\documentclass[a4paper,11pt]{article}
\pdfoutput=1 

\usepackage[utf8]{inputenc}
\usepackage{amsmath}
\usepackage{amsfonts}
\usepackage{amssymb}
\usepackage{mathtools}
\usepackage{amsthm}
\usepackage{dsfont}
\usepackage{hyperref}
\usepackage[capitalise,sort]{cleveref}
\usepackage{pgfplots}
\usepackage{pgfplotstable}
\usepackage{tabularx}
\usepackage{environ}
\usepackage[inline]{enumitem}
\usepackage{comment}
\usepackage{tikz}
\usetikzlibrary{decorations.pathmorphing,calc,fit,matrix,cd,shapes,arrows,positioning, intersections,chains,positioning,arrows.meta,hobby,patterns,shadings,shadows,pgfplots.groupplots}
\usepackage{blkarray}
\usepackage{shadow}
\usepackage{fancybox}
\usepackage{bm}
\usepackage{lscape}
\usepackage{verbatim}
\usepackage[normalem]{ulem}
\usepackage{mdframed}
\usepackage{orcidlink}
\usepackage{pdflscape} 
\usepackage{longtable}
\usepackage{hhline}
\usepackage{array}
\usepackage{multirow}
\usepackage{graphicx}
\usepackage{subcaption}
\usepackage[tmargin=2cm,bmargin=2cm,lmargin=2cm,rmargin=2cm,footskip=1cm]{geometry}
\usepackage{nowidow}
\usepackage[numbers, sort&compress]{natbib}

\numberwithin{equation}{section}

\makeatletter
\def\myrotate{\ifodd\c@page\else+\fi 90}
\g@addto@macro{\landscape}{\PLS@Rotate{\myrotate}}
\makeatother

\newcommand{\Lpagenumber}{\ifdim\textwidth=\linewidth\else\bgroup
	\dimendef\margin=0 
	\ifodd\value{page}\margin=\oddsidemargin
	\else\margin=\evensidemargin
	\fi
	\raisebox{\dimexpr -3\topmargin-\headheight-\headsep-0.5\linewidth}[0pt][0pt]{%
		\rlap{\hspace{\dimexpr \margin+\textheight+3\footskip}%
			\llap{\rotatebox{90}{\hspace{-6.5cm}\normalsize\thepage\hfill}}}}%
	\egroup\fi}
\AddToHook{shipout/background}{\Lpagenumber}%

\def\fnote#1#2{\begingroup\def\thefootnote{#1}\footnote{#2}
     \addtocounter{footnote}{-1}\endgroup}

\pgfplotsset{
        compat=1.18,
        compat/bar nodes=1.18,
    }
    
\renewcommand{\tabularxcolumn}[1]{m{#1}}

\makeatletter
\def\@xfootnote[#1]{%
	\protected@xdef\@thefnmark{#1}%
	\@footnotemark\@footnotetext}
\makeatother

\definecolor{prhigh}{HTML}{ff0000}
\definecolor{sechigh}{HTML}{e0fbfc}
\definecolor{prcolor}{HTML}{1d3557}
\definecolor{seccolor}{HTML}{457b9d}
\definecolor{tercolor}{HTML}{98c1d9}
\definecolor{blueplot}{HTML}{58468e}

\newcommand{\AM}[1]{{\color{blue}{\bf{AM}}: zzzz #1 }}
\newcommand{\GS}[1]{{\color{red}{\bf{GS}}: zzzz #1}}
\newcommand{\JY}[1]{{\color{orange}{\bf{JY}}: zzzz #1}}
\newcommand\cC{\mathcal{C}}
\newcommand{\hJ}{\hat{J}}

\newcommand\bea{\begin{eqnarray}}
\newcommand\eea{\end{eqnarray}}

\newcommand\sS{{\bf S}}

\theoremstyle{plain}

\theoremstyle{definition}

\newtheorem*{conjecture*}{Conjecture}

\newtheorem{remark*}{Remark}



\DeclareMathOperator{\Res}{Res}
\DeclareMathOperator{\Disc}{Disc}
\DeclareMathOperator{\U}{U}
\DeclareMathOperator{\SU}{SU}
\DeclareMathOperator{\SO}{SO}
\DeclareMathOperator{\SL}{SL}
\DeclareMathOperator{\GL}{GL}
\DeclareMathOperator{\Sp}{Sp}
\DeclareMathOperator{\USp}{USp}
\DeclareMathOperator{\str}{STr}
\DeclareMathOperator{\rank}{rank}
\DeclareMathOperator{\PE}{PE}
\DeclareMathOperator{\PL}{PL}
\DeclareMathOperator{\spa}{Span}
\DeclareMathOperator{\adj}{Ad}
\DeclareMathOperator{\spec}{Spec}
\DeclareMathOperator{\sign}{sign}
\DeclareMathOperator{\perm}{perm}
\DeclareMathOperator{\re}{Re}
\DeclareMathOperator{\im}{Im}
\DeclareMathOperator{\tr}{tr}
\DeclareMathOperator{\Tr}{Tr}
\newcommand{\de}{\partial}
\newcommand{\CC}{\mathbb{C}}
\newcommand{\PP}{\mathbb{P}}
\newcommand{\RR}{\mathbb{R}}
\newcommand{\ZZ}{\mathbb{Z}}
\newcommand{\FF}{\mathbb{F}}
\newcommand{\ID}{\mathds{1}}
\newcommand{\tder}[2]{\frac{d#1}{d#2}}
\newcommand{\tderr}[1]{\frac{d}{d#1}}
\newcommand{\ntder}[3]{\frac{d^{#3}#1}{d#2^{#3}}}
\newcommand{\ntderr}[2]{\frac{d^{#2}}{d #1^{#2}}}
\newcommand{\pder}[2]{\frac{\de#1}{\de#2}}
\newcommand{\pderr}[1]{\frac{\de}{\de #1}}
\newcommand{\npder}[3]{\frac{\de^{#3}#1}{\de#2^{#3}}}
\newcommand{\npderr}[2]{\frac{\de^{#2}}{\de #1^{#2}}}
\newcommand{\coma}{\, , \quad}
\newcommand{\fstop}{\, .}
\newcommand{\hc}{\quad\text{ h.c.}}
\newcommand{\with}{\quad\text{with}\quad}
\newcommand{\e}{\quad\text{and}\quad}
\newcommand{\AdS}{\text{AdS}}
\newcommand{\dS}{\text{dS}}
\newcommand{\dP}{\text{dP}}
\newcommand{\UV}{\text{UV}}
\newcommand{\IR}{\text{IR}}
\newcommand{\CY}{\text{CY}}
\newcommand{\YM}{\text{\tiny YM}}
\newcommand{\cV}{\mathcal{V}}
\newcommand{\cK}{\mathcal{K}}
\newcommand{\ii}{{\rm i}}
\newcommand{\IX}{{\bf X}}
\newcommand{\IY}{{\bf Y}}
\newcommand{\Pl}{\text{\tiny Pl}}
\newcommand{\KK}{\text{\tiny KK}}
\newcommand{\spKK}{\text{\tiny sp,KK}}
\newcommand{\spc}{\text{\tiny sp}}
\newcommand{\Kt}{\text{\tiny K3}}
\newcommand{\WGC}{\text{\tiny WGC}}
\newcommand{\QG}{\text{\tiny QG}}
\newcommand{\T}{\text{T}}
\newcommand{\het}{\text{\tiny het.}}
\newcommand{\IIA}{\text{\tiny IIA}}
\newcommand{\IIB}{\text{\tiny IIB}}
\newcommand{\PlD}[1]{\text{\tiny Pl,\,#1}}
\newcommand{\Plinf}{\text{\tiny Pl,\,$\infty$}}
\newcommand{\spD}[1]{\text{\tiny sp,\,#1}}
\newcommand{\YMD}[1]{\text{\tiny YM,\,#1}}
\newcommand{\UoD}[1]{\text{\tiny U(1),\,#1}}
\newcommand{\ttiny}[1]{\text{\tiny #1}}
\newcommand{\WGCD}[1]{\text{\tiny WGC,\,#1}}
\newcommand{\AdSD}[1]{\text{\tiny AdS,\,#1}}
\newcommand{\MinkD}[1]{\text{\tiny Mink,\,#1}}
\newcommand{\fn}{\mathfrak{n}}

\renewcommand{\epsilon}{\varepsilon}

\makeatletter
\newsavebox{\measure@tikzpicture}
\NewEnviron{scaletikzpicturetowidth}[1]{%
  \def\tikz@width{#1}%
  \def\tikzscale{1}\begin{lrbox}{\measure@tikzpicture}%
  \BODY
  \end{lrbox}%
  \pgfmathparse{#1/\wd\measure@tikzpicture}%
  \edef\tikzscale{\pgfmathresult}%
  \BODY
}
\makeatother

\makeatletter
\newcommand{\inlineitem}[1][]{%
\ifnum\enit@type=\tw@
    {\descriptionlabel{#1}}
  \hspace{0pt}%
\else
  \ifnum\enit@type=\z@
      \hspace{-15pt} \refstepcounter{\@listctr}\fi
    \quad\@itemlabel\hspace{0pt}%
\fi}
\makeatother
\let\mathcaldefault\mathcal
\DeclareMathAlphabet{\mathdutchcal}{U}{dutchcal}{m}{n}

\def\fnote#1#2{\begingroup\def\thefootnote{#1}\footnote{#2}
     \addtocounter{footnote}{-1}\endgroup}
  
\tikzset{
    partial ellipse/.style args={#1:#2:#3}{
        insert path={+ (#1:#3) arc (#1:#2:#3)}
    }
}

\tikzset{cross/.style={cross out, draw=black, fill=none, minimum size=2*(#1-\pgflinewidth), inner sep=0pt, outer sep=0pt}, cross/.default={2pt}}

\tikzset{
	pics/torus/.style n args={3}{
		code = {
			\providecolor{pgffillcolor}{rgb}{1,1,1}
			\begin{scope}[
				yscale=cos(#3),
				outer torus/.style = {draw,line width/.expanded={\the\dimexpr2\pgflinewidth+#2*2},line join=round},
				inner torus/.style = {draw=pgffillcolor,line width={#2*2}}
				]
				\draw[outer torus] circle(#1);\draw[inner torus] circle(#1);
				\draw[outer torus] (180:#1) arc (180:360:#1);\draw[inner torus,line cap=round] (180:#1) arc (180:360:#1);
			\end{scope}
		}
	}
}

\tikzset{
	pics/hole/.style n args={2}{
		code = {
			\draw[fill=white] (0,0) arc(120:60:#1 and #2)  arc(-60:-120:#1 and #2);
            \draw (0,0) arc(-120:-130:#1 and #2) (#1,0) arc(-60:-50:#1 and #2);
		}
	}
}

\makeatletter
\newcommand*{\itemequation}[3][]{%
  \item
  \begingroup
    \refstepcounter{equation}%
    \ifx\\#1\\%
    \else  
      \label{#1}%
    \fi
    \sbox0{#2}%
    \sbox2{$\displaystyle#3\m@th$}%
    \sbox4{\@eqnnum}%
    \dimen@=.5\dimexpr\linewidth-\wd2\relax
    \ifcase
        \ifdim\wd0>\dimen@
          \z@
        \else
          \ifdim\wd4>\dimen@
            \z@
          \else 
            \@ne
          \fi 
        \fi
      \@latex@warning{Equation is too large}%
    \fi
    \noindent   
    \rlap{\copy0}%
    \rlap{\hbox to \linewidth{\hfill\copy2\hfill}}%
    \hbox to \linewidth{\hfill\copy4}%
    \hspace{0pt}
  \endgroup
  \ignorespaces 
}
\makeatother  

\hypersetup{
	pdftitle={Exploring Line Bundle Standard Models with Transformers},    
	pdfauthor={\textcopyright\ Jacky H. T. Yip, Alessandro Mininno, Gary Shiu},     
	pdfsubject={hep-th},   
	pdfcreator={pdfLaTex},   
	pdfproducer={LaTex}, 
	pdfkeywords={},
	colorlinks=true
  ,urlcolor=blue
  ,anchorcolor=blue
  ,citecolor=blue
  ,filecolor=blue
  ,linkcolor=blue
  ,menucolor=blue
  ,linktocpage=true
}

\allowdisplaybreaks[1] 

\crefname{figure}{Figure}{Figures}
\crefname{table}{Table}{Tables}
\crefname{definition}{Definition}{Definitions}
\crefname{proposition}{Proposition}{Propositions}
\crefname{claim}{Claim}{Claims}
\crefname{conjecture}{Conjecture}{Conjectures}

\def\beq{\begin{equation}}
\def\eeq{\end{equation}}
\def\bc{\begin{cases}}
\def\ec{\end{cases}}
\def\bal{\begin{aligned}}
\def\eal{\end{aligned}}
\newcommand{\mc}{\mathcal}
\newcommand{\mf}{\mathfrak}

\allowdisplaybreaks[1]

\begin{document}

\begin{titlepage}

\vspace*{3cm} 

\begin{center}
{\LARGE\bfseries Exploring Line Bundle Standard Models with Transformers}
\vspace{0.4cm}

{\large  Jacky H. T. Yip\,\orcidlink{0009-0002-1921-524X},$^{\spadesuit}$ Alessandro Mininno\,\orcidlink{0000-0002-9593-0440},$^{\spadesuit}$ Gary Shiu\,\orcidlink{0000-0003-1308-5202}}\\
\vspace{.6cm}

{Department of Physics, University of Wisconsin--Madison,\\1150 University Avenue, Madison, WI 53706, USA}\par
\vspace{.2cm}

\scalebox{0.8}{\tt \href{mailto:jacky.ht.yip@gmail.com}{jacky.ht.yip@gmail.com}, \href{mailto:mininno@physics.wisc.edu}{mininno@physics.wisc.edu}, \href{mailto:shiu@physics.wisc.edu}{shiu@physics.wisc.edu}}
\vspace{0.3cm}

\renewcommand{\thefootnote}{} 
\footnotemark\footnotetext{$^\spadesuit$ Equal contribution.}
\renewcommand{\thefootnote}{\arabic{footnote}} 

\setcounter{footnote}{0} 

\textbf{Abstract}
\end{center}

\noindent We propose a Transformer-based Reinforcement Learning architecture, ``LB-Explorer'', to search for heterotic line bundle standard models arising from compactifications on smooth Calabi--Yau (CY) threefolds. We focus on $E_8\times E_8$ heterotic string theory compactifications on CY with abelian line bundles to produce $\SU(5)\times \text{S}(\U(1)^5)$ symmetry, whose $\SU(5)$ can be further broken to an MSSM-like gauge group using appropriate discrete Wilson lines. We test the LB-Explorer environment on complete intersection Calabi--Yau (CICY) manifolds, though the neural network architecture naturally generalizes to any CY admitting a simplicial Mori cone and a freely-acting discrete symmetry. The LB-Explorer efficiently learns constraints on the line bundle sums, guaranteeing the $E_8$ gauge embedding, anomaly cancellation, poly-stability (supersymmetry), chirality of the spectrum, and the absence of exotic matter. Valid configurations can be subsequently filtered by imposing the missing constraints, such as the equivariant structure of the line bundle sum and further requirements on the particle spectrum. In this direction, we introduce a hybrid architecture incorporating CP-SAT solvers that aims to impose some of the conditions exactly by perturbing solutions found by the LB-Explorer. The versatility and scalability of the LB-Explorer make it a powerful tool for navigating the string landscape with a large number of moduli. The code and tools necessary to reproduce our findings are available at \href{https://github.com/alexmininno/LB-Explorer}{GitHub repository}.

\vspace{1cm}
\vfill 
\end{titlepage}

\tableofcontents
\bigskip\medskip
\hrule
\bigskip\bigskip

\section{Introduction}
One of the main purposes of string phenomenology is the construction of four-dimensional effective field theories (EFTs) from string theory compactifications that contain the gauge symmetry and the particle content of the Minimal Supersymmetric Standard Model (MSSM) (see \cite{Marchesano:2024gul, Cvetic:2022fnv} for recent reviews). However, even if significantly smaller than the entire string theory landscape, the number of EFTs that have a SM-like particle spectrum is still large \cite{Constantin:2018xkj,Cvetic:2019gnh}, making heuristic and exhaustive scans realistically impossible. For these reasons, in recent years, machine learning (ML) techniques have been used to guide and systematize the search of solutions. 

In this paper, we focus on $E_8\times E_8$ heterotic string theory compactifications on CY \cite{Gross:1984dd, Gross:1985fr, Gross:1985rr,Candelas:1985en,Greene:1986ar,Greene:1986bm,Greene:1986jb} with abelian internal gauge fields \cite{Anderson:2011ns,Anderson:2012yf,Anderson:2013xka} to produce $\SU(5)\times \text{S}(\U(1)^5)$ symmetry, whose $\SU(5)$ can be further broken to an MSSM-like gauge group using appropriate discrete Wilson lines. In the literature, these models have been dubbed ``line bundle standard models'' \cite{Anderson:2012yf} and are four-dimensional EFTs with an MSSM-type spectrum and additional $\text{S}(\U(1)^5)$ symmetries. These extra $\U(1)$ factors are anomalous, but such anomalies are canceled by the Green--Schwarz (GS) mechanism \cite{Green:1984sg} via axion-like fields mixing with the gauge bosons. 

These models have already been subjected to both systematic scans \cite{Anderson:2013xka,He:2013ofa,Constantin:2018xkj,Anderson:2026eyl} and ML searches \cite{Larfors:2020ugo,Constantin:2021for,Abel:2023zwg,Berglund:2024reu}. The former have reached modern computational limits, as the number of configurations grows exponentially with the number of K\"ahler moduli, limiting the scan to relatively small values of $h^{1,1}(\IX_3)$ or of the first Chern class of the line bundles. The latter have shown promising results, proposing Reinforcement Learning (RL) and Genetic Algorithm (GA) algorithms to overcome the large number of possible configurations and the complexity of the problems. 

\subsubsection*{Summary of the Results}

We decided to improve in the latter direction by proposing ``LB-Explorer'', a novel neural network (NN) environment that combines Transformers \cite{Vaswani:2017xxx} with RL. LB-Explorer, together with scripts to reproduce and test the results, is provided in a \href{https://github.com/alexmininno/LB-Explorer}{Github repository}. Our purpose is to design LB-Explorer to search for configurations yielding an $\SU(5)\times \text{S}(\U(1)^5)$ gauge symmetry. The LB-Explorer can identify models with the correct chiral spectrum across a broad range of CY manifolds under minimal assumptions.
In fact, the current version of LB-Explorer assumes that the CY admits a simplicial Mori cone and the existence of a freely-acting symmetry $\Gamma$. Under these assumptions, the NN will learn to impose the conditions that guarantee the embedding of the $\SU(5)\times\text{S}(\U(1)^5)$ gauge group into one of the two $E_8$ gauge groups, the cancellation of the GS anomaly, the preservation of minimal supersymmetry in four dimensions, the presence of a chiral spectrum, and the absence of exotic matter. These conditions do not guarantee that the resulting configuration leads to a MSSM-like spectrum, leaving the user to filter the solutions. The reason for this choice is to allow a more general application of LB-Explorer to CYs for which the action of the freely-acting symmetry $\Gamma$ is not known or classified, or if the computations of line bundle cohomologies require the use of external tools when closed formulas are not known.

We tested the LB-Explorer on around $50$ distinct favorable complete intersection CY (CICYs) \cite{Candelas:1987kf,Anderson:2017aux} with different values of freely-acting symmetries $\Gamma$ \cite{Braun:2010vc,Lukas:2017vqp}. We list the geometric properties for the CICYs we considered in Table \ref{tab:CICYdataset}. For these manifolds, we performed the exploration over 10 million episodes across five different seeds that initialized the Transformers, with line bundle entries $|\mathbf{k}_a|\leq 8$. This choice allowed us to make a systematic analysis of the performance of LB-Explorer over different geometries under the same conditions, counting the configurations found and filtering them with the conditions not learned by LB-Explorer. Moreover, for each CICY, we also computed the number of solutions modded out by the $S_5$ symmetry associated to the relabeling of the line bundles composing the vector bundle and by the symmetry $G_{\IX_3}$ associated to the relabeling of the divisors basis.

We analyze the status of the exploration, identifying three different stages of LB-Explorer that can be used to probe how well the search for line bundle sums is proceeding. We find that increasing the number of seeds in the transformer initialization guaranties a better search in the space of solutions. LB-Explorer can, hence, be parallelized on different seeds and run for the reasonable number of episodes (we found 10 million episodes to be such an amount) to guarantee the best performance.

One of the most important analyzes we conducted in this work is about the capacity of the LB-Explorer for transfer learning across CICYs. We find that many of the constraints learned by the LB-Explorer for a given CICY are retained, which makes training on a different CY more efficient without complete restructuring of the learned policies. This result points in the direction that it is, in principle, possible to train the LB-Explorer on a set of CYs and use that training to extract solutions for other geometries, making our architecture suitable for large scans. Although there is always transfer learning in terms of finding solutions in fewer episodes than training from scratch, we also discuss a directional asymmetry that we have observed. In order to obtain a stable and consistent search for solutions in the transfer geometry, we recommend choosing the source CY to pre-train the LB-Explorer to be one with a small $G_{\IX_3}$, large $|\Gamma|$, and large $h^{1,1}(\IX_3)$. Any combination of these requirements leads to optimal transfer learning. However, if the purpose of transfer learning is to find solutions for the transfer CICY in significantly fewer steps than training from scratch, and there is no interest in finding all possible solutions, we find that transfer learning also works if the source geometry does not satisfy the conditions stated above. 

LB-Explorer does not learn to impose the equivariant structure of the line bundles, nor the exact conditions on the line bundle cohomologies. We need, instead, to filter the solutions afterwards, using scripts included in the \href{https://github.com/alexmininno/LB-Explorer}{GitHub repository}. Imposing the equivariance reduces the number of solutions found more or less dramatically, depending on whether the action of $\Gamma$ on the divisors is trivial or not. We believe that one way to improve in this direction is to let the LB-Explorer run over multiple seeds so that it can find more solutions. The check of the spectrum is done by using known modules, such as \texttt{pyCICY} \cite{Larfors:2019sie}. Since our purpose was not to be exhaustive in finding solutions satisfying all the MSSM conditions but to design an efficient and flexible explorer, we restricted the check of the spectrum to the equivariant solutions that also had $|\mathbf{k}_a|\leq 2$, so that the checks with \texttt{pyCICY} were fast enough. We chose \texttt{pyCICYs} because it is designed to compute line bundle cohomologies for CICYs.\footnote{During the completion of this work, the \texttt{CIPro} package \cite{Anderson:2026eyl} in Mathematica has been released that also allows for the computation of line bundle cohomologies for CICYs, and much more.} For the future extension of LB-Explorer to other CYs, there are tools like \texttt{cohomcalg} \cite{Blumenhagen:2010pv} or recent extensions of \texttt{CYTools} \cite{Demirtas:2022hqf,Gendler:2026uux} that could also be used for such computations.

However, we also propose a way in which, theoretically, it is possible to impose extra conditions on the LB-Explorer without renouncing its general applicability. In fact, we discuss a hybrid architecture of the LB-Explorer where extra conditions can be solved exactly using the CP-SAT module \cite{cpsatlp}, which perturbs the solutions found by the LB-Explorer to satisfy other conditions. Instead of testing it by imposing the equivariance that would have made the conditions specific for each CICYs, for which we would not have had a benchmark to test the efficacy of this method, we decided to take CICYs with small $h^{1,1}(\IX_3)$ over which the LB-Explorer did not find valid solutions due to their sparsity, and remove the chirality among the conditions learned by the LB-Explorer. The chirality is instead satisfied with CP-SAT by perturbing what the LB-Explorer found. We believe that this can be a way to impose more granular conditions on the LB-Explorer without relying completely on the filtering of the solutions \textit{a posteriori}.

\subsubsection*{Structure of the Paper}

The paper is structured as follows. In Section \ref{sec:reviewHet}, we provide a brief review of how to construct line bundle standard models from $E_8\times E_8$ heterotic string theory compactifications on Calabi--Yau threefolds, detailing both the physical moduli stabilization at the maximal splitting locus and the algebraic constraints on the bundle. In particular, Section \ref{sec:GUTHetIngred} explains how these requirements translate into specific constraints on our LB-Explorer NN, summarized in Table \ref{tab:conditions}. In Section \ref{sec:CICYstatistics}, we discuss the redundancies of the configurations under the relabeling of line bundles or divisor bases. This leads directly to the definition of our training dataset in Section \ref{sec:dataset_selection}, with data listed in Table \ref{tab:CICYdataset}. In Section \ref{sec:nn_architecture}, we define the LB-Explorer framework, detailing the Transformer architecture, reinforcement learning loss functions, and specific implementation details used during training. In Section \ref{sec:results}, we present our primary exploration results, providing detailed solution counts (Table \ref{tab:detailed_counts}), analyzing the learning patterns over 10 million episodes across five different seeds, and exploring transfer learning across different geometries (Section \ref{sec:transferlearning}). In Section \ref{sec:spectrumanalysis}, we proceed to the physics validation, filtering the generated solutions by first enforcing the equivariant structure of the line bundle sum and then, using existing cohomology computation tools like \texttt{pyCICY} \cite{Larfors:2019sie}, checking for the MSSM spectrum. Building on the insights from our baseline model, Section \ref{sec:HybridRL} introduces a hybrid architecture incorporating CP-SAT to check whether one can impose even fewer conditions from Table \ref{tab:conditions} and solve for the remaining ones exactly. This is used as a benchmark to test if CP-SAT can, in principle, be used to impose the missing conditions we decided not to teach to the LB-Explorer. Finally, we discuss our conclusions and potential future directions in Section \ref{sec:conclusions}. Appendix \ref{app:CICYreview} reviews the construction and symmetries of CICYs, and all comprehensive result tables are compiled in Appendix \ref{app:tables}. The LB-Explorer code, along with the tools necessary to reproduce our findings, is publicly available in our \href{https://github.com/alexmininno/LB-Explorer}{GitHub repository}.

\section{A Light Review of Heterotic Compactifications on Calabi--Yau}
\label{sec:reviewHet}

In this section, we are going to review the main ingredients for $E_8\times E_8$ heterotic string theory compactifications on CY threefolds \cite{Gross:1984dd, Gross:1985fr, Gross:1985rr,Candelas:1985en,Greene:1986ar,Greene:1986bm,Greene:1986jb} with general embeddings. This section is based on more comprehensive and detailed reviews that can be found in the literature (see e.g. \cite{Anderson:2008uw,Anderson:2007nc,Anderson:2008ex,Anderson:2011ns,Anderson:2012yf,Anderson:2013xka} and references therein).

The starting point is $E_8\times E_8$ heterotic string theory compactified on $\mathcal{M}_4\times \IX_3$, where $\IX_3$ is a compact CY manifold. To each $E_8$ group, we associate vector bundles $V$ and $\widetilde{V}$ such that the bundle associated with the $E_8\times E_8$ field strength is the direct product of $V$ and $\widetilde{V}$. We take $V$ to be a general holomorphic vector bundle over $\IX_3$ with structure group $H$. In this way, the structure group of the ten-dimensional bundle decomposes into $G\times H\subset E_8$, where $G$ is the commutant of $H$ in $E_8$. A four-dimensional EFT with gauge group $G$ is obtained by choosing $V$ accordingly. This is what is known as ``general embeddings" in heterotic compactifications \cite{Distler:1987ee, Distler:1993mk, Kachru:1995em}. However, in order for such general vector bundles to lead to supersymmetric vacua, they must satisfy (at least) two properties: anomaly cancellation and poly-stability of the line bundles \cite{Donaldson:1985zz, Uhlenbeck:1986ntc}. Moreover, the CY must be non-simply connected to allow for the existence of Wilson lines that will further break the Grand Unified Theory (GUT) group $G$ to the Standard Model gauge group. We are now going to review these three properties in detail, while extra conditions on the particle spectrum will be discussed in Section \ref{sec:GUTHetIngred}.

\paragraph{Anomaly Cancellation:} To guarantee the anomaly cancellation imposed by the Green--Schwarz (GS) mechanism, i.e.,
\begin{equation}\label{eq:dH3condition}
    dH_3 = \Tr (R\wedge R) - \Tr (F_2 \wedge F_2)\coma
\end{equation}
we note that it can be expressed in terms of the second Chern character of the $V\times \widetilde{V}$ bundle and the tangent bundle $T\IX_3$ of the CY, such that $dH_3$ is trivial in homology. The Chern class for $V$ with curvature $F_2$ is defined as
\begin{equation}
    c(V) = \det \left(\ID+\frac{i}{2\pi}F_2\right)\coma
\end{equation}
yielding the first and second Chern classes to be
\begin{equation}
    c_1(V) = \frac{i}{2\pi}\Tr(F_2)\coma c_2(V) = \frac{1}{2}\left(c_1(V)^2-\Tr(F_2\wedge F_2)\right)\fstop
\end{equation}
Similarly, for the tangent bundle $T\IX_3$ of the CY $\IX_3$, the Chern classes are constructed from the Riemann curvature $R$:
\begin{equation}
    c_1(T\IX_3) = \frac{i}{2\pi}\Tr(R) \stackrel{!}{=} 0 \coma c_2(T\IX_3) = -\frac{1}{2}\Tr(R\wedge R)\coma 
\end{equation}
where we have already imposed the CY condition that enforces the vanishing of the first Chern class of the tangent bundle. In order to guarantee the existence of a spin structure on $V$, we impose a similar constraint on the $V\times \widetilde{V}$ bundle, i.e.
\begin{equation}
    c_1(V) = c_1(\widetilde{V}) = 0 \, \mod \, 2\fstop\label{eq:c1V=0}
\end{equation}
Coming back to \eqref{eq:dH3condition}, the condition that $dH_3$ must be trivial in homology can be expressed in terms of the second Chern classes of the bundles and the tangent bundle of the CY, i.e.,
\begin{equation}
    c_2(T\IX_3) - c_2(V) - c_2(\widetilde{V}) = 0\fstop\label{eq:sndChernChar=0}
\end{equation}
However, \eqref{eq:sndChernChar=0} can be extended to include the presence of NS5-branes wrapping 2-cycles in $\IX_3$ of class $W_5$, such that the anomaly cancellation condition becomes
\begin{equation}\label{eq:anomalycancellation}
    c_2(T\IX_3) - c_2(V) - c_2(\widetilde{V}) = W_5\fstop
\end{equation}
In the following, we consider compactifications where the dynamics of the theory are entirely encoded in one of the two $E_8$ factors, while the other represents the hidden sector. Setting $\widetilde{V}$ to be a trivial bundle, the anomaly cancellation condition reduces to
\begin{equation}\label{eq:curveconditionW5}
    c_2(T\IX_3) - c_2(V) = W_5\coma
\end{equation} 
which is satisfied by including a consistent number of NS5-branes wrapping effective curves in $\IX_3$. In our paper, we assume that the Mori cone of effective curves of $\IX_3$ is simplicial. Under this assumption, we can choose a basis of divisors $\{J_i\}_{i=1}^{h^{1,1}(\IX_3)}$ for $H^2(\IX_3, \mathbb{Z})$ that generates the K\"ahler cone and is dual to the generators $\mathcal{C}^j$ of the Mori cone, satisfying $\int_{\mathcal{C}^j} J_i = \delta_i^j$. 

The K\"ahler form $J$, residing in the interior of the K\"ahler cone, can then be expanded as $J = t^i J_i$, where the K\"ahler moduli $t^i>0$ correspond precisely to the volumes of the Mori cone generating curves. With this choice of basis, demanding that the curve class $W_5 = c_2(T\IX_3) - c_2(V)$ is effective (i.e., it can be written as a non-negative integer linear combination of the Mori cone generators) translates directly into the requirement that its intersection with each K\"ahler cone generator is non-negative:
\begin{equation}
    \int_{\IX_3} \left(c_2(T\IX_3) - c_2(V)\right) \wedge J_i \geq 0\coma \quad i = 1,\ldots, h^{1,1}(\IX_3)\fstop
\end{equation}

\paragraph{Poly-stability Property of Line Bundles:} We also demand that the resulting EFT is supersymmetric. This is related to the requirement that the gauge connection on $V$ satisfies the hermitian Yang--Mills (YM) equations at zero slope. This condition is usually called the ``poly-stability property" since, by the Donaldson--Uhlenbeck--Yau theorem, the existence and uniqueness of such a connection satisfying the hermitian YM equation is guaranteed for each ``poly-stable" holomorphic vector bundle $V$. 

The slope for a coherent sheaf $F$ on $\IX_3$ is defined as
\begin{equation}
    \mu(F) = \frac{1}{\text{rank}(F)}\int_{\IX_3} c_1(F) \wedge J \wedge J = \frac{1}{\text{rank}(F)}\sum_{i,j,k=1}^{h^{1,1}(\IX_3)} \kappa_{ijk}c_1^i(F)t^j t^k\coma
\end{equation}
where we have introduced the triple intersection numbers of $\IX_3$ as
\begin{equation}
    \kappa_{ijk} = \int_{\IX_3} J_i \wedge J_j \wedge J_k\fstop
\end{equation}
A holomorphic vector bundle $V$ is ``slope-stable" (resp. ``slope-semi-stable") if 
\begin{equation}
    \mu(F)<\mu(V) \quad (\text{resp. } \mu(F)\leq\mu(V)) \coma \forall F\subset V \text{ with } 0< \text{rank}(F) < \text{rank}(V)\fstop
\end{equation}
In our construction, we restrict $V$ to be a direct sum of line bundles, i.e.
\begin{equation}
    V = \bigoplus_{a=1}^N L_a\coma
\end{equation}
hence, because line bundles have rank $1$, they contain no proper sub-sheaves and are trivially stable. A direct sum of stable bundles is ``poly-stable" if and only if all components share the same slope. Given the $E_8$ embedding requirement \eqref{eq:c1V=0}, the overall slope is $\mu(V)=0$; therefore, poly-stability requires that the slope of each individual line bundle vanishes simultaneously, i.e.
\begin{equation}
    \mu(L_a) = 0 \coma \forall a\fstop
\end{equation}
It follows that any poly-stable bundle is also semi-stable.

\paragraph{Equivariant Structure:} The constraints discussed thus far ensure a stable supersymmetric vacuum with an $\SU(5)$ GUT group. However, obtaining a realistic four-dimensional Standard Model requires breaking this GUT group down to $\SU(3)_C \times \SU(2)_L \times \U(1)_Y$. In heterotic compactifications, this is typically achieved by turning on discrete Wilson lines, which is only possible if the internal manifold is non-simply connected.

To construct such a geometry, one starts with a simply connected CY threefold $\IX_3$ that admits a freely acting discrete symmetry group $\Gamma$ of order $|\Gamma|$. By quotienting the manifold, one obtains a smooth, non-simply connected physical threefold $\widetilde{\IX}_3 = \IX_3/\Gamma$. For the resulting four-dimensional EFT to be well-defined, the vector bundle $V$ constructed on the upstream covering space $\IX_3$ must be fully compatible with this discrete symmetry. Such a bundle is said to admit an ``equivariant structure'' under $\Gamma$. A necessary condition for equivariance is that all relevant topological invariants, including the Euler characteristic of the manifold and the bundle partitions, must be integer multiples of the group order $|\Gamma|$ \cite{Anderson:2008uw,He:2009wi,Anderson:2011ns,Anderson:2012yf,He:2013ofa}.

\subsection{Ingredients for GUT Heterotic Compactifications}
\label{sec:GUTHetIngred}

The focus of this paper is to consider CY manifolds constructed as complete intersections (CICY) in products of projective spaces \cite{Candelas:1987kf} or as hypersurfaces in toric fourfolds \cite{Kreuzer:2000xy}. We will also assume that the CY admits a simplicial (i.e., the number of generators is $h^{1,1}(\IX_3)$) Mori cone. Under these assumptions, we can consider the K\"ahler cone basis $\{J_i\}$ as a basis of divisors in which we can expand the K\"ahler form $J$ of the CY. Although this is not a necessary requirement for model building of GUT theories in heterotic compactifications, it makes the machine learning we are going to construct easier since we can interpret the K\"ahler moduli $t^i$ as positive volumes associated to Mori cone curves, and the triple intersections of the CY are all non-negative. 

As mentioned in the previous section, we focus on a single $E_8$ group of the $E_8\times E_8$ heterotic string, so that we associate a trivial line bundle $\widetilde{V}$ to the hidden sector. On the other hand, we will need an equivariant bundle $V$ such that the ten-dimensional gauge group breaks into a split structure group of the form
\begin{equation}
    H = \text{S}\left(\U(1)^N\right)\fstop
\end{equation}
The choice of splitting above is called ``maximal splitting".\footnote{This locus leaves some moduli unstabilized. However, deforming the bundle into a non-Abelian extension does not modify the net number of chiral families in the spectrum, and it generates the required terms for the stabilization of these moduli. For further discussion, see, e.g., \cite{Anderson:2010mh, Anderson:2011ty, Anderson:2013qca}.} Eventually, we will consider $N=5$, but for the moment, we keep $N$ generic. In the case of $N=5$, the $E_8$ group is broken down into $G\times H = \SU(5)\times \text{S}\left(\U(1)^5\right)$. 

In order to realize this split structure, we construct line bundles on $\IX_3$ by restricting line bundles from the ambient space. Let $H_i$ be the hyperplane classes of the ambient space $\mathcal{A}$. We define ambient space line bundles \[\mathcal{L}_a = \mathcal{O}_{\mathcal{A}}\left(\sum_{i=1}^{h^{1,1}(\IX_3)} k_a^i H_i\right)\] and define $L_a$ on the CY threefold via the restriction map $\iota^*$:
\begin{equation}
L_a = \iota^* \mathcal{L}_a = \mathcal{O}_{\IX_3}\left(\sum_{i=1}^{h^{1,1}(\IX_3)} k_a^i J_i\right)\, , \quad \text{with } k_a^i \in \mathbb{Z}\, ,
\end{equation}
where $J_i = \iota^* H_i$ forms the basis of divisors for $H^2(\IX_3, \mathbb{Z})$ descending from the ambient space, which we assume coincides with the K\"ahler cone basis of $\IX_3$. In the following, we will use a vector shorthand $\mathbf{k}_a = (k_1,\ldots, k_{h^{1,1}(\IX_3)})\in \ZZ^{h^{1,1}(\IX_3)}$, such that the first Chern class of the line bundle can be written as $c_1(L_a) = \sum_{i=1}^{h^{1,1}(\IX_3)}k^{i}_a J_i$, and we will denote it as $L_a = \mathcal{O}_{\IX_3}(\mathbf{k}_a)$. Each line bundle has a structure group $\U(1)$, and we set
\begin{equation}
    V = \bigoplus_{a=1}^N L_a\fstop
\end{equation}
The requirement that the line bundle sum has an embedding into $E_8$ imposes
\begin{equation}\label{eq:VembeddinginE8}
    c_1(V) = \sum_{a=1}^N c_1(L_a) \equiv \sum_{a=1}^N \mathbf{k}_a= 0\fstop
\end{equation}
However, \eqref{eq:VembeddinginE8} is only a necessary but not a sufficient condition to guarantee that there are no obstructions in having $H = \text{S}\left(\U(1)^N\right)$, given $V$ as the sum of $N$ line bundles \cite[Appendix A]{Anderson:2013xka}. In fact, we need to impose that
\begin{equation}\label{eq:Vembeddingsubsets}
    \sum_{a\in S} c_1(L_a) \neq 0\coma 
\end{equation}
for all proper subsets $S$ of $\{1,\ldots, N\}$.\footnote{When considering $N=5$, since \eqref{eq:VembeddinginE8} must hold, it is sufficient that \eqref{eq:Vembeddingsubsets} holds for subsets of dimension 1 and 2, requiring that all line bundles are non-trivial and none is equal to minus another.} Specifically, for a sum of an odd number of line bundles, the presence of a trivial line bundle can lead to an accidental isomorphism $V\simeq V^*$. This forces the transition functions into $\SO(5)$ or $\text{Sp}(4)$, yielding a non-unitary structure group and breaking the required $\SU(5)$ GUT embedding \cite{Anderson:2013xka}.

We impose the anomaly cancellation condition \eqref{eq:anomalycancellation} by assuming that there are enough 5-branes to cancel the anomaly. Hence, this can be achieved by requiring that the curve $c_2(T\IX_3)-c_2(V)$ is effective. Since we are working with the K\"ahler cone basis as a basis of divisors for the CY, it reduces to the requirement that
\begin{equation}
    c_2(T\IX_3) - c_2(V)  \geq 0\fstop
\end{equation}
In components, we impose $h^{1,1}(\IX_3)$ conditions as
\begin{equation}
    c_{2,i}(T\IX_3) \geq \frac{1}{2}\sum_{j,k= 1}^{h^{1,1}(\IX_3)}\sum_{a=1}^N\left(\kappa_{ijk}c_1^j(L_a)c_1^k(L_a)\right) \fstop
\end{equation}

Analogously, the poly-stability property of the line bundles can be expressed as 
\begin{equation}\label{eq:mucondition}
    \mu(L_a) = 0 \Longrightarrow \sum_{i,j,k= 1}^{h^{1,1}(\IX_3)}\kappa_{ijk}c_1^i(L_a)t^j t^k = 0\coma
\end{equation}
somewhere in the interior of the K\"ahler cone. The assumption of working in the K\"ahler cone basis simplifies the condition, since, as already noted in \cite{Abel:2023zwg}, we can replace \eqref{eq:mucondition} with a weaker condition that each of the matrices
\begin{equation}\label{eq:Majk-poly}
    (M_a)_{jk} = \kappa_{ijk}\mathbf{k}^i_a
\end{equation}
has at least one positive and one
negative entry. Moreover, the same should hold for
every linear combination $v^aM_a$.\footnote{In practice, we consider all vectors $v_a$ with integer entries between
$-2$ and $2$. This bound was found heuristically in \cite{Abel:2023zwg}, and while this finite check does not constitute a formal proof of poly-stability, we believe that it still serves as a sufficient computational proxy.}

The last condition in Section \ref{sec:reviewHet} is about the equivariant structure of the line bundles, i.e., requiring the existence of a freely acting $\Gamma$ on $\IX_3$, such that $\widetilde{\IX}_3 = \IX_3/\Gamma$ is non-simply connected. In order for a vector bundle $V$ to descend to the smooth quotient manifold $\widetilde{\IX}_3$, the bundle must admit an equivariant structure under the freely-acting discrete symmetry $\Gamma$. A formal verification of this structure requires that the dimensions of the cohomology groups $H^i(\IX_3,V)$ transform as specific representations of $\Gamma$. For symmetries $\Gamma$ acting trivially on the divisor basis $J_i$, one should require that the Euler characteristic of every maximal partial sum of line bundles in $V$ (i.e., line bundles with identical first Chern classes) is divisible by $|\Gamma|$ \cite{Anderson:2011ns, Anderson:2012yf, Anderson:2013xka}, yielding
\begin{equation}\label{eq:equivariantcond_trivial}
m(L) \chi(\IX_3,L) = 0 \mod |\Gamma| \quad \forall \text{ distinct } L \subset V \coma
\end{equation}
where $m(L)$ is the multiplicity of $L$ in the line bundle sum $V$.

If the symmetries possess a non-trivial action on the divisor basis, they induce a permutation subgroup $H_\Gamma$ on the K\"ahler generators $J_i$. For CICYs, this has been classified in \cite{Braun:2010vc}, with a database available, e.g., \href{https://www-thphys.physics.ox.ac.uk/projects/CalabiYau/cicylist/index.html}{here}. In this scenario, we enforce the equivariant structure through a three-step verification \cite{Abel:2023zwg}. First, we require the invariance of the bundle $V$ under the induced action of $H_\Gamma$. For every line bundle $L \in V$ and every permutation $g \in H_\Gamma$, the pulled-back bundle $g^*L$ must be present in $V$ with a multiplicity:
\begin{equation}\label{eq:equivariantcond_multiset}
m(g^* L) = m(L) \quad \forall g \in H_\Gamma \fstop
\end{equation}
Second, we verify that the Euler characteristic is preserved, i.e., $\chi(\IX_3, g^* L) = \chi(\IX_3, L)$. Finally, $V$ must admit a partition into $H_\Gamma$-invariant partial sums, where the Euler characteristic of each partial sum is divisible by $|\Gamma|$. Under the non-trivial action of $H_\Gamma$, the line bundles in $V$ are partitioned into disjoint orbits (equivariant blocks). Because the discrete group $\Gamma$ does not mix distinct orbits, each orbit must descend to the quotient manifold to preserve the split nature of the bundle. Thus, the divisibility constraint applies to the total Euler characteristic of each individual orbit, rather than the entire bundle:
\begin{equation}\label{eq:equivariantcond_nontrivial}
\sum_{L \in \text{Orbit}_i} m(L) \chi(\IX_3,L) = 0 \mod |\Gamma| \quad \forall \text{ distinct orbits } i \coma
\end{equation}
where the sum runs over the distinct line bundles comprising the $i$-th orbit. These conditions filter our solutions to ensure $\Gamma$-invariance. 

To keep our LB-Explorer framework general and computationally efficient during the search phase, we do not perform explicit cohomology computations on-the-fly. Instead, we enforce the equivariance structure of the line bundles by filtering the generated configurations against the constraints of \cref{eq:equivariantcond_trivial,eq:equivariantcond_nontrivial}. For the particle spectrum, we impose weaker necessary conditions derived from the Atiyah--Singer index theorem. Specifically, we demand that the chiral indices of the bundle and its relevant tensor products scale appropriately with the group order $|\Gamma|$ and remain bounded. This isolates configurations with the correct net chiral asymmetry needed to produce the Standard Model spectrum upon quotienting. Full, automated cohomology computations (requiring tools such as \texttt{pyCICYs} \cite{Larfors:2019sie}, \texttt{CIPro} \cite{Anderson:2026eyl}, \texttt{cohomcalg} \cite{Blumenhagen:2010pv}, or recent extensions of \texttt{CYTools} \cite{Demirtas:2022hqf,Gendler:2026uux}) are then relegated to a subsequent validation step.\footnote{It would be interesting to further explore the possibility of computing line bundle cohomologies using ML techniques, as pioneered in \cite{Constantin:2018hvl, Klaewer:2018sfl} and further developed in \cite{Brodie:2019dfx, Brodie:2021nit, Constantin:2021for, Constantin:2024ulu}.} The conditions we impose during the search are formalized in the next section and summarized in Table \ref{tab:conditions}.

\subsubsection{Conditions on the Particle Spectrum}

Finally, we need to impose conditions on the spectrum. Usually, this involves the computation of cohomology groups of $\IX_3$ with values in the line bundles. Algorithms to compute these cohomologies exist for CICYs \cite{Larfors:2019sie} and for CYs in toric ambient space \cite{Blumenhagen:2010pv,Gendler:2026uux}. However, for our purposes, we decided to train LB-Explorer under weaker conditions, leaving the possibility of filtering the found solutions to the user's necessity. The conditions we impose are the existence of three chiral families and the absence of exotic representations, whose constraints we explain in the following.

For simplicity, let us focus on $H = \text{S}\left(\U(1)^5\right)$ so that the resulting four-dimensional spectrum consists of multiplets charged under the $\SU(5)\times \text{S}\left(\U(1)^5\right)$ gauge group with representations \cite{Anderson:2013xka}:
\begin{enumerate}
    \item $\mathbf{10}_a$ (resp. $\overline{\mathbf{10}}_a$) that carries charge 1 (resp. $-1$) under the a-th $\U(1)$ of $H$ and is uncharged under the others. The associated cohomology counting these multiplets is $H^1(\IX_3,L_a)$ (resp. $H^1(\IX_3,L_a^*)$).
    \item $\overline{\mathbf{5}}_{a,b}$ (resp. $\mathbf{5}_{a,b}$) for $a<b$ that carries charge $1$ (resp. $-1$) under the a-th and b-th $\U(1)$. The associated cohomology counting these multiplets is $H^1(\IX_3,L_a\otimes L_b)$ (resp. $H^1(\IX_3,L_a^*\otimes L_b^*)$).
    \item $\mathbf{1}_{a,b}$ for $a\neq b$ that has charge $1$ under the a-th $\U(1)$ and $-1$ under the b-th $\U(1)$. The associated cohomology is $H^1(\IX_3,L_a\otimes L_b^*)$.
\end{enumerate}

Instead of computing the cohomologies, we enforce the net chirality of the spectrum. First, we want the spectrum to be chiral and to have three families of multiplets in the $\mathbf{10}-\overline{\mathbf{10}}$ representation. This is accounted for by imposing  
\begin{equation}\label{eq:indV3Gamma}
    \text{ind}(V) =  \int_X \left(\text{ch}_3(V)+\frac{1}{12}c_2(T\IX_3)\wedge c_1(V)\right) \stackrel{!}{=} -3|\Gamma|\coma
\end{equation} 
which translates, in terms of $L_a$, under the condition that
\begin{equation}
    \text{ind}(V) = \sum_{a=1}^N \text{ind}(L_a) = -3|\Gamma|\coma
\end{equation}
with
\begin{equation}
    \text{ind}(L_a) 
     = \frac{1}{6}\sum_{i,j,k= 1}^{h^{1,1}(\IX_3)}\kappa_{ijk}c_1^i(L_a)c_1^j(L_a)c_1^k(L_a) + \frac{1}{12}\sum_{i=1}^{h^{1,1}(\IX_3)}c_{2,i}(T\IX_3)c_1^i(L_a)\fstop
\end{equation}
Interestingly, it was shown in \cite{Anderson:2013xka} that for $\SU(5)$ GUT, $\text{ind}(V) = \text{ind}(\wedge^2V)$; thus, the same constraint also imposes that $\mathbf{5}-\overline{\mathbf{5}}$ is chiral. Second, we impose bounds on the individual index contributions. To prevent anti-generations, we demand \cite{Anderson:2013xka}
\begin{equation}
    -3|\Gamma|\leq \text{ind}(L_a) \leq 0 \coma a = 1,\ldots ,5\coma
\end{equation}
for the $\mathbf{10}-\overline{\mathbf{10}}$, a similar constraint applies to the $\mathbf{5}-\overline{\mathbf{5}}$ sector. To address the doublet/triplet splitting problem, we impose
\begin{equation}
     -3|\Gamma|\leq \text{ind}(L_a\otimes L_b) \leq 0\coma a<b = 1,\ldots, 5\fstop
\end{equation}
If this index were positive on $\IX_3$, it would remain positive on $\widetilde{\IX}_3$, leading to Higgs color triplets that cannot be projected out by Wilson lines \cite{Anderson:2011ns,Anderson:2012yf}.

In Table \ref{tab:conditions}, we summarize the conditions that LB-Explorer will learn to find $\SU(5)$ GUT theories from heterotic compactifications.

\begin{table}[!htp]
    \centering
    \renewcommand*{\arraystretch}{1.3}
    \caption{Conditions for $\SU(5)$ GUT theories from heterotic compactifications on CY that LB-Explorer has learned.}
    \label{tab:conditions}
    \begin{tabular}{c|c}
     &  Conditions\\ \hline
       $E_8$ embedding  & $\displaystyle c_1(V) = 0\coma \sum_{a\in S} c_1(L_a) \neq 0$ \\
        Anomaly cancellation & $c_2(T\IX_3) - c_2(V)\geq 0$\\
        Poly-stability & $\mu(L_a) = 0$\\
        Three chiral families & $\text{ind}(V)=-3|\Gamma|$\\
        No exotic representations & $-3|\Gamma|\leq \text{ind}(L_a) \leq 0 \coma -3|\Gamma|\leq \text{ind}(L_a\otimes L_b) \leq 0$
    \end{tabular}
\end{table}

\section{Statistics on the CICYs}
\label{sec:CICYstatistics}

\begin{figure}[!htp]
    \centering
    \caption{Distribution of the maximal freely acting symmetry group orders across the favorable CICY landscape. There exist CICYs that admit multiple realization of symmetries with the same orders and they are counted once in the histogram. However, the same CICY may appear in multiple columns because it has more than one freely-acting groups.}
    \label{fig:gamma_histogram}
    \begin{tikzpicture}
        \begin{axis}[
            ybar, 
            bar width=10pt,
            width=10cm, 
            height=8cm,
            xlabel={Order of the Freely-Acting Group ($|\Gamma|$)},
            ylabel={Number CICYs},
            ymin=0,
            symbolic x coords={2, 3, 4, 5, 6, 8, 9, 10, 12, 16, 20, 25, 32},
            xtick=data, 
            nodes near coords, 
            nodes near coords style={
                    font=\tiny, 
                    text=black,
                    inner sep=1pt 
                },
            nodes near coords align={vertical},
            axis lines*=left,
            enlarge x limits=0.08,
            ymajorgrids=true,
            grid style=dashed
        ]
        
        \addplot[fill=prcolor!40, draw=prcolor!80, thick] coordinates {
            (2, 149)
            (3, 24)
            (4, 38)
            (5, 5)
            (6, 4)
            (8, 8)
            (9, 5)
            (10, 3)
            (12, 2)
            (16, 4)
            (20, 1)
            (25, 1)
            (32, 1)
        };
        \legend{$|\Gamma|$}
        \end{axis}
    \end{tikzpicture}
\end{figure}
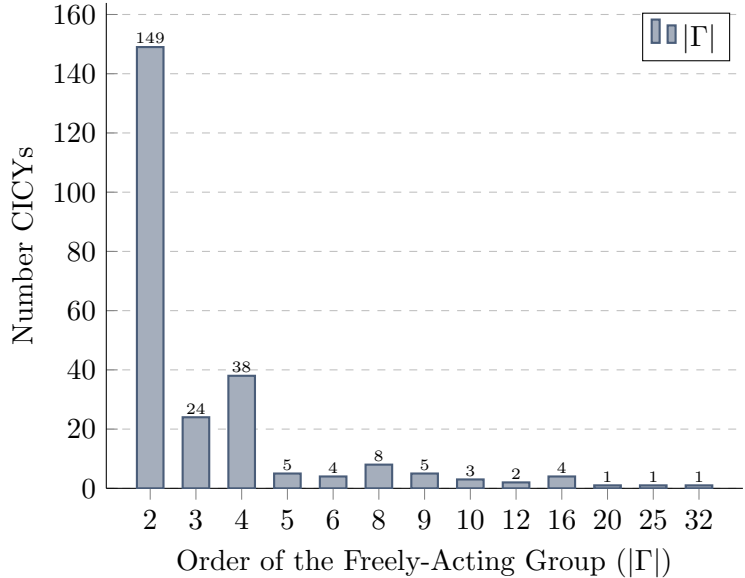

In this work, we focus on CICY manifolds \cite{Candelas:1987kf}, which we briefly review in Appendix \ref{app:CICYreview}. These are manifolds defined as the zero-locus of a set of holomorphic polynomials in an ambient space given by $\mathcal{A} = \prod_i \PP^{n_i}$. As shown in \cite{Anderson:2017aux}, all CICYs, apart from 70, admit a favorable realization, meaning the second cohomology descends entirely from the embedding space (see Appendix \ref{app:CICYreview} for a more precise definition), leading to $h^{1,1}(\IX_3) = h^{1,1}(\mathcal{A})$. Thus, in the following, we will first restrict ourselves to working with favorable CICYs only. In our \href{https://github.com/alexmininno/LB-Explorer}{GitHub repository}, we provide an updated database for these favorable CICYs. This version includes the intersection ring in the K\"ahler cone basis (following \cite{Carta:2021sms}), along with the symmetry groups of the configuration matrix ($G_\text{conf}$) and the CICY itself ($G_{\IX_3}$), and a list of freely acting symmetries that updates the existing database (found \href{https://www-thphys.physics.ox.ac.uk/projects/CalabiYau/cicylist/index.html}{here}) to the entries of the CICYs in \cite{Anderson:2017aux}.\footnote{The \texttt{CIPro} package \cite{Anderson:2026eyl} in Mathematica allows for the computation of such groups as well. A related CICY database containing Coxeter symmetries induced by isomorphic flops on the K\"ahler moduli space \cite{Alvarez-Garcia:2026vwq} is available \href{https://www.algarafa.com/cicy-coxeter/}{here}.} However, we have seen in the previous section that we require the existence of a freely acting symmetry group $\Gamma$ to define a smooth quotient space $\widetilde{\IX}_3=\IX_3/\Gamma$, such that the resulting CY manifold has a non-trivial fundamental group. In this way, Wilson lines are possible, and we can further break the GUT group $G$ down to the Standard Model. As we review in Appendix \ref{app:CoordinateSymmetries}, freely-acting symmetries in CICY manifolds have been studied in \cite{Candelas:1987du,Candelas:2008wb,Candelas:2010ve}, and a complete classification of all freely-acting symmetries that descend from linear actions on the projective ambient space has been completed in \cite{Braun:2010vc}. The result of the classification is that only 171 CICYs admit (at least) a freely-acting symmetry. Among these CICYs, there are CICYs that admit multiple freely-acting groups, so that we can consider the same CICY but different values of $|\Gamma|$. In Figure \ref{fig:gamma_histogram}, we show a distribution of the order of the freely-acting groups among these 171 CICYs. As it is clear from Figure \ref{fig:gamma_histogram}, the same CICY can appear multiple times among the various columns of the histogram, but we only count it once if it admits multiple freely-acting groups of the same order. An example is the mirror quintic, i.e., CICY $7890$, which admits multiple $\Gamma = \ZZ_5$ groups, but also a $\Gamma = \ZZ_5\times \ZZ_5$. The distribution heavily favors small cyclic groups, with $\mathbb{Z}_2$ being the most prevalent.

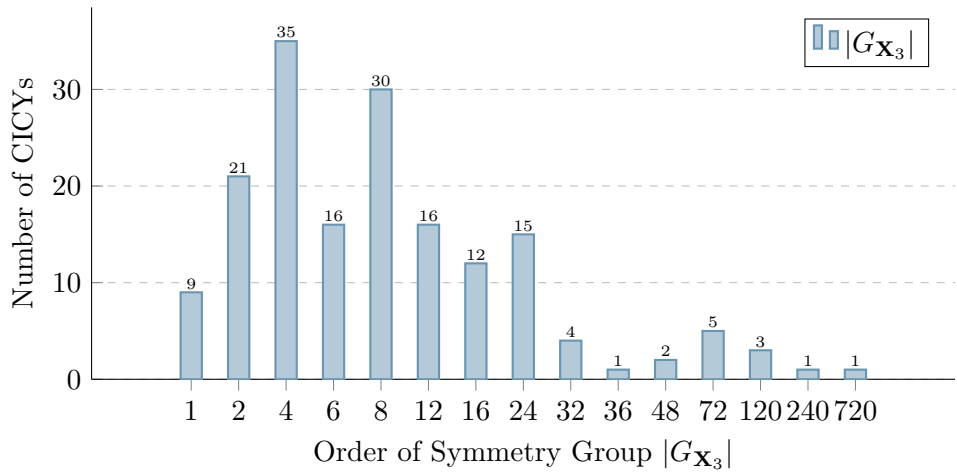
\begin{figure}[!htp]
        \centering
\caption{Distribution of symmetry group orders restricted to the 171 CICYs that admit freely-acting quotients.}
    \label{fig:conf_vs_geom_subplots_filtered}
        \begin{tikzpicture}
            \begin{axis}[
                ybar=3pt,
                bar width=8pt,
                width=13cm,
                height=6.5cm,
                ylabel={Number of CICYs},
                xlabel={Order of Symmetry Group $|G_{\IX_3}|$},
                ymin=0,
                symbolic x coords={1, 2, 4, 6, 8, 12, 16, 24,32,36,48,72,120,240,720},
                xtick=data,
                axis lines*=left,
                enlarge x limits=0.15,
                ymajorgrids=true,
                grid style=dashed,
                legend pos=north east,
                nodes near coords,
                nodes near coords style={
                    font=\tiny, 
                    text=black,
                    inner sep=1pt 
                }
            ]
            
            \addplot[fill=seccolor!40, draw=seccolor!80, thick] coordinates {
                (1, 9)
                (2, 21)
                (4, 35)
                (6, 16)
                (8, 30)
                (12, 16)
                (16, 12)
                (24, 15)
                (32, 4)
                (36, 1)
                (48, 2)
                (72, 5)
                (120, 3)
                (240, 1)
                (720, 1)
            };

            \legend{$|G_{\IX_3}|$
            }
            \end{axis}
        \end{tikzpicture}
\end{figure}

In exploring the landscape of possible vector bundle configurations, we want to mod out redundant solutions, i.e., bundles $V$ that are equivalent up to the relabeling of the line bundles $L_a$ or of the basis of divisors $J_i$. In fact, $V$ can be interpreted as an $(h^{1,1}\times 5)$-dimensional matrix that is constrained by the conditions explained in  \cref{sec:reviewHet,sec:GUTHetIngred}. Since we are considering only maximal splitting, the first redundancy is an $S_5$ permutation symmetry associated to the relabeling of the five line bundles $L_a$. The scan finds vector bundles $V$, where we will have already modded out this redundancy. However, there is also another redundancy, which is the group of automorphisms of the intersection ring that preserve the second Chern class $c_2(T\IX_3)$ of $\IX_3$. As we explain in Appendix \ref{app:CICYsymmetries}, this is a subgroup of the symmetric group $S_{h^{1,1}(\IX_3)}$ defined in \eqref{eq:GX3def}. In Appendix \ref{app:CICYsymmetries}, we explain the algorithm that we used to compute such a group for the whole CICY database. We will, then, also provide the number of solutions for which we will have modded out by $G_{\IX_3}$. However, rather than analyzing the general automorphisms of the entire dataset, we restrict our evaluation $G_{\IX_3}$, exclusively to the 171 manifolds that admit a freely-acting quotient. Figure \ref{fig:conf_vs_geom_subplots_filtered} displays the distribution of the group orders $|G_{\IX_3}|$ over the dataset of 171 CICY manifolds for which a freely-acting quotient $\Gamma$ has been classified \cite{Braun:2010vc}.

\subsection{Dataset Selection and Specific Geometries}
\label{sec:dataset_selection}

While the landscape of favorable CICYs admitting freely-acting discrete symmetries comprises 171 manifolds (with CICYs admitting multiple freely-acting discrete symmetries as well), training LB-Explorer across the entire database simultaneously is prohibitive and not particularly physically meaningful. Instead, we select specific subsets of this landscape to evaluate both the baseline efficiency of the NN architecture and its capacity for transfer learning.

\paragraph{Baseline Training Set.}
To establish the baseline performance of LB-Explorer, we consider a set of 54 CICYs for various $h^{1,1}(\IX_3)$ and $|\Gamma|$. We considered $h^{1,1}(\IX_3)\in [4,15]$ and $|\Gamma|\in [2,4]$. For each pair $(h^{1,1}(\IX_3),|\Gamma|)$, we picked a set of CICYs that span a representative range of $|G_{\IX_3}|$ from its minimal to its maximal value for fixed $h^{1,1}(\IX_3)$. We list them with their properties in Table \ref{tab:CICYdataset}. For all these CICYs, we have run a training of LB-Explorer over five different seeds for 10 million episodes. LB-Explorer has been asked to learn conditions in Table \ref{tab:conditions}. We will divide the solutions into four categories, as follows. By $N$, we call all the solutions found by LB-Explorer satisfying the conditions in Table \ref{tab:conditions}. We then filter these solutions by taking only one representative among those that are related by relabeling of the five line bundles $L_a$. The remaining solutions will be called $N_{S_5}$. Analogously, by acting with the symmetry group $G_{\IX_3}$, we define $N_{G_{\IX_3}}$ as the number of unique solutions obtained by modding out redundancies related to the relabeling of the divisor basis $J_i$. Finally, $N_{\text{full}}$ will be the combinations of the two previous filters. The discussion of the scan is presented in Section \ref{sec:results}, with data shown in Table \ref{tab:detailed_counts}. The solutions are further filtered by imposing \cref{eq:equivariantcond_trivial,eq:equivariantcond_nontrivial}, and we refer to Table \ref{tab:CICYdataset} to determine whether the action of $\Gamma$ is trivial or not on the divisor basis. In Table \ref{tab:detailed_counts}, this is shown with a \checkmark. 

Finally, in Section \ref{sec:spectrumanalysis}, we will consider some of the solutions associated with CICY in Table \ref{tab:detailed_counts} that have a \checkmark, and we will compute the line bundle cohomologies to count how many will satisfy the constraints on the spectrum, relying on \texttt{pyCICY} \cite{Larfors:2019sie} for the computations of the cohomologies of line bundles. We show the results in Table \ref{tab:spectrum_stats}.

\paragraph{Transfer Learning.} The generality and flexibility of the LB-Explorer naturally lead to the question of whether it is possible to pre-train the NN over a CICY and see if the pre-trained NN can also learn to find solutions for a different CICY. We decided, then, to conduct some experiments on transfer learning of three possible kinds: the first is among CICYs that share the same $h^{1,1}(\IX_3)$ and $\Gamma$ but are not related by any transformation that makes them equivalent according to Wall's theorem.\footnote{It is, of course, known that all CICYs can be obtained by performing a finite chain of conifold transitions that sometimes require an unfavorable realization of the CICY \cite{Candelas:1989ug}. Here, by unrelated, we mean that they are CICYs for which it is not possible to find a matrix transformation that allows one to rewrite the second Chern class and the intersection numbers of one in terms of the other. Redundancies in CICYs have been known since their first formulation \cite{Candelas:1990pi,Avram:1995pu,Candelas:1989ug,Candelas:2007ac}, and they have been classified over the years, e.g., in \cite{Anderson:2008uw,Carta:2021sms}.} We call this transfer learning ``transfer learning within bucket". The second is considering a CICY admitting more freely-acting symmetries and checking transfer learning between a pre-trained NN over a $|\Gamma|$ and using it to find solutions over another $|\Gamma|$. This transfer learning is dubbed ``transfer learning across $\Gamma$. Finally, we considered CICYs related by a conifold transition that increases or decreases $h^{1,1}(\IX_3)$ by 1, and we studied the ``transfer learning across $h^{1,1}(\IX_3)$". These experiments will be discussed in Section \ref{sec:transferlearning}. It is important to stress that these were just examples of transfer learning to show the potentiality of the LB-Explorer in learning to satisfy the conditions in Table \ref{tab:conditions} without relying too much on the specific geometric data of the CICY.

\paragraph{Hybrid Approach.} As we have explained in Section \ref{sec:GUTHetIngred}, the LB-Explorer is not learning all the conditions that are necessary to obtain a MSSM-like spectrum from the line bundles. In fact, we only impose the conditions in Table \ref{tab:conditions}, delegating the check on equivariance and particle spectrum to a further filtering of the solutions. The reason is to keep the number of exact conditions to be learned by the LB-Explorer to the minimum, together with the difficulties in computing line bundle cohomologies for general values of $\mathbf{k}_a$ or large $h^{1,1}(\IX_3)$. However, it is, in principle, possible to let the LB-Explorer impose these extra conditions using a Hybrid LB-Explorer, where the missing conditions could, in principle, be found exactly by means of the Constraint Programming and Boolean Satisfiability (CP-SAT) module \cite{cpsatlp}, which perturbs the solutions found by the LB-Explorer while trying to satisfy the remaining constraints. In order to have some idea that this approach is feasible, instead of trying to impose the extra conditions, we attempted to remove one condition among those in Table \ref{tab:conditions}, i.e., the check on the index of $V$, and let CP-SAT perturb the solutions of the LB-Explorer in order to find solutions satisfying the check on the index exactly. As a benchmark, we used CICYs with low $h^{1,1}(\IX_3)$ that usually have sparse solutions, aiming to improve the exploration of the LB-Explorer. We discuss more in Section \ref{sec:HybridRL}.

\section{LB-Explorer Architecture}
\label{sec:nn_architecture}

In this section, we describe the LB-Explorer architecture for the search of vector line bundles. The network search is formulated as an autoregressive sequence generation task and solved using Proximal Policy Optimization (PPO) \cite{Schulman:2017xxx}. The aim is to construct the vector bundle $V$, represented as an $h^{1,1}(\IX_3)\times 5$ matrix $\mathbf{K}=\{k_a^i\}_{i=1,\ldots,h^{1,1}}^{a = 1,\ldots,5}$, token-by-token, imposing the conditions in Table \ref{tab:conditions} on the entries. 

The first Chern class condition, i.e., $c_1(V) = 0$, is imposed by construction by requiring that
\begin{equation}
    \mathbf{k}_5 = -\sum_{a=1}^{4}\mathbf{k}_a\coma
\end{equation}
while the remaining entries of $\mathbf{K}$ are flattened in row-major order into a token sequence $\mathbf{S} = \{s_t\}_{t=0,\ldots, S-1}$, with $S = h^{1,1}(\IX_3)\times 4$. For instance, $s_{10}$ corresponds to the entry $k^3_3$ in $\mathbf{K}$.

To fill the sequence vector, we define an action space of integers $a_t\in [0,2k_{\text{max}}]$, that is mapped to the values of $s_t$ by $s_t = a_t - k_{\text{max}}$.

\begin{figure}[!htp]
    \centering
    \caption{Schematic representation of the LB-Explorer architecture.}
    \label{fig:nn_schema}
    \begin{tikzpicture}[
        node distance = 0.6cm and 1.5cm,
        box/.style = {draw, rectangle, rounded corners, minimum width=3.5cm, minimum height=0.8cm, align=center, fill=gray!20, font=\small},
        splitbox/.style = {draw, rectangle, rounded corners, minimum width=2cm, minimum height=0.8cm, align=center, fill=blue!20, font=\small},
        envbox/.style = {draw, rectangle, rounded corners, minimum width=6cm, minimum height=1.2cm, align=center, fill=green!30, font=\small, dashed},
        arrow/.style = {-Latex, thick}
    ]
    
    \node (input) [box, fill=white] {Input Sequence $\mathbf{S}_t$ \\ $(a_0, a_1, \dots, a_{t-1})$};
    \node (embed) [box, below=of input] {Token Embedding};
    \node (pos) [box, right=of embed] {Positional Encoding};
    
    \node (attn) [box, below=1.0cm of embed] {Masked Multi-Head \\ Self-Attention (8 Heads)};
    \node (addnorm1) [box, below=0.8cm of attn] {Residual Connection \& LayerNorm};
    \node (ffn) [box, below=0.8cm of addnorm1] {Feed-Forward Network \\ (ReLU activation)};
    \node (addnorm2) [box, below=0.8cm of ffn] {Residual Connection \& LayerNorm};
    
    \node (actor) [splitbox, below left=1cm and -0.5cm of addnorm2] {Actor Head \\ (Linear \& Softmax)};
    \node (critic) [splitbox, below right=1cm and -0.5cm of addnorm2] {Critic Head \\ (Linear)};
    
    \node (pi) [below=0.4cm of actor] {Action $a_t \sim \pi_\theta(a_t | \mathbf{S}_t)$};
    \node (val) [below=0.4cm of critic] {Value $V_\theta(\mathbf{S}_t)$};
    
    \node (env) [envbox, below=4cm of addnorm2] {Physics Environment \& Reward Engine \\ \textit{Triviality Mask $\cdot$ Log-Barrier Constraints $\cdot$ Novelty Buffer}};
    
    \draw [arrow] (input) -- (embed);
    \draw [arrow] (pos) -- (embed);
    \draw [arrow] (embed) -- (attn);
    \draw [arrow] (attn) -- (addnorm1);
    \draw [arrow] (addnorm1) -- (ffn);
    \draw [arrow] (ffn) -- (addnorm2);
    
    \draw [arrow] (embed.south) ++(0,-0.4) coordinate(a) -- ++(3.5,0) coordinate(res1) |- (addnorm1.east);
    \draw [arrow] (addnorm1.south) ++(0,-0.4) coordinate(b) -- ++(3.5,0) |- (addnorm2.east);
    
    \draw [arrow] (addnorm2.south) -- ++(0,-0.4) -| (actor.north);
    \draw [arrow] (addnorm2.south) -- ++(0,-0.4) -| (critic.north);
    
    \draw [arrow] (actor) -- (pi);
    \draw [arrow] (critic) -- (val);
    
    \draw [arrow] (pi.south) -- ++(0,-0.4) -| (env.north);
    \draw [arrow, dashed, color=red!70!black] (env.west) -- ++(-1.5,0) |- node[pos=0.25, left, align=center, font=\footnotesize] {Reward $r_S$ \\ \& \\ Advantage (GAE)} (actor.west);
    
    \node[draw, dashed, inner sep=0.3cm, fit=(attn) (addnorm2) (res1), label={[anchor=north west, inner sep=3pt]north west:$\times 4$ Layers}] {};

    \end{tikzpicture}
\end{figure}
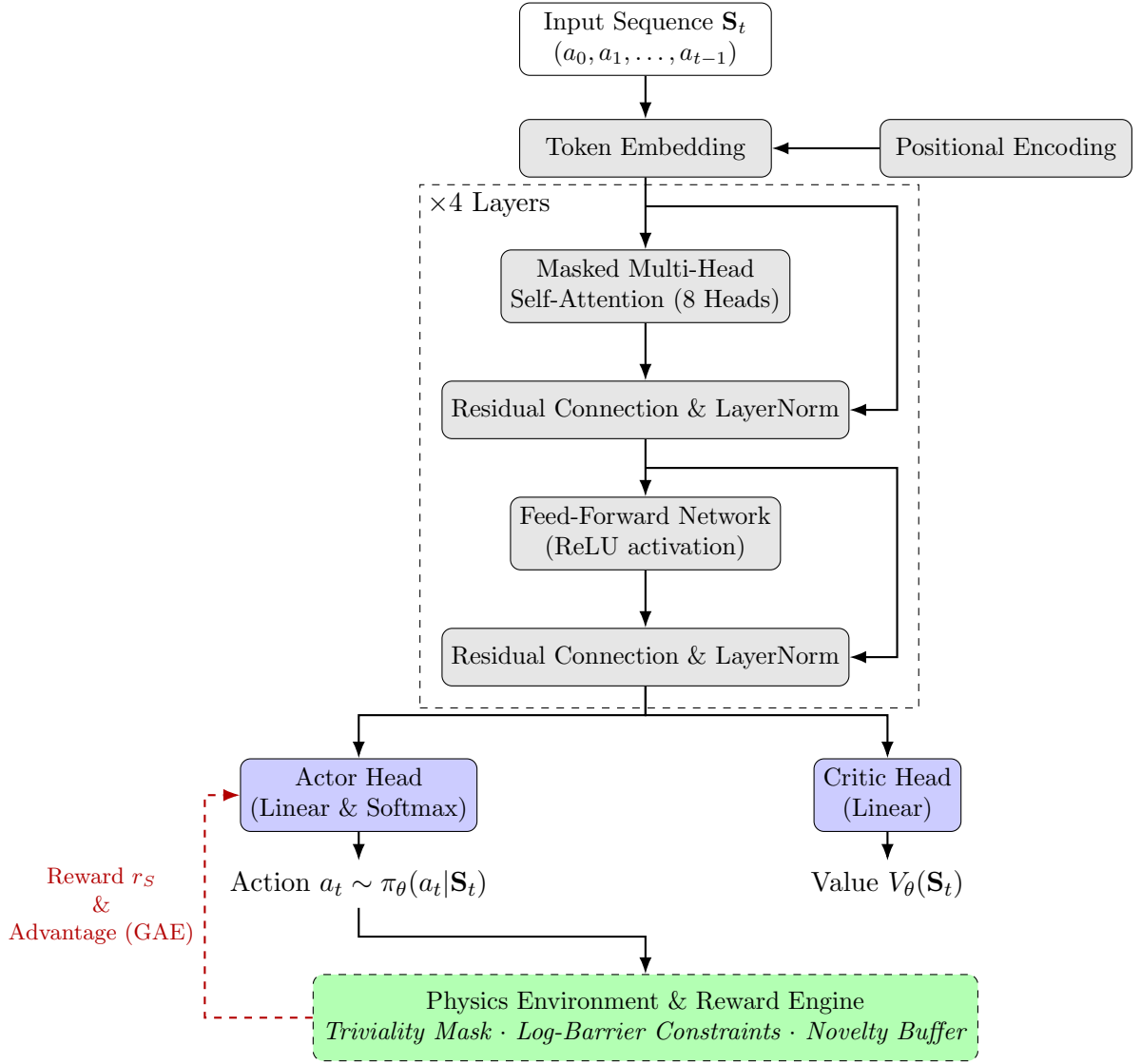

We depict the full schematic representation of the architecture in Figure \ref{fig:nn_schema}, which we will explain in detail in the following section. The input sequence at step $t$ is denoted by $\mathbf{S}_t = (a_0, a_1, \dots, a_{t-1})$.

\subsection{Network Architecture}

The policy and value functions are parameterized by a shared, decoder-only Transformer network \cite{Vaswani:2017xxx}. The LB-Explorer has $540,946$ trainable parameters in total. At step $t$, the state is the sequence of previously generated tokens $\mathbf{S}_t = (a_0, a_1, \dots, a_{t-1})$. 

\paragraph{Input Embedding.} Given a bound $k_{\text{max}}$ on the line bundle values,\footnote{In our work we considered $k_{\text{max}}= 8$.} a discrete vocabulary $A$ of size $|A|=2k_{\text{max}}+1$ is constructed, where $a_t\in A$ are tokens in this vocabulary.\footnote{Precisely speaking, the input vocabulary also contains a padding token to initialize the autoregressive generation. However, the action space excludes the padding token, ensuring it is never sampled.} The input to LB-Explorer consists of $a_t$ tokens mapped to a dense vector via a learnable embedding matrix $\mathbf{E}\in \RR^{(|A|+1)\times d_{\text{model}}}$, with $d_{\text{model}} = 128$. Since token order matters for the conversion back to $\mathbf{K}$, we also introduce a learnable positional encoding matrix $\mathbf{P}\in \RR^{S\times d_{\text{model}}}$. This allows us to break the permutation equivariance of the attention mechanism. The input at a given position of the sequence is then 
\begin{equation}
    x_t = E_{a_t}+P_t\coma
\end{equation}
where $E_{a_t}$ is the row of $\mathbf{E}$ corresponding to token $a_t$, and $P_t$ is the $t$-th row of $\mathbf{P}$.

\paragraph{Transformer Decoder.} The input sequence $X^{(0)} = [x_0, \dots, x_{t-1}] \in \mathbb{R}^{t \times d_{\text{model}}}$ is processed by $L=4$ identical transformer blocks. For each layer $l=1,\ldots,4$, the Masked Multi-Head Self-Attention (MHA) operation computes $H=8$ independent attention heads. The projections for the $h$-th head are parameterized by $W^Q_h, W^K_h, W^V_h \in \mathbb{R}^{d_{\text{model}} \times d_k}$, where $d_k = d_{\text{model}}/H$. The attention mechanism is enforced by an upper-triangular mask $\mathbf{M} \in \mathbb{R}^{t \times t}$ with entries $M_{ij} = 0$ for $j \leq i$ and $M_{ij} = -\infty$ for $j > i$:
\begin{equation}
    \text{head}_h = \text{Softmax}\left(\frac{X^{(l-1)} W^Q_h (X^{(l-1)} W^K_h)^\top}{\sqrt{d_k}} + M\right) X^{(l-1)} W^V_h \fstop
\end{equation}
The outputs of the individual heads are concatenated and projected via an output weight matrix $W^O \in \mathbb{R}^{d_{\text{model}} \times d_{\text{model}}}$. The primary role of $W^O$ is to linearly combine the diverse, specialized representations learned by the independent attention heads back into a cohesive $d_{\text{model}}$-dimensional vector. We also introduce a residual connection followed by a Layer Normalization (LN) \cite{Ba:2016xxx} block to improve stability. Since the conditions imposed on the vector bundles are non-linear in the line bundle values, we apply a Feed-Forward Network (FFN) \cite{Glorot:2010xxx} to each token's representation. Effectively, it expands the dimensionality to $d_{\text{FF}} = 2 d_{\text{model}}$ using a ReLU activation before projecting back down, i.e.,
\begin{align}
    Y^{(l)} &= \text{LN}\left(X^{(l-1)} + \text{Concat}(\text{head}_1, \dots, \text{head}_H) W^O\right) \coma \\
    X^{(l)} &= \text{LN}\left(Y^{(l)} + \text{ReLU}(Y^{(l)} W^1 + b^1) W^2 + b^2\right) \coma
\end{align}
with $W^1 \in \mathbb{R}^{d_{\text{model}} \times d_{\text{FF}}}$ and $W^2 \in \mathbb{R}^{d_{\text{FF}} \times d_{\text{model}}}$.\footnote{Dropout is applied at 4 standard sites per transformer block (on the attention weights, after the attention projection, mid-FFN, and post-FFN) to guard against policy collapse and value-function overfitting by inducing exploration noise.}

\paragraph{Actor-Critic Architecture.} To predict the next action $a_t$ and evaluate the quality of the current state, LB-Explorer extracts the final hidden state vector $h_t = X^{(L)}_{t-1} \in \mathbb{R}^{d_{\text{model}}}$ corresponding to the last processed token. This vector serves as a summary of the generated matrix prefix $(a_0, \dots, a_{t-1})$. $h_t$ is routed through two independent linear layers:
\begin{enumerate}
    \item \textit{Actor Head (Policy)}: Maps $h_t$ to a logit across the entire action space via a weight matrix $W_\pi \in \mathbb{R}^{|A| \times d_{\text{model}}}$ and a bias vector $b_\pi \in \mathbb{R}^{|A|}$. A softmax activation converts these logits into a probability distribution:
    \begin{equation}
        \pi_\theta(a_t | \mathbf{S}_t) = \text{Softmax}(W_\pi h_t + b_\pi) \fstop
    \end{equation}
    Here, $\pi_\theta(a_t | \mathbf{S}_t)$ is the probability of the network selecting the token $a_t$ given the current sequence prefix $\mathbf{S}_t$.
    \item \textit{Critic Head (State-Value Estimate)}: Maps the same hidden state $h_t$ to a scalar value via $W_v \in \mathbb{R}^{1 \times d_{\text{model}}}$ and $b_v \in \mathbb{R}$:
    \begin{equation}
        V_\theta(\mathbf{S}_t) = W_v h_t + b_v \coma
    \end{equation}
    where $\theta$ represents a collective name for all learnable parameters. $V_\theta(\mathbf{S}_t)$ predicts the expected cumulative reward the agent anticipates accumulating from the sequence prefix $\mathbf{S}_t$.
\end{enumerate}

\subsection{Optimization and Loss Functions}

In this section, we explain in more detail how the PPO-RL is updated. While in the previous section we have seen that the Actor-Critic architecture constructs the vector bundle matrix $V$ token-by-token, the conditions in Table \ref{tab:conditions} can be checked only once the full vector bundle configuration has been specified. Hence, the environment assigns a dense reward $r_S$ only at the terminal step $t=S$, leaving all intermediate steps with zero reward ($r_t = 0$ for $t < S$). In the following, we explain how we assign such a reward.

To construct the final dense reward $r_S$, the environment evaluates the constraints using a composite scoring system:
\begin{enumerate}
    \item \textit{Triviality Masking}: We first apply a binary mask that zeros the final score if the bundle is degenerate (e.g., contains an all-zero column or two columns that cancel each other).
    \item \textit{Continuous and Binary Scoring}: During learning, we compute a continuous score for each constraint using a log-barrier function, 
    \begin{equation}\label{eq:Sicont}
        S_{i, \text{cont}} = \frac{1}{1 + w_i \ln(1 + e_i)}\coma
    \end{equation}
    providing a gradient signal even when constraints are violated. The error $e_i$ is formulated for each constraint as follows:
    \begin{itemize}
        \item \textit{Anomaly Cancellation:} The error $e_{\text{anom}}$ is defined as the mean positive violation across the $h^{1,1}(\IX_3)$ components of the line bundles, $$e_{\text{anom}} = \frac{1}{h^{1,1}(\IX_3)}\sum_j \max(0, c_2^j(V) - c_2^j(T\IX_3))\fstop$$
        \item \textit{Poly-stability:}  The error $e_{\text{stab}}$ aggregates the extent to which the matrix in \eqref{eq:Majk-poly} fails to possess both positive and negative entries across the grid of test vectors $v^a\in [-2,2]$.
        \item \textit{Chirality:} The error is the absolute difference $$e_{\text{sum}} = \left| \left(\sum_a \text{ind}(L_a)\right) - (-3|\Gamma|) \right|\fstop$$
        For the range conditions on the index of the single line bundles, the error $e_{\text{rng}}$ sums the clamped positive deviations outside the window $[-3|\Gamma|,0]$.
    \end{itemize}
    Note that each $e_i$ is an average error normalized by the number of constraint equations. Therefore, the choice of $S_{i,\text{cont}}$ in \eqref{eq:Sicont} leads to stable PPO training for three main reasons: (i) its range is in $(0,1]$, regardless of the magnitude of the single errors $e_i$, (ii) with the logarithm preserving the signal even for large values of errors that would lead to a suppressed reward otherwise. (iii) Finally, the reward is highest only when $e_i$ is near zero, without leading to a flattened loss function even for moderate errors. We define the continuous base score $B = \sum_i\mathfrak{w}_iS_{i, \text{cont}}$, where LB-Explorer defaults to uniform weights $\mathfrak{w}_i=w_i=1$. As hyperparameters, these weights may be tuned to achieve an artificial learning curriculum.\footnote{Before settling on the uniform choice of weights, we tested two adaptive schemes for the reward weights $\mathfrak{w}_i$ as alternatives: 1. \textit{Random Loss Weighting (RLW)}~\cite{lin2022reasonableeffectivenessrandomweighting}, where, at each batch, a new weight vector is sampled from a Dirichlet distribution and re-normalized. 2. \textit{EMA-adaptive weighting}~\cite{moralesbrotons2024exponentialmovingaverageweights}, where a running Exponential Moving Average (EMA) of each constraint's passing rate is maintained, and weights are set inversely so that the most heavily failing constraint receives the largest learning pressure. These two alternatives have been tested on the CICYs in the bucket with $h^{1,1}=5$ and $|\Gamma|=4$, and we compared their performance with the uniform choice of weights. RLW added pure noise, completely unable to identify any constraint. EMA often exhibited runaway feedback on one constraint from a single batch that caused the explorer to struggle to learn the other conditions. Eventually, the uniform distribution of the weights produced the most consistent learning across all test cases. There exist more multi-task techniques, such as those operating at the gradient level (e.g., FAMO~\cite{liu2023famofastadaptivemultitask}, GradVac~\cite{wang2020gradientvaccineinvestigatingimproving}), that were not tested in this work.}
    \item \textit{Perfection Bonus}: A valid matrix satisfying all constraints simultaneously receives an additive spike ($+5$ bonus) to reinforce valid configurations.
\end{enumerate}

To avoid mode collapse into a single known solution, we implement a rolling First-In-First-Out (FIFO) novelty buffer of previously generated matrices. When the agent generates a matrix that is equivalent under $S_5$ (the relabeling permutation symmetry of the line bundles) to any entry in the buffer, its reward is scaled down by a penalty factor $p_{\text{pen}} = 0.5$.

Synthesizing the triviality masking, the base score $B$, the perfection bonus, and the novelty penalty, the terminal reward $r_S$ assigned by the environment at step $t=S$ for matrix $\mathbf{K}$ is:
\begin{equation}\label{eq:terminal_reward}
    r_S = \begin{cases}
    0 & \text{if degenerate}\coma \\
    B + 5 & \text{if all constraints are met and $\mathbf{K}$ is novel}\coma \\
    p_{\text{pen}} (B + 5) & \text{if all constraints are met but $\mathbf{K}$ is a duplicate}\coma \\
    B & \text{if some constraints fail, but $\mathbf{K}$ is non-degenerate and novel}\coma \\
    p_{\text{pen}} B & \text{if some constraints fail, but $\mathbf{K}$ is non-degenerate and a duplicate}\fstop
    \end{cases}
\end{equation}

\paragraph{Generalized Advantage Estimation (GAE).} Because the environment provides no immediate feedback for the intermediate tokens ($r_t = 0$ for $t < S$), we propagate the terminal reward $r_S$ backwards through the sequence using Generalized Advantage Estimation \cite{Schulman:2015xxx}. At each step $t$, we compute the Temporal Difference (TD) error $\delta_t$. The advantage function $\hat{A}_t$ aggregates these temporal errors:
\begin{equation}
    \hat{A}_t = \sum_{l=0}^{S-t-1} (\gamma \lambda)^l \delta_{t+l}\coma \text{with } \delta_t = r_t + \gamma V_{\theta_{\text{old}}}(\mathbf{S}_{t+1}) - V_{\theta_{\text{old}}}(\mathbf{S}_t) \fstop
\end{equation}
Here, $\gamma = 0.99$ is the discount factor determining the importance of future rewards, and $\lambda = 0.95$ is the GAE trace-decay parameter controlling the bias-variance tradeoff of the advantage estimates. The terminal reward $r_S$ is thereby distributed across the intermediate steps, assigning credit to the specific choices made during generation. We further z-score normalize $\hat{A}_t$ by its mean and standard deviation across the sequence and batch dimensions to keep the gradient step size on a consistent scale, decoupling the gradient magnitude from the reward scale. 

\paragraph{PPO Loss Function and Update.} Each batch of generated matrices is reused over $E_{\text{ppo}}=4$ Episodes, with the network parameters $\theta$ updated by minimizing a total loss function that combines three distinct components:
\begin{equation}
    \mathcal{L}^{\text{total}}(\theta) = \mathcal{L}^{\text{CLIP}}(\theta) + c_1 \mathcal{L}^{\text{VF}}(\theta) - c_2 \mathcal{S}[\pi_\theta] \coma
\end{equation}
where the first two loss functions are the actor and critic loss, respectively, while the last is an entropy regulator, explained in the following.
\begin{enumerate}
    \item \textit{Clipped Surrogate Objective (Actor Loss)}: The purpose of this loss function is to avoid destructive updates during the training process. The PPO mitigates this by bounding how much the policy can change in a single update. Let $\rho_t(\theta) = \frac{\pi_\theta(a_t|\mathbf{S}_t)}{\pi_{\theta_{\text{old}}}(a_t|\mathbf{S}_t)}$ denote the importance sampling ratio between the newly updated policy and the old policy, specifically for the sampled token $a_t$.\footnote{Note that $\rho_t=1$ for the first mini-batch of the first PPO episode.} The loss is:
    \begin{equation}
        \mathcal{L}^{\text{CLIP}}(\theta) = -\frac{1}{B} \sum_{b=1}^{B} \sum_{t=1}^{S} \min\left( \rho_t(\theta) \hat{A}_t, \text{clip}(\rho_t(\theta), 1-\epsilon, 1+\epsilon) \hat{A}_t \right) \coma
    \end{equation}
    where $\epsilon=0.2$ is the clipping hyperparameter, and $B=8192$ is the batch size.\footnote{In practice, each PPO Episode is split into mini-batches of size $B_{\text{mini}}=1024$ to manage memory constraints.}
    \item \textit{Value Function Loss (Critic Loss)}: 
    \begin{equation}
        \mathcal{L}^{\text{VF}}(\theta) = \frac{1}{B} \sum_{b=1}^{B} \sum_{t=1}^{S} \left( V_\theta(\mathbf{S}_t) - \hat{R}_t \right)^2 \coma
    \end{equation}
    where $\hat{R}_t = \hat{A}_t + V_{\theta_{\text{old}}}(\mathbf{S}_t)$ represents the target return, computed as the sum of the estimated advantage and the previous value prediction.
    \item \textit{Entropy Regularization}: To prevent the policy from prematurely converging on trivial local minima, we maximize the Shannon entropy of the policy distribution:
    \begin{equation}
        \mathcal{S}[\pi_\theta] = \frac{1}{B} \sum_{b=1}^{B} \sum_{t=1}^{S} \left( - \sum_{a \in A} \pi_\theta(a|\mathbf{S}_t) \ln \pi_\theta(a|\mathbf{S}_t) \right) \fstop
    \end{equation}
    This entropy loss contrasts with the Actor Loss's tendency to concentrate probability on high-advantage actions. The coefficient $c_2$ may be reduced during the training process to first encourage broad exploration and only later fine-tune valid solutions.\footnote{In our experiments, we held $c_2$ constant at $0.05$ throughout training, allowing us to compare results without per-CICY hyperparameter tuning.}
\end{enumerate}

\section{Results}
\label{sec:results}

In this section, we will discuss the results of the LB-Explorer training, commenting on the learning pattern and the analysis of the exploration status. In Section \ref{sec:transferlearning}, we analyze the transfer learning properties of the LB-Explorer.

\subsection{Analysis of Solutions}
\label{sec:generalexploration}

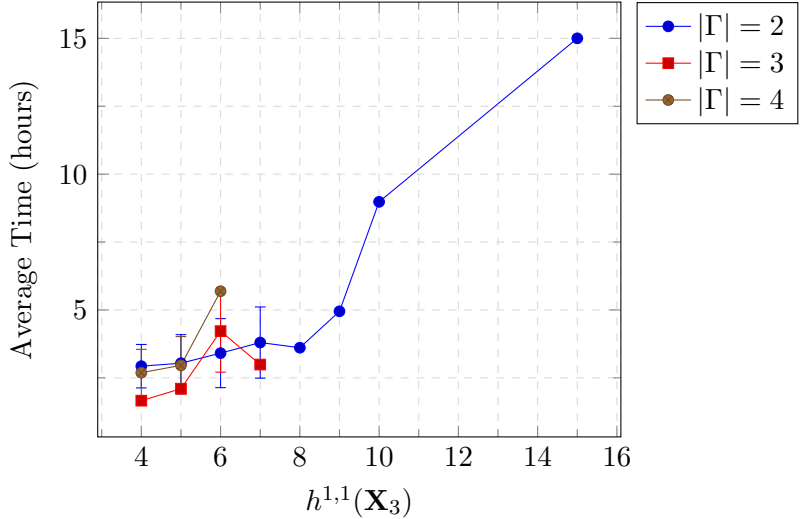
\begin{figure}[!htp]
		\centering
		\begin{tikzpicture}
			\begin{axis}[
				width=0.5\textwidth,
				xlabel={$h^{1,1}(\IX_3)$},
				ylabel={Average Time (hours)},
				legend pos=outer north east,
				grid=both,
				grid style={dashed, gray!30},
				minor tick num=1,
				mark=none
				]
				\addplot+[error bars/.cd, y dir=both, y explicit] coordinates {
					(4, 2.93) +- (0, 0.80) 
					(5, 3.04) +- (0, 1.05) 
					(6, 3.41) +- (0, 1.27) 
					(7, 3.80) +- (0, 1.31) 
					(8, 3.61) +- (0, 0.00) 
					(9, 4.95) +- (0, 0.00) 
					(10, 8.98) +- (0, 0.00) 
					(15, 15.00) +- (0, 0.00)
				};
				\addlegendentry{$|\Gamma|=2$}
				\addplot+[error bars/.cd, y dir=both, y explicit] coordinates {
					(4, 1.66) +- (0, 0.00) 
					(5, 2.10) +- (0, 0.01) 
					(6, 4.22) +- (0, 1.51) 
					(7, 2.99) +- (0, 0.00)
				};
				\addlegendentry{$|\Gamma|=3$}
				\addplot+[error bars/.cd, y dir=both, y explicit] coordinates {
					(4, 2.69) +- (0, 0.86) 
					(5, 2.96) +- (0, 1.06) 
					(6, 5.69) +- (0, 0.00)
				};
				\addlegendentry{$|\Gamma|=4$}
			\end{axis}
		\end{tikzpicture}
		\caption{Average time of LB-Explorer for each seed.}
		\label{fig:timerun}
\end{figure}

LB-Explorer has been running for 10 million episodes on five different seeds\footnote{All training was performed on \href{https://www.runpod.io/}{RunPod} community pods. Each LB-Explorer instance can run on a single RTX $4090$ ($24$~GB) at the default batch configuration ($B=8192$, $B_\text{mini}=1024$). Each pod typically runs two to six GPU in parallel and uses at least $8$ vCPUs. Across all experiments reported in this work, the total compute footprint is approximately $1000$ GPU-hours.} for the CICYs in Table \ref{tab:CICYdataset}, with the purpose of searching for solutions that satisfy the conditions in Table \ref{tab:conditions}. How long the run lasts depends on the CICY, as we show in Figure \ref{fig:timerun}. Up to $h^{1,1}(\IX_3)= 9$, the average time of a run is between 2 to 6 hours. The increase in time for larger CICYs is mainly due to time needed to save the found solutions and for the repetition buffer.

The results of the training are listed in Table \ref{tab:detailed_counts}, as we will explain in the following. Each row of Table \ref{tab:detailed_counts} contains the counts of solutions for a given CICY and the associated freely-acting symmetry $\Gamma$. We report, as a reference for the reader, the $h^{1,1}(\IX_3)$ and the order of the symmetry group $G_{\IX_3}$ for each CICY, from Table \ref{tab:CICYdataset}. The solutions are listed in four sets: $N$ counts the solutions found during the exploration. $N_{S_5}$ is the set of the $N$ solutions that are unique under the relabeling of the line bundles, while $N_{G_{\IX_3}}$ is the set of the $N$ solutions that are unique up to the action of $G_{\IX_3}$ for a given CICY. Finally, $N_{\text{full}}$ is the set of solutions that are unique under both of the previous symmetries. 

We report those numbers by seeds, meaning that they are the solutions found by those five initializing seeds for the Transformers. While, in general, the reason for choosing different seeds is to ensure the exploration of different regions of the solution space, it is possible for different seeds to find the same solution, and we report the number of solutions found by at least two seeds in the `Dup.' column. The column `Total' reports the total number of solutions summed across the seeds, where the duplicated solutions are counted only once. This is the reason why, when summing the solutions per seed, we generally get a larger total number of solutions: the duplicated solutions across seeds are included in the counting of the solutions for single seeds, but they are counted only once when summed in the total. In order to see the percentage of solutions uniquely found by each seed, we refer to Figure \ref{fig:percentage_solutions}.

Finally, in `Equivariance', we show the vector bundles that admit an equivariant structure. Equivariance is not imposed at the level of LB-Explorer and we have filtered the solutions in post-processing. If the number in the column is followed by a \checkmark, it means that the action of $\Gamma$ we considered for that CICY acts trivially on the basis of divisors $J_i$, therefore we imposed only \eqref{eq:equivariantcond_trivial}. Otherwise, the solutions have been filtered imposing also \cref{eq:equivariantcond_nontrivial}.

\subsubsection{Learning Pattern}

\begin{figure}[!htp]
    \centering
    \caption{Training metrics for three examples where learning clearly progresses in stages.}
    \label{fig:staged_learning_examples}
    
    \begin{subfigure}[b]{0.49\textwidth}
        \centering
        \begin{tikzpicture}[font=\footnotesize,baseline=0]
            \node (plot) {\includegraphics[width=0.9\textwidth,keepaspectratio]{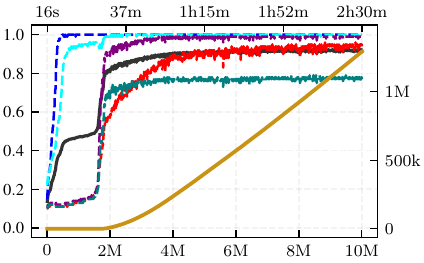}};
            \node (xaxis) at (plot.south) [below=0mm] {Episode};
            \node (yaxis) at (plot.west) [left=-2mm, rotate=90, anchor=south] {Success Rate};
            \node (xtop) at (plot.north) [above=0mm] {Wall time};
        \end{tikzpicture}
        \caption{CICY 3413, $|\Gamma|=3$, Seed 42}
        \label{sfig:plateau_log}
    \end{subfigure}\hfill
    \begin{subfigure}[b]{0.49\textwidth}
        \centering
        \begin{tikzpicture}[font=\footnotesize,baseline=0]
            \node (plot) {\includegraphics[width=0.9\textwidth,keepaspectratio]{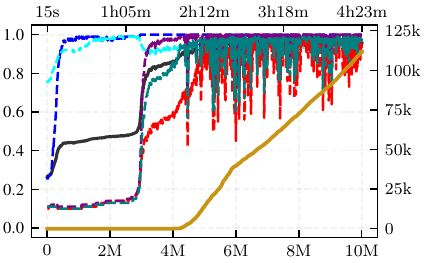}};
            \node (xaxis) at (plot.south) [below=0mm] {Episode};
            \node (yright) at (plot.east) [right=3mm, rotate=90, anchor=south] {Solutions};
            \node (xtop) at (plot.north) [above=0mm] {Wall time};
        \end{tikzpicture}
        \caption{CICY 2639, $|\Gamma|=2$, Seed 45}
        \label{sfig:abrupt_log}
    \end{subfigure}
    
    \vspace{1em}
    
    \begin{subfigure}[b]{0.49\textwidth}
        \centering

        \begin{tikzpicture}[font=\footnotesize,baseline=0]
            \node (plot) {\includegraphics[width=0.9\textwidth,keepaspectratio]{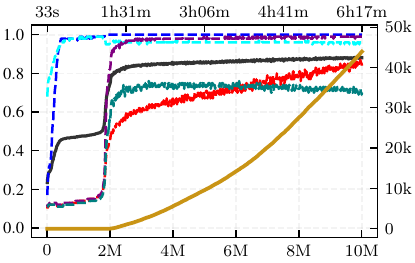}};
            \node (xaxis) at (plot.south) [below=0mm] {Episode};
            \node (yaxis) at (plot.west) [left=0mm, rotate=90, anchor=south] {Success Rate};
            \node (yright) at (plot.east) [right=3mm, rotate=90, anchor=south] {Solutions};
            \node (xtop) at (plot.north) [above=0mm] {Wall time};
        \end{tikzpicture}
        \caption{CICY 2544, $|\Gamma|=2$, Seed 42}
        \label{sfig:hierarchical_log}
    \end{subfigure}\hfill
    \begin{minipage}[b]{0.49\textwidth}
        \centering
        \begin{tikzpicture}
         \definecolor{cReward}{HTML}{000000}
\definecolor{cAnom}{HTML}{0000FF}
\definecolor{cStab}{HTML}{00FFFF}
\definecolor{cSum}{HTML}{FF0000}
\definecolor{cRng}{HTML}{800080}
\definecolor{cPair}{HTML}{008080}
\definecolor{cFound}{HTML}{C99415}
            \matrix [
                draw, 
                fill=white, 
                inner sep=3pt, 
                nodes={inner sep=3pt, anchor=west},
                column 1/.style={nodes={anchor=center}},
                font=\small,
                row sep=-3pt
            ] (legend) {
               \draw[cReward, line width=1pt] (0,0) -- (0.3,0); & \node {Total reward}; \\
    \draw[cAnom, line width=1pt, dashed] (0,0) -- (0.3,0); & \node {Anomaly cancellation}; \\
    \draw[cStab, line width=1pt, dashed] (0,0) -- (0.3,0); & \node {Poly-stability}; \\
    \draw[cSum, line width=1pt, dashed] (0,0) -- (0.3,0); & \node {Chirality}; \\
    \draw[cRng, line width=1pt, dashed] (0,0) -- (0.3,0); & \node {$-3|\Gamma|\leq\text{ind}(L_a)\leq 0$}; \\
    \draw[cPair, line width=1pt, dashed] (0,0) -- (0.3,0); & \node {$-3|\Gamma|\leq\text{ind}(L_a\oplus L_b)\leq 0$}; \\
    \draw[cFound, line width=1pt] (0,0) -- (0.3,0); & \node {Solutions}; \\
            };
        \end{tikzpicture}
        \vspace{1.75cm} 
    \end{minipage}
\end{figure}
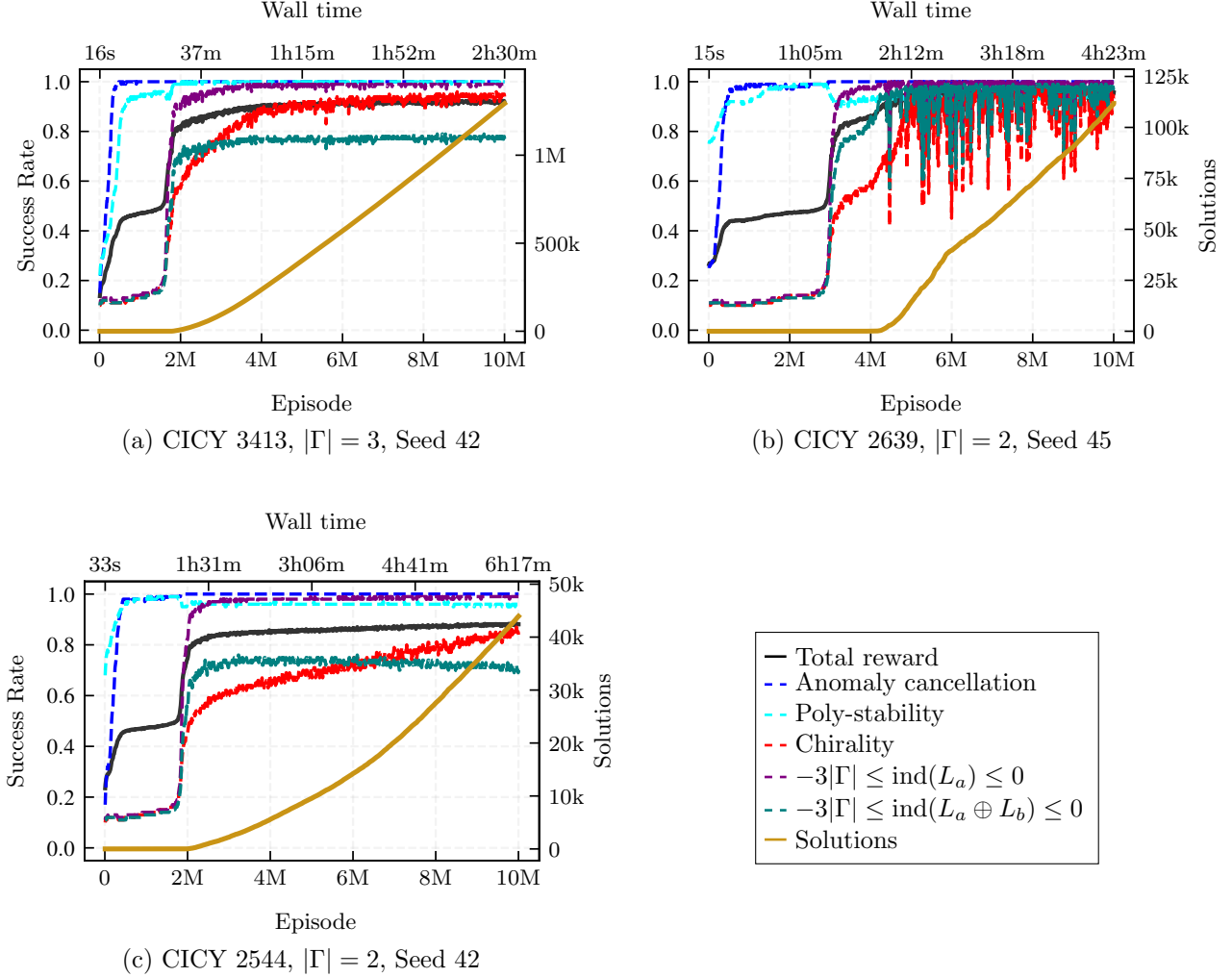

We noticed that the conditions in Table \ref{tab:conditions} are learned in stages by the LB-Explorer, as we show, e.g., in Figure \ref{fig:staged_learning_examples}. The plots represent the success rate of satisfying each of the conditions in Table \ref{tab:conditions}.\footnote{Recall that the $E_8$ embedding is imposed by construction on the LB-Explorer.} Clearly, the first conditions that the LB-Explorer learns are anomaly cancellation and poly-stability. This is interesting because, as we discussed in Section \ref{sec:nn_architecture}, the weights of each reward for each condition are the same. This pattern of learning has also been observed in \cite{Loges:2021hvn}, which focused on the search for realistic D-brane models.
However, these two conditions are not enough to guarantee the finding of solutions satisfying all the conditions in Table \ref{tab:conditions}. In fact, the solutions are found roughly when the chirality condition starts to be rewarded. The absence of exotic representations is then learned as a fine-tuning of the solutions satisfying the chirality condition, and all three guide the LB-Explorer to accumulate valid solutions. However, as we will see in the next section, it is not guaranteed that the solutions found are unrelated by permutations of columns or by relabeling of the divisor basis,\footnote{Similar permutation symmetries and relabeling of the basis of cycles play a role in the precise counting of the number of gauge-inequivalent D-brane configurations \cite{Loges:2022mao}.} but those redundancies can be used to determine, in real time, the health of the exploration.

\subsubsection{Comments on Exploration Status}

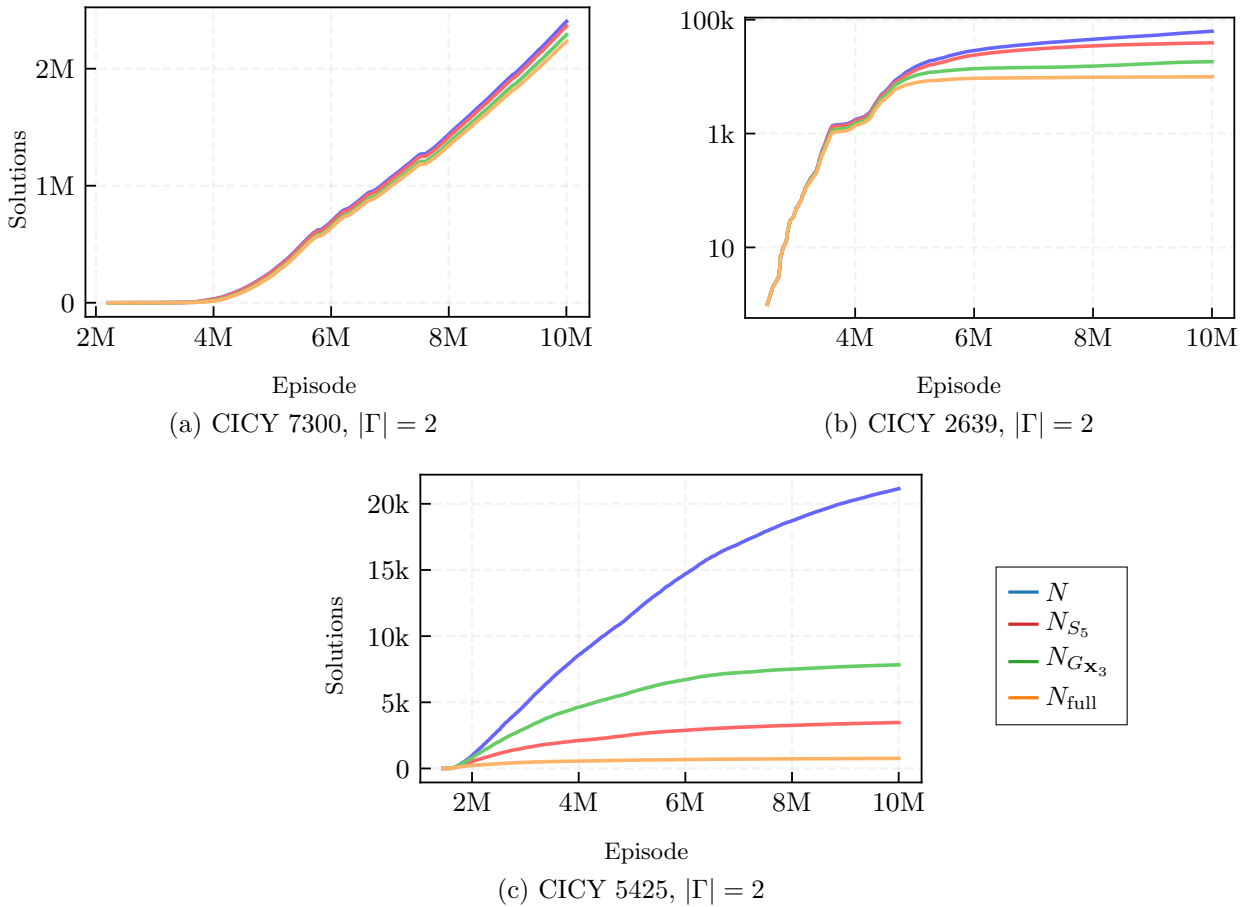
\begin{figure}[!htp]
        \centering
\caption{We show the accumulation of solutions for three examples of explorations. Figure \ref{sfig:exploring} after 10 million episodes is still finding solutions, Figure \ref{sfig:plataued} is plateauing, while Figure \ref{sfig:exhausted} looks like that is finding new solutions but they mostly are redundant under the $S_5$ and $G_{\IX_3}$ of the CICY. }\label{fig:exploration_examples}
     \begin{subfigure}[b]{0.49\textwidth}
    \centering
    \begin{tikzpicture}[font=\footnotesize,baseline=0]
        \node (plot) {\includegraphics[width=0.9\textwidth,keepaspectratio]{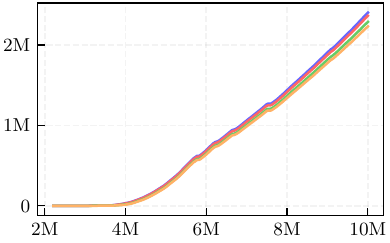}};
        \node (xaxis) at (plot.south) [below=0mm] {Episode};
    \node (yaxis) at (plot.west) [left=-2mm, rotate=90, anchor=south] {Solutions};
    \end{tikzpicture}
    \caption{CICY 7300, $|\Gamma|=2$}
    \label{sfig:exploring}
     \end{subfigure}\hfill
     \begin{subfigure}[b]{0.49\textwidth}
    \centering
     \begin{tikzpicture}[font=\footnotesize,baseline=0]
        \node (plot) {\includegraphics[width=0.9\textwidth,keepaspectratio]{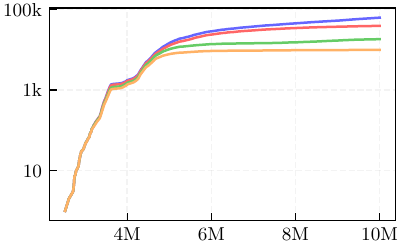}};
    \node (xaxis) at (plot.south) [below=0mm] {Episode};
    \end{tikzpicture}
    \caption{CICY 2639, $|\Gamma|=2$}
    \label{sfig:plataued}
     \end{subfigure}

     \begin{subfigure}[b]{0.49\textwidth}
    \centering
     \begin{tikzpicture}[font=\footnotesize,baseline=0]
        \node (plot) {\includegraphics[width=0.9\textwidth,keepaspectratio]{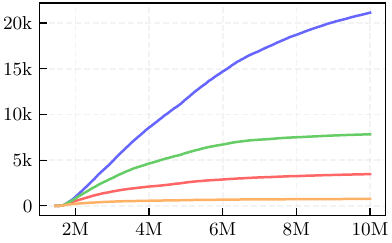}};
        \node (xaxis) at (plot.south) [below=0mm] {Episode};
    \node (yaxis) at (plot.west) [left=0mm, rotate=90, anchor=south] {Solutions};
        \node (legend) [right=6mm of plot] {
        \begin{tikzpicture}
        \definecolor{C0}{HTML}{1f77b4}
\definecolor{C1}{HTML}{ff7f0e}
\definecolor{C2}{HTML}{2ca02c}
\definecolor{C3}{HTML}{d62728}
    \matrix [
        draw, 
        fill=white, 
        inner sep=4pt, 
        nodes={inner sep=2pt, anchor=west},
        column 1/.style={nodes={anchor=center}},
        font=\small,
        row sep=0pt
    ] (legend) at (0,0) { 
        \draw[C0, line width=1.2pt] (0,0) -- (0.4,0); & \node {$N$}; \\
        \draw[C3, line width=1.2pt] (0,0) -- (0.4,0);    & \node {$N_{S_5}$}; \\
        \draw[C2, line width=1.2pt] (0,0) -- (0.4,0);  & \node {$N_{G_{\mathbf{X}_3}}$}; \\
        \draw[C1, line width=1.2pt] (0,0) -- (0.4,0);  & \node {$N_{\text{full}}$}; \\
    };
\end{tikzpicture}};
    \end{tikzpicture}
    \caption{CICY 5425, $|\Gamma|=2$}
    \label{sfig:exhausted}
     \end{subfigure}
\end{figure}

From Table \ref{tab:detailed_counts}, it is unclear whether the exploration of LB-Explorer was exhaustive or if a longer exploration would have led to more solutions. In fact, as we show in Figure \ref{fig:exploration_examples}, there can be three possible scenarios. In Figure \ref{sfig:exploring}, associated to CICY 7300 and $|\Gamma|=2$, it was shown that LB-Explorer was consistently finding new solutions, mostly not related by any symmetry of the CICY or of the line bundles sum $V$, meaning that a longer exploration would have increased the number of solutions. On the other hand, Figure \ref{sfig:plataued} shows that for CICY 2639 and $|\Gamma|=2$, a slowdown in the accumulation of solutions was starting, and more solutions were found; however, these were related by permutations of line bundles or divisors. This is even clearer from Figure \ref{sfig:exhausted} for CICY 5425 and $|\Gamma|=2$, where the number of solutions appears to be increasing, but they are mostly related by permutations of line bundles, drastically reducing the effective number of solutions.\footnote{This fact is also related to the observation that multiple seeds found the same solutions, as shown in Table \ref{tab:detailed_counts}.}

We believe that to demonstrate the power of LB-Explorer, the democratic choice of fixing the number of Episodes to 10 million and the number of seeds to five is enough. In order to find all solutions for a given CY, one can leverage either an increase in the length of exploration or an increase in the number of seeds. In fact, we decided to pick two CYs in Table \ref{tab:detailed_counts} to show that increasing the number of steps or changing the seeds helps in finding more solutions. 

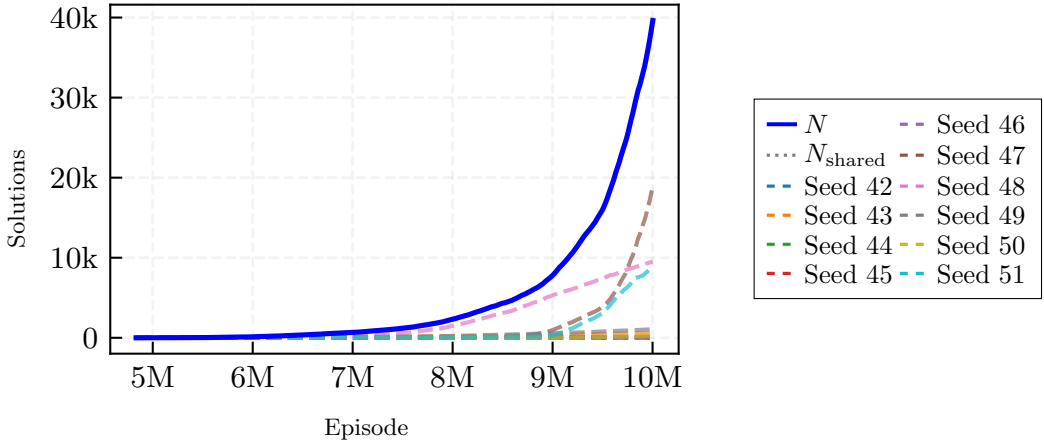
\begin{figure}[!htp]
    \centering
 \caption{Number of total solutions found for CICY 6 over the ten seeds considered.}
    \label{fig:total_sol_cy6}
    \begin{tikzpicture}[font=\footnotesize,baseline=0]
        \node (plot) {\includegraphics[width=0.5\textwidth,keepaspectratio]{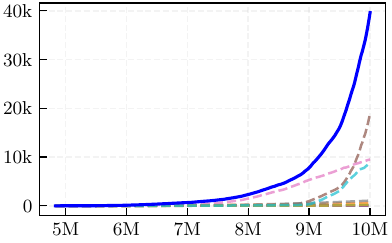}};
        \node (xaxis) at (plot.south) [below=0mm] {Episode};
    \node (yaxis) at (plot.west) [left=0mm, rotate=90, anchor=south] {Solutions};
        \node (legend) [right=6mm of plot] {\begin{tikzpicture}
    \definecolor{seed42}{HTML}{1F77B4}
    \definecolor{seed43}{HTML}{FF7F0E}
    \definecolor{seed44}{HTML}{2CA02C}
    \definecolor{seed45}{HTML}{D62728}
    \definecolor{seed46}{HTML}{9467BD}
    \definecolor{seed47}{HTML}{8C564B}
    \definecolor{seed48}{HTML}{E377C2}
    \definecolor{seed49}{HTML}{7F7F7F}
    \definecolor{seed50}{HTML}{BCBD22}
    \definecolor{seed51}{HTML}{17BECF}
    
    \matrix [
        draw, 
        fill=white, 
        inner sep=4pt, 
        nodes={inner sep=2pt, anchor=west},
        column 1/.style={nodes={anchor=center}},
        column 3/.style={nodes={anchor=center}},
        font=\small,
        row sep=0pt
    ] (legend) at (0,0) { 
        \draw[blue, line width=1.5pt] (0,0) -- (0.4,0); & \node {$N$}; &
        \draw[seed46, line width=1.2pt, dashed] (0,0) -- (0.4,0); & \node {Seed 46}; \\
        
        \draw[gray, line width=1.2pt, dotted] (0,0) -- (0.4,0); & \node {$N_{\text{shared}}$}; &
        \draw[seed47, line width=1.2pt, dashed] (0,0) -- (0.4,0); & \node {Seed 47}; \\
        
        \draw[seed42, line width=1.2pt, dashed] (0,0) -- (0.4,0); & \node {Seed 42}; &
        \draw[seed48, line width=1.2pt, dashed] (0,0) -- (0.4,0); & \node {Seed 48}; \\
        
        \draw[seed43, line width=1.2pt, dashed] (0,0) -- (0.4,0); & \node {Seed 43}; &
        \draw[seed49, line width=1.2pt, dashed] (0,0) -- (0.4,0); & \node {Seed 49}; \\
        
        \draw[seed44, line width=1.2pt, dashed] (0,0) -- (0.4,0); & \node {Seed 44}; &
        \draw[seed50, line width=1.2pt, dashed] (0,0) -- (0.4,0); & \node {Seed 50}; \\
        
        \draw[seed45, line width=1.2pt, dashed] (0,0) -- (0.4,0); & \node {Seed 45}; &
        \draw[seed51, line width=1.2pt, dashed] (0,0) -- (0.4,0); & \node {Seed 51}; \\
    };
\end{tikzpicture}
};
    \end{tikzpicture}
\end{figure}

For instance, CICY 6, despite having $h^{1,1}=15$, found only 846 solutions across the five seeds we considered in Table \ref{tab:detailed_counts}. Hence, we trained the LB-Explorer on seeds from 47 to 51 for 10 million episodes, finding the solutions we report in Table \ref{tab:detailed_counts_cy6} or shown in Figure \ref{fig:total_sol_cy6}. We believe that the increase of two orders of magnitude in the number of solutions for this choice of seeds is a good indicator that, in order to perform an exhaustive search of all possible solutions, one needs to run the LB-Explorer over multiple seeds. We also believe that the larger $h^{1,1}(\IX_3)$ is, the less overlap there is in the solutions found among the seeds.

\setlength{\LTleft}{\fill}
\setlength{\LTright}{\fill}
\setlength{\LTcapwidth}{\linewidth}
\renewcommand{\arraystretch}{1}
\begin{longtable}{c | c | c |  c | c | c | c | c | c | c}
\caption{Solutions for CICY 6 over five more seeds than those considered in Table \ref{tab:detailed_counts}. The meaning of each columns is the same as in Table \ref{tab:detailed_counts}. The total number of solutions, and the equivariant solutions are also considering those in Table \ref{tab:detailed_counts}.} 
\label{tab:detailed_counts_cy6} \\
\hhline{=|=|=====|=|=|=}
\multirow{2}{*}{\#} & \multirow{2}{*}{Sol.} & \multicolumn{5}{|c|}{Seeds} & \multirow{2}{*}{Dup. $\left(\frac{N_\text{dup}}{N_\text{tot}}\%\right)$} & \multirow{2}{*}{Total} & \multirow{2}{*}{Equivariance} \\ \cline{3-7}
 &  & 47 & 48 & 49 & 50 & 51 &  &  & \\
\hhline{=|=|=|=|=|=|=|=|=|=}
\endfirsthead
\multicolumn{10}{c}{{\tablename\ \thetable\ -- \textit{Continued from previous page}}} \\
\hhline{=|=|=====|=|=|=}
\multirow{2}{*}{\#} & \multirow{2}{*}{Sol.} & \multicolumn{5}{|c|}{Seeds} & \multirow{2}{*}{Dup. $\left(\frac{N_\text{dup}}{N_\text{tot}}\%\right)$} & \multirow{2}{*}{Total} & \multirow{2}{*}{Equivariance} \\ \cline{3-7}
 &  & 47 & 48 & 49 & 50 & 51 &  &  & \\
\hhline{=|=|=|=|=|=|=|=|=|=}
\endhead
\multicolumn{10}{r}{\textit{Continued on next page}} \\
\endfoot
\hhline{=|=|=|=|=|=|=|=|=|=}
\endlastfoot
\multirow{4}{*}{6} & $N$ & 18943 & 9526 & 1070 & 208 & 9035 & 0 (0.0\%) & 39628 & 5166 \checkmark \\
 &  $N_{S_5}$ &  18849 & 8682 & 1045 & 199 & 8289 & 1 (0.003\%) & 37891 & 4941 \checkmark \\
 &  $N_{G_{\mathbf{X}_3}}$ &  18943 & 9525 & 1070 & 206 & 9034 & 0 (0.0\%) & 39623 & 5164 \checkmark \\
 &  $N_{\text{full}}$ &  18849 & 8680 & 1045 & 199 & 8284 & 2 (0.005\%) & 37882 & 4940 \checkmark \\ \hline
\end{longtable}

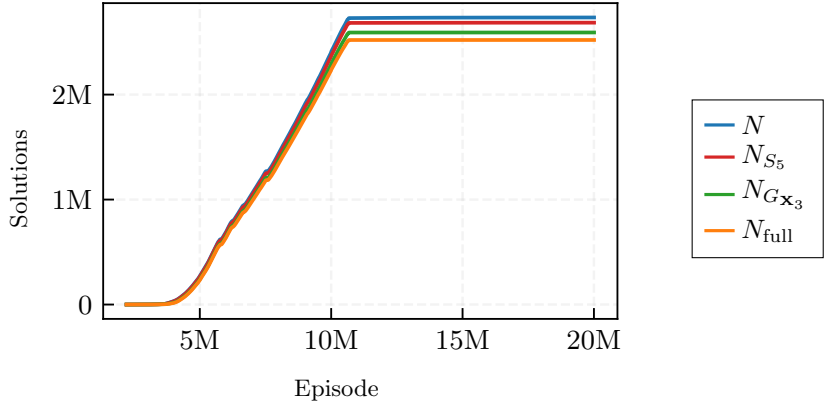
\begin{figure}[!htp]
    \centering
 \caption{Exploration status of CICY 7300 after 20 million episodes.}
    \label{fig:exploration_cy7300}
    \begin{tikzpicture}[font=\footnotesize,baseline=0]
        \node (plot) {\includegraphics[width=0.45\textwidth,keepaspectratio]{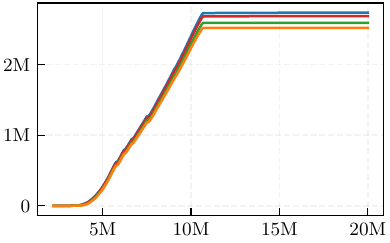}};
        \node (xaxis) at (plot.south) [below=0mm] {Episode};
    \node (yaxis) at (plot.west) [left=0mm, rotate=90, anchor=south] {Solutions};
        \node (legend) [right=6mm of plot] {\begin{tikzpicture}
        \definecolor{C0}{HTML}{1f77b4}
\definecolor{C1}{HTML}{ff7f0e}
\definecolor{C2}{HTML}{2ca02c}
\definecolor{C3}{HTML}{d62728}
    \matrix [
        draw, 
        fill=white, 
        inner sep=4pt, 
        nodes={inner sep=2pt, anchor=west},
        column 1/.style={nodes={anchor=center}},
        font=\small,
        row sep=0pt
    ] (legend) at (0,0) { 
        \draw[C0, line width=1.2pt] (0,0) -- (0.4,0); & \node {$N$}; \\
        \draw[C3, line width=1.2pt] (0,0) -- (0.4,0);    & \node {$N_{S_5}$}; \\
        \draw[C2, line width=1.2pt] (0,0) -- (0.4,0);  & \node {$N_{G_{\mathbf{X}_3}}$}; \\
        \draw[C1, line width=1.2pt] (0,0) -- (0.4,0);  & \node {$N_{\text{full}}$}; \\
    };
\end{tikzpicture}};
    \end{tikzpicture}
\end{figure}

To prove instead that allowing us to explore the LB-Explorer for more episodes is sometimes a way to find more solutions, we decided to run the LB-Explorer over CICY 7300 shown in \ref{sfig:exploring} for 10 million additional episodes. The results are in Table \ref{tab:detailed_counts_cy7300}. We observe roughly an increase of 14\% in the number of solutions for each seed, with the LB-Explorer finally plateauing, as we expected. We show this in Figure \ref{fig:exploration_cy7300}. Noting the small overlap in the solutions found, we believe that running the LB-Explorer on different seeds could also improve the total number of solutions that can be obtained.

\setlength{\LTleft}{\fill}
\setlength{\LTright}{\fill}
\setlength{\LTcapwidth}{\linewidth}
\renewcommand{\arraystretch}{1}
\begin{longtable}{c | c | c |  c | c | c | c | c | c | c}
\caption{Total number of solutions for CICY 7300 continuing the exploration that produced the solutions in Table \ref{tab:detailed_counts} for 10 million more episodes. The number of solutions is the one accumulated over the whole 20 million episodes.} 
\label{tab:detailed_counts_cy7300} \\
\hhline{=|=|=====|=|=|=}
\multirow{2}{*}{\#} & \multirow{2}{*}{Sol.} & \multicolumn{5}{|c|}{Seeds} & \multirow{2}{*}{Dup. $\left(\frac{N_\text{dup}}{N_\text{tot}}\%\right)$} & \multirow{2}{*}{Total} & \multirow{2}{*}{Equivariance} \\ \cline{3-7}
 &  & 42 & 43 & 44 & 45 & 46 &  &  & \\
\hhline{=|=|=|=|=|=|=|=|=|=}
\endfirsthead
\multicolumn{10}{c}{{\tablename\ \thetable\ -- \textit{Continued from previous page}}} \\
\hhline{=|=|=====|=|=|=}
\multirow{2}{*}{\#} & \multirow{2}{*}{Sol.} & \multicolumn{5}{|c|}{Seeds} & \multirow{2}{*}{Dup. $\left(\frac{N_\text{dup}}{N_\text{tot}}\%\right)$} & \multirow{2}{*}{Total} & \multirow{2}{*}{Equivariance} \\ \cline{3-7}
 &  & 42 & 43 & 44 & 45 & 46 &  &  & \\
\hhline{=|=|=|=|=|=|=|=|=|=}
\endhead
\multicolumn{10}{r}{\textit{Continued on next page}} \\
\endfoot
\hhline{=|=|=|=|=|=|=|=|=|=}
\endlastfoot
\multirow{4}{*}{7300} & $N$ & 2474 & 3805 & 4797 & 2682954 & 40604 & 3 (0.000\%) & 2734631 & 427908 \checkmark \\
 & $N_{S_5}$ & 2454 & 3794 & 3351 & 2655028 & 20581 & 21 (0.001\%) & 2685187 & 423916 \checkmark \\
 & $N_{G_{\mathbf{X}_3}}$ & 325 & 265 & 1489 & 2588140 & 1357 & 28 (0.001\%) & 2591548 & 407864 \checkmark \\
 & $N_{\text{full}}$ & 306 & 240 & 683 & 2518309 & 768 & 67 (0.003\%) & 2520239 & 396819 \checkmark \\ \hline
\end{longtable}

\subsection{Transfer Learning}
\label{sec:transferlearning}

In this section, we are going to discuss the experiments on transfer learning that we performed. Since we are not aiming to be exhaustive, we decided to use pre-trained NN over other CICYs, running the LB-Explorer for 5 million episodes over three different seeds. The transfer learning we tested is ``within bucket", i.e., CICYs with the same $h^{1,1}(\IX_3)$ and $\Gamma$ but unrelated by any linear transformation on the second Chern class and triple intersection numbers. We also considered transfer learning ``across $\Gamma$", considering the same CICY but admitting multiple freely-acting symmetries. Finally, we explored transfer learning ``across $h^{1,1}(\IX_3)$", considering CICYs related by a conifold transition that increases/decreases $h^{1,1}(\IX_3)$ by 1. In order to quantify the transfer learning, we performed various tests that we are going to describe in Section \ref{sec:TL_methods}. Generally, we believe that LB-Explorer has proven to pass the tests we designed to quantify transfer learning for the various sets of theories we considered. All the results of the training and of the tests can be found in the following tables:
\begin{itemize}
    \item \cref{tab:detailed_counts_within_bucket,tab:efficacy_within_bucket,tab:extra_tests_within_bucket,tab:forgetting_within_bucket} contain the detailed counting of the solutions found and the results of the tests of transfer learning within the bucket.
    \item \cref{tab:detailed_counts_cross_gamma,tab:efficacy_cross_gamma,tab:extra_tests_cross_gamma,tab:forgetting_cross_gamma} contain the detailed counting of the solutions found and the results of the tests of transfer learning across $\Gamma$.
    \item \cref{tab:detailed_counts_cross_h11,tab:efficacy_cross_h11,tab:extra_tests_cross_h11,tab:forgetting_cross_h11} contain the detailed counting of the solutions found and the results of the tests of transfer learning across $h^{1,1}(\IX_3)$.
\end{itemize}
Moreover, in \cref{sec:TL_withinbucket,sec:TL_acrossgamma,sec:TL_acrossh11}, we will consider representative examples for a more immediate visualization of the results of the transfer learning tests we have performed. 

\subsubsection{Transfer Learning Methods}
\label{sec:TL_methods}

To test transfer learning in our LB-Explorer model, we decided to perform some evaluations. At the level of macroscopic measurements, we considered two metrics:
\begin{itemize}
    \item \textit{Zero-shot Reward:} The reward obtained by the pre-trained source policy $\pi_{\text{src}}$ when evaluated on the new target geometry at episode $E=0$, before any parameter updates occur. This measures the extent to which the model has generalized the conditions in Table \ref{tab:conditions} independently of the specific CICY.
    \item \textit{Episodes to First Solution:} As the name suggests, it is the episode at which the LB-Explorer has found the first line bundle sum $V$ that satisfies all the conditions in Table \ref{tab:conditions}. This serves as an indication that the transfer policy requires fewer steps to identify a valid solution.
\end{itemize}
Besides these empirical tests, we wanted to test how the NN on the target CICY adapted when input weights were determined by training on a different source CICY. We then performed five additional tests: 
\begin{itemize}
   \item \textit{Layer-wise Weight Distance:} To quantify the extent of parameter restructuring during transfer learning, we compute the mean cosine distance between the parameter tensors of the source and target networks at each layer. For a given transformer decoder layer $L$, the distance is defined as:
    \begin{equation}
        d_L = \frac{1}{|W_L|} \sum_{w \in W_L} \left( 1 - \frac{w_{\text{trg}} \cdot w_{\text{src}}}{\|w_{\text{trg}}\| \|w_{\text{src}}\|} \right)\coma
    \end{equation}
    where $w_{\text{trg}}$ and $w_{\text{src}}$ represent the flattened corresponding parameter tensors (e.g., attention weights and feed-forward matrices) within the parameter set $W_L$. The use of cosine distance measures directional parameter shifts while remaining invariant to scalar magnitude scaling. 

    This metric evaluates whether the network required a complete structural overhaul or simply adjusted terminal action probability distributions. Specifically, it measures how much foundational representation is preserved in early layers versus how much later policy layers adapt to the new target CICY. A uniformly high distance across all layers implies the model destructively restructured, effectively relearning constraints from scratch. Conversely, near-zero distances across all layers carry a dual interpretation depending on the task performance. If transfer performance is poor, it suggests the network failed to escape the local minimum of the source geometry, struggling to adapt to the target parameters. However, if transfer performance is high, near-zero distances indicate highly successful, near-zero-shot transfer, where the features learned on the source CICY were robust and directly applicable to the target CICY without requiring meaningful weight updates. Because this direct comparison does not account for internal network permutation symmetries, it serves as a diagnostic heuristic rather than a standalone proof of successful transfer.
    \item \textit{Representation Alignment (Kernel CKA):} To test if the target model retains the internal logic of the source model, we compare their hidden layer activations. We evaluate this by feeding both networks a batch of $n$ valid line bundle sums $V$ and measuring the similarity of their internal responses. The alignment is initially computed using a Radial Basis Function (RBF) kernel, which measures the similarity between two activation vectors $x_i$ and $x_j$:
\begin{equation}
k(x_i, x_j) = \exp\left(-\frac{\|x_i - x_j\|^2}{2\sigma^2}\right)\coma
\end{equation}
where the bandwidth parameter $\sigma^2$ is set via the median heuristic. This means $\sigma^2$ is chosen as the median of all squared distances ($\|x_i - x_j\|^2$) between all pairs of activation vectors in the current batch. Let $K$ and $L$ denote these kernel matrices derived from a specific source layer and target layer, respectively.\footnote{It is clear that these kernels are sensitive to the batch size and the dynamic bandwidth estimation, which introduces variance in alignment scores across different sampled subsets of valid states.} Before comparing $K$ and $L$, we must remove statistical bias by centering them, which maps their feature means to zero. The centered kernel matrix $K_c$ is obtained via:
\begin{equation}
K_c = H K H \quad \text{with} \quad H = \ID - \frac{1}{n} \mathbf{1} \mathbf{1}^T\coma
\end{equation}
where $H$ is the centering matrix, $\ID$ is the $n \times n$ identity matrix, and $\mathbf{1}$ is an $n$-dimensional vector of ones. The target matrix $L_c$ is centered similarly. With the matrices centered, we measure the statistical dependence between the two layers using the Hilbert-Schmidt Independence Criterion (HSIC):
\begin{equation}
\text{HSIC}(K, L) = \frac{1}{(n-1)^2} \text{Tr}(K_c L_c)\fstop
\end{equation}
Finally, the Centered Kernel Alignment (CKA) normalizes this overlap to a value between $0$ and $1$:
\begin{equation}
\text{CKA}(K, L) = \frac{\text{HSIC}(K, L)}{\sqrt{\text{HSIC}(K, K) \text{HSIC}(L, L)}}\fstop
\end{equation}

The resulting CKA scores are plotted as a heatmap comparing the source layers ($y$-axis) to the target layers ($x$-axis). A unadapted transfer has CKA scores near $1.0$ along the diagonal, indicating the NN preserved the source architecture and it did not adjust to the target geometry. Conversely, a successful transfer has high alignment in early layers that degrades and disperses asymmetrically off-diagonal in later layers, demonstrating that the network restructured specific internal logic to accommodate the conditions for the target CICY. If alignment scores are low and lack a coherent pattern, the model completely discarded the source representations and learned the target CICY from scratch.

\item \textit{Temporal Policy Divergence:} To evaluate the token-by-token behavioral shift of the autoregressive policy, we compute the step-wise Kullback--Leibler (KL) divergence between the transfer model ($\pi_{\theta_\text{trg}}$) and the corresponding model of the target geometry when trained from scratch ($\pi_{\theta_\text{scratch}}$). Measured over generation trajectories, the divergence at step $t$ is defined as
    \begin{equation}
        D_{\text{KL}}^{(t)} = \mathbb{E}_{\mathbf{s}_{<t} \sim \mathcal{D}_{\text{trg}}} \left[ \sum_{a \in \mathcal{A}} \pi_{\theta_\text{scratch}}(a \mid \mathbf{s}_{<t}) \log \frac{\pi_{\theta_\text{scratch}}(a \mid \mathbf{s}_{<t})}{\pi_{\theta_\text{trg}}(a \mid \mathbf{s}_{<t})} \right]\coma
    \end{equation}
    where $\mathbf{s}_{<t}$ denotes the partial state sequence up to step $t$, sampled from a dataset of known valid configurations of the line bundle sum $V$, denoted $\mathcal{D}_{\text{trg}}$, and $\mathcal{A}$ represents the vocabulary. A successful transfer strategy typically displays low divergence during the initial steps, namely indicating that the NN has learned universal heuristics for the conditions in Table \ref{tab:conditions}, with isolated spikes occurring when the model is forced to make decisions specific to the target CICY.
    \item \textit{Sample Efficiency:} Expanding upon the \textit{Episodes to First Solution} metric, we evaluate the number of episodes required for the LB-Explorer to discover $N$ unique solutions\footnote{For this analysis, we set $N=1000$.} during transfer learning, comparing it directly to how many episodes the same geometry needed when trained from scratch. This will be the most recurring metric in the following. While transfer learning typically accelerates the discovery of the initial solution, this metric tests the policy's sustained exploration capacity. Specifically, it verifies that the network is able to discover diverse configurations, rather than suffering from mode collapse and repeatedly generating duplicates of the initially discovered solutions.
    \item \textit{Catastrophic Forgetting:} This test assesses whether the NN has learned the general conditions to find the line bundle sums $V$, or if it has overwritten its weights to overfit the target CICY parameters. To quantify catastrophic forgetting, we compute two quantities: the Continuous Reward Score (CRS) and the Validation Success Rate (VSR). We consider the target policy $\pi_{\theta_\text{trg}}$ after training concludes and evaluate it on the source CICY $\mathcal{E}_\text{src}$. Considering the constraints $\mathcal{C}$ in Table \ref{tab:conditions}, we generate a sequence $\mathbf{s}\sim \pi_{\theta_\text{trg}}$ from the probability distribution defined by the target policy weights $\theta_\text{trg}$. CRS is defined as the expectation value of the weighted average of the constraint satisfaction scores:
\begin{equation}\label{eq:CRS}
    \text{CRS} = \mathbb{E}_{\mathbf{s}\sim \pi_{\theta_\text{trg}}}\left[ \frac{1}{\sum_{c\in \mathcal{C}} \alpha_c}\sum_{c\in \mathcal{C}} \alpha_c \left( \frac{1}{1 + w_c \ln(1 + \epsilon_c(\mathbf{s}|\mathcal{E}_\text{src}))} \right) \right]\coma
\end{equation}
where $\epsilon_c$ is the absolute error associated with constraint $c$, $w_c$ is its corresponding log-barrier penalty weight, and $\alpha_c$ is its linear combination coefficient for the final reward calculation. The expectation value is computed by averaging this reward across a batch of sequences generated by the target model. The VSR is defined as the probability that a generated sequence satisfies all constraints:
\begin{equation}\label{eq:VSR}
\text{VSR} = \mathbb{E}_{\mathbf{s} \sim \pi_{\theta_\text{trg}}} \left[ \prod_{c \in \mathcal{C}} \mathbb{I}(\epsilon_c(\mathbf{s} \mid \mathcal{E}_\text{src}) < \tau) \right]\coma
\end{equation}
where $\tau= 10^{-5}$ is the satisfaction threshold. The indicator function $\mathbb{I}$ yields $1$ (pass) if the error $\epsilon_c$ is below this tolerance, and $0$ (fail) otherwise. The product $\prod_{c \in \mathcal{C}}$ ensures a configuration is valid if and only if every condition is satisfied. Successful transfer learning is indicated by high CRS and $\text{VSR}$ when evaluating $\pi_{\theta_\text{trg}}$ on $\mathcal{E}_\text{src}$.
\end{itemize}

\subsubsection{Transfer Learning within Bucket}
\label{sec:TL_withinbucket}

In this section, we consider transfer learning for geometries with the same $h^{1,1}(\IX_3)$ and $|\Gamma|$, but not trivially related. This transfer learning analysis aims to understand the NN's ability to generalize among theories that are more or less symmetric under the $G_{\IX_3}$ group.

Looking at Table \ref{tab:efficacy_within_bucket}, we note an overall positive improvement in the \textit{Zero-shot Reward} evaluated at episode $E=0$, with the \textit{Episodes to First Solution} also dropping significantly compared to baseline models trained from scratch. As explained in Section \ref{sec:TL_methods}, these are the first signals that the NN did not simply overfit the source geometries, but that it has retained the fundamental information necessary to satisfy the conditions in Table \ref{tab:conditions}.

Despite reporting all the data in \cref{tab:efficacy_within_bucket,tab:extra_tests_within_bucket,tab:forgetting_within_bucket}, here we isolate representative CICYs that better pass the tests in Section \ref{sec:TL_methods}.

\begin{figure}[!htp]
    \centering
  \caption{Weight distance and CKA heatmap for the transfer learning analysis from CICY 7447 to CICY 5256 in Seed 43.}
    \label{fig:CICY5256seed43-LCKA}
     \begin{subfigure}[b]{0.49\textwidth}
     \centering
      \begin{tikzpicture}[font=\footnotesize,baseline=0]
        \node (plot) {\includegraphics[width=0.9\textwidth,keepaspectratio]{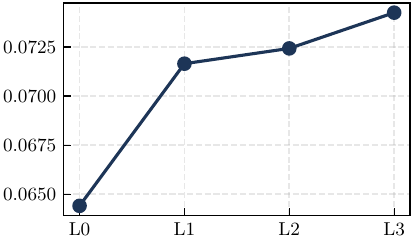}};
        \node (xaxis) at (plot.south) [below=0mm] {Layers};
    \node (yaxis) at (plot.west) [left=0mm, rotate=90, anchor=south] {Distance};
    \end{tikzpicture}
     \caption{Weight Distance}
     \label{sfig:CICY5256seed43-WD}
     \end{subfigure}\hfill
     \begin{subfigure}[b]{0.49\textwidth}
     \centering
     \begin{tikzpicture}[font=\footnotesize,baseline=0]
        \node (plot) {\includegraphics[width=0.9\textwidth,keepaspectratio]{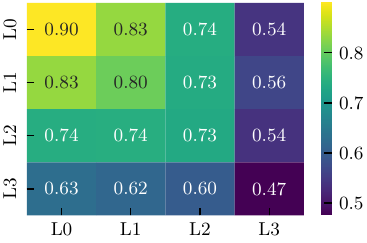}};
        \node (xaxis) at (plot.south) [below=0mm] {CICY 5256 Layers};
    \node (yaxis) at (plot.west) [left=0mm, rotate=90, anchor=south] {CICY 7447 Layers};
    \end{tikzpicture}
     \caption{CKA Heatmap}
     \label{sfig:CICY5256seed43-CKA}
     \end{subfigure}
\end{figure}
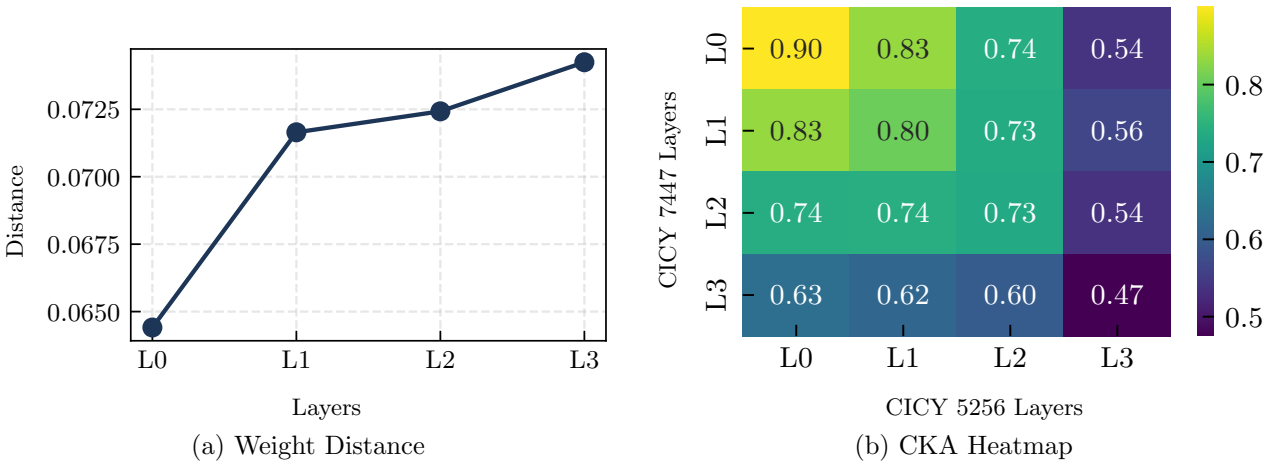

\paragraph{Best Transfer Learning Seed: CICY 7447 to CICY 5256 in Seed 43.}

We identify the transfer from source CICY 7447 to target CICY 5256 (Seed 43) as the most balanced across all transfer learning tests we performed. This does not mean it is the absolute extremum in any single metric, but it has been shown to perform better than other geometries on average across all tests. The Zero-shot Reward improved by $+177.8\%$ ($0.257$ to $0.714$), and the episodes required to find the first solution fell from $1343488$ to $8192$ ($+99.4\%$ improvement). 

However, evaluating its \textit{sample efficiency}, we find that, despite discovering the first solution rapidly, its exploration capacity is average since it found the first $1000$ unique solutions only $12.1\%$ faster ($5087232$ to $4472832$ episodes) than the training performed from scratch on that geometry. 

At the level of transformer layers, we can see from Figure \ref{fig:CICY5256seed43-LCKA} that the NN needs some restructuring, but the CKA exhibits the typical behavior of successful transfer, with degradation in the deeper layers while keeping a coherent pattern overall. This means that the layers are mapped from the source to the target CICY without any severe permutation of the NN's internal logic for learning the conditions in Table \ref{tab:conditions}, clearly showing the transfer of policies and not a training from scratch on the target CICY.

\paragraph{Average Best Transfer Learning: CICY 5967 to CICY 3413.}

\begin{figure}[!htp]
    \centering
  \caption{Sample Efficiency and Weight distance for the average transfer learning analysis from CICY 5967 to CICY 3413.}
    \label{fig:CICY3413mean-SEWD}
     \begin{subfigure}[b]{0.49\textwidth}
     \centering
      \begin{tikzpicture}[font=\footnotesize,baseline=0]
        \node (plot) {\includegraphics[width=0.9\textwidth,keepaspectratio]{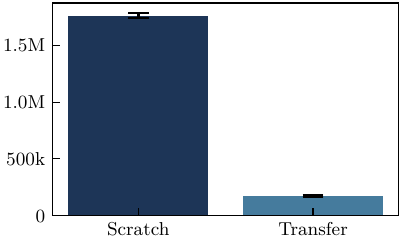}};
        \node (xaxis) at (plot.south) [below=0mm] {CICY 3413 training};
    \node (yaxis) at (plot.west) [left=0mm, rotate=90, anchor=south] {Episodes to Solution};
    \end{tikzpicture}
     \caption{Sample Efficiency}
     \label{sfig:CICY3413mean-SE}
     \end{subfigure}\hfill
     \begin{subfigure}[b]{0.49\textwidth}
     \centering
     \begin{tikzpicture}[font=\footnotesize,baseline=0]
        \node (plot) {\includegraphics[width=0.9\textwidth,keepaspectratio]{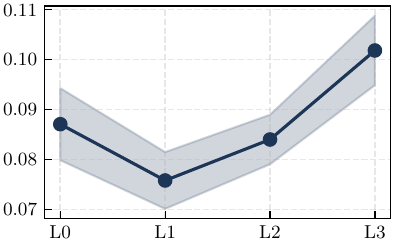}};
        \node (xaxis) at (plot.south) [below=0mm] {Layers};
    \node (yaxis) at (plot.west) [left=0mm, rotate=90, anchor=south] {Distance};
    \end{tikzpicture}
     \caption{Weight Distance}
     \label{sfig:CICY3413mean-WD}
     \end{subfigure}
\end{figure}
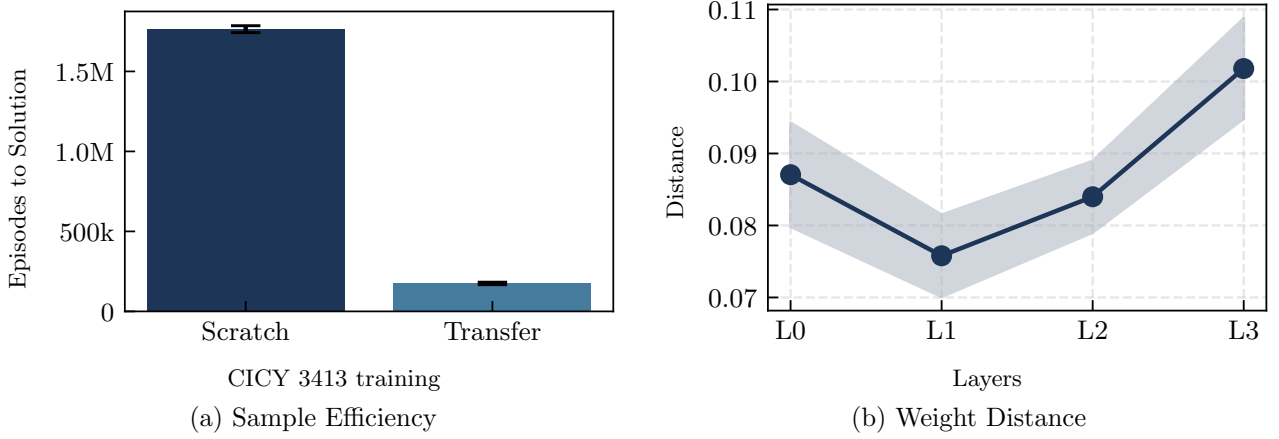

While CICY 5256 had the best performance on Seed 43, we also decided to examine which transfer learning performed best across the seeds. We find that the transfer learning from CICY 5967 to CICY 3413 had the most balanced transfer learning across all seeds. Macroscopically, its average Zero-Shot Reward improved by $+260.6\%$ ($0.155$ to $0.558$), and the average number of episodes to discover the first solution dropped from $1608362$ to $90112$ ($+94.4\%$ improvement). Furthermore, as we show in Figure \ref{fig:CICY3413mean-SEWD}, we find an $+86.4\%$ improvement in \textit{sample efficiency}, with only $245760$ episodes necessary to find $1000$ solutions versus the original $1807701$ ones. The transfer learning is also successful concerning the other metrics, as both the policy divergence (averaged to $ D_{\text{KL}} = 12.91$) and the weight distance (averaged to $d_L = 0.09$) are relatively low. 

\paragraph{Directional Asymmetries in Transfer Learning.}

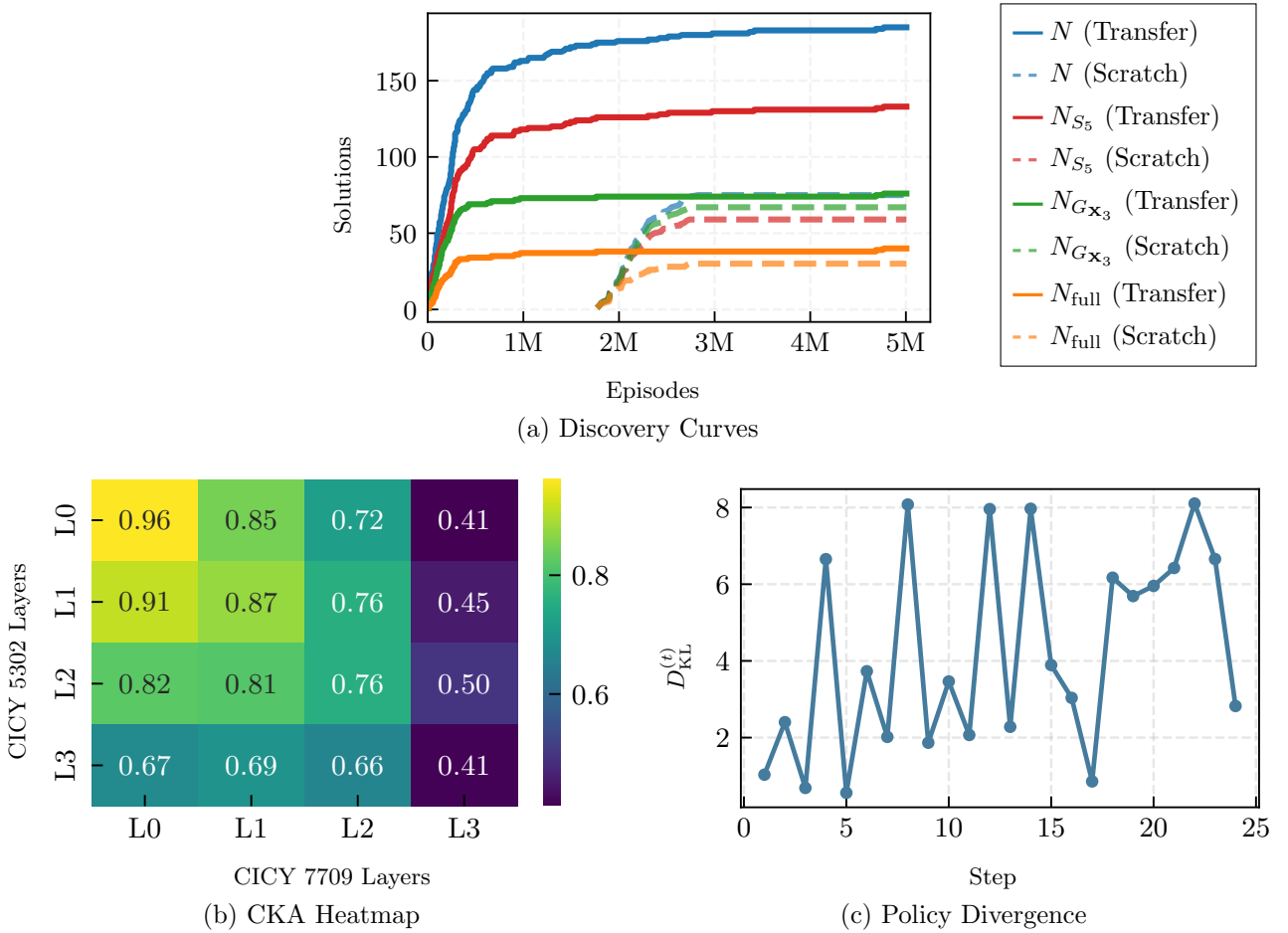
\begin{figure}[!htp]
    \centering
  \caption{Discovery Curve, CKA Heatmap and Policy Divergence for transfer learning analysis from CICY 5302 to CICY 7709 in Seed 42.}
    \label{fig:CICY7709seed42-DCPD}
     \begin{subfigure}[b]{0.49\textwidth}
     \centering
      \begin{tikzpicture}[font=\footnotesize,baseline=0]
        \node (plot) {\includegraphics[width=0.9\textwidth,keepaspectratio]{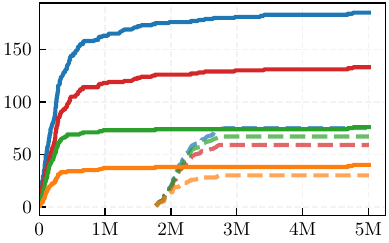}};
        \node (xaxis) at (plot.south) [below=0mm] {Episodes};
    \node (yaxis) at (plot.west) [left=0mm, rotate=90, anchor=south] {Solutions};
    \node (legend) [right=6mm of plot] {\begin{tikzpicture}
    \definecolor{C0}{HTML}{1f77b4}
\definecolor{C1}{HTML}{ff7f0e}
\definecolor{C2}{HTML}{2ca02c}
\definecolor{C3}{HTML}{d62728}
    \matrix [
        draw, 
        fill=white, 
        inner sep=4pt, 
        nodes={inner sep=2pt, anchor=west},
        column 1/.style={nodes={anchor=center}},
        font=\small,
        row sep=2pt
    ] (legend) at (0,0) {
        \draw[C0, line width=1.2pt] (0,0) -- (0.4,0);                 & \node {$N$ (Transfer)}; \\
        \draw[C0, line width=1.2pt, dashed, opacity=0.7] (0,0) -- (0.4,0); & \node {$N$ (Scratch)}; \\
        
        \draw[C3, line width=1.2pt] (0,0) -- (0.4,0);                  & \node {$N_{S_5}$ (Transfer)}; \\
        \draw[C3, line width=1.2pt, dashed, opacity=0.7] (0,0) -- (0.4,0);  & \node {$N_{S_5}$ (Scratch)}; \\
        
        \draw[C2, line width=1.2pt] (0,0) -- (0.4,0);                & \node {$N_{G_{\mathbf{X}_3}}$ (Transfer)}; \\
        \draw[C2, line width=1.2pt, dashed, opacity=0.7] (0,0) -- (0.4,0);& \node {$N_{G_{\mathbf{X}_3}}$ (Scratch)}; \\
        
        \draw[C1, line width=1.2pt] (0,0) -- (0.4,0);               & \node {$N_{\text{full}}$ (Transfer)}; \\
        \draw[C1, line width=1.2pt, dashed, opacity=0.7] (0,0) -- (0.4,0);& \node {$N_{\text{full}}$ (Scratch)}; \\
    };
\end{tikzpicture}};
    \end{tikzpicture}
     \caption{Discovery Curves}
     \label{sfig:CICY7709seed42-DC}
     \end{subfigure}\\
   \begin{subfigure}[b]{0.49\textwidth}
     \centering
     \begin{tikzpicture}[font=\footnotesize,baseline=0]
        \node (plot) {\includegraphics[width=0.9\textwidth,keepaspectratio]{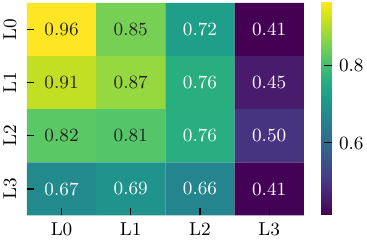}};
        \node (xaxis) at (plot.south) [below=0mm] {CICY 7709 Layers};
    \node (yaxis) at (plot.west) [left=0mm, rotate=90, anchor=south] {CICY 5302 Layers};
    \end{tikzpicture}
     \caption{CKA Heatmap}
     \label{sfig:CICY7709seed42-CKA}
     \end{subfigure}\hfill
     \begin{subfigure}[b]{0.49\textwidth}
     \centering
     \begin{tikzpicture}[font=\footnotesize,baseline=0]
        \node (plot) {\includegraphics[width=0.9\textwidth,keepaspectratio]{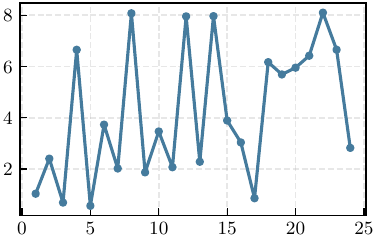}};
        \node (xaxis) at (plot.south) [below=0mm] {Step};
    \node (yaxis) at (plot.west) [left=0mm, rotate=90, anchor=south] {$D^{(t)}_\text{KL}$};
    \end{tikzpicture}
     \caption{Policy Divergence}
     \label{sfig:CICY7709seed42-PD}
     \end{subfigure}
\end{figure}

Evaluating the transfer learning performance reveals an asymmetry depending on the direction of the transfer with respect to the order of the geometric symmetry group $G_{\IX_3}$. Looking at the average speedup in finding solutions from CICY with large $|G_{\IX_3}|$ to small $|G_{\IX_3}|$, we notice a mean speedup of $67.7\times$ to the first valid solution while still maintaining a high understanding of the conditions in Table \ref{tab:conditions} (for instance, the reward associated to the anomaly cancellation condition is, on average, around $96\%$). However, the sample efficiency for this transfer learning is similar to that of training from scratch. This means that the policy quickly finds a local minimum, but it navigates the space of solutions less efficiently. A representative example is the transfer from CICY 5302 ($|G_{\IX_3}|=720$) to CICY 7709 ($|G_{\IX_3}|=12$) at Seed 42, for which we show discovery curves and policy divergence in Figure \ref{fig:CICY7709seed42-DCPD}. It achieved a remarkable $216.0\times$ speedup to the first solution and a $+0.436$ asymptotic reward gain, but its discovery curve does not beat the one of the training from scratch.

\begin{figure}[!htp]
    \centering
  \caption{Discovery Curve, CKA Heatmap and Mean CRS for transfer learning analysis from CICY 5967 to CICY 3413 in Seed 43.}
    \label{fig:CICY3413seed43-DCCRS}
     \begin{subfigure}[b]{0.49\textwidth}
     \centering
      \begin{tikzpicture}[font=\footnotesize,baseline=0]
        \node (plot) {\includegraphics[width=0.9\textwidth,keepaspectratio]{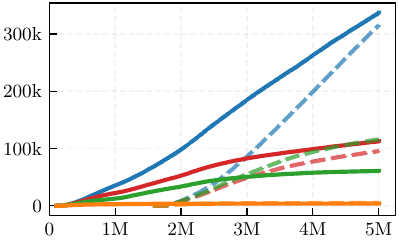}};
        \node (xaxis) at (plot.south) [below=0mm] {Episodes};
    \node (yaxis) at (plot.west) [left=0mm, rotate=90, anchor=south] {Solutions};
    \node (legend) [right=6mm of plot] {\begin{tikzpicture}
    \definecolor{C0}{HTML}{1f77b4}
\definecolor{C1}{HTML}{ff7f0e}
\definecolor{C2}{HTML}{2ca02c}
\definecolor{C3}{HTML}{d62728}

    \matrix [
        draw, 
        fill=white, 
        inner sep=4pt, 
        nodes={inner sep=2pt, anchor=west},
        column 1/.style={nodes={anchor=center}},
        font=\small,
        row sep=2pt
    ] (legend) at (0,0) {
        \draw[C0, line width=1.2pt] (0,0) -- (0.4,0);                 & \node {$N$ (Transfer)}; \\
        \draw[C0, line width=1.2pt, dashed, opacity=0.7] (0,0) -- (0.4,0); & \node {$N$ (Scratch)}; \\
        
        \draw[C3, line width=1.2pt] (0,0) -- (0.4,0);                  & \node {$N_{S_5}$ (Transfer)}; \\
        \draw[C3, line width=1.2pt, dashed, opacity=0.7] (0,0) -- (0.4,0);  & \node {$N_{S_5}$ (Scratch)}; \\
        
        \draw[C2, line width=1.2pt] (0,0) -- (0.4,0);                & \node {$N_{G_{\mathbf{X}_3}}$ (Transfer)}; \\
        \draw[C2, line width=1.2pt, dashed, opacity=0.7] (0,0) -- (0.4,0);& \node {$N_{G_{\mathbf{X}_3}}$ (Scratch)}; \\
        
        \draw[C1, line width=1.2pt] (0,0) -- (0.4,0);               & \node {$N_{\text{full}}$ (Transfer)}; \\
        \draw[C1, line width=1.2pt, dashed, opacity=0.7] (0,0) -- (0.4,0);& \node {$N_{\text{full}}$ (Scratch)}; \\
    };
\end{tikzpicture}};
    \end{tikzpicture}
     \caption{Discovery Curves}
     \label{sfig:CICY3413seed43-DC}
     \end{subfigure}\\
     \begin{subfigure}[b]{0.49\textwidth}
     \centering
     \begin{tikzpicture}[font=\footnotesize,baseline=0]
        \node (plot) {\includegraphics[width=0.9\textwidth,keepaspectratio]{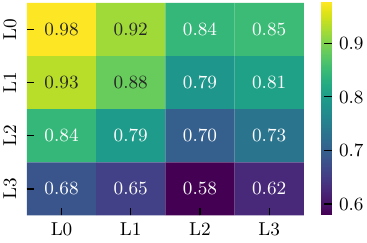}};
        \node (xaxis) at (plot.south) [below=0mm] {CICY 3413 Layers};
    \node (yaxis) at (plot.west) [left=0mm, rotate=90, anchor=south] {CICY 5967 Layers};
    \end{tikzpicture}
     \caption{CKA Heatmap}
     \label{sfig:CICY3413seed43-CKA}
     \end{subfigure}\hfill
     \begin{subfigure}[b]{0.49\textwidth}
     \centering
     \begin{tikzpicture}[font=\footnotesize,baseline=0]
        \node (plot) {\includegraphics[width=0.9\textwidth,keepaspectratio]{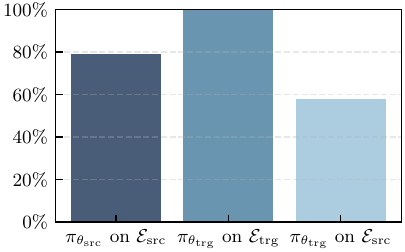}};
    \node (yaxis) at (plot.west) [left=0mm, rotate=90, anchor=south] {Mean CRS};
    \end{tikzpicture}
     \caption{Mean CRS}
     \label{sfig:CICY3413seed43-CRS}
     \end{subfigure}
\end{figure}
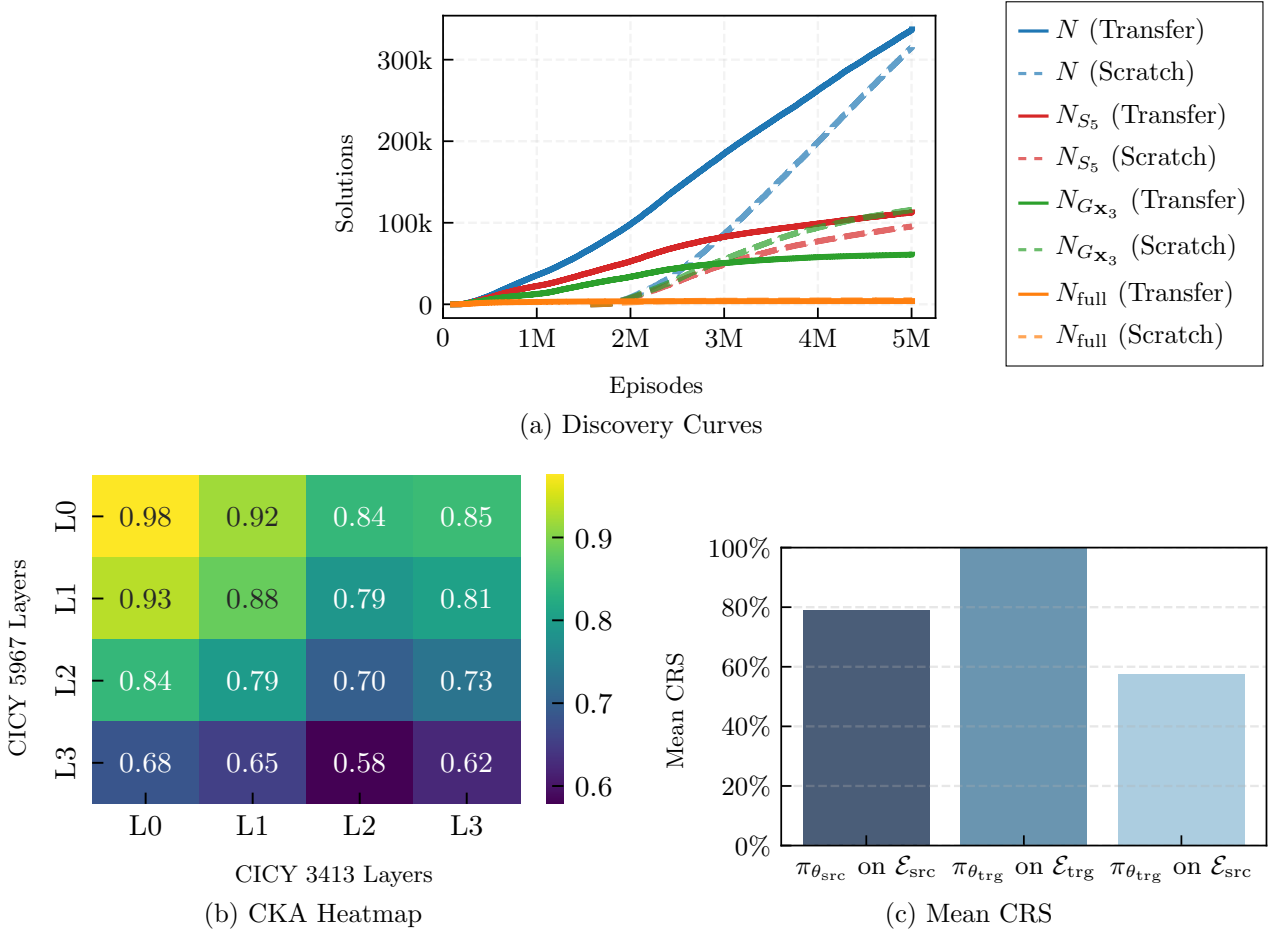

Conversely, transferring from a small $|G_{\IX_3}|$ to a large $|G_{\IX_3}|$ leads to more sustained exploration, but it requires more restructuring of the NN. While the initial discovery mean speedup is moderate ($25.7\times$), the transfer led to a $3.9\times$ speedup in discovering $1000$ solutions. This adaptation, however, reduces the retention of the NN structure obtained during training on the source CICY. Mean reward retention falls to $63.8\%$, and some other constraint rewards also decrease. For instance, transferring from CICY 5967 ($|G_{\IX_3}|=12$) to CICY 3413 ($|G_{\IX_3}|=72$) at Seed 43 yielded a $14.7\times$ initial speedup, and we show the discovery curves and the mean CRS in Figure \ref{fig:CICY3413seed43-DCCRS}. Ultimately, these results highlight a complementary trade-off: transferring from high to low symmetry offers rapid convergence and retention of constraints, whereas transferring from low to high symmetry encourages broader exploration at the expense of retaining information about source conditions and modifying the source policies. 

\subsubsection{Transfer Learning across \texorpdfstring{$\Gamma$}{Gamma}}
\label{sec:TL_acrossgamma}

In this section, we consider some examples of transfer learning for geometries that admit more freely-acting symmetries. This transfer learning analysis aims to understand the NN's ability to generalize when the constraints regarding the particle spectrum of the LBSMs can be either strengthened or loosened, depending on $|\Gamma|$.

\paragraph{Best Transfer Learning Seed: CICY 5302 ($|\Gamma|=4$) to CICY 5302 ($|\Gamma|=2$) in Seed 43.}

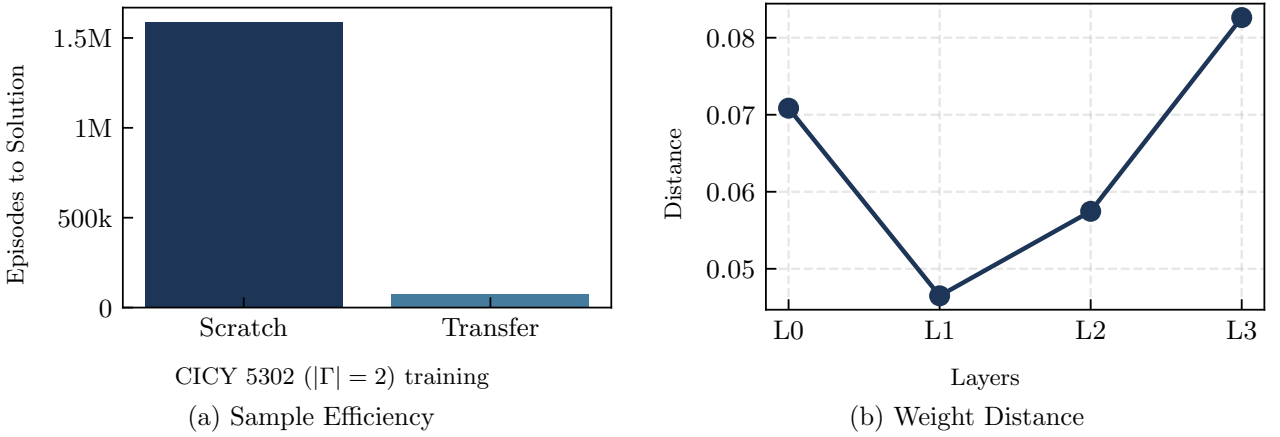
\begin{figure}[!htp]
    \centering
  \caption{Sample Efficiency and Weight distance for the transfer learning analysis from CICY 5302 ($|\Gamma|=4$) to CICY 5302 ($|\Gamma|=2$) in Seed 43.}
    \label{fig:CICY5302seed43-SEWD}
     \begin{subfigure}[b]{0.49\textwidth}
     \centering
      \begin{tikzpicture}[font=\footnotesize,baseline=0]
        \node (plot) {\includegraphics[width=0.9\textwidth,keepaspectratio]{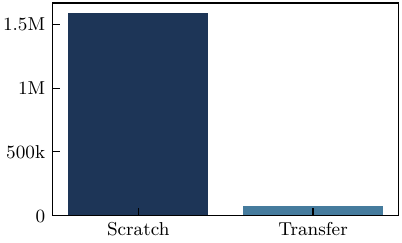}};
        \node (xaxis) at (plot.south) [below=0mm] {CICY 5302 ($|\Gamma|=2$) training};
    \node (yaxis) at (plot.west) [left=0mm, rotate=90, anchor=south] {Episodes to Solution};
    \end{tikzpicture}
     \caption{Sample Efficiency}
     \label{sfig:CICY5302seed43-SE}
     \end{subfigure}\hfill
     \begin{subfigure}[b]{0.49\textwidth}
     \centering
     \begin{tikzpicture}[font=\footnotesize,baseline=0]
        \node (plot) {\includegraphics[width=0.9\textwidth,keepaspectratio]{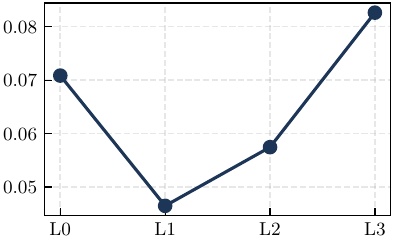}};
        \node (xaxis) at (plot.south) [below=0mm] {Layers};
    \node (yaxis) at (plot.west) [left=0mm, rotate=90, anchor=south] {Distance};
    \end{tikzpicture}
     \caption{Weight Distance}
     \label{sfig:CICY5302seed43-WD}
     \end{subfigure}
\end{figure}

We identify the transfer from source CICY 5302 ($|\Gamma|=4$) to target CICY 5302 ($|\Gamma|=2$) at Seed 43 as the most balanced and performant transfer overall. It exhibited a remarkable improvement in initial discovery: the Zero-shot Reward evaluated at episode $E=0$ increased by $+203.0\%$ ($0.268$ to $0.812$), and the episodes required to find the first valid line bundle configuration fell from $1425408$ to $16384$, reflecting an $87.0\times$ improvement. This configuration also maintained good exploration capability since it has the best sample efficiency of all considered transfer learning: It discovered $1000$ solutions in only $188416$ episodes compared to the $1712128$ that training from scratch required, meaning it had a $9.1\times$ sample efficiency speedup. This means that by having the NN learn to find solutions in a larger range, it has not had much difficulty restricting them to the smaller range. In fact, looking at Figure \ref{fig:CICY5302seed43-SEWD}, we see minimal structural disruption in the weight distance alongside excellent performance.

\paragraph{Average Best Transfer Learning: CICY 5302 ($|\Gamma|=4$) to CICY 5302 ($|\Gamma|=2$).}

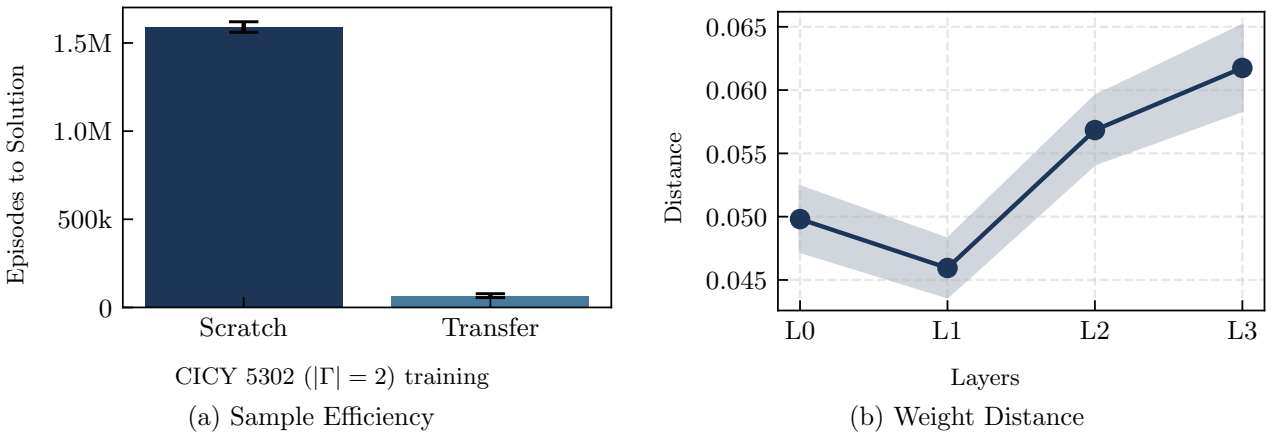
\begin{figure}[!htp]
    \centering
  \caption{Sample Efficiency and Weight distance for the average transfer learning analysis from CICY 5302 ($|\Gamma|=4$) to CICY 5302 ($|\Gamma|=2$).}
    \label{fig:CICY5302mean-SEWD}
     \begin{subfigure}[b]{0.49\textwidth}
     \centering
      \begin{tikzpicture}[font=\footnotesize,baseline=0]
        \node (plot) {\includegraphics[width=0.9\textwidth,keepaspectratio]{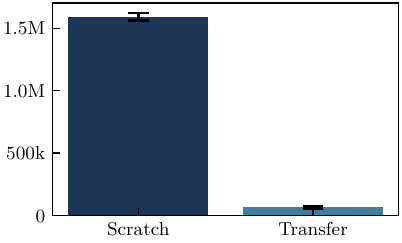}};
        \node (xaxis) at (plot.south) [below=0mm] {CICY 5302 ($|\Gamma|=2$) training};
    \node (yaxis) at (plot.west) [left=0mm, rotate=90, anchor=south] {Episodes to Solution};
    \end{tikzpicture}
     \caption{Sample Efficiency}
     \label{sfig:CICY5302mean-SE}
     \end{subfigure}\hfill
     \begin{subfigure}[b]{0.49\textwidth}
     \centering
     \begin{tikzpicture}[font=\footnotesize,baseline=0]
        \node (plot) {\includegraphics[width=0.9\textwidth,keepaspectratio]{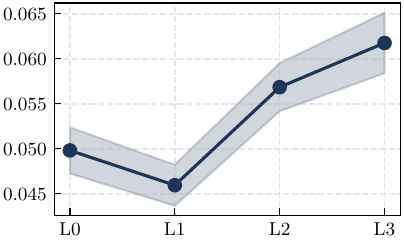}};
        \node (xaxis) at (plot.south) [below=0mm] {Layers};
    \node (yaxis) at (plot.west) [left=0mm, rotate=90, anchor=south] {Distance};
    \end{tikzpicture}
     \caption{Weight Distance}
     \label{sfig:CICY5302mean-WD}
     \end{subfigure}
\end{figure}

This success is remarkably consistent. Across all seeds, the transfer from CICY 5302 ($|\Gamma|=4$) to CICY 5302 ($|\Gamma|=2$) is also the best overall transfer learning across $\Gamma$. On average across all seeds, its Zero-Shot Reward improved by $+197.2\%$ (from a mean of $0.275$ to $0.811$), and the average episodes to discover the first solution dropped to $\sim 19$K (a $77.6\times$ improvement over the mean  number of episodes that were necessary when trained from scratch). Furthermore, it maintains excellent exploration, improving its average sample efficiency for $1000$ unique solutions by $8.7\times$ (down to $196608$ episodes from $\sim 1.7$M). This confirms that the policy successfully adapts its prior knowledge of the constraints in Table \ref{tab:conditions} to simply find solutions in the smaller range allowed by the smaller $|\Gamma|$. The average sample efficiency and weight distances are shown in Figure \ref{fig:CICY5302mean-SEWD}.

\paragraph{Directional Asymmetries in Transfer Learning.}

\begin{figure}[!htp]
    \centering
  \caption{Discovery Curve and Mean CRS for transfer learning analysis from CICY 5302 ($|\Gamma|=2$) to CICY 5302 ($|\Gamma|=4$) at Seed 42.}
    \label{fig:CICY5302seed42-DCCRS}
     \begin{subfigure}[b]{0.49\textwidth}
     \centering
      \begin{tikzpicture}[font=\footnotesize,baseline=0]
        \node (plot) {\includegraphics[width=0.9\textwidth,keepaspectratio]{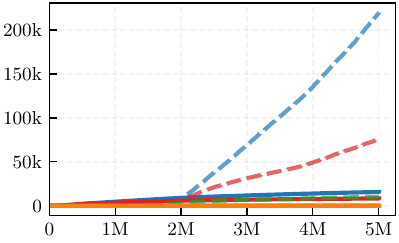}};
        \node (xaxis) at (plot.south) [below=0mm] {Episodes};
    \node (yaxis) at (plot.west) [left=0mm, rotate=90, anchor=south] {Solutions};
    \node (legend) [right=6mm of plot] {\begin{tikzpicture}
    \definecolor{C0}{HTML}{1f77b4}
\definecolor{C1}{HTML}{ff7f0e}
\definecolor{C2}{HTML}{2ca02c}
\definecolor{C3}{HTML}{d62728}

    \matrix [
        draw, 
        fill=white, 
        inner sep=4pt, 
        nodes={inner sep=2pt, anchor=west},
        column 1/.style={nodes={anchor=center}},
        font=\small,
        row sep=2pt
    ] (legend) at (0,0) {
        \draw[C0, line width=1.2pt] (0,0) -- (0.4,0);                 & \node {$N$ (Transfer)}; \\
        \draw[C0, line width=1.2pt, dashed, opacity=0.7] (0,0) -- (0.4,0); & \node {$N$ (Scratch)}; \\
        
        \draw[C3, line width=1.2pt] (0,0) -- (0.4,0);                  & \node {$N_{S_5}$ (Transfer)}; \\
        \draw[C3, line width=1.2pt, dashed, opacity=0.7] (0,0) -- (0.4,0);  & \node {$N_{S_5}$ (Scratch)}; \\
        
        \draw[C2, line width=1.2pt] (0,0) -- (0.4,0);                & \node {$N_{G_{\mathbf{X}_3}}$ (Transfer)}; \\
        \draw[C2, line width=1.2pt, dashed, opacity=0.7] (0,0) -- (0.4,0);& \node {$N_{G_{\mathbf{X}_3}}$ (Scratch)}; \\
        
        \draw[C1, line width=1.2pt] (0,0) -- (0.4,0);               & \node {$N_{\text{full}}$ (Transfer)}; \\
        \draw[C1, line width=1.2pt, dashed, opacity=0.7] (0,0) -- (0.4,0);& \node {$N_{\text{full}}$ (Scratch)}; \\
    };
\end{tikzpicture}};
    \end{tikzpicture}
     \caption{Discovery Curves}
     \label{sfig:CICY5302seed42-DC}
     \end{subfigure}\\
      \begin{subfigure}[b]{0.49\textwidth}
     \centering
     \begin{tikzpicture}[font=\footnotesize,baseline=0]
        \node (plot) {\includegraphics[width=0.9\textwidth,keepaspectratio]{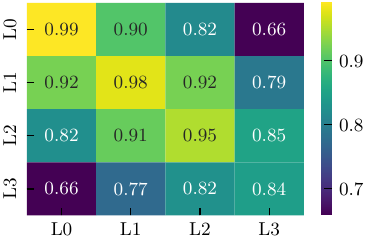}};
        \node (xaxis) at (plot.south) [below=0mm] {CICY 5302 ($|\Gamma|=4$) Layers};
    \node (yaxis) at (plot.west) [left=0mm, rotate=90, anchor=south] {CICY 5302 ($|\Gamma|=2$) Layers};
    \end{tikzpicture}
     \caption{CKA Heatmap}
     \label{sfig:CICY5302seed42-CKA}
     \end{subfigure}\hfill
     \begin{subfigure}[b]{0.49\textwidth}
     \centering
     \begin{tikzpicture}[font=\footnotesize,baseline=0]
        \node (plot) {\includegraphics[width=0.9\textwidth,keepaspectratio]{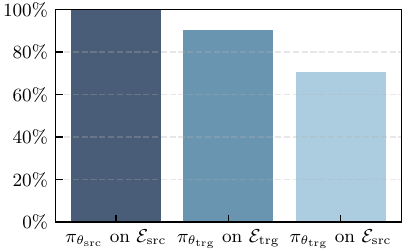}};
    \node (yaxis) at (plot.west) [left=0mm, rotate=90, anchor=south] {Mean CRS};
    \end{tikzpicture}
     \caption{Mean CRS}
     \label{sfig:CICY5302seed42-CRS-Gamma}
     \end{subfigure}
\end{figure}
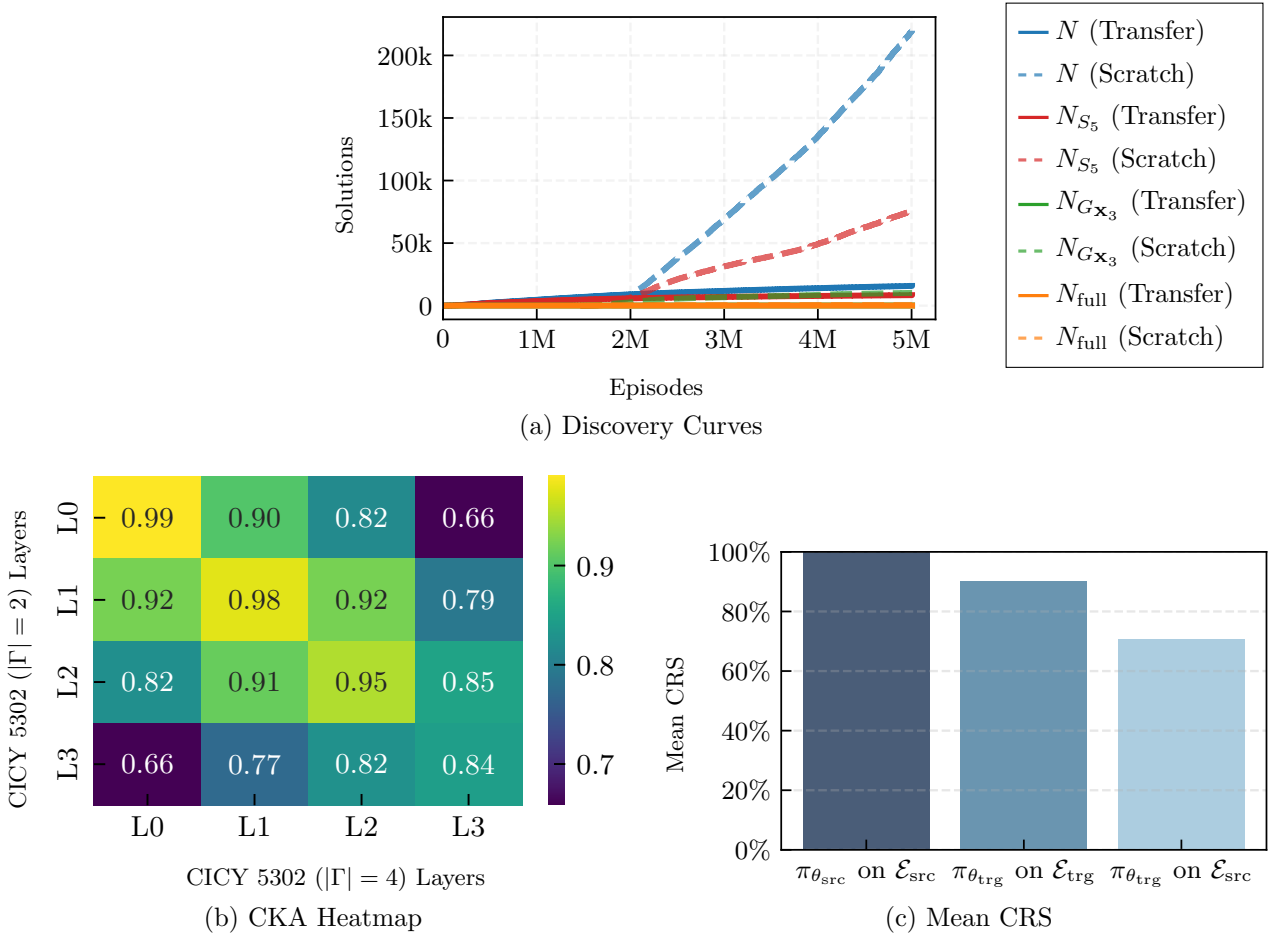

We can, once again, check for directional asymmetries in the transfer learning. We have seen in the previous section that transferring from larger $|\Gamma|$ to smaller ones yields good transfer learning performance. In transferring from a configuration with a smaller freely-acting symmetry to a larger one (e.g., from $|\Gamma|=2$ to $|\Gamma|=4$), it also leads to a high mean speedup of $73.4\times$ to the first solution. However, sample efficiency shows that this transfer direction is not as efficient as the other: the policy rapidly becomes trapped in a deep local minimum, discovering fewer solutions asymptotically than a model trained from scratch. This is a similar situation to what we encountered in the previous section (but opposite) whenever we considered transfer learning from larger $|G_{\IX_3}|$ to smaller $|G_{\IX_3}|$. A representative example is the transfer from CICY 5302 ($|\Gamma|=2$) to CICY 5302 ($|\Gamma|=4$) at Seed 42, shown in Figure \ref{fig:CICY5302seed42-DCCRS}.

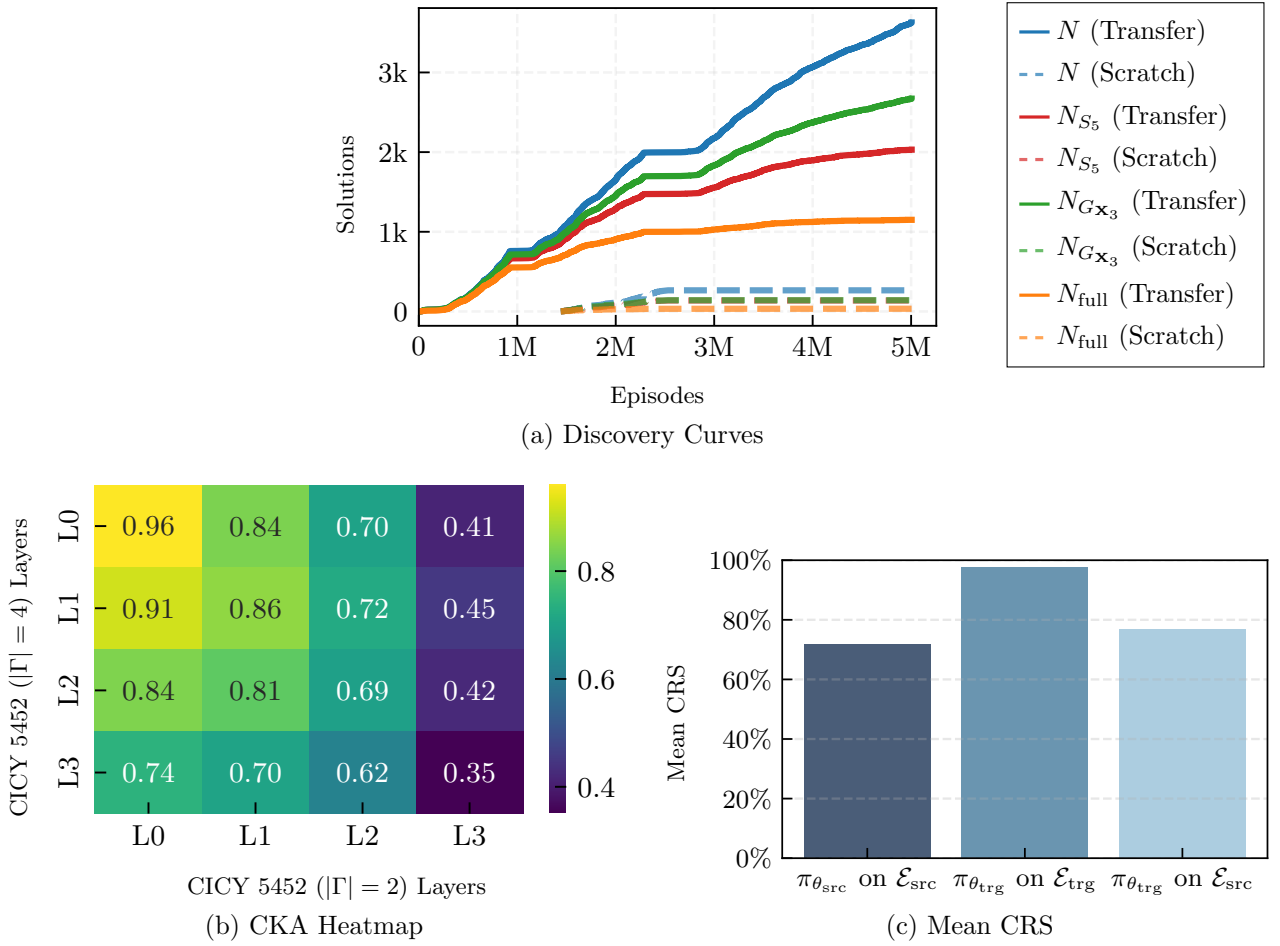
\begin{figure}[!htp]
    \centering
  \caption{Discovery Curve and Mean CRS for transfer learning analysis from CICY 5452 ($|\Gamma|=4$) to CICY 5452 ($|\Gamma|=2$) at Seed 43.}
    \label{fig:CICY5452seed43-DCCRS}
     \begin{subfigure}[b]{0.49\textwidth}
     \centering
      \begin{tikzpicture}[font=\footnotesize,baseline=0]
        \node (plot) {\includegraphics[width=0.9\textwidth,keepaspectratio]{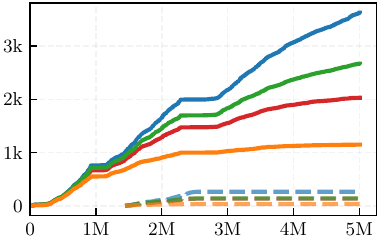}};
        \node (xaxis) at (plot.south) [below=0mm] {Episodes};
    \node (yaxis) at (plot.west) [left=0mm, rotate=90, anchor=south] {Solutions};
    \node (legend) [right=6mm of plot] {\begin{tikzpicture}
    \definecolor{C0}{HTML}{1f77b4}
\definecolor{C1}{HTML}{ff7f0e}
\definecolor{C2}{HTML}{2ca02c}
\definecolor{C3}{HTML}{d62728}

    \matrix [
        draw, 
        fill=white, 
        inner sep=4pt, 
        nodes={inner sep=2pt, anchor=west},
        column 1/.style={nodes={anchor=center}},
        font=\small,
        row sep=2pt
    ] (legend) at (0,0) {
        \draw[C0, line width=1.2pt] (0,0) -- (0.4,0);                 & \node {$N$ (Transfer)}; \\
        \draw[C0, line width=1.2pt, dashed, opacity=0.7] (0,0) -- (0.4,0); & \node {$N$ (Scratch)}; \\
        
        \draw[C3, line width=1.2pt] (0,0) -- (0.4,0);                  & \node {$N_{S_5}$ (Transfer)}; \\
        \draw[C3, line width=1.2pt, dashed, opacity=0.7] (0,0) -- (0.4,0);  & \node {$N_{S_5}$ (Scratch)}; \\
        
        \draw[C2, line width=1.2pt] (0,0) -- (0.4,0);                & \node {$N_{G_{\mathbf{X}_3}}$ (Transfer)}; \\
        \draw[C2, line width=1.2pt, dashed, opacity=0.7] (0,0) -- (0.4,0);& \node {$N_{G_{\mathbf{X}_3}}$ (Scratch)}; \\
        
        \draw[C1, line width=1.2pt] (0,0) -- (0.4,0);               & \node {$N_{\text{full}}$ (Transfer)}; \\
        \draw[C1, line width=1.2pt, dashed, opacity=0.7] (0,0) -- (0.4,0);& \node {$N_{\text{full}}$ (Scratch)}; \\
    };
\end{tikzpicture}};
    \end{tikzpicture}
     \caption{Discovery Curves}
     \label{sfig:CICY5452seed43-DC-Gamma}
     \end{subfigure}\\
     \begin{subfigure}[b]{0.49\textwidth}
     \centering
     \begin{tikzpicture}[font=\footnotesize,baseline=0]
        \node (plot) {\includegraphics[width=0.9\textwidth,keepaspectratio]{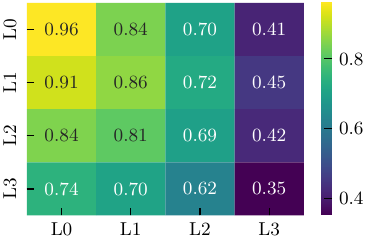}};
        \node (xaxis) at (plot.south) [below=0mm] {CICY 5452 ($|\Gamma|=2$) Layers};
    \node (yaxis) at (plot.west) [left=0mm, rotate=90, anchor=south] {CICY 5452 ($|\Gamma|=4$) Layers};
    \end{tikzpicture}
     \caption{CKA Heatmap}
     \label{sfig:CICY5452seed43-CKA-Gamma}
     \end{subfigure}\hfill
     \begin{subfigure}[b]{0.49\textwidth}
     \centering
     \begin{tikzpicture}[font=\footnotesize,baseline=0]
        \node (plot) {\includegraphics[width=0.9\textwidth,keepaspectratio]{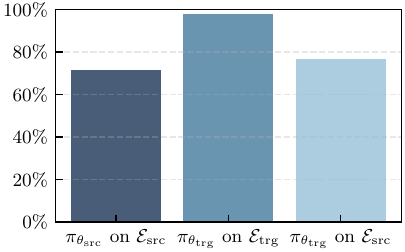}};
    \node (yaxis) at (plot.west) [left=0mm, rotate=90, anchor=south] {Mean CRS};
    \end{tikzpicture}
     \caption{Mean CRS}
     \label{sfig:CICY5452seed43-CRS-Gamma}
     \end{subfigure}
\end{figure}

While achieving a massive $179.0\times$ speedup to the first solution, the transfer model ultimately found only $15786$ solutions over $5$ million episodes, compared to $676184$ discovered by training from scratch. Conversely, transferring from a larger $|\Gamma|$ to a smaller $|\Gamma|$ results in slower initial convergence but more exploration. The mean speedup to the first solution drops to $42.3\times$, but the model continues to find new solutions. For instance, transferring from CICY 5452 ($|\Gamma|=4$) to CICY 5452 ($|\Gamma|=2$) at Seed 43 (in Figure \ref{fig:CICY5452seed43-DCCRS}) yielded an $88.0\times$ initial speedup, and the transfer policy successfully navigated the solution space to find $3627$ valid matrices, well above the $265$ found from scratch. 

Furthermore, the retention of conditions learned by the source training is also high. This leads us to the conclusion that transferring to a larger $\Gamma$ allows for finding solutions faster, but the NN is not able to find new solutions easily. On the other hand, transferring to a smaller $\Gamma$, despite a moderate speedup, has the potential to explore and find more unique solutions.

\newpage
\subsubsection{Transfer Learning Across \texorpdfstring{$h^{1,1}(\IX_3)$}{h11}}
\label{sec:TL_acrossh11}

In this section, we finally consider some examples of transfer learning for geometries related by a conifold transition. This transfer learning analysis aims to understand the NN's ability to scale to larger or smaller setups, thus demonstrating the capacity to adapt when constraints are added or removed. 

\paragraph{Best Transfer Learning Seed: CICY 3413 to CICY 6024 in Seed 44.}

\begin{figure}[!htp]
    \centering
  \caption{Discovery Curve and Policy Divergence for the transfer learning analysis from CICY 3413 to CICY 6024 in Seed 44.}
    \label{fig:CICY6024seed44-DCPD}
     \begin{subfigure}[b]{0.49\textwidth}
     \centering
      \begin{tikzpicture}[font=\footnotesize,baseline=0]
        \node (plot) {\includegraphics[width=0.9\textwidth,keepaspectratio]{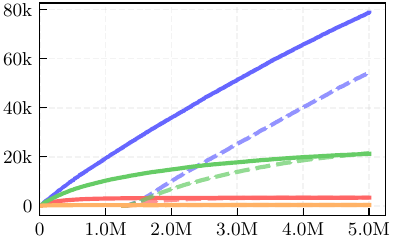}};
        \node (xaxis) at (plot.south) [below=0mm] {Episodes};
    \node (yaxis) at (plot.west) [left=0mm, rotate=90, anchor=south] {Solutions};
    \end{tikzpicture}
     \caption{Discovery Curves}
     \label{sfig:CICY6024seed44-DC}
     \end{subfigure}\hfill
     \begin{subfigure}[b]{0.49\textwidth}
     \centering
     \begin{tikzpicture}[font=\footnotesize,baseline=0]
        \node (plot) {\includegraphics[width=0.9\textwidth,keepaspectratio]{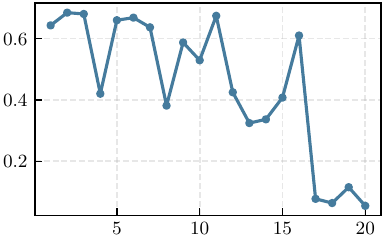}};
        \node (xaxis) at (plot.south) [below=0mm] {Step};
    \node (yaxis) at (plot.west) [left=0mm, rotate=90, anchor=south] {$D^{(t)}_\text{KL}$};
    \end{tikzpicture}
     \caption{Policy Divergence}
     \label{sfig:CICY6024seed44-PD}
     \end{subfigure}
\end{figure}
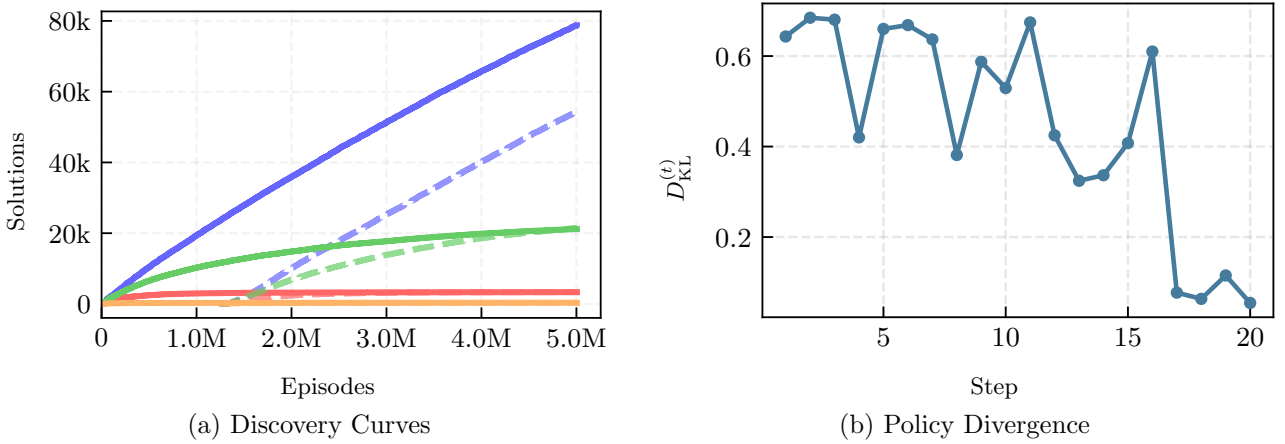

The best transfer learning across $h^{1,1}(\IX_3)$ is from source CICY 3413 ($h^{1,1}(\IX_3)=6$) to target CICY 6024 ($h^{1,1}(\IX_3)=5$) at Seed 44. For this transfer, we see a Zero-shot reward of $+253.0\%$ ($0.185$ to $0.653$) and the first solution was found after only $8192$ episodes compared to the original $1236992$ episodes required by the training from scratch.

We also see that it found $1000$ unique solutions only after $49152$, improving the original training by $+96.6\%$. Looking at Figure \ref{fig:CICY6024seed44-DCPD}, the policy divergence displays an initial structural shift, which then stabilizes. This suggests that the NN adapted to the new geometry without completely forgetting the conditions learned during the training of the source.

\paragraph{Average Best Transfer Learning: CICY 3413 to CICY 6024.}

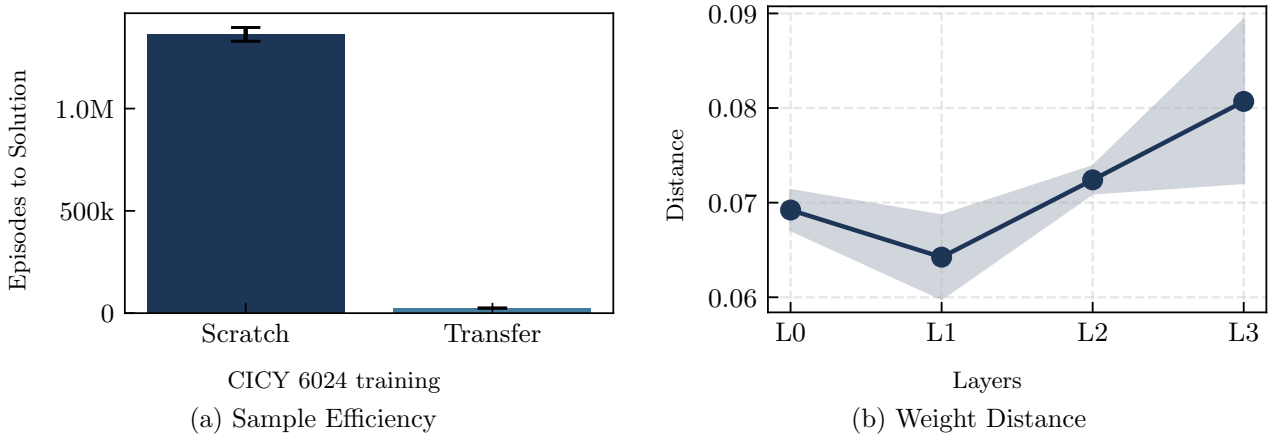
\begin{figure}[!htp]
    \centering
  \caption{Sample Efficiency and Weight distance for the average transfer learning analysis from CICY 3413 to CICY 6024.}
    \label{fig:CICY6024mean-SEWD}
     \begin{subfigure}[b]{0.49\textwidth}
     \centering
      \begin{tikzpicture}[font=\footnotesize,baseline=0]
        \node (plot) {\includegraphics[width=0.9\textwidth,keepaspectratio]{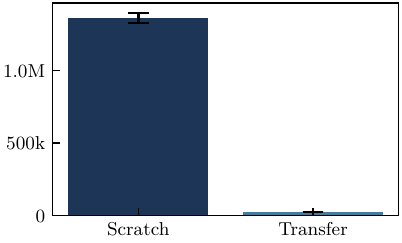}};
        \node (xaxis) at (plot.south) [below=0mm] {CICY 6024 training};
    \node (yaxis) at (plot.west) [left=0mm, rotate=90, anchor=south] {Episodes to Solution};
    \end{tikzpicture}
     \caption{Sample Efficiency}
     \label{sfig:CICY6024mean-SE}
     \end{subfigure}\hfill
     \begin{subfigure}[b]{0.49\textwidth}
     \centering
     \begin{tikzpicture}[font=\footnotesize,baseline=0]
        \node (plot) {\includegraphics[width=0.9\textwidth,keepaspectratio]{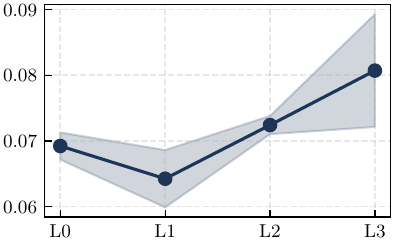}};
        \node (xaxis) at (plot.south) [below=0mm] {Layers};
    \node (yaxis) at (plot.west) [left=0mm, rotate=90, anchor=south] {Distance};
    \end{tikzpicture}
     \caption{Weight Distance}
     \label{sfig:CICY6024mean-WD}
     \end{subfigure}
\end{figure}

Once again, the pair that performed best on a single seed is also the one that had the best transfer learning overall. On average, the Zero-Shot Reward improved by $+303.7\%$ (from a mean of $0.162$ to $0.654$), and the average number of episodes required to discover the first solution dropped from $\sim 1.25$M of the training from scratch to $8192$. Furthermore, the sample efficiency is at $+96.3\%$, since the NN required only $54613$ to find $1000$ solutions over the $1466368$ episodes of the scratch training. Similar to what we have seen for the transfer learning across $\Gamma$, the NN is able to restrict the solutions for the smaller CICY after it has learned the conditions on the larger one. These results are confirmed in Figure \ref{fig:CICY6024mean-SEWD}.

\paragraph{Directional Asymmetries in Transfer Learning.}

\begin{figure}[!htp]
    \centering
  \caption{Discovery Curve, CKA Heatmap and Mean CRS for transfer learning analysis from CICY 6024 ($h^{1,1}(\IX_3)=5$) to CICY 3413 ($h^{1,1}(\IX_3)=6$) at Seed 42.}
    \label{fig:CICY3413seed42-DCCRS}
     \begin{subfigure}[b]{0.49\textwidth}
     \centering
      \begin{tikzpicture}[font=\footnotesize,baseline=0]
        \node (plot) {\includegraphics[width=0.9\textwidth,keepaspectratio]{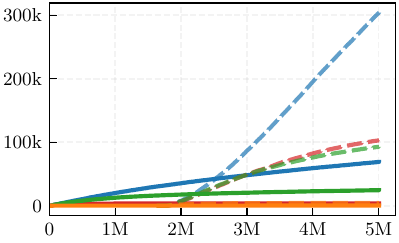}};
        \node (xaxis) at (plot.south) [below=0mm] {Episodes};
    \node (yaxis) at (plot.west) [left=0mm, rotate=90, anchor=south] {Solutions};
    \node (legend) [right=6mm of plot] {\begin{tikzpicture}
    \definecolor{C0}{HTML}{1f77b4}
\definecolor{C1}{HTML}{ff7f0e}
\definecolor{C2}{HTML}{2ca02c}
\definecolor{C3}{HTML}{d62728}

    \matrix [
        draw, 
        fill=white, 
        inner sep=4pt, 
        nodes={inner sep=2pt, anchor=west},
        column 1/.style={nodes={anchor=center}},
        font=\small,
        row sep=2pt
    ] (legend) at (0,0) {
        \draw[C0, line width=1.2pt] (0,0) -- (0.4,0);                 & \node {$N$ (Transfer)}; \\
        \draw[C0, line width=1.2pt, dashed, opacity=0.7] (0,0) -- (0.4,0); & \node {$N$ (Scratch)}; \\
        
        \draw[C3, line width=1.2pt] (0,0) -- (0.4,0);                  & \node {$N_{S_5}$ (Transfer)}; \\
        \draw[C3, line width=1.2pt, dashed, opacity=0.7] (0,0) -- (0.4,0);  & \node {$N_{S_5}$ (Scratch)}; \\
        
        \draw[C2, line width=1.2pt] (0,0) -- (0.4,0);                & \node {$N_{G_{\mathbf{X}_3}}$ (Transfer)}; \\
        \draw[C2, line width=1.2pt, dashed, opacity=0.7] (0,0) -- (0.4,0);& \node {$N_{G_{\mathbf{X}_3}}$ (Scratch)}; \\
        
        \draw[C1, line width=1.2pt] (0,0) -- (0.4,0);               & \node {$N_{\text{full}}$ (Transfer)}; \\
        \draw[C1, line width=1.2pt, dashed, opacity=0.7] (0,0) -- (0.4,0);& \node {$N_{\text{full}}$ (Scratch)}; \\
    };
\end{tikzpicture}};
    \end{tikzpicture}
     \caption{Discovery Curves}
     \label{sfig:CICY3413seed42-DC}
     \end{subfigure}\\
      \begin{subfigure}[b]{0.49\textwidth}
     \centering
     \begin{tikzpicture}[font=\footnotesize,baseline=0]
        \node (plot) {\includegraphics[width=0.9\textwidth,keepaspectratio]{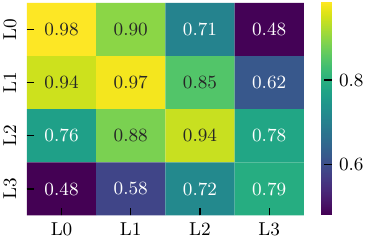}};
        \node (xaxis) at (plot.south) [below=0mm] {CICY 3413 ($h^{1,1}(\IX_3)=6$) Layers};
    \node (yaxis) at (plot.west) [left=0mm, rotate=90, anchor=south] {CICY 6024 ($h^{1,1}(\IX_3)=5$) Layers};
    \end{tikzpicture}
     \caption{CKA Heatmap}
     \label{sfig:CICY3413seed42-CKA}
     \end{subfigure}\hfill
     \begin{subfigure}[b]{0.49\textwidth}
     \centering
     \begin{tikzpicture}[font=\footnotesize,baseline=0]
        \node (plot) {\includegraphics[width=0.9\textwidth,keepaspectratio]{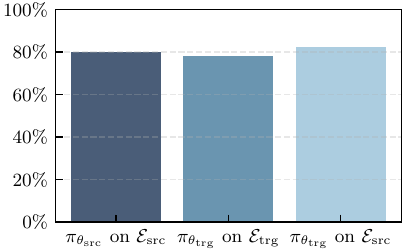}};
    \node (yaxis) at (plot.west) [left=0mm, rotate=90, anchor=south] {Mean CRS};
    \end{tikzpicture}
     \caption{Mean CRS}
     \label{sfig:CICY3413seed42-CRS-h11}
     \end{subfigure}
\end{figure}
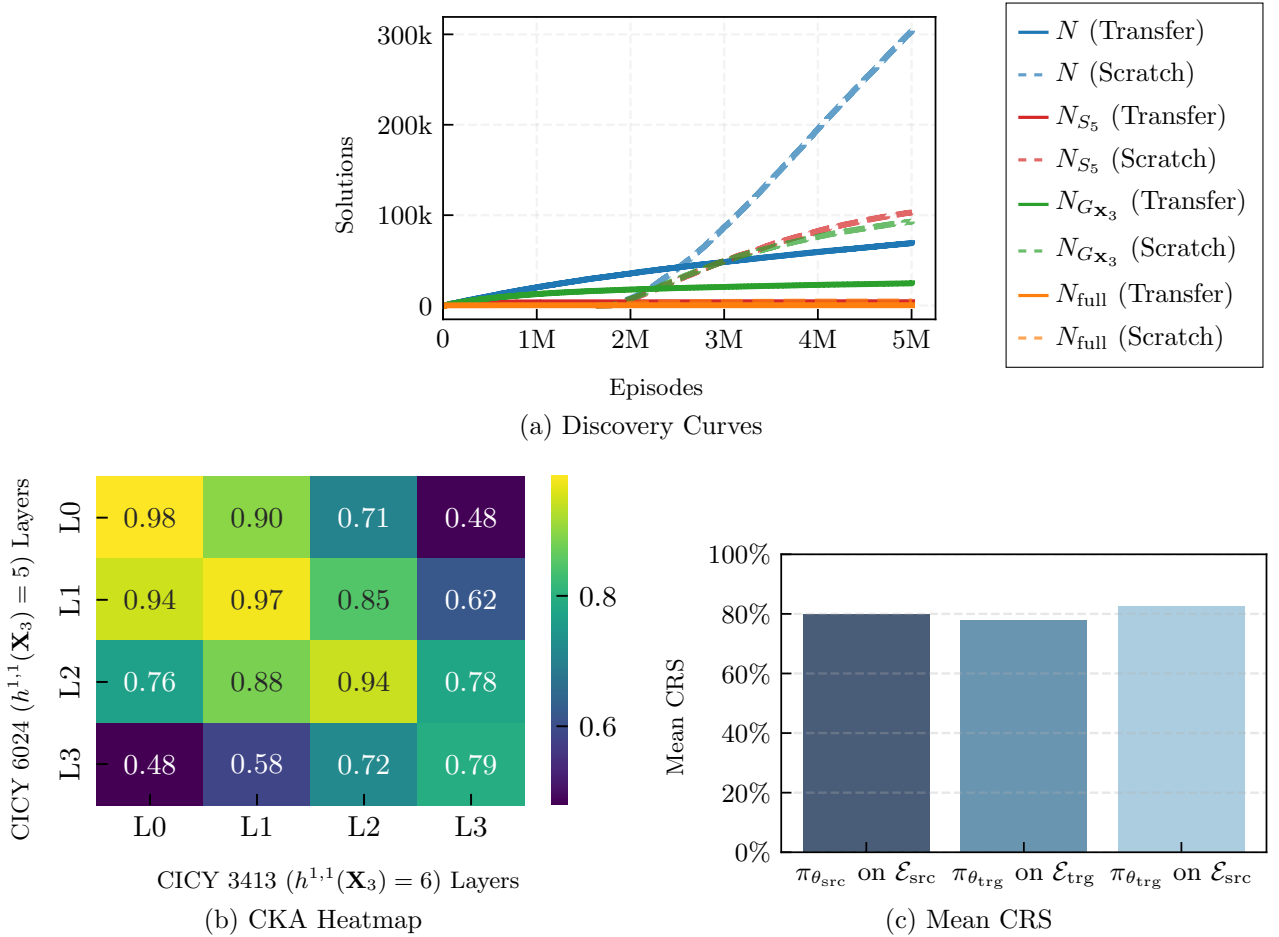

The similarities with the transfer learning across $\Gamma$ point in the direction that there could be a directional asymmetry also in the transfer learning across $h^{1,1}(\IX_3)$. In fact, in transferring from a configuration with a smaller $h^{1,1}(\IX_3)$ to a larger one (e.g., from $h^{1,1}(\IX_3)=5$ to $h^{1,1}(\IX_3)=6$), we observe an extreme mean speedup of $171.5\times$ to the first valid solution. However, the policy is forced to restructure, leading to a drop in retention of the conditions (e.g., retaining only $82.4\%$ of the anomaly cancellation and $78.3\%$ of the index sum conditions). Consequently, while the initial exploitation is extremely fast, the sample efficiency shows that the model discovers far fewer unique solutions asymptotically. A representative example is the transfer from CICY 6024 ($h^{1,1}(\IX_3)=5$) to CICY 3413 ($h^{1,1}(\IX_3)=6$) at Seed 42, shown in Figure \ref{fig:CICY3413seed42-DCCRS}. While achieving a $199.0\times$ speedup to the first solution, the transfer model ultimately found $69118$ unique solutions asymptotically, compared to the $1293813$ solutions ($300$k already at 5 million episodes) discovered by training from scratch. 

Conversely, transferring from a larger $h^{1,1}(\IX_3)$ to a smaller $h^{1,1}(\IX_3)$ results in a less extreme initial convergence, despite being high at $85.0\times$; however, the model preserves a much larger fraction of the condition rewards (e.g., $97.1\%$ for the anomaly cancellation condition and $91.3\%$ for stability). For instance, transferring from CICY 3413 ($h^{1,1}(\IX_3)=6$) to CICY 6024 ($h^{1,1}(\IX_3)=5$) at Seed 43 (in Figure \ref{fig:CICY6024seed43-DCCRS}) yielded a $157.0\times$ initial speedup, and the transfer policy led to finding $79953$ solutions (compared to $\sim 40$k from that episode). Ultimately, scaling across conifold transitions uncovers a distinct asymmetry: transferring to a higher dimensional topology forces destructive adaptation to maximize early solutions but sacrifices long-term breadth, whereas transferring to a lower dimensional topology focuses on retaining source conditions seamlessly with more sustained exploration.

\begin{figure}[!htp]
    \centering
  \caption{Discovery Curve, CKA Heatmap and Mean CRS for transfer learning analysis from CICY 3413 ($h^{1,1}(\IX_3)=6$) to CICY 6024 ($h^{1,1}(\IX_3)=5$) at Seed 43.}
    \label{fig:CICY6024seed43-DCCRS}
     \begin{subfigure}[b]{0.49\textwidth}
     \centering
      \begin{tikzpicture}[font=\footnotesize,baseline=0]
        \node (plot) {\includegraphics[width=0.9\textwidth,keepaspectratio]{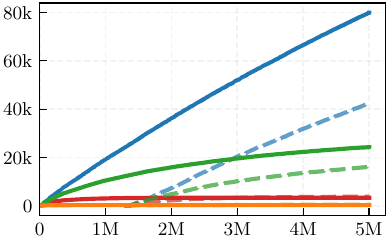}};
        \node (xaxis) at (plot.south) [below=0mm] {Episodes};
    \node (yaxis) at (plot.west) [left=0mm, rotate=90, anchor=south] {Solutions};
    \node (legend) [right=6mm of plot] {\begin{tikzpicture}
    \definecolor{C0}{HTML}{1f77b4}
\definecolor{C1}{HTML}{ff7f0e}
\definecolor{C2}{HTML}{2ca02c}
\definecolor{C3}{HTML}{d62728}

    \matrix [
        draw, 
        fill=white, 
        inner sep=4pt, 
        nodes={inner sep=2pt, anchor=west},
        column 1/.style={nodes={anchor=center}},
        font=\small,
        row sep=2pt
    ] (legend) at (0,0) {
        \draw[C0, line width=1.2pt] (0,0) -- (0.4,0);                 & \node {$N$ (Transfer)}; \\
        \draw[C0, line width=1.2pt, dashed, opacity=0.7] (0,0) -- (0.4,0); & \node {$N$ (Scratch)}; \\
        
        \draw[C3, line width=1.2pt] (0,0) -- (0.4,0);                  & \node {$N_{S_5}$ (Transfer)}; \\
        \draw[C3, line width=1.2pt, dashed, opacity=0.7] (0,0) -- (0.4,0);  & \node {$N_{S_5}$ (Scratch)}; \\
        
        \draw[C2, line width=1.2pt] (0,0) -- (0.4,0);                & \node {$N_{G_{\mathbf{X}_3}}$ (Transfer)}; \\
        \draw[C2, line width=1.2pt, dashed, opacity=0.7] (0,0) -- (0.4,0);& \node {$N_{G_{\mathbf{X}_3}}$ (Scratch)}; \\
        
        \draw[C1, line width=1.2pt] (0,0) -- (0.4,0);               & \node {$N_{\text{full}}$ (Transfer)}; \\
        \draw[C1, line width=1.2pt, dashed, opacity=0.7] (0,0) -- (0.4,0);& \node {$N_{\text{full}}$ (Scratch)}; \\
    };
\end{tikzpicture}};
    \end{tikzpicture}
     \caption{Discovery Curves}
     \label{sfig:CICY6024seed43-DC-h11}
     \end{subfigure}\\
     \begin{subfigure}[b]{0.49\textwidth}
     \centering
     \begin{tikzpicture}[font=\footnotesize,baseline=0]
        \node (plot) {\includegraphics[width=0.9\textwidth,keepaspectratio]{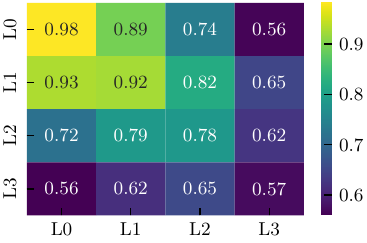}};
        \node (xaxis) at (plot.south) [below=0mm] {CICY 6024 ($h^{1,1}(\IX_3)=5$) Layers};
    \node (yaxis) at (plot.west) [left=0mm, rotate=90, anchor=south] {CICY 3413 ($h^{1,1}(\IX_3)=6$) Layers};
    \end{tikzpicture}
     \caption{CKA Heatmap}
     \label{sfig:CICY6024seed43-CKA-h11}
     \end{subfigure}\hfill
     \begin{subfigure}[b]{0.49\textwidth}
     \centering
     \begin{tikzpicture}[font=\footnotesize,baseline=0]
        \node (plot) {\includegraphics[width=0.9\textwidth,keepaspectratio]{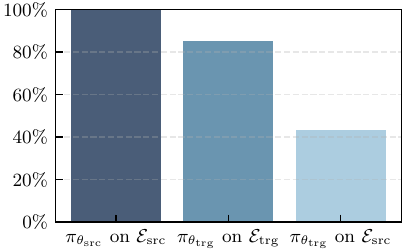}};
    \node (yaxis) at (plot.west) [left=0mm, rotate=90, anchor=south] {Mean CRS};
    \end{tikzpicture}
     \caption{Mean CRS}
     \label{sfig:CICY6024seed43-CRS-h11}
     \end{subfigure}
\end{figure}
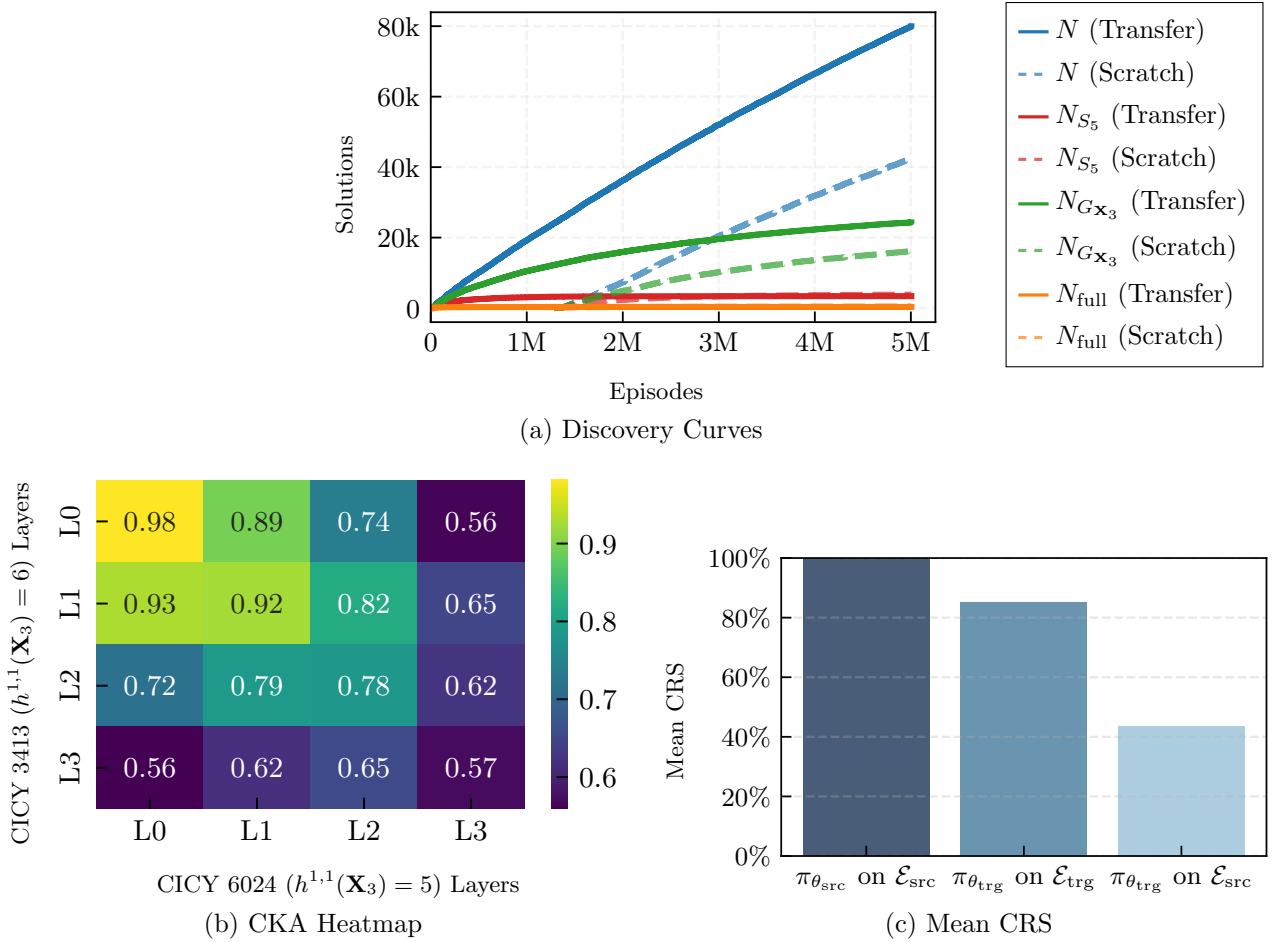

\section{Equivariance and Spectrum Analysis}
\label{sec:spectrumanalysis}

In Section \ref{sec:GUTHetIngred}, we listed the spectrum constraints to obtain heterotic $\SU(5)$ GUT; however, in Table \ref{tab:conditions}, we only imposed some necessary but not sufficient conditions on the spectrum of particles. In this section, we discuss the filtering of the solutions by imposing the equivariant structure of the vector bundles and further conditions on the spectrum of particles.

\subsection{Comments on Equivariant Structure}

As evident from Table \ref{tab:detailed_counts}, imposing the equivariance conditions \cref{eq:equivariantcond_trivial,eq:equivariantcond_nontrivial} drastically truncates the set of viable solutions discovered by the LB-Explorer. In this work, the equivariant structure check was intentionally left out of the LB-Explorer training loop to maintain a framework capable of finding line bundles for $\IX_3$, independent of specific quotient actions.

We find that when the action of the freely-acting symmetry is trivial on the divisors, i.e., if the equivariance can be checked using \eqref{eq:equivariantcond_trivial}, a larger fraction of the configurations found by the LB-Explorer survives the filtering. However, a non-trivial group action on the divisors, requiring \eqref{eq:equivariantcond_nontrivial}, dramatically reduces the number of solutions found by the LB-Explorer.

Requiring that the LB-Explorer be able to learn \cref{eq:equivariantcond_trivial,eq:equivariantcond_nontrivial} would have introduced many extra constraints that would have made the exploration inefficient. Moreover, it would have made the transfer learning analysis more difficult because the freely-acting symmetries might have different ways of acting on the CICYs. While it would be interesting to consider a future extension of the LB-Explorer that also includes the equivariant structure constraints, in Section \ref{sec:HybridRL}, we propose an alternative way to impose conditions in an exact manner that could be an efficient way to encode extra constraints without modifying the current architecture.

\subsection{Comments on Spectrum Analysis}

In this section, instead, we comment on checking how many of the found solutions satisfy the conditions on the spectrum that can be imposed by computing the cohomologies of the found line bundles. In particular, after imposing the equivariance constraints \cref{eq:equivariantcond_trivial,eq:equivariantcond_nontrivial}, one requires that the vector bundle $V$ satisfies the conditions \cite{Anderson:2011ns, Anderson:2013xka}:
\begin{equation}\label{eq:fullspectrum}
\begin{array}{rclrcl}
h^1(\IX_3,V) &=& 3|\Gamma| \coma & h^2(\IX_3,V) & = & 0 \coma \\
h^1(\IX_3,\wedge^2 V) & = & 3|\Gamma| + n_h \coma & h^2(\IX_3,\wedge^2 V) & = & n_h\coma
\end{array}
\end{equation}
where $n_h$ is the number of Higgs doublets. For some CICY manifolds, there exist closed formulas to compute line cohomologies \cite{Constantin:2018hvl}, but in general, it is possible to compute them using algorithmic tools such as \texttt{cohomcalg} \cite{Blumenhagen:2010pv} and \texttt{CIPro} \cite{Anderson:2026eyl}, which utilizes Koszul sequence techniques, or \texttt{pyCICY} \cite{Larfors:2019sie}, or recent extensions of \texttt{CYTools} \cite{Demirtas:2022hqf,Gendler:2026uux}. However, these algorithms scale poorly with $h^{1,1}(\IX_3)$ and the magnitude of the line bundle integer charges $|\mathbf{k}_a|$. For this reason, as with the equivariance conditions, we have decided not to include these computations among the conditions that the LB-Explorer had to learn, and to filter the solutions \textit{a posteriori}. However, it would be interesting in the future to find a way to include these computations in a NN environment by letting the network predict the line bundle cohomologies.\footnote{See, e.g., \cite{Klaewer:2018sfl, Brodie:2019dfx, Brodie:2021nit, Constantin:2021for, Constantin:2024ulu} for efforts in this direction} 

For the moment, we have decided to consider the CICYs that admit a symmetry $\Gamma$ acting trivially on the basis divisors $J_i$ (indicated by a \checkmark in Table \ref{tab:detailed_counts}), and we have restricted the solutions to the equivariant ones with $|\mathbf{k}_a|\leq 2$. This restriction already reduces the number of solutions drastically since, by construction, $\mathbf{k}_5$ was left free to be the sum of all the entries in order to satisfy $c_1(V) = 0$. As shown in Table \ref{tab:spectrum_stats}, requiring that all the line bundles have entries between $-2$ and $2$ cuts the space of solutions to a few thousand in the best case scenario. Nevertheless, for completeness in the analysis of the solutions found by LB-Explorer, we used \texttt{pyCICY} to compute the line bundle cohomologies and report the surviving solutions in Table \ref{tab:spectrum_stats} as well.

\section{Hybrid LB-Explorer Architecture}
\label{sec:HybridRL}

In this section, we discuss a way to augment the LB-Explorer with further constraints that can be solved exactly. The main idea is that, while the PPO agent is efficient at navigating the space of configurations, it might struggle with solving the conditions that involve equalities, e.g., the chirality in Table \ref{tab:conditions}.

In general, if one wants to impose the equivariant structure, or even the line bundle cohomology constraints, it is no longer sufficient for the LB-Explorer to find solutions in a given range; it must find components of the vector bundle $V$ that precisely sum up to certain constraints. 

In our work, we have decided to impose the extra constraints by filtering the solutions afterward. However, one could, in principle, try to exploit numerical methods to see if perturbing the found solutions with the LB-Explorer can satisfy more conditions. Instead of trying it on the equivariance or the spectrum conditions, which would have required solving many equations and, in particular, would have been specific to each CICYs, we have decided to remove the chirality condition among the constraints learned by the LB-Explorer and check if those conditions can be solved exactly afterward with a method we are now going to describe.

We implemented a ``Phase 0" trigger that halts the training and saves a checkpoint when the scores for all the conditions except for the chirality of the spectrum are above a specific threshold. In this way, we declare that the LB-Explorer has learned all the conditions, and the matrices found are those satisfying everything except for the chiral condition. These predictions $\mathbf{K}_0$ are then almost solutions of Table \ref{tab:conditions}, and we can perturb them so that they satisfy the missing condition using the CP-SAT module \cite{cpsatlp}.

\subsection{Solving Constraints via CP-SAT}
\label{sec:cpsat_closure}

CP-SAT is an optimization solver specifically designed for discrete optimization problems. Given a set of constraints for a given problem, CP-SAT uses those constraints to perform a logic and tree search of the solution space instead of relying on gradient descent. Moreover, when the solver makes a series of guesses that lead to a violation of these constraints, it generates new logical rules (called ``clauses") that summarize those failures and adds them to the already existing constraints to prevent the solver from exploring those invalid branches again. Finally, once a valid solution is found, it adds a hidden constraint that guides the search for the next solution and continues until no better solution exists.

The objective of the CP-SAT module is to find a perturbed matrix $\mathbf{K}'$ that satisfies all constraints in Table \ref{tab:conditions}, while minimizing the deviation from the predicted $\mathbf{K}_0$. In formulas, this is achieved by minimizing the distance between the two matrix configurations, i.e.
\begin{equation}
    \min_{\mathbf{K}'} \|\mathbf{K}' - \mathbf{K}_0\|_1 = \min_{\mathbf{K}'} \sum_{i=1}^{h^{1,1}(\IX_3)} \sum_{a=1}^{5} |\mathbf{K}^{'\,i}_{a} - \mathbf{K}_{0|a}^i| \coma
\end{equation}
subject to the condition that $\mathbf{K}'$ is also a valid vector bundle. Bounding the distance ensures that the solver searches the local neighborhood discovered by the LB-Explorer rather than restarting the search from scratch.

\subsection{Results with CP-SAT}

We applied the Hybrid LB-Explorer to those CICYs in Table \ref{tab:CICYdataset} for which we could not find any solutions using LB-Explorer. The reason why the LB-Explorer was not able to find any solutions for these CICYs is most probably due to the sparsity of the solutions, since they are CICYs with small $h^{1,1}(\IX_3)$. These CICYs have been the object of exact scans \cite{Anderson:2013xka} in the past, while our LB-Explorer, as we have proven in the previous sections, is better suited to explore a large space of solutions. However, precisely these CICYs can represent a good benchmark to check whether solving exact conditions with CP-SAT can be a solution to augment the LB-Explorer with conditions we have not imposed, such as the equivariant structure.  

The results of our use of CP-SAT for these CICYs are shown in Table \ref{tab:detailed_counts_CP} and Figure \ref{fig:percentage_solutions_CP}. We clearly see that, even if CP-SAT is able to find solutions satisfying the chirality conditions, it shows promising outcomes if one decides to use CP-SAT to impose other kinds of exact conditions. In fact, we see that without imposing the equivariant structure, the perturbed solutions produced by CP-SAT are not enough to guarantee that they also satisfy \cref{eq:equivariantcond_trivial,eq:equivariantcond_nontrivial}, meaning that they should have been added among the constraints of CP-SAT. The situation is even clearer once we consider the only CICY in Table \ref{tab:detailed_counts_CP} that has some equivariant solutions (i.e., CICY 7862), and we compute the line bundle cohomologies. As we show in Table \ref{tab:spectrum_stats_CP}, for solutions with entries $|\mathbf{k}_a|\leq 2$, using \texttt{pyCICY}, all the solutions are excluded.\footnote{It is worth noting that the computation of the line bundle cohomologies has been done over only $10\%$ of the actual equivariant solutions found by CP-SAT, but the remaining solutions have entries $|\mathbf{k}_a|> 2$.} 

Nevertheless, CP-SAT proved to be able to find perturbed solutions from those found by the LB-Explorer, meaning that it can be considered a more efficient way to filter solutions satisfying conditions not learned by the LB-Explorer while also being able to find more solutions. It would be interesting to test CP-SAT on some selected CICYs to see if it can be used to impose the equivariant structure instead of relying on filtering the solutions in post-processing. We plan to return to this in the future.

\begin{longtable}{c | c | c | c | c | c | c}
\caption{Spectrum analysis summary. The meaning of $N$, $N_{S_5}$, $N_{G_{\mathbf{X}_3}}$ and $N_{\text{full}}$ is the same as in Table \ref{tab:detailed_counts}.  In the `Solutions' column we show only the solutions with entries $|\mathbf{k}_a|\leq 2$. `Equivariance' shows solutions in the range $|\mathbf{k}_a|\leq 2$ that also satisfy \cref{eq:equivariantcond_trivial,eq:equivariantcond_nontrivial} for the equivariance of line bundles under the action of $\Gamma$. `Spectrum' counts the solutions satisfying the spectrum constraints in \eqref{eq:fullspectrum}. The line bundle cohomologies have been computed with \texttt{pyCICY} \cite{Larfors:2019sie}. We show in `Errors', the number of solutions for which \texttt{pyCICY} reported a warning that the computation was not reliable.} \label{tab:spectrum_stats_CP} \\
\hhline{=|=|=|=|=|=|=}
\# & $h^{1,1}(\IX_3)$ & $|\Gamma|$ & & Sol ($|\mathbf{k}_a|\leq 2$) & Equivariant & Spectrum (Errors) \\
\hhline{=|=|=|=|=|=|=}
\endfirsthead
\multirow{4}{*}{7862} & \multirow{4}{*}{4} & \multirow{4}{*}{4} & $N$ & 2525 & 231 & 0 (0) \\
 &  &  & $N_{S_5}$ & 24 & 2 & 0 (0) \\
 &  &  & $N_{G_{\mathbf{X}_3}}$ & 120 & 120 & 0 (0) \\
 &  &  & $N_{\text{full}}$ & 1 & 1 & 0 (0) \\ \hline
\end{longtable}

\section{Conclusions and Outlooks}
\label{sec:conclusions}

In this work, we present a novel approach to systematically explore heterotic line bundle standard models using advanced machine learning techniques. We demonstrate the scalability of LB-Explorer by successfully identifying thousands of valid configurations on CICY manifolds admitting freely-acting symmetries. In particular, Section \ref{sec:transferlearning} is devoted to an exhaustive transfer learning analysis that truly demonstrates the power of the LB-Explorer. The conclusions that can be drawn from the transfer learning are that, generally, the LB-Explorer is able to learn the fundamental conditions in Table \ref{tab:conditions}, adapting itself to the new geometries to which it is trained. The asymmetric directions over which the transfer learning performs better naturally lead one to believe that training the LB-Explorer on large CICYs with large freely-acting symmetry and small $|G_{\IX_3}|$ can be used to create a generic explorer that will adapt to specific CICYs in a few episodes. This result and the scalability of the LB-Explorer are the most prominent novelties of our work. While we restricted our analysis to CICYs, the framework and the conditions imposed are general enough to be readily applied to other CY geometries, such as those realized as toric hypersurface singularities \cite{Kreuzer:2000xy}, provided they admit a simplicial Mori cone and a known freely-acting symmetry.

Looking forward, there are several promising extensions to this work. From a computational perspective, the bottleneck posed by line bundle cohomology computations could be alleviated by integrating our LB-Explorer with emerging ML techniques designed to compute these cohomologies directly \cite{Klaewer:2018sfl, Brodie:2019dfx, Brodie:2021nit, Constantin:2021for, Constantin:2024ulu}. Furthermore, while our hybrid CP-SAT approach proved successful, a natural next step would be to embed the equivariant structure constraints built-into the RL reward environment to eliminate post-generation filtering entirely.

From a phenomenological perspective, the valid Abelian configurations identified by the LB-Explorer at the split locus serve as robust topological stepping stones. A physical extension will be to explicitly construct the non-Abelian deformations of these bundles, which are required to break residual global symmetries, generate realistic Yukawa couplings, and achieve full complex structure moduli stabilization.

Finally, we believe that the architectures proposed here can be adapted to tackle numerous analogous problems in the string landscape. One immediate extension is applying a similar RL framework to the search for perturbatively flat flux vacua in Type IIB compactifications \cite{Demirtas:2019sip, Alvarez-Garcia:2020pxd, Demirtas:2020ffz}, which has thus far been approached only via exact scans \cite{Carta:2021kpk, Carta:2022oex}.
It would be even more interesting to see if our autoregressive RL explorer can discover new algorithms to find vacua with small cosmological constants that have not been previously designed by humans.
Another immediate application of our RL explorer is in finding realistic D-brane models (see
\cite{Blumenhagen:2005mu} for a review). Machine learning guided searches for such models have so far been done only for a single orientifold background \cite{Halverson:2019tkf, Loges:2021hvn,Loges:2022mao}. It is conceivable that the versatile architecture we developed here can tackle the string landscape at scale.

We anticipate that this work, alongside contemporary efforts such as \cite{Yip:2025hon,Arnal:2026zyo}, establishes a solid foundation for the modern integration of Transformers in string phenomenology, proving that these sequence-modeling architectures are uniquely suited for exploring vast topological landscapes and making unprecedented levels of scalability accessible.

\subsubsection*{Acknowledgments}

The authors thank Federico Carta, Cesar Fierro Cota, Martina Danese, Bj\"orn Hassfeld, Magdalena Larfors, Jakob Moritz, Andreas Schachner, Moritz Walden, and Timo Weigand for helpful discussions and comments. The authors also thank Moritz M\"unchmeyer for access to his group's multi-GPU server during the initial exploratory phase of the project. This work is supported in part by the DOE grants DE-SC0017647 and DE-SC-0023719. Software development was assisted by Gemini Pro 3.1 (Google) and Claude Opus 4.7 (Anthropic). This work was finalized while A. M. and G. S. were at the Aspen Center for Physics, which is supported by National Science Foundation grant PHY-2210452.

\appendix

\section{Complete Intersection Calabi--Yau Threefolds}
\label{app:CICYreview}

In this appendix, we review the basic concepts and properties of CICYs \cite{Candelas:1987kf}. These manifolds are described as algebraic varieties obtained as intersections of the zero loci of $k$ homogeneous polynomials $p_j(k)$ in an ambient space $\mathcal{A}$, which is taken to be a product of projective spaces $\PP^{n_1}\times \ldots \times \PP^{n_m}$. The CY $\IX_3$ is called ``complete intersection" if the complex dimension of $\IX_3$ is equal to the dimension of $\mathcal{A}$ minus the number of polynomials. Since we are interested in threefolds, the constraint is
\begin{equation}
    \sum_{i=1}^m n_i - k = 3\fstop
\end{equation}
Each polynomial $p_j(k)$ is characterized by its multi-degree $q_j^i$, with $j = 1,\ldots, k$ and $i = 1,\ldots, m$ specifying the degrees in the homogeneous coordinates of each $\PP^{n_i}$. This information is encoded in a ``configuration matrix":
\begin{equation}
\left[
\begin{tabular}{c|cccc}
$\PP^{n_1}$ &   $q_1^1$ & $\cdots$  & $q_k^1 $ \\
$\PP^{n_2}$  &   $q_1^2$ & $\cdots$  &$ q_k^2$  \\
$\vdots$ &   $\vdots$ & $\ddots$ & $\vdots$  \\ 
$\PP^{n_m} $&   $q_1^m $& $\cdots$ & $q_k^m $
\end{tabular}
\right] \fstop
\label{eq:configuration}
\end{equation}
The resulting manifold is CY if we impose that $c_1(T\IX_3)$ vanishes, i.e.
\begin{equation}\label{eq:CYconditionCICY}
    \sum_{j=1}^k q_j^i = n_i +1 \coma \forall i = 1,\ldots,m\fstop
\end{equation}
Each homogeneous polynomial $p_j(x)$ can be thought of as a holomorphic section of a line bundle over the ambient space $\mathcal{A}$, defined as:
\begin{equation}
\mathcal{O}_{\mathcal{A}}(q_j) = \bigotimes_{i=1}^{m} \pi_{i}^{*} \mathcal{O}_{\mathbb{P}^{n_i}}(q_j^i)\, ,
\end{equation}
where $\pi_i: \mathcal{A} \to \mathbb{P}^{n_i}$ is the projection onto the $i$-th factor, and $\mathcal{O}_{\mathbb{P}^{n_i}}(1)$ is the standard hyperplane line bundle. Let $H_i = c_1(\pi_{i}^{*} \mathcal{O}_{\mathbb{P}^{n_i}}(1))$ denote the hyperplane classes generating $H^2(\mathcal{A}, \mathbb{Z})$. With $\mathcal{O}_{\mathcal{A}}(q_j)$, we define the normal bundle $\mathcal{N}_{\IX_3/\mathcal{A}}$ to $\IX_3$ as the direct sum of the $k$ line bundles:
\begin{equation}
\mathcal{N}_{\IX_3/\mathcal{A}} = \bigoplus_{j=1}^{k} \mathcal{O}_{\mathcal{A}}(q_j)\, .
\end{equation}

We restrict our attention to CICYs for which the second cohomology descends entirely from the ambient space. Let $\iota: \IX_3 \hookrightarrow \mathcal{A}$ be the embedding of the CY manifold. Strictly speaking, defining $S=\bigoplus_{i=1}^m \mathcal{O}_{\IX_3}(\mathbf{e}_i)^{\oplus(n_i+1)}$ (where $\mathbf{e}_i$ are the standard unit vectors in $m$ dimensions) and noting $N=\mathcal{N}_{\IX_3/\mathcal{A}}$, a CICY is called ``favorable'' if ${\rm Coker}(H^1(\IX_3,S)\rightarrow H^1(\IX_3,N))=0$ and $H^2(\IX_3,S)=0$ \cite{Anderson:2007nc, Anderson:2008uw}. Under these conditions, the restriction map $\iota^* : H^2(\mathcal{A}, \mathbb{Z}) \to H^2(\IX_3, \mathbb{Z})$ is surjective. As a consequence, $h^{1,1}(\IX_3) = h^{1,1}(\mathcal{A}) = m$.

We define the basis of divisors $J_i$ on $\IX_3$ as the pullback of the ambient hyperplanes:
\begin{equation}
J_i = \iota^* H_i\, , \quad i = 1, \dots, m\, .
\end{equation}
Assuming these divisors span the K\"ahler cone of $\IX_3$, we can expand the K\"ahler form on $\IX_3$ as:
\begin{equation}
J = t^i J_i\, , \quad t^i > 0\, ,
\end{equation}
where $t^i$ are the K\"ahler moduli corresponding to the volumes of the generators of the Mori cone. We define the triple intersection number $\kappa_{abc}$ as
\begin{equation}
    \kappa_{abc} = \int_{\IX_3} J_a \wedge J_b \wedge J_c\fstop
\end{equation}
Since the divisors descend from the hyperplane class of the $\PP^{n_i}$ in the ambient space, we can lift the integral to an integral over the ambient space, i.e.
\begin{equation}\label{eq:TripleIntersectionsCICY}
     \kappa_{abc} = \int_{\mathcal{A}} J_a \wedge J_b \wedge J_c \wedge c_k(\mathcal{N}_{\IX_3/\mathcal{A}})=\int_{\mathcal{A}} J_a \wedge J_b \wedge J_c \bigwedge_{j=1}^k\left(\sum_{i=1}^mq_j^i J_i\right)\fstop
\end{equation}
The advantage comes from the ambient space $\mathcal{A}$ being a Cartesian product, so that the integral factorizes as an integral over the single $\PP^{n_i}$ factors. As a consequence, the only time when such integrals are non-zero is when the integrand exactly contains the volume form of $\PP^{n_i}$; that is, for every $\PP^{n_i}$, the integrand is non zero if it involves a wedge product of $n_i$ $J_i$s. The triple intersection number corresponds to the overall coefficient multiplying the volume form of $\mathcal{A}$. Moreover, using $J_i$, we can compute the total Chern class of $\IX_3$ by adjunction:
\begin{equation}\label{eq:adjunctionCICY}
    c(T\IX_3) = \frac{c(T\mathcal{A}|_{\IX_3})}{c(\mathcal{N}_{\IX_3/\mathcal{A}})} = \dfrac{\displaystyle\prod_{i=1}^m(1+J_i)^{n_i+1}}{\displaystyle\prod_{j=1}^k\left(1+\sum_{a=1}^mq_j^aJ_a\right)}\fstop
\end{equation}
Note that \eqref{eq:CYconditionCICY} comes from requiring $c_1(T\IX_3) = 0$ in \eqref{eq:adjunctionCICY}. Moreover, from \eqref{eq:adjunctionCICY}, one can compute the Euler characteristic of $\IX_3$ and, for favorable CICYs, extract  $h^{2,1}(\IX_3)$.

\subsection{Symmetries of CICYs}
\label{app:CICYsymmetries}

By construction, the ambient space $\mathcal{A}$ exhibits permutation symmetries when it contains multiple identical projective factors $\PP^{n_i}$. At the level of the configuration matrix \eqref{eq:configuration}, this manifest symmetry group $G_{\text{conf}}$, corresponds to the set of row permutations (swapping identical $\PP^{n_i}$ factors) that leave the matrix invariant, up to compensating column permutations (swapping polynomials). Elements of $G_{\text{conf}}$ represent geometric automorphisms of the defining equations in the ambient space.

However, $G_{\text{conf}}$ is generally a subgroup of the full symmetry group of the resulting CY manifold $G_{\IX_3}$. By Wall's theorem, the diffeomorphism class of a simply connected CY threefold is determined by its Hodge numbers, the intersection ring $\kappa_{abc}$, and the second Chern class evaluated on the basis divisors $c_{2,a}$. Thus, the symmetry group of $\IX_3$ corresponds to the automorphisms of the intersection ring that simultaneously preserve $c_{2,a}$. In the basis of divisors that corresponds to the generators of the K\"ahler cone, $G_{\IX_3}$ is the subgroup of the symmetric group $S_{h^{1,1}(\IX_3)}$ defined by
\begin{equation}\label{eq:GX3def}
    G_{\IX_3} = \left\{ \sigma \in S_{h^{1,1}(\IX_3)} \mid \kappa_{abc} = \kappa_{\sigma(a)\sigma(b)\sigma(c)} \text{ and } c_{2,a} = c_{2,\sigma(a)} \coma \forall a,b,c \right\}\fstop
\end{equation}
In the main text and in Table \ref{tab:CICYdataset}, when we write $G_{\IX_3}$, it denotes the group generated by \eqref{eq:GX3def}. We computed this group by mapping each divisor to a node in a graph. These nodes are divided into groups based on their values of $c_2(T\IX_3)$, and we restrict the algorithm to swap only nodes within the same group. We also introduce ``tensor nodes" for every non-zero intersection number, similarly grouped by their numerical values. These tensor nodes connect to their corresponding divisor nodes using labels denoting their power (e.g., a term $J_a^2J_b$ creates a connection to $J_a$ with label $2$ and a connection to $J_b$ with label $1$). Finally, we used \texttt{GAP} \cite{GAP4} within \texttt{SageMath} \cite{sagemath} to compute the graph symmetries that preserve these groupings of the divisor and tensor nodes. The code is included in the \href{https://github.com/alexmininno/LB-Explorer}{GitHub repository}. 

To illustrate the difference between $G_{\text{conf}}$ and $G_{\IX_3}$, consider, e.g., CICY 6281 in the enumeration of \cite{Anderson:2017aux}. This CICY has $h^{1,1}=6$ and $h^{2,1}=34$, defined by the following configuration matrix:
\begin{equation}
\left[
\begin{tabular}{c|cccccccc}
$\PP^2$ & 1 & 1 & 1 & 0 & 0 & 0 & 0 & 0 \\
$\PP^1$ & 0 & 0 & 1 & 0 & 0 & 1 & 0 & 0 \\
$\PP^1$ & 0 & 0 & 0 & 1 & 1 & 0 & 0 & 0 \\
$\PP^1$ & 0 & 1 & 0 & 0 & 0 & 0 & 1 & 0 \\
$\PP^1$ & 1 & 0 & 0 & 0 & 0 & 0 & 1 & 0 \\
$\PP^5$ & 0 & 0 & 0 & 1 & 1 & 1 & 1 & 2
\end{tabular}
\right] \fstop
\end{equation}
The configuration symmetry group is $G_{\text{conf}} \cong \mathbb{Z}_2$, generated by the row permutation $\langle(45)\rangle$, corresponding to swapping the fourth and fifth rows of the configuration matrix. The original matrix, after swapping row 4 and row 5, is restored by swapping columns 1 and 2. Because column permutations reorder the polynomial equations without altering the CY locus, $\langle(45)\rangle$ represents a valid symmetry. 

We can extract the triple intersection numbers and the second Chern class of the $\IX_3$ using the method described in Section \ref{app:CICYreview}. The triple intersection ring, encoded in the intersection ring $\mathcal{I} = \kappa_{abc} J_a J_b J_c$, is given by:
\begin{equation}
\begin{split}
    \mathcal{I} =& \,\, 2 J_1^2 J_3 + 2 J_1 J_2 J_3 + 2 J_1 J_3 J_4 + 2 J_2 J_3 J_4 + 2 J_1 J_3 J_5 + 2 J_2 J_3 J_5 + 2 J_3 J_4 J_5  \\
    &+ 4 J_1^2 J_6 + 4 J_1 J_2 J_6 + 6 J_1 J_3 J_6 + 4 J_2 J_3 J_6 + 4 J_1 J_4 J_6 + 4 J_2 J_4 J_6 + 4 J_3 J_4 J_6 \\
    &+ 4 J_1 J_5 J_6 + 4 J_2 J_5 J_6 + 4 J_3 J_5 J_6 + 4 J_4 J_5 J_6 + 12 J_1 J_6^2 + 8 J_2 J_6^2 + 4 J_3 J_6^2 \\
    &+ 8 J_4 J_6^2 + 8 J_5 J_6^2 + 8 J_6^3 \coma
\end{split}
\end{equation}
while the second Chern class, evaluated on the basis of divisors, is 
\begin{equation}
    c_2(T\IX_3)\cdot J = \{36, 24, 24, 24, 24, 56\}\fstop
\end{equation}
By constructing the graph for this CICY, we find that the symmetry group is $G_{\IX_3} \cong S_3$ (of order 6), generated by the permutations $\langle(45)\rangle$ and $\langle(24)\rangle$. 

We see, then, that the generator $\langle(24)\rangle$ arises at the level of the CY by $J_2$ and $J_4$, leaving $\mathcal{I}$ and $c_{2,a}$ invariant. Yet, at the level of the configuration matrix, swapping row 2 and row 4 cannot be resolved by any column permutation.

\subsubsection{Coordinate Symmetries and Quotient Manifolds}
\label{app:CoordinateSymmetries}

While $G_{\text{conf}}$ and $G_{\IX_3}$ describe automorphisms of the configuration matrix and the intersection ring, respectively, the explicit construction of non-simply connected CY manifolds requires analyzing symmetries at the level of the ambient space coordinates. 

For our purposes, we are interested in freely-acting symmetries in CICY manifolds \cite{Candelas:1987du,Candelas:2008wb,Candelas:2010ve}. A complete classification of all freely-acting symmetries that descend from linear actions on the projective ambient space has been completed in \cite{Braun:2010vc}.\footnote{There has been less work on freely-acting discrete symmetries of CY manifolds defined as hyper-surfaces in toric four-folds. The underlying reflexive polytopes were fully classified in \cite{Batyrev:2005jc}, where they also identified 16 reflexive polytopes admitting a freely-acting symmetry. However, the identification of freely-acting discrete symmetries on these spaces was initiated in \cite{Braun:2017juz}.} Crucially, a given CY can be embedded in multiple different products of projective spaces; whether a symmetry action is realized linearly or non-linearly depends on this specific choice of embedding \cite{Braun:2010vc,Candelas:2008wb}.

As we explained in the main text, such freely-acting symmetries of CY manifolds are useful for constructing a non-trivial fundamental group by forming quotients (see e.g.  \cite{Candelas:2008wb,Candelas:2010ve,Candelas:2015amz,Candelas:2016fdy,Constantin:2016xlj} for work in that direction). The existence of such quotients of CY manifolds is important for the existence of Wilson lines that would break the GUT group to the Standard Model.

Following the convention of \cite{Lukas:2017vqp}, let $\mathbf{x}_i = [x_{i,0} : \cdots : x_{i,n_i}]$ denote the homogeneous coordinates of the $i$-th projective factor $\PP^{n_i}$ in the ambient space $\mathcal{A}$. A discrete symmetry group $\Gamma$ acts linearly on $\mathcal{A}$. An element $g \in \Gamma$ acts via a combination of a permutation $\sigma \in S_m$ of projective factors and a linear transformation on the coordinates of each factor. Specifically, the action takes the form:
\begin{equation}
    g \cdot \mathbf{x}_i = M_{i} \, \mathbf{x}_{\sigma^{-1}(i)} \coma
\end{equation}
where $M_i \in \text{PGL}(n_i + 1, \mathbb{C})$ is a non-singular transformation matrix, typically restricted to a diagonal matrix of roots of unity. 

For $\IX_3 \subset \mathcal{A}$ to be invariant under $\Gamma$, the zero locus of the defining polynomials $\{p_j(\mathbf{x})\}_{j=1}^k$ must be mapped to itself. This means that the ideal generated by the polynomials must be invariant, and the polynomials transform into linear combinations of themselves of the same multi-degree:
\begin{equation}
    p_j(g \cdot \mathbf{x}) = \sum_{l=1}^k {C_j}^l(g) \, p_l(\mathbf{x}) \coma
\end{equation}
where $C(g)$ is an invertible $k \times k$ constant matrix forming a representation of the group $\Gamma$. Because $g$ must map equations to equations of the same multi-degree, the permutation $\sigma$ must leave the configuration matrix \eqref{eq:configuration} invariant (up to column permutations). Therefore, the permutation part of any valid coordinate symmetry must be an element of the configuration symmetry group: $\sigma \in G_{\text{conf}}$.

Once the invariant coordinate actions are identified, they are classified by their fixed-point structure:
\begin{enumerate}
    \item \textit{Freely Acting Symmetries:} If $\Gamma$ has no fixed points on $\IX_3$ (i.e., for all non-trivial $g \in \Gamma$ and $g \cdot x \neq x \,\, \forall x \in \IX_3$), the symmetry can be used to define a smooth quotient space $\widetilde{\IX}_3 = \IX_3 / \Gamma$. This quotient is a CY manifold with fundamental group $\pi_1(\widetilde{\IX}_3) \cong \Gamma$ and a reduced Euler characteristic $\chi(\widetilde{\IX}_3) = \chi(\IX_3)/|\Gamma|$.
    \item \textit{Non-Freely Acting Symmetries:} If fixed points exist, the symmetry cannot be used to quotient the manifold without introducing singularities. However, these non-freely acting groups frequently survive as regular symmetries \cite{Lukas:2017vqp}, leading to discrete flavor symmetries in the four-dimensional EFT.
\end{enumerate}

These symmetries can be related to $G_{\text{conf}}$ and $G_{\IX_3}$ discussed in Appendix \ref{app:CICYsymmetries}. The action of $\Gamma$ on the coordinates induces a pullback $g^*$ on the cohomology group $H^{1,1}(\IX_3)$. In the basis of divisors given by $J_i$ corresponding to the hyperplanes of $\PP^{n_i}$, the phase matrices $M_i$ act trivially on the divisor classes, meaning the induced action on the K\"ahler generators is dictated by the permutation $\sigma$:
\begin{equation}
    g^* J_i = J_{\sigma(i)} \fstop
\end{equation}
Consequently, there is a natural group homomorphism projecting the coordinate symmetry group $\Gamma$ onto a subgroup of the configuration symmetry group $G_{\text{conf}}$. It is using this information that we have updated the database of freely-acting symmetries provided \href{https://www-thphys.physics.ox.ac.uk/projects/CalabiYau/cicylist/index.html}{here} with the entries from the database of the CICYs in \cite{Anderson:2017aux}.

\begin{landscape}
\section{Tables}
\label{app:tables}

\pagestyle{empty}

\setlength{\LTleft}{\fill}
\setlength{\LTright}{\fill}
\setlength{\LTcapwidth}{\linewidth}
\renewcommand{\arraystretch}{1}
\begin{longtable}{c | c | c | c | c | c |  c | c | c | c | c | c | c}
\caption{Counting of solutions found by LB-Explorer. $N$ denotes the total number of solutions found, $N_{S_5}$ are the solutions unique under columns permutations, $N_{G_{\mathbf{X}_3}}$ are the solutions unique under the group action $G_{\mathbf{X}_3}$, and $N_{\text{full}}$ are the solutions unique under both $S_5$ and $G_{\mathbf{X}_3}$. `Dup.' indicates solutions found by multiple seeds. In fact, the counting of solutions for each seed is absolute, and some solutions can appear multiple times across seeds. `Total' is the count of solutions with the duplicated ones counted once. For a visualization of the distributions of solutions across seeds, we refer to Figure \ref{fig:percentage_solutions}. Finally, `Equivariance'  contains the number of solutions satisfying \cref{eq:equivariantcond_trivial,eq:equivariantcond_nontrivial} for the equivariance of line bundles under the action of $\Gamma$. If $\Gamma$ is acting trivially on the divisor basis, we put a \checkmark.} \label{tab:detailed_counts} \\
\hhline{=|=|=|=|=|=====|=|=|=}
\multirow{2}{*}{\#} & \multirow{2}{*}{$h^{1,1}$} & \multirow{2}{*}{$|\Gamma|$} & \multirow{2}{*}{$|G_{\mathbf{X}_3}|$} & \multirow{2}{*}{Sol.} & \multicolumn{5}{|c|}{Seeds} & \multirow{2}{*}{Dup. $\left(\frac{N_\text{dup}}{N_\text{tot}}\%\right)$} & \multirow{2}{*}{Total} & \multirow{2}{*}{Equivariance} \\ \cline{6-10}
 &   &  &  &  & 42 & 43 & 44 & 45 & 46 &  &  & \\
\hhline{=|=|=|=|=|=|=|=|=|=|=|=|=}
\endfirsthead
\multicolumn{13}{c}{{\tablename\ \thetable\ -- \textit{Continued from previous page}}} \\
\hhline{=|=|=|=|=|=====|=|=|=}
\multirow{2}{*}{\#} & \multirow{2}{*}{$h^{1,1}$} & \multirow{2}{*}{$|\Gamma|$} & \multirow{2}{*}{$|G_{\mathbf{X}_3}|$} & \multirow{2}{*}{Sol.} & \multicolumn{5}{|c|}{Seeds} & \multirow{2}{*}{Dup. $\left(\frac{N_\text{dup}}{N_\text{tot}}\%\right)$} & \multirow{2}{*}{Total} & \multirow{2}{*}{Equivariance} \\ \cline{6-10}
 &   &  &  &  & 42 & 43 & 44 & 45 & 46 &  &  & \\
\hhline{=|=|=|=|=|=|=|=|=|=|=|=|=}
\endhead
\multicolumn{13}{r}{\textit{Continued on next page}} \\
\endfoot
\hhline{=|=|=|=|=|=|=|=|=|=|=|=|=}
\endlastfoot
\multirow{4}{*}{7862} & \multirow{4}{*}{4} & \multirow{4}{*}{4} & \multirow{4}{*}{24} & $N$ & 884 & 893 & 859 & 954 & 888 & 1153 (71.0\%) & 1623 & 169 \checkmark \\
 &  &  &  & $N_{S_5}$ & 30 & 28 & 28 & 32 & 28 & 30 (81.1\%) & 37 & 15 \checkmark \\
 &  &  &  & $N_{G_{\mathbf{X}_3}}$ & 115 & 89 & 94 & 128 & 110 & 126 (86.9\%) & 145 & 29 \checkmark \\
 &  &  &  & $N_{\text{full}}$ & 2 & 2 & 3 & 2 & 2 & 2 (66.7\%) & 3 & 2 \checkmark \\ \hline
\multirow{4}{*}{7403} & \multirow{4}{*}{4} & \multirow{4}{*}{2} & \multirow{4}{*}{4} & $N$ & 576 & 286 & 0 & 287 & 1175 & 288 (16.4\%) & 1751 & 0 \\
 &  &  &  & $N_{S_5}$ & 98 & 48 & 0 & 48 & 196 & 48 (16.3\%) & 294 & 0 \\
 &  &  &  & $N_{G_{\mathbf{X}_3}}$ & 290 & 144 & 0 & 144 & 887 & 144 (12.2\%) & 1177 & 0 \\
 &  &  &  & $N_{\text{full}}$ & 50 & 24 & 0 & 24 & 148 & 48 (32.0\%) & 150 & 0 \\
\hline
\end{longtable}

\newpage

\begin{figure}[!htp]
\caption{Percentage of solutions as in Table \ref{tab:detailed_counts} by seeds. Solutions found uniquely by Seeds 42, 43, 44, 45 46 are, respectively, shown in {\color{blue!60}{blue}}, {\color{red!60}{red}}, {\color{green!60}{green}}, {\color{orange!60}{orange}} and {\color{purple!60}{purple}}. Solutions that have been found by multiple seeds are shown in {\color{gray!60}{gray}}.}\label{fig:percentage_solutions}
    \centering
    \pgfplotsset{
        compat=1.18,
        every axis/.append style={
            width=\textwidth, 
            height=4.5cm, 
            ymin=0, ymax=110, ybar stacked, bar width=12pt,
            xtick={0,1,2,3}, xticklabels={$N$, $N_{S_5}$, $N_{G_{\mathbf{X}_3}}$, $N_{\text{full}}$},
            enlarge x limits=0.2, 
            legend style={cells={anchor=west}, legend pos=north east, font=\tiny},
            legend image code/.code={\draw[#1, draw=none] (0cm,-0.1cm) rectangle (0.2cm,0.1cm);},
            tick label style={font=\scriptsize},
            label style={font=\footnotesize}
        }
    }
    \pgfplotstableread{
Label S42 S43 S44 S45 S46 Dup
N 5.9 4.1 3.2 10.1 5.7 71.0
NS5 2.7 0.0 2.7 10.8 2.7 81.1
NGX3 2.8 1.4 2.8 5.5 0.7 86.9
Nfull 0.0 0.0 33.3 0.0 0.0 66.7
    }\dataCB
    \begin{subfigure}{0.33\textwidth}
        \centering
        \begin{tikzpicture}
            \begin{axis}[ylabel={Solutions (\%)},]
                \addplot [fill=blue!60, draw=black!80] table [y=S42, x expr=\coordindex] {\dataCB};
                \addplot [fill=red!60, draw=black!80] table [y=S43, x expr=\coordindex] {\dataCB};
                \addplot [fill=green!60, draw=black!80] table [y=S44, x expr=\coordindex] {\dataCB};
                \addplot [fill=orange!60, draw=black!80] table [y=S45, x expr=\coordindex] {\dataCB};
                \addplot [fill=purple!60, draw=black!80] table [y=S46, x expr=\coordindex] {\dataCB};
                \addplot [fill=gray!60, draw=black!80] table [y=Dup, x expr=\coordindex] {\dataCB};
            \end{axis}
        \end{tikzpicture}
        \caption*{CICY 7862, $|\Gamma|=4$}
    \end{subfigure}\hfill
    \pgfplotstableread{
Label S42 S43 S44 S45 S46 Dup
N 16.4 0.0 0.0 0.0 67.1 16.4
NS5 17.0 0.0 0.0 0.0 66.7 16.3
NGX3 12.4 0.0 0.0 0.0 75.4 12.2
Nfull 1.3 0.0 0.0 0.0 66.7 32.0
    }\dataBU
    \begin{subfigure}{0.33\textwidth}
        \centering
        \begin{tikzpicture}
            \begin{axis}[]
                \addplot [fill=blue!60, draw=black!80] table [y=S42, x expr=\coordindex] {\dataBU};
                \addplot [fill=red!60, draw=black!80] table [y=S43, x expr=\coordindex] {\dataBU};
                \addplot [fill=green!60, draw=black!80] table [y=S44, x expr=\coordindex] {\dataBU};
                \addplot [fill=orange!60, draw=black!80] table [y=S45, x expr=\coordindex] {\dataBU};
                \addplot [fill=purple!60, draw=black!80] table [y=S46, x expr=\coordindex] {\dataBU};
                \addplot [fill=gray!60, draw=black!80] table [y=Dup, x expr=\coordindex] {\dataBU};
            \end{axis}
        \end{tikzpicture}
        \caption*{CICY 7403, $|\Gamma|=2$}
    \end{subfigure}\hfill
    \pgfplotstableread{
Label S42 S43 S44 S45 S46 Dup
N 2.9 2.4 2.8 13.6 11.3 67.0
NS5 0.0 0.0 1.4 0.0 0.0 98.6
NGX3 0.6 0.5 0.8 7.2 4.4 86.4
Nfull 0.0 0.0 7.7 0.0 0.0 92.3
    }\dataBR
    \begin{subfigure}{0.33\textwidth}
        \centering
        \begin{tikzpicture}
            \begin{axis}[]
                \addplot [fill=blue!60, draw=black!80] table [y=S42, x expr=\coordindex] {\dataBR};
                \addplot [fill=red!60, draw=black!80] table [y=S43, x expr=\coordindex] {\dataBR};
                \addplot [fill=green!60, draw=black!80] table [y=S44, x expr=\coordindex] {\dataBR};
                \addplot [fill=orange!60, draw=black!80] table [y=S45, x expr=\coordindex] {\dataBR};
                \addplot [fill=purple!60, draw=black!80] table [y=S46, x expr=\coordindex] {\dataBR};
                \addplot [fill=gray!60, draw=black!80] table [y=Dup, x expr=\coordindex] {\dataBR};
            \end{axis}
        \end{tikzpicture}
        \caption*{CICY 7247, $|\Gamma|=3$}
    \end{subfigure}\hfill
    \pgfplotstableread{
Label S42 S43 S44 S45 S46 Dup
N 0.0 0.0 0.0 5.7 94.2 0.1
NS5 0.0 0.0 0.0 13.6 82.6 3.8
NGX3 0.0 0.0 0.0 13.3 86.3 0.4
Nfull 0.0 0.0 0.0 23.7 72.4 3.9
    }\dataBQ
    \begin{subfigure}{0.33\textwidth}
        \centering
        \begin{tikzpicture}
            \begin{axis}
                \addplot [fill=blue!60, draw=black!80] table [y=S42, x expr=\coordindex] {\dataBQ};
                \addplot [fill=red!60, draw=black!80] table [y=S43, x expr=\coordindex] {\dataBQ};
                \addplot [fill=green!60, draw=black!80] table [y=S44, x expr=\coordindex] {\dataBQ};
                \addplot [fill=orange!60, draw=black!80] table [y=S45, x expr=\coordindex] {\dataBQ};
                \addplot [fill=purple!60, draw=black!80] table [y=S46, x expr=\coordindex] {\dataBQ};
                \addplot [fill=gray!60, draw=black!80] table [y=Dup, x expr=\coordindex] {\dataBQ};
            \end{axis}
        \end{tikzpicture}
        \caption*{CICY 7245, $|\Gamma|=2$}
    \end{subfigure}

    \pgfplotstableread{
Label S42 S43 S44 S45 S46 Dup
N 66.2 23.7 0.0 0.0 0.0 10.0
NS5 51.0 0.0 0.0 0.0 0.0 49.0
NGX3 26.3 47.4 0.0 0.0 0.0 26.3
Nfull 0.0 0.0 0.0 0.0 0.0 100.0
    }\dataBL
    \begin{subfigure}{0.33\textwidth}
        \centering
        \begin{tikzpicture}
            \begin{axis}[ylabel={Solutions (\%)},]
                \addplot [fill=blue!60, draw=black!80] table [y=S42, x expr=\coordindex] {\dataBL};
                \addplot [fill=red!60, draw=black!80] table [y=S43, x expr=\coordindex] {\dataBL};
                \addplot [fill=green!60, draw=black!80] table [y=S44, x expr=\coordindex] {\dataBL};
                \addplot [fill=orange!60, draw=black!80] table [y=S45, x expr=\coordindex] {\dataBL};
                \addplot [fill=purple!60, draw=black!80] table [y=S46, x expr=\coordindex] {\dataBL};
                \addplot [fill=gray!60, draw=black!80] table [y=Dup, x expr=\coordindex] {\dataBL};
            \end{axis}
        \end{tikzpicture}
        \caption*{CICY 6831, $|\Gamma|=2$}
    \end{subfigure}\hfill
    \pgfplotstableread{
Label S42 S43 S44 S45 S46 Dup
N 0.0 0.0 100.0 0.0 0.0 0.0
NS5 0.0 0.0 100.0 0.0 0.0 0.0
NGX3 0.0 0.0 100.0 0.0 0.0 0.0
Nfull 0.0 0.0 100.0 0.0 0.0 0.0
    }\dataCA
    \begin{subfigure}{0.33\textwidth}
        \centering
        \begin{tikzpicture}
            \begin{axis}[]
                \addplot [fill=blue!60, draw=black!80] table [y=S42, x expr=\coordindex] {\dataCA};
                \addplot [fill=red!60, draw=black!80] table [y=S43, x expr=\coordindex] {\dataCA};
                \addplot [fill=green!60, draw=black!80] table [y=S44, x expr=\coordindex] {\dataCA};
                \addplot [fill=orange!60, draw=black!80] table [y=S45, x expr=\coordindex] {\dataCA};
                \addplot [fill=purple!60, draw=black!80] table [y=S46, x expr=\coordindex] {\dataCA};
                \addplot [fill=gray!60, draw=black!80] table [y=Dup, x expr=\coordindex] {\dataCA};
            \end{axis}
        \end{tikzpicture}
        \caption*{CICY 7800, $|\Gamma|=3$}
    \end{subfigure}\hfill
    \pgfplotstableread{
Label S42 S43 S44 S45 S46 Dup
N 16.6 7.6 7.3 14.2 11.0 43.3
NS5 5.6 3.5 4.9 4.7 4.2 77.0
NGX3 20.9 3.3 3.2 7.2 5.2 60.1
Nfull 5.2 2.3 2.6 2.6 2.6 84.6
    }\dataBX
    \begin{subfigure}{0.33\textwidth}
        \centering
        \begin{tikzpicture}
            \begin{axis}
                \addplot [fill=blue!60, draw=black!80] table [y=S42, x expr=\coordindex] {\dataBX};
                \addplot [fill=red!60, draw=black!80] table [y=S43, x expr=\coordindex] {\dataBX};
                \addplot [fill=green!60, draw=black!80] table [y=S44, x expr=\coordindex] {\dataBX};
                \addplot [fill=orange!60, draw=black!80] table [y=S45, x expr=\coordindex] {\dataBX};
                \addplot [fill=purple!60, draw=black!80] table [y=S46, x expr=\coordindex] {\dataBX};
                \addplot [fill=gray!60, draw=black!80] table [y=Dup, x expr=\coordindex] {\dataBX};
            \end{axis}
        \end{tikzpicture}
        \caption*{CICY 7487, $|\Gamma|=4$}
    \end{subfigure}\hfill
    \pgfplotstableread{
Label S42 S43 S44 S45 S46 Dup
N 3.9 4.1 6.3 30.2 35.7 19.8
NS5 0.5 0.6 1.1 1.5 36.2 60.1
NGX3 1.2 2.5 2.1 31.5 28.6 34.1
Nfull 0.4 0.5 0.4 0.5 0.4 97.8
    }\dataBV
    \begin{subfigure}{0.33\textwidth}
        \centering
        \begin{tikzpicture}
            \begin{axis}[]
                \addplot [fill=blue!60, draw=black!80] table [y=S42, x expr=\coordindex] {\dataBV};
                \addplot [fill=red!60, draw=black!80] table [y=S43, x expr=\coordindex] {\dataBV};
                \addplot [fill=green!60, draw=black!80] table [y=S44, x expr=\coordindex] {\dataBV};
                \addplot [fill=orange!60, draw=black!80] table [y=S45, x expr=\coordindex] {\dataBV};
                \addplot [fill=purple!60, draw=black!80] table [y=S46, x expr=\coordindex] {\dataBV};
                \addplot [fill=gray!60, draw=black!80] table [y=Dup, x expr=\coordindex] {\dataBV};
            \end{axis}
        \end{tikzpicture}
        \caption*{CICY 7447, $|\Gamma|=2$}
    \end{subfigure}

    \pgfplotstableread{
Label S42 S43 S44 S45 S46 Dup
N 15.3 12.8 6.5 14.5 10.7 40.2
NS5 3.5 5.0 3.0 3.3 36.5 48.8
NGX3 16.6 9.0 2.8 7.6 3.1 60.9
Nfull 6.5 5.0 2.5 2.2 19.2 64.5
    }\dataBW
    \begin{subfigure}{0.33\textwidth}
        \centering
        \begin{tikzpicture}
            \begin{axis}[ylabel={Solutions (\%)},]
                \addplot [fill=blue!60, draw=black!80] table [y=S42, x expr=\coordindex] {\dataBW};
                \addplot [fill=red!60, draw=black!80] table [y=S43, x expr=\coordindex] {\dataBW};
                \addplot [fill=green!60, draw=black!80] table [y=S44, x expr=\coordindex] {\dataBW};
                \addplot [fill=orange!60, draw=black!80] table [y=S45, x expr=\coordindex] {\dataBW};
                \addplot [fill=purple!60, draw=black!80] table [y=S46, x expr=\coordindex] {\dataBW};
                \addplot [fill=gray!60, draw=black!80] table [y=Dup, x expr=\coordindex] {\dataBW};
            \end{axis}
        \end{tikzpicture}
        \caption*{CICY 7447, $|\Gamma|=4$}
    \end{subfigure}\hfill
  \pgfplotstableread{
Label S42 S43 S44 S45 S46 Dup
N 0.7 2.5 0.5 2.3 92.9 1.0
NS5 0.0 0.1 0.0 0.1 98.5 1.3
NGX3 0.5 1.7 0.4 1.6 93.6 2.2
Nfull 0.0 0.0 0.0 0.1 99.2 0.7
    }\dataBS
    \begin{subfigure}{0.33\textwidth}
        \centering
        \begin{tikzpicture}
            \begin{axis}
                \addplot [fill=blue!60, draw=black!80] table [y=S42, x expr=\coordindex] {\dataBS};
                \addplot [fill=red!60, draw=black!80] table [y=S43, x expr=\coordindex] {\dataBS};
                \addplot [fill=green!60, draw=black!80] table [y=S44, x expr=\coordindex] {\dataBS};
                \addplot [fill=orange!60, draw=black!80] table [y=S45, x expr=\coordindex] {\dataBS};
                \addplot [fill=purple!60, draw=black!80] table [y=S46, x expr=\coordindex] {\dataBS};
                \addplot [fill=gray!60, draw=black!80] table [y=Dup, x expr=\coordindex] {\dataBS};
            \end{axis}
        \end{tikzpicture}
        \caption*{CICY 7279, $|\Gamma|=2$}
    \end{subfigure}\hfill
    \pgfplotstableread{
Label S42 S43 S44 S45 S46 Dup
N 29.0 2.0 7.9 21.6 24.4 15.2
NS5 2.7 0.6 0.8 14.4 17.1 64.3
NGX3 23.7 1.7 6.9 15.7 16.7 35.3
Nfull 1.7 3.4 1.7 7.6 10.1 75.6
    }\dataBP
    \begin{subfigure}{0.33\textwidth}
        \begin{tikzpicture}
            \begin{axis}[]
                \addplot [fill=blue!60, draw=black!80] table [y=S42, x expr=\coordindex] {\dataBP};
                \addplot [fill=red!60, draw=black!80] table [y=S43, x expr=\coordindex] {\dataBP};
                \addplot [fill=green!60, draw=black!80] table [y=S44, x expr=\coordindex] {\dataBP};
                \addplot [fill=orange!60, draw=black!80] table [y=S45, x expr=\coordindex] {\dataBP};
                \addplot [fill=purple!60, draw=black!80] table [y=S46, x expr=\coordindex] {\dataBP};
                \addplot [fill=gray!60, draw=black!80] table [y=Dup, x expr=\coordindex] {\dataBP};
            \end{axis}
        \end{tikzpicture}
        \caption*{CICY 6947, $|\Gamma|=4$}
    \end{subfigure}\hfill
    \pgfplotstableread{
Label S42 S43 S44 S45 S46 Dup
N 10.3 2.0 18.7 3.9 45.3 19.9
NS5 0.6 1.0 19.3 1.1 47.5 30.5
NGX3 6.5 1.0 15.1 2.1 47.6 27.7
Nfull 0.8 1.7 9.2 2.5 39.5 46.2
    }\dataBO
    \begin{subfigure}{0.33\textwidth}
        \centering
        \begin{tikzpicture}
            \begin{axis}[]
                \addplot [fill=blue!60, draw=black!80] table [y=S42, x expr=\coordindex] {\dataBO};
                \addplot [fill=red!60, draw=black!80] table [y=S43, x expr=\coordindex] {\dataBO};
                \addplot [fill=green!60, draw=black!80] table [y=S44, x expr=\coordindex] {\dataBO};
                \addplot [fill=orange!60, draw=black!80] table [y=S45, x expr=\coordindex] {\dataBO};
                \addplot [fill=purple!60, draw=black!80] table [y=S46, x expr=\coordindex] {\dataBO};
                \addplot [fill=gray!60, draw=black!80] table [y=Dup, x expr=\coordindex] {\dataBO};
            \end{axis}
        \end{tikzpicture}
        \caption*{CICY 6927, $|\Gamma|=4$}
    \end{subfigure}
\caption*{\hfill\textit{Continued on next page}}
\end{figure}
\newpage
\begin{figure*}[!htp]
\ContinuedFloat
    \centering
    \pgfplotsset{
        compat=1.18,
        every axis/.append style={
            width=\textwidth, 
            height=4.5cm, 
            ymin=0, ymax=110, ybar stacked, bar width=12pt,
            xtick={0,1,2,3}, xticklabels={$N$, $N_{S_5}$, $N_{G_{\mathbf{X}_3}}$, $N_{\text{full}}$},
            enlarge x limits=0.2, 
            legend style={cells={anchor=west}, legend pos=north east, font=\tiny},
            legend image code/.code={\draw[#1, draw=none] (0cm,-0.1cm) rectangle (0.2cm,0.1cm);},
            tick label style={font=\scriptsize},
            label style={font=\footnotesize}
        }
    }
    \pgfplotstableread{
Label S42 S43 S44 S45 S46 Dup
N 13.9 12.6 10.2 20.4 21.7 21.4
NS5 3.9 6.4 4.4 7.4 5.9 72.1
NGX3 11.1 10.6 6.4 15.4 25.8 30.7
Nfull 1.4 11.3 2.8 5.6 4.2 74.6
    }\dataBN
    \begin{subfigure}{0.33\textwidth}
        \centering
        \begin{tikzpicture}
            \begin{axis}[ylabel={Solutions (\%)},]
                \addplot [fill=blue!60, draw=black!80] table [y=S42, x expr=\coordindex] {\dataBN};
                \addplot [fill=red!60, draw=black!80] table [y=S43, x expr=\coordindex] {\dataBN};
                \addplot [fill=green!60, draw=black!80] table [y=S44, x expr=\coordindex] {\dataBN};
                \addplot [fill=orange!60, draw=black!80] table [y=S45, x expr=\coordindex] {\dataBN};
                \addplot [fill=purple!60, draw=black!80] table [y=S46, x expr=\coordindex] {\dataBN};
                \addplot [fill=gray!60, draw=black!80] table [y=Dup, x expr=\coordindex] {\dataBN};
            \end{axis}
        \end{tikzpicture}
        \caption*{CICY 6890, $|\Gamma|=2$}
    \end{subfigure}\hfill
    \pgfplotstableread{
Label S42 S43 S44 S45 S46 Dup
N 24.0 3.4 6.7 7.2 16.9 41.9
NS5 9.9 2.1 0.9 1.1 9.1 76.9
NGX3 17.7 1.8 3.6 5.0 6.0 65.9
Nfull 4.7 0.9 0.9 0.9 3.8 88.7
    }\dataBM
    \begin{subfigure}{0.33\textwidth}
        \centering
        \begin{tikzpicture}
            \begin{axis}[]
                \addplot [fill=blue!60, draw=black!80] table [y=S42, x expr=\coordindex] {\dataBM};
                \addplot [fill=red!60, draw=black!80] table [y=S43, x expr=\coordindex] {\dataBM};
                \addplot [fill=green!60, draw=black!80] table [y=S44, x expr=\coordindex] {\dataBM};
                \addplot [fill=orange!60, draw=black!80] table [y=S45, x expr=\coordindex] {\dataBM};
                \addplot [fill=purple!60, draw=black!80] table [y=S46, x expr=\coordindex] {\dataBM};
                \addplot [fill=gray!60, draw=black!80] table [y=Dup, x expr=\coordindex] {\dataBM};
            \end{axis}
        \end{tikzpicture}
        \caption*{CICY 6836, $|\Gamma|=4$}
    \end{subfigure}\hfill
    \pgfplotstableread{
Label S42 S43 S44 S45 S46 Dup
N 30.8 5.6 2.4 2.7 48.4 10.2
NS5 0.6 1.9 1.6 0.2 76.1 19.6
NGX3 24.5 2.3 0.0 0.3 57.7 15.2
Nfull 1.2 3.8 0.0 0.0 85.0 10.0
    }\dataBJ
    \begin{subfigure}{0.33\textwidth}
        \centering
        \begin{tikzpicture}
            \begin{axis}[]
                \addplot [fill=blue!60, draw=black!80] table [y=S42, x expr=\coordindex] {\dataBJ};
                \addplot [fill=red!60, draw=black!80] table [y=S43, x expr=\coordindex] {\dataBJ};
                \addplot [fill=green!60, draw=black!80] table [y=S44, x expr=\coordindex] {\dataBJ};
                \addplot [fill=orange!60, draw=black!80] table [y=S45, x expr=\coordindex] {\dataBJ};
                \addplot [fill=purple!60, draw=black!80] table [y=S46, x expr=\coordindex] {\dataBJ};
                \addplot [fill=gray!60, draw=black!80] table [y=Dup, x expr=\coordindex] {\dataBJ};
            \end{axis}
        \end{tikzpicture}
        \caption*{CICY 6788, $|\Gamma|=2$}
    \end{subfigure}\hfill
    \pgfplotstableread{
Label S42 S43 S44 S45 S46 Dup
N 11.8 3.2 11.6 9.0 32.3 32.1
NS5 9.0 0.4 0.3 1.0 18.7 70.5
NGX3 3.7 3.6 8.7 6.1 21.9 55.9
Nfull 6.1 1.0 0.0 1.0 12.1 79.8
    }\dataBK
    \begin{subfigure}{0.33\textwidth}
        \centering
        \begin{tikzpicture}
            \begin{axis}
                \addplot [fill=blue!60, draw=black!80] table [y=S42, x expr=\coordindex] {\dataBK};
                \addplot [fill=red!60, draw=black!80] table [y=S43, x expr=\coordindex] {\dataBK};
                \addplot [fill=green!60, draw=black!80] table [y=S44, x expr=\coordindex] {\dataBK};
                \addplot [fill=orange!60, draw=black!80] table [y=S45, x expr=\coordindex] {\dataBK};
                \addplot [fill=purple!60, draw=black!80] table [y=S46, x expr=\coordindex] {\dataBK};
                \addplot [fill=gray!60, draw=black!80] table [y=Dup, x expr=\coordindex] {\dataBK};
            \end{axis}
        \end{tikzpicture}
        \caption*{CICY 6788, $|\Gamma|=4$}
    \end{subfigure}

    \pgfplotstableread{
Label S42 S43 S44 S45 S46 Dup
N 1.0 36.3 11.4 33.1 1.3 16.8
NS5 1.9 0.0 0.0 5.8 3.8 88.5
NGX3 0.4 23.3 11.0 18.8 1.6 44.9
Nfull 16.7 0.0 0.0 0.0 0.0 83.3
    }\dataBI
    \begin{subfigure}{0.33\textwidth}
        \centering
        \begin{tikzpicture}
            \begin{axis}[ylabel={Solutions (\%)},]
                \addplot [fill=blue!60, draw=black!80] table [y=S42, x expr=\coordindex] {\dataBI};
                \addplot [fill=red!60, draw=black!80] table [y=S43, x expr=\coordindex] {\dataBI};
                \addplot [fill=green!60, draw=black!80] table [y=S44, x expr=\coordindex] {\dataBI};
                \addplot [fill=orange!60, draw=black!80] table [y=S45, x expr=\coordindex] {\dataBI};
                \addplot [fill=purple!60, draw=black!80] table [y=S46, x expr=\coordindex] {\dataBI};
                \addplot [fill=gray!60, draw=black!80] table [y=Dup, x expr=\coordindex] {\dataBI};
            \end{axis}
        \end{tikzpicture}
        \caption*{CICY 6724, $|\Gamma|=2$}
    \end{subfigure}\hfill
    \pgfplotstableread{
Label S42 S43 S44 S45 S46 Dup
N 19.6 5.1 3.7 10.3 27.3 34.1
NS5 16.1 0.3 0.2 10.6 29.7 43.1
NGX3 1.8 3.0 0.7 19.7 40.5 34.3
Nfull 0.0 0.4 0.0 20.6 47.5 31.4
    }\dataBG
    \begin{subfigure}{0.33\textwidth}
        \centering
        \begin{tikzpicture}
            \begin{axis}[]
                \addplot [fill=blue!60, draw=black!80] table [y=S42, x expr=\coordindex] {\dataBG};
                \addplot [fill=red!60, draw=black!80] table [y=S43, x expr=\coordindex] {\dataBG};
                \addplot [fill=green!60, draw=black!80] table [y=S44, x expr=\coordindex] {\dataBG};
                \addplot [fill=orange!60, draw=black!80] table [y=S45, x expr=\coordindex] {\dataBG};
                \addplot [fill=purple!60, draw=black!80] table [y=S46, x expr=\coordindex] {\dataBG};
                \addplot [fill=gray!60, draw=black!80] table [y=Dup, x expr=\coordindex] {\dataBG};
            \end{axis}
        \end{tikzpicture}
        \caption*{CICY 6715, $|\Gamma|=2$}
    \end{subfigure}\hfill
    \pgfplotstableread{
Label S42 S43 S44 S45 S46 Dup
N 33.6 3.3 7.2 7.1 29.1 19.7
NS5 7.1 0.9 1.3 0.7 18.6 71.5
NGX3 27.3 4.1 6.0 3.8 15.8 43.0
Nfull 1.8 4.5 1.8 0.9 12.6 78.4
    }\dataBH
    \begin{subfigure}{0.33\textwidth}
        \centering
        \begin{tikzpicture}
            \begin{axis}
                \addplot [fill=blue!60, draw=black!80] table [y=S42, x expr=\coordindex] {\dataBH};
                \addplot [fill=red!60, draw=black!80] table [y=S43, x expr=\coordindex] {\dataBH};
                \addplot [fill=green!60, draw=black!80] table [y=S44, x expr=\coordindex] {\dataBH};
                \addplot [fill=orange!60, draw=black!80] table [y=S45, x expr=\coordindex] {\dataBH};
                \addplot [fill=purple!60, draw=black!80] table [y=S46, x expr=\coordindex] {\dataBH};
                \addplot [fill=gray!60, draw=black!80] table [y=Dup, x expr=\coordindex] {\dataBH};
            \end{axis}
        \end{tikzpicture}
        \caption*{CICY 6715, $|\Gamma|=4$}
    \end{subfigure}\hfill
    \pgfplotstableread{
Label S42 S43 S44 S45 S46 Dup
N 6.2 21.9 5.3 31.0 12.1 23.4
NS5 0.0 0.0 0.0 0.0 0.0 100.0
NGX3 3.8 20.2 3.1 25.8 11.8 35.2
Nfull 0.0 0.0 0.0 0.0 0.0 100.0
    }\dataBC
    \begin{subfigure}{0.33\textwidth}
        \centering
        \begin{tikzpicture}
            \begin{axis}[]
                \addplot [fill=blue!60, draw=black!80] table [y=S42, x expr=\coordindex] {\dataBC};
                \addplot [fill=red!60, draw=black!80] table [y=S43, x expr=\coordindex] {\dataBC};
                \addplot [fill=green!60, draw=black!80] table [y=S44, x expr=\coordindex] {\dataBC};
                \addplot [fill=orange!60, draw=black!80] table [y=S45, x expr=\coordindex] {\dataBC};
                \addplot [fill=purple!60, draw=black!80] table [y=S46, x expr=\coordindex] {\dataBC};
                \addplot [fill=gray!60, draw=black!80] table [y=Dup, x expr=\coordindex] {\dataBC};
            \end{axis}
        \end{tikzpicture}
        \caption*{CICY 6204, $|\Gamma|=2$}
    \end{subfigure}
    
    \pgfplotstableread{
Label S42 S43 S44 S45 S46 Dup
N 6.0 5.3 10.1 11.0 10.9 56.7
NS5 9.4 3.0 0.9 0.7 1.0 85.0
NGX3 3.1 1.0 1.0 1.1 1.0 92.8
Nfull 16.5 6.4 3.3 1.7 2.2 70.0
    }\dataAY
    \begin{subfigure}{0.33\textwidth}
        \centering
        \begin{tikzpicture}
            \begin{axis}[ylabel={Solutions (\%)},]
                \addplot [fill=blue!60, draw=black!80] table [y=S42, x expr=\coordindex] {\dataAY};
                \addplot [fill=red!60, draw=black!80] table [y=S43, x expr=\coordindex] {\dataAY};
                \addplot [fill=green!60, draw=black!80] table [y=S44, x expr=\coordindex] {\dataAY};
                \addplot [fill=orange!60, draw=black!80] table [y=S45, x expr=\coordindex] {\dataAY};
                \addplot [fill=purple!60, draw=black!80] table [y=S46, x expr=\coordindex] {\dataAY};
                \addplot [fill=gray!60, draw=black!80] table [y=Dup, x expr=\coordindex] {\dataAY};
            \end{axis}
        \end{tikzpicture}
        \caption*{CICY 6024, $|\Gamma|=3$}
    \end{subfigure}\hfill
    \pgfplotstableread{
Label S42 S43 S44 S45 S46 Dup
N 20.0 12.8 14.8 14.1 16.9 21.4
NS5 5.9 5.6 5.6 5.9 6.2 70.8
NGX3 12.3 13.1 9.0 16.9 10.1 38.5
Nfull 5.7 3.8 5.7 1.9 1.9 81.1
    }\dataAT
    \begin{subfigure}{0.33\textwidth}
        \centering
        \begin{tikzpicture}
            \begin{axis}
                \addplot [fill=blue!60, draw=black!80] table [y=S42, x expr=\coordindex] {\dataAT};
                \addplot [fill=red!60, draw=black!80] table [y=S43, x expr=\coordindex] {\dataAT};
                \addplot [fill=green!60, draw=black!80] table [y=S44, x expr=\coordindex] {\dataAT};
                \addplot [fill=orange!60, draw=black!80] table [y=S45, x expr=\coordindex] {\dataAT};
                \addplot [fill=purple!60, draw=black!80] table [y=S46, x expr=\coordindex] {\dataAT};
                \addplot [fill=gray!60, draw=black!80] table [y=Dup, x expr=\coordindex] {\dataAT};
            \end{axis}
        \end{tikzpicture}
        \caption*{CICY 5452, $|\Gamma|=2$}
    \end{subfigure}\hfill
    \pgfplotstableread{
Label S42 S43 S44 S45 S46 Dup
N 14.0 14.2 6.0 16.2 23.1 26.5
NS5 5.2 4.0 1.6 33.7 16.0 39.6
NGX3 9.9 9.3 5.2 12.5 24.2 38.8
Nfull 3.8 2.4 1.2 39.4 20.2 33.0
    }\dataAU
    \begin{subfigure}{0.33\textwidth}
        \centering
        \begin{tikzpicture}
            \begin{axis}[]
                \addplot [fill=blue!60, draw=black!80] table [y=S42, x expr=\coordindex] {\dataAU};
                \addplot [fill=red!60, draw=black!80] table [y=S43, x expr=\coordindex] {\dataAU};
                \addplot [fill=green!60, draw=black!80] table [y=S44, x expr=\coordindex] {\dataAU};
                \addplot [fill=orange!60, draw=black!80] table [y=S45, x expr=\coordindex] {\dataAU};
                \addplot [fill=purple!60, draw=black!80] table [y=S46, x expr=\coordindex] {\dataAU};
                \addplot [fill=gray!60, draw=black!80] table [y=Dup, x expr=\coordindex] {\dataAU};
            \end{axis}
        \end{tikzpicture}
        \caption*{CICY 5452, $|\Gamma|=4$}
    \end{subfigure}\hfill
    \pgfplotstableread{
Label S42 S43 S44 S45 S46 Dup
N 9.8 23.2 21.6 14.4 12.4 18.6
NS5 3.3 17.5 21.9 8.4 3.3 45.7
NGX3 6.0 20.6 21.6 13.2 6.2 32.5
Nfull 0.9 14.8 28.6 2.8 1.6 51.4
    }\dataAP
    \begin{subfigure}{0.33\textwidth}
        \centering
        \begin{tikzpicture}
            \begin{axis}[]
                \addplot [fill=blue!60, draw=black!80] table [y=S42, x expr=\coordindex] {\dataAP};
                \addplot [fill=red!60, draw=black!80] table [y=S43, x expr=\coordindex] {\dataAP};
                \addplot [fill=green!60, draw=black!80] table [y=S44, x expr=\coordindex] {\dataAP};
                \addplot [fill=orange!60, draw=black!80] table [y=S45, x expr=\coordindex] {\dataAP};
                \addplot [fill=purple!60, draw=black!80] table [y=S46, x expr=\coordindex] {\dataAP};
                \addplot [fill=gray!60, draw=black!80] table [y=Dup, x expr=\coordindex] {\dataAP};
            \end{axis}
        \end{tikzpicture}
        \caption*{CICY 5301, $|\Gamma|=4$}
    \end{subfigure}
\caption*{\hfill\textit{Continued on next page}}
\end{figure*}
\newpage
\begin{figure*}[!htp]
\ContinuedFloat
    \centering
    \pgfplotsset{
        compat=1.18,
        every axis/.append style={
            width=\textwidth, 
            height=4.5cm, 
            ymin=0, ymax=110, ybar stacked, bar width=12pt,
            xtick={0,1,2,3}, xticklabels={$N$, $N_{S_5}$, $N_{G_{\mathbf{X}_3}}$, $N_{\text{full}}$},
            enlarge x limits=0.2, 
            legend style={cells={anchor=west}, legend pos=north east, font=\tiny},
            legend image code/.code={\draw[#1, draw=none] (0cm,-0.1cm) rectangle (0.2cm,0.1cm);},
            tick label style={font=\scriptsize},
            label style={font=\footnotesize}
        }
    }
    \pgfplotstableread{
Label S42 S43 S44 S45 S46 Dup
N 32.8 8.1 12.0 15.8 17.5 13.8
NS5 14.5 2.3 5.0 7.3 5.7 65.3
NGX3 33.0 3.5 5.7 10.3 12.5 35.1
Nfull 7.4 0.0 0.0 7.4 1.9 83.3
    }\dataAL
    \begin{subfigure}{0.33\textwidth}
        \centering
        \begin{tikzpicture}
            \begin{axis}[ylabel={Solutions (\%)},]
                \addplot [fill=blue!60, draw=black!80] table [y=S42, x expr=\coordindex] {\dataAL};
                \addplot [fill=red!60, draw=black!80] table [y=S43, x expr=\coordindex] {\dataAL};
                \addplot [fill=green!60, draw=black!80] table [y=S44, x expr=\coordindex] {\dataAL};
                \addplot [fill=orange!60, draw=black!80] table [y=S45, x expr=\coordindex] {\dataAL};
                \addplot [fill=purple!60, draw=black!80] table [y=S46, x expr=\coordindex] {\dataAL};
                \addplot [fill=gray!60, draw=black!80] table [y=Dup, x expr=\coordindex] {\dataAL};
            \end{axis}
        \end{tikzpicture}
        \caption*{CICY 5256, $|\Gamma|=2$}
    \end{subfigure}\hfill
    \pgfplotstableread{
Label S42 S43 S44 S45 S46 Dup
N 7.8 52.4 6.5 15.7 6.2 11.4
NS5 1.7 69.4 1.3 4.9 1.3 21.3
NGX3 5.2 54.3 4.0 13.7 3.9 19.0
Nfull 0.5 76.9 0.3 2.5 0.3 19.6
    }\dataAM
    \begin{subfigure}{0.33\textwidth}
        \centering
        \begin{tikzpicture}
            \begin{axis}[]
                \addplot [fill=blue!60, draw=black!80] table [y=S42, x expr=\coordindex] {\dataAM};
                \addplot [fill=red!60, draw=black!80] table [y=S43, x expr=\coordindex] {\dataAM};
                \addplot [fill=green!60, draw=black!80] table [y=S44, x expr=\coordindex] {\dataAM};
                \addplot [fill=orange!60, draw=black!80] table [y=S45, x expr=\coordindex] {\dataAM};
                \addplot [fill=purple!60, draw=black!80] table [y=S46, x expr=\coordindex] {\dataAM};
                \addplot [fill=gray!60, draw=black!80] table [y=Dup, x expr=\coordindex] {\dataAM};
            \end{axis}
        \end{tikzpicture}
        \caption*{CICY 5256, $|\Gamma|=4$}
    \end{subfigure}\hfill
    \pgfplotstableread{
Label S42 S43 S44 S45 S46 Dup
N 20.7 18.5 17.7 23.6 14.6 5.0
NS5 26.2 11.8 8.2 14.6 9.1 30.0
NGX3 12.0 17.8 14.3 27.7 15.7 12.4
Nfull 25.2 5.3 5.6 16.2 7.5 40.2
    }\dataBZ
    \begin{subfigure}{0.33\textwidth}
        \centering
        \begin{tikzpicture}
            \begin{axis}[]
                \addplot [fill=blue!60, draw=black!80] table [y=S42, x expr=\coordindex] {\dataBZ};
                \addplot [fill=red!60, draw=black!80] table [y=S43, x expr=\coordindex] {\dataBZ};
                \addplot [fill=green!60, draw=black!80] table [y=S44, x expr=\coordindex] {\dataBZ};
                \addplot [fill=orange!60, draw=black!80] table [y=S45, x expr=\coordindex] {\dataBZ};
                \addplot [fill=purple!60, draw=black!80] table [y=S46, x expr=\coordindex] {\dataBZ};
                \addplot [fill=gray!60, draw=black!80] table [y=Dup, x expr=\coordindex] {\dataBZ};
            \end{axis}
        \end{tikzpicture}
        \caption*{CICY 7731, $|\Gamma|=2$}
    \end{subfigure}\hfill
    \pgfplotstableread{
Label S42 S43 S44 S45 S46 Dup
N 2.0 40.7 13.8 12.6 16.0 15.0
NS5 1.1 57.3 3.5 5.1 7.4 25.6
NGX3 3.8 44.0 9.2 11.0 15.6 16.5
Nfull 0.9 50.0 3.0 4.4 7.4 34.3
    }\dataBY
    \begin{subfigure}{0.33\textwidth}
        \centering
        \begin{tikzpicture}
            \begin{axis}
                \addplot [fill=blue!60, draw=black!80] table [y=S42, x expr=\coordindex] {\dataBY};
                \addplot [fill=red!60, draw=black!80] table [y=S43, x expr=\coordindex] {\dataBY};
                \addplot [fill=green!60, draw=black!80] table [y=S44, x expr=\coordindex] {\dataBY};
                \addplot [fill=orange!60, draw=black!80] table [y=S45, x expr=\coordindex] {\dataBY};
                \addplot [fill=purple!60, draw=black!80] table [y=S46, x expr=\coordindex] {\dataBY};
                \addplot [fill=gray!60, draw=black!80] table [y=Dup, x expr=\coordindex] {\dataBY};
            \end{axis}
        \end{tikzpicture}
        \caption*{CICY 7709, $|\Gamma|=2$}
    \end{subfigure}

  \pgfplotstableread{
Label S42 S43 S44 S45 S46 Dup
N 0.0 5.1 0.0 94.5 0.4 0.0
NS5 0.0 1.6 0.0 91.9 0.5 5.9
NGX3 0.0 11.6 0.0 87.5 0.0 0.9
Nfull 0.0 1.7 0.0 81.4 0.0 16.9
    }\dataBE
    \begin{subfigure}{0.33\textwidth}
        \centering
        \begin{tikzpicture}
            \begin{axis}[ylabel={Solutions (\%)},]
                \addplot [fill=blue!60, draw=black!80] table [y=S42, x expr=\coordindex] {\dataBE};
                \addplot [fill=red!60, draw=black!80] table [y=S43, x expr=\coordindex] {\dataBE};
                \addplot [fill=green!60, draw=black!80] table [y=S44, x expr=\coordindex] {\dataBE};
                \addplot [fill=orange!60, draw=black!80] table [y=S45, x expr=\coordindex] {\dataBE};
                \addplot [fill=purple!60, draw=black!80] table [y=S46, x expr=\coordindex] {\dataBE};
                \addplot [fill=gray!60, draw=black!80] table [y=Dup, x expr=\coordindex] {\dataBE};
            \end{axis}
        \end{tikzpicture}
        \caption*{CICY 6281, $|\Gamma|=2$}
    \end{subfigure}\hfill
    \pgfplotstableread{
Label S42 S43 S44 S45 S46 Dup
N 11.0 0.2 5.7 14.1 41.7 27.4
NS5 8.6 0.4 0.8 11.1 11.9 67.2
NGX3 7.5 0.3 10.4 10.1 40.3 31.4
Nfull 9.2 0.9 0.9 9.2 8.3 71.6
    }\dataBD
    \begin{subfigure}{0.33\textwidth}
        \centering
        \begin{tikzpicture}
            \begin{axis}[]
                \addplot [fill=blue!60, draw=black!80] table [y=S42, x expr=\coordindex] {\dataBD};
                \addplot [fill=red!60, draw=black!80] table [y=S43, x expr=\coordindex] {\dataBD};
                \addplot [fill=green!60, draw=black!80] table [y=S44, x expr=\coordindex] {\dataBD};
                \addplot [fill=orange!60, draw=black!80] table [y=S45, x expr=\coordindex] {\dataBD};
                \addplot [fill=purple!60, draw=black!80] table [y=S46, x expr=\coordindex] {\dataBD};
                \addplot [fill=gray!60, draw=black!80] table [y=Dup, x expr=\coordindex] {\dataBD};
            \end{axis}
        \end{tikzpicture}
        \caption*{CICY 6231, $|\Gamma|=2$}
    \end{subfigure}\hfill
    \pgfplotstableread{
Label S42 S43 S44 S45 S46 Dup
N 14.5 13.1 0.5 8.0 0.0 63.9
NS5 12.9 11.4 0.0 8.6 0.0 67.1
NGX3 12.7 10.2 1.0 4.1 0.0 72.1
Nfull 11.7 9.0 0.0 1.8 0.0 77.5
    }\dataBB
    \begin{subfigure}{0.33\textwidth}
        \centering
        \begin{tikzpicture}
            \begin{axis}
                \addplot [fill=blue!60, draw=black!80] table [y=S42, x expr=\coordindex] {\dataBB};
                \addplot [fill=red!60, draw=black!80] table [y=S43, x expr=\coordindex] {\dataBB};
                \addplot [fill=green!60, draw=black!80] table [y=S44, x expr=\coordindex] {\dataBB};
                \addplot [fill=orange!60, draw=black!80] table [y=S45, x expr=\coordindex] {\dataBB};
                \addplot [fill=purple!60, draw=black!80] table [y=S46, x expr=\coordindex] {\dataBB};
                \addplot [fill=gray!60, draw=black!80] table [y=Dup, x expr=\coordindex] {\dataBB};
            \end{axis}
        \end{tikzpicture}
        \caption*{CICY 6202, $|\Gamma|=2$}
    \end{subfigure}\hfill
    \pgfplotstableread{
Label S42 S43 S44 S45 S46 Dup
N 0.0 1.1 97.2 0.0 0.6 1.1
NS5 0.0 1.6 96.1 0.0 0.8 1.6
NGX3 0.0 2.6 93.5 0.0 0.0 3.9
Nfull 0.0 2.2 91.3 0.0 0.0 6.5
    }\dataBA
    \begin{subfigure}{0.33\textwidth}
        \centering
        \begin{tikzpicture}
            \begin{axis}[]
                \addplot [fill=blue!60, draw=black!80] table [y=S42, x expr=\coordindex] {\dataBA};
                \addplot [fill=red!60, draw=black!80] table [y=S43, x expr=\coordindex] {\dataBA};
                \addplot [fill=green!60, draw=black!80] table [y=S44, x expr=\coordindex] {\dataBA};
                \addplot [fill=orange!60, draw=black!80] table [y=S45, x expr=\coordindex] {\dataBA};
                \addplot [fill=purple!60, draw=black!80] table [y=S46, x expr=\coordindex] {\dataBA};
                \addplot [fill=gray!60, draw=black!80] table [y=Dup, x expr=\coordindex] {\dataBA};
            \end{axis}
        \end{tikzpicture}
        \caption*{CICY 6187, $|\Gamma|=2$}
    \end{subfigure}

\pgfplotstableread{
Label S42 S43 S44 S45 S46 Dup
N 0.0 0.0 65.3 0.0 34.7 0.0
NS5 0.0 0.0 50.1 0.0 49.9 0.0
NGX3 0.0 0.0 68.4 0.0 31.6 0.0
Nfull 0.0 0.0 52.4 0.0 47.6 0.0
    }\dataAW
    \begin{subfigure}{0.33\textwidth}
        \centering
        \begin{tikzpicture}
            \begin{axis}[ylabel={Solutions (\%)},]
                \addplot [fill=blue!60, draw=black!80] table [y=S42, x expr=\coordindex] {\dataAW};
                \addplot [fill=red!60, draw=black!80] table [y=S43, x expr=\coordindex] {\dataAW};
                \addplot [fill=green!60, draw=black!80] table [y=S44, x expr=\coordindex] {\dataAW};
                \addplot [fill=orange!60, draw=black!80] table [y=S45, x expr=\coordindex] {\dataAW};
                \addplot [fill=purple!60, draw=black!80] table [y=S46, x expr=\coordindex] {\dataAW};
                \addplot [fill=gray!60, draw=black!80] table [y=Dup, x expr=\coordindex] {\dataAW};
            \end{axis}
        \end{tikzpicture}
        \caption*{CICY 5982, $|\Gamma|=3$}
    \end{subfigure}\hfill
    \pgfplotstableread{
Label S42 S43 S44 S45 S46 Dup
N 62.4 0.0 35.9 0.2 0.0 1.4
NS5 59.1 0.0 35.9 0.1 0.0 5.0
NGX3 62.0 0.0 32.0 0.1 0.0 5.9
Nfull 54.0 0.0 31.5 0.0 0.0 14.5
    }\dataAV
    \begin{subfigure}{0.33\textwidth}
        \centering
        \begin{tikzpicture}
            \begin{axis}
                \addplot [fill=blue!60, draw=black!80] table [y=S42, x expr=\coordindex] {\dataAV};
                \addplot [fill=red!60, draw=black!80] table [y=S43, x expr=\coordindex] {\dataAV};
                \addplot [fill=green!60, draw=black!80] table [y=S44, x expr=\coordindex] {\dataAV};
                \addplot [fill=orange!60, draw=black!80] table [y=S45, x expr=\coordindex] {\dataAV};
                \addplot [fill=purple!60, draw=black!80] table [y=S46, x expr=\coordindex] {\dataAV};
                \addplot [fill=gray!60, draw=black!80] table [y=Dup, x expr=\coordindex] {\dataAV};
            \end{axis}
        \end{tikzpicture}
        \caption*{CICY 5967, $|\Gamma|=3$}
    \end{subfigure}\hfill
    \pgfplotstableread{
Label S42 S43 S44 S45 S46 Dup
N 14.1 11.7 21.3 8.8 19.9 24.1
NS5 10.3 7.9 11.6 4.1 2.8 63.4
NGX3 16.4 9.3 19.9 5.8 14.6 34.0
Nfull 14.7 6.9 4.3 3.5 2.4 68.2
    }\dataAS
    \begin{subfigure}{0.33\textwidth}
        \centering
        \begin{tikzpicture}
            \begin{axis}[]
                \addplot [fill=blue!60, draw=black!80] table [y=S42, x expr=\coordindex] {\dataAS};
                \addplot [fill=red!60, draw=black!80] table [y=S43, x expr=\coordindex] {\dataAS};
                \addplot [fill=green!60, draw=black!80] table [y=S44, x expr=\coordindex] {\dataAS};
                \addplot [fill=orange!60, draw=black!80] table [y=S45, x expr=\coordindex] {\dataAS};
                \addplot [fill=purple!60, draw=black!80] table [y=S46, x expr=\coordindex] {\dataAS};
                \addplot [fill=gray!60, draw=black!80] table [y=Dup, x expr=\coordindex] {\dataAS};
            \end{axis}
        \end{tikzpicture}
        \caption*{CICY 5425, $|\Gamma|=2$}
    \end{subfigure}\hfill
    \pgfplotstableread{
Label S42 S43 S44 S45 S46 Dup
N 11.0 8.8 13.7 8.5 13.0 45.0
NS5 6.0 6.0 5.0 5.9 3.4 73.6
NGX3 6.2 9.3 13.4 5.9 7.0 58.3
Nfull 2.1 5.2 4.1 4.1 6.2 78.4
    }\dataAQ
    \begin{subfigure}{0.33\textwidth}
        \centering
        \begin{tikzpicture}
            \begin{axis}[]
                \addplot [fill=blue!60, draw=black!80] table [y=S42, x expr=\coordindex] {\dataAQ};
                \addplot [fill=red!60, draw=black!80] table [y=S43, x expr=\coordindex] {\dataAQ};
                \addplot [fill=green!60, draw=black!80] table [y=S44, x expr=\coordindex] {\dataAQ};
                \addplot [fill=orange!60, draw=black!80] table [y=S45, x expr=\coordindex] {\dataAQ};
                \addplot [fill=purple!60, draw=black!80] table [y=S46, x expr=\coordindex] {\dataAQ};
                \addplot [fill=gray!60, draw=black!80] table [y=Dup, x expr=\coordindex] {\dataAQ};
            \end{axis}
        \end{tikzpicture}
        \caption*{CICY 5302, $|\Gamma|=2$}
    \end{subfigure}
\caption*{\hfill\textit{Continued on next page}}
\end{figure*}
\newpage
\begin{figure*}[!htp]
\ContinuedFloat
    \centering
    \pgfplotsset{
        compat=1.18,
        every axis/.append style={
            width=\textwidth, 
            height=4.5cm, 
            ymin=0, ymax=110, ybar stacked, bar width=12pt,
            xtick={0,1,2,3}, xticklabels={$N$, $N_{S_5}$, $N_{G_{\mathbf{X}_3}}$, $N_{\text{full}}$},
            enlarge x limits=0.2, 
            legend style={cells={anchor=west}, legend pos=north east, font=\tiny},
            legend image code/.code={\draw[#1, draw=none] (0cm,-0.1cm) rectangle (0.2cm,0.1cm);},
            tick label style={font=\scriptsize},
            label style={font=\footnotesize}
        }
    }
    \pgfplotstableread{
Label S42 S43 S44 S45 S46 Dup
N 10.7 9.8 17.7 8.9 14.6 38.3
NS5 2.7 2.2 2.7 2.6 5.3 84.5
NGX3 1.9 8.6 3.8 5.6 15.2 64.9
Nfull 3.5 6.5 4.9 4.4 6.0 74.6
    }\dataAR
    \begin{subfigure}{0.33\textwidth}
        \centering
        \begin{tikzpicture}
            \begin{axis}[ylabel={Solutions (\%)},]
                \addplot [fill=blue!60, draw=black!80] table [y=S42, x expr=\coordindex] {\dataAR};
                \addplot [fill=red!60, draw=black!80] table [y=S43, x expr=\coordindex] {\dataAR};
                \addplot [fill=green!60, draw=black!80] table [y=S44, x expr=\coordindex] {\dataAR};
                \addplot [fill=orange!60, draw=black!80] table [y=S45, x expr=\coordindex] {\dataAR};
                \addplot [fill=purple!60, draw=black!80] table [y=S46, x expr=\coordindex] {\dataAR};
                \addplot [fill=gray!60, draw=black!80] table [y=Dup, x expr=\coordindex] {\dataAR};
            \end{axis}
        \end{tikzpicture}
        \caption*{CICY 5302, $|\Gamma|=4$}
    \end{subfigure}\hfill
    \pgfplotstableread{
Label S42 S43 S44 S45 S46 Dup
N 5.2 12.5 11.7 52.8 5.5 12.3
NS5 0.8 2.8 2.9 76.5 1.4 15.7
NGX3 7.0 14.5 17.8 35.4 6.5 18.9
Nfull 1.3 3.0 3.6 65.9 1.8 24.4
    }\dataAO
    \begin{subfigure}{0.33\textwidth}
        \centering
        \begin{tikzpicture}
            \begin{axis}[]
                \addplot [fill=blue!60, draw=black!80] table [y=S42, x expr=\coordindex] {\dataAO};
                \addplot [fill=red!60, draw=black!80] table [y=S43, x expr=\coordindex] {\dataAO};
                \addplot [fill=green!60, draw=black!80] table [y=S44, x expr=\coordindex] {\dataAO};
                \addplot [fill=orange!60, draw=black!80] table [y=S45, x expr=\coordindex] {\dataAO};
                \addplot [fill=purple!60, draw=black!80] table [y=S46, x expr=\coordindex] {\dataAO};
                \addplot [fill=gray!60, draw=black!80] table [y=Dup, x expr=\coordindex] {\dataAO};
            \end{axis}
        \end{tikzpicture}
        \caption*{CICY 5273, $|\Gamma|=2$}
    \end{subfigure}\hfill
    \pgfplotstableread{
Label S42 S43 S44 S45 S46 Dup
N 10.7 11.1 10.3 10.9 15.9 41.0
NS5 3.3 3.7 3.3 3.5 10.2 76.1
NGX3 3.5 7.2 6.0 8.1 3.6 71.6
Nfull 1.8 3.7 2.9 3.2 2.5 85.9
    }\dataAC
    \begin{subfigure}{0.33\textwidth}
        \centering
        \begin{tikzpicture}
            \begin{axis}[]
                \addplot [fill=blue!60, draw=black!80] table [y=S42, x expr=\coordindex] {\dataAC};
                \addplot [fill=red!60, draw=black!80] table [y=S43, x expr=\coordindex] {\dataAC};
                \addplot [fill=green!60, draw=black!80] table [y=S44, x expr=\coordindex] {\dataAC};
                \addplot [fill=orange!60, draw=black!80] table [y=S45, x expr=\coordindex] {\dataAC};
                \addplot [fill=purple!60, draw=black!80] table [y=S46, x expr=\coordindex] {\dataAC};
                \addplot [fill=gray!60, draw=black!80] table [y=Dup, x expr=\coordindex] {\dataAC};
            \end{axis}
        \end{tikzpicture}
        \caption*{CICY 3413, $|\Gamma|=3$}
    \end{subfigure}\hfill
    \pgfplotstableread{
Label S42 S43 S44 S45 S46 Dup
N 20.4 0.0 32.2 12.4 35.0 0.0
NS5 20.2 0.0 32.3 12.4 35.1 0.0
NGX3 20.4 0.0 32.2 12.4 35.0 0.0
Nfull 20.2 0.0 32.3 12.4 35.1 0.0
    }\dataBF
    \begin{subfigure}{0.33\textwidth}
        \centering
        \begin{tikzpicture}
            \begin{axis}
                \addplot [fill=blue!60, draw=black!80] table [y=S42, x expr=\coordindex] {\dataBF};
                \addplot [fill=red!60, draw=black!80] table [y=S43, x expr=\coordindex] {\dataBF};
                \addplot [fill=green!60, draw=black!80] table [y=S44, x expr=\coordindex] {\dataBF};
                \addplot [fill=orange!60, draw=black!80] table [y=S45, x expr=\coordindex] {\dataBF};
                \addplot [fill=purple!60, draw=black!80] table [y=S46, x expr=\coordindex] {\dataBF};
                \addplot [fill=gray!60, draw=black!80] table [y=Dup, x expr=\coordindex] {\dataBF};
            \end{axis}
        \end{tikzpicture}
        \caption*{CICY 6502, $|\Gamma|=3$}
    \end{subfigure}

    \pgfplotstableread{
Label S42 S43 S44 S45 S46 Dup
N 1.3 0.6 4.3 91.9 1.0 0.8
NS5 3.6 1.7 9.4 78.7 2.6 4.0
NGX3 6.4 3.0 15.5 64.7 4.8 5.6
Nfull 14.0 6.4 27.3 35.6 7.2 9.4
    }\dataAZ
    \begin{subfigure}{0.33\textwidth}
        \centering
        \begin{tikzpicture}
            \begin{axis}[ylabel={Solutions (\%)},]
                \addplot [fill=blue!60, draw=black!80] table [y=S42, x expr=\coordindex] {\dataAZ};
                \addplot [fill=red!60, draw=black!80] table [y=S43, x expr=\coordindex] {\dataAZ};
                \addplot [fill=green!60, draw=black!80] table [y=S44, x expr=\coordindex] {\dataAZ};
                \addplot [fill=orange!60, draw=black!80] table [y=S45, x expr=\coordindex] {\dataAZ};
                \addplot [fill=purple!60, draw=black!80] table [y=S46, x expr=\coordindex] {\dataAZ};
                \addplot [fill=gray!60, draw=black!80] table [y=Dup, x expr=\coordindex] {\dataAZ};
            \end{axis}
        \end{tikzpicture}
        \caption*{CICY 6178, $|\Gamma|=2$}
    \end{subfigure}\hfill
    \pgfplotstableread{
Label S42 S43 S44 S45 S46 Dup
N 59.3 16.5 1.7 4.3 18.0 0.1
NS5 68.6 19.7 2.0 2.3 7.0 0.4
NGX3 68.8 18.1 2.0 2.6 6.9 1.6
Nfull 72.8 18.9 2.1 1.0 2.1 3.1
    }\dataAK
    \begin{subfigure}{0.33\textwidth}
        \centering
        \begin{tikzpicture}
            \begin{axis}[]
                \addplot [fill=blue!60, draw=black!80] table [y=S42, x expr=\coordindex] {\dataAK};
                \addplot [fill=red!60, draw=black!80] table [y=S43, x expr=\coordindex] {\dataAK};
                \addplot [fill=green!60, draw=black!80] table [y=S44, x expr=\coordindex] {\dataAK};
                \addplot [fill=orange!60, draw=black!80] table [y=S45, x expr=\coordindex] {\dataAK};
                \addplot [fill=purple!60, draw=black!80] table [y=S46, x expr=\coordindex] {\dataAK};
                \addplot [fill=gray!60, draw=black!80] table [y=Dup, x expr=\coordindex] {\dataAK};
            \end{axis}
        \end{tikzpicture}
        \caption*{CICY 5248, $|\Gamma|=2$}
    \end{subfigure}\hfill
    \pgfplotstableread{
Label S42 S43 S44 S45 S46 Dup
N 16.7 29.6 8.8 16.0 7.7 21.2
NS5 12.5 14.8 5.1 12.5 6.3 48.8
NGX3 13.5 31.6 8.9 12.0 9.3 24.7
Nfull 10.6 14.4 3.6 9.3 4.4 57.8
    }\dataAJ
    \begin{subfigure}{0.33\textwidth}
        \centering
        \begin{tikzpicture}
            \begin{axis}
                \addplot [fill=blue!60, draw=black!80] table [y=S42, x expr=\coordindex] {\dataAJ};
                \addplot [fill=red!60, draw=black!80] table [y=S43, x expr=\coordindex] {\dataAJ};
                \addplot [fill=green!60, draw=black!80] table [y=S44, x expr=\coordindex] {\dataAJ};
                \addplot [fill=orange!60, draw=black!80] table [y=S45, x expr=\coordindex] {\dataAJ};
                \addplot [fill=purple!60, draw=black!80] table [y=S46, x expr=\coordindex] {\dataAJ};
                \addplot [fill=gray!60, draw=black!80] table [y=Dup, x expr=\coordindex] {\dataAJ};
            \end{axis}
        \end{tikzpicture}
        \caption*{CICY 4738, $|\Gamma|=2$}
    \end{subfigure}\hfill
    \pgfplotstableread{
Label S42 S43 S44 S45 S46 Dup
N 15.7 21.1 10.3 12.5 26.6 13.8
NS5 9.9 17.5 7.6 6.8 36.1 22.1
NGX3 23.3 9.3 12.6 14.2 22.1 18.5
Nfull 16.9 6.0 4.2 4.8 30.7 37.3
    }\dataAI
    \begin{subfigure}{0.33\textwidth}
        \centering
        \begin{tikzpicture}
            \begin{axis}[]
                \addplot [fill=blue!60, draw=black!80] table [y=S42, x expr=\coordindex] {\dataAI};
                \addplot [fill=red!60, draw=black!80] table [y=S43, x expr=\coordindex] {\dataAI};
                \addplot [fill=green!60, draw=black!80] table [y=S44, x expr=\coordindex] {\dataAI};
                \addplot [fill=orange!60, draw=black!80] table [y=S45, x expr=\coordindex] {\dataAI};
                \addplot [fill=purple!60, draw=black!80] table [y=S46, x expr=\coordindex] {\dataAI};
                \addplot [fill=gray!60, draw=black!80] table [y=Dup, x expr=\coordindex] {\dataAI};
            \end{axis}
        \end{tikzpicture}
        \caption*{CICY 4335, $|\Gamma|=2$}
    \end{subfigure}

    \pgfplotstableread{
Label S42 S43 S44 S45 S46 Dup
N 45.0 0.1 46.2 3.3 5.5 0.0
NS5 45.0 0.1 46.2 3.3 5.5 0.0
NGX3 45.0 0.1 46.2 3.3 5.5 0.0
Nfull 44.9 0.1 46.2 3.3 5.5 0.1
    }\dataAH
    \begin{subfigure}{0.33\textwidth}
        \centering
        \begin{tikzpicture}
            \begin{axis}[ylabel={Solutions (\%)},]
                \addplot [fill=blue!60, draw=black!80] table [y=S42, x expr=\coordindex] {\dataAH};
                \addplot [fill=red!60, draw=black!80] table [y=S43, x expr=\coordindex] {\dataAH};
                \addplot [fill=green!60, draw=black!80] table [y=S44, x expr=\coordindex] {\dataAH};
                \addplot [fill=orange!60, draw=black!80] table [y=S45, x expr=\coordindex] {\dataAH};
                \addplot [fill=purple!60, draw=black!80] table [y=S46, x expr=\coordindex] {\dataAH};
                \addplot [fill=gray!60, draw=black!80] table [y=Dup, x expr=\coordindex] {\dataAH};
            \end{axis}
        \end{tikzpicture}
        \caption*{CICY 4185, $|\Gamma|=2$}
    \end{subfigure}\hfill
   \pgfplotstableread{
Label S42 S43 S44 S45 S46 Dup
N 16.4 10.4 19.9 19.9 33.5 0.0
NS5 16.4 10.4 19.9 19.9 33.5 0.0
NGX3 16.4 10.4 19.9 19.9 33.5 0.0
Nfull 16.4 10.4 19.9 19.9 33.5 0.0
    }\dataAG
    \begin{subfigure}{0.33\textwidth}
        \centering
        \begin{tikzpicture}
            \begin{axis}
                \addplot [fill=blue!60, draw=black!80] table [y=S42, x expr=\coordindex] {\dataAG};
                \addplot [fill=red!60, draw=black!80] table [y=S43, x expr=\coordindex] {\dataAG};
                \addplot [fill=green!60, draw=black!80] table [y=S44, x expr=\coordindex] {\dataAG};
                \addplot [fill=orange!60, draw=black!80] table [y=S45, x expr=\coordindex] {\dataAG};
                \addplot [fill=purple!60, draw=black!80] table [y=S46, x expr=\coordindex] {\dataAG};
                \addplot [fill=gray!60, draw=black!80] table [y=Dup, x expr=\coordindex] {\dataAG};
            \end{axis}
        \end{tikzpicture}
        \caption*{CICY 4108, $|\Gamma|=2$}
    \end{subfigure}\hfill
    \pgfplotstableread{
Label S42 S43 S44 S45 S46 Dup
N 26.6 11.1 43.7 2.9 15.8 0.0
NS5 26.6 11.1 43.7 2.9 15.7 0.0
NGX3 26.7 11.1 43.6 2.9 15.8 0.0
Nfull 26.6 11.1 43.7 2.9 15.7 0.1
    }\dataAF
    \begin{subfigure}{0.33\textwidth}
        \centering
        \begin{tikzpicture}
            \begin{axis}[]
                \addplot [fill=blue!60, draw=black!80] table [y=S42, x expr=\coordindex] {\dataAF};
                \addplot [fill=red!60, draw=black!80] table [y=S43, x expr=\coordindex] {\dataAF};
                \addplot [fill=green!60, draw=black!80] table [y=S44, x expr=\coordindex] {\dataAF};
                \addplot [fill=orange!60, draw=black!80] table [y=S45, x expr=\coordindex] {\dataAF};
                \addplot [fill=purple!60, draw=black!80] table [y=S46, x expr=\coordindex] {\dataAF};
                \addplot [fill=gray!60, draw=black!80] table [y=Dup, x expr=\coordindex] {\dataAF};
            \end{axis}
        \end{tikzpicture}
        \caption*{CICY 4078, $|\Gamma|=2$}
    \end{subfigure}\hfill
    \pgfplotstableread{
Label S42 S43 S44 S45 S46 Dup
N 28.0 23.7 0.0 23.8 24.5 0.0
NS5 28.0 23.7 0.0 23.8 24.5 0.0
NGX3 28.0 23.7 0.0 23.8 24.5 0.0
Nfull 28.0 23.7 0.0 23.8 24.5 0.0
    }\dataAE
    \begin{subfigure}{0.33\textwidth}
        \centering
        \begin{tikzpicture}
            \begin{axis}[]
                \addplot [fill=blue!60, draw=black!80] table [y=S42, x expr=\coordindex] {\dataAE};
                \addplot [fill=red!60, draw=black!80] table [y=S43, x expr=\coordindex] {\dataAE};
                \addplot [fill=green!60, draw=black!80] table [y=S44, x expr=\coordindex] {\dataAE};
                \addplot [fill=orange!60, draw=black!80] table [y=S45, x expr=\coordindex] {\dataAE};
                \addplot [fill=purple!60, draw=black!80] table [y=S46, x expr=\coordindex] {\dataAE};
                \addplot [fill=gray!60, draw=black!80] table [y=Dup, x expr=\coordindex] {\dataAE};
            \end{axis}
        \end{tikzpicture}
        \caption*{CICY 4071, $|\Gamma|=2$}
    \end{subfigure}
\caption*{\hfill\textit{Continued on next page}}
\end{figure*}
\newpage
\begin{figure*}[!htp]
\ContinuedFloat
    \centering
    \pgfplotsset{
        compat=1.18,
        every axis/.append style={
            width=\textwidth, 
            height=4.5cm, 
            ymin=0, ymax=110, ybar stacked, bar width=12pt,
            xtick={0,1,2,3}, xticklabels={$N$, $N_{S_5}$, $N_{G_{\mathbf{X}_3}}$, $N_{\text{full}}$},
            enlarge x limits=0.2, 
            legend style={cells={anchor=west}, legend pos=north east, font=\tiny},
            legend image code/.code={\draw[#1, draw=none] (0cm,-0.1cm) rectangle (0.2cm,0.1cm);},
            tick label style={font=\scriptsize},
            label style={font=\footnotesize}
        }
    }
    \pgfplotstableread{
Label S42 S43 S44 S45 S46 Dup
N 10.9 31.6 19.2 17.0 9.3 12.1
NS5 8.0 17.8 3.5 23.6 3.5 43.6
NGX3 13.9 30.1 11.2 22.1 6.3 16.5
Nfull 11.2 14.2 1.9 20.2 2.2 50.4
    }\dataAD
    \begin{subfigure}{0.33\textwidth}
        \centering

        \caption*{CICY 3929, $|\Gamma|=2$}
    \end{subfigure}\hfill
    \pgfplotstableread{
Label S42 S43 S44 S45 S46 Dup
N 15.4 17.4 16.6 28.0 7.1 15.5
NS5 7.9 22.4 5.7 10.6 6.5 47.0
NGX3 14.0 20.0 12.2 26.0 5.6 22.2
Nfull 7.2 24.4 4.0 8.1 5.8 50.5
    }\dataAA
    \begin{subfigure}{0.33\textwidth}
        \centering
        %
        \caption*{CICY 2544, $|\Gamma|=2$}
    \end{subfigure}\hfill
    \pgfplotstableread{
Label S42 S43 S44 S45 S46 Dup
N 0.1 0.1 0.2 98.1 1.5 0.0
NS5 0.1 0.1 0.1 98.8 0.9 0.0
NGX3 0.0 0.0 0.1 99.9 0.1 0.0
Nfull 0.0 0.0 0.0 99.9 0.0 0.0
    }\dataBT
    \begin{subfigure}{0.33\textwidth}
        \centering
        %
        \caption*{CICY 7300, $|\Gamma|=2$}
    \end{subfigure}\hfill
    \pgfplotstableread{
Label S42 S43 S44 S45 S46 Dup
N 4.8 0.4 1.1 90.9 2.6 0.2
NS5 7.6 0.5 1.6 85.7 4.1 0.5
NGX3 11.5 1.2 2.6 81.1 1.9 1.7
Nfull 20.2 1.8 4.0 66.6 3.4 3.9
    }\dataAB
    \begin{subfigure}{0.33\textwidth}
        \centering
        %
        \caption*{CICY 2639, $|\Gamma|=2$}
    \end{subfigure}

    \pgfplotstableread{
Label S42 S43 S44 S45 S46 Dup
N 13.4 12.3 10.0 44.9 19.3 0.0
NS5 13.5 12.3 10.0 44.9 19.3 0.0
NGX3 13.5 12.4 10.0 44.9 19.3 0.0
Nfull 13.6 12.4 10.0 44.5 19.5 0.0
    }\dataAN
\begin{subfigure}{0.33\textwidth}
        \centering
        %
        \caption*{CICY 5257, $|\Gamma|=2$}
    \end{subfigure}
    \pgfplotstableread{
Label S42 S43 S44 S45 S46 Dup
N 5.3 71.3 11.2 4.3 7.9 0.0
NS5 5.4 70.8 11.4 4.3 8.1 0.0
NGX3 5.3 71.2 11.2 4.3 7.9 0.0
Nfull 5.4 70.7 11.4 4.4 8.1 0.0
    }\dataAX
    \begin{subfigure}{0.33\textwidth}
        \centering
        %
        \caption*{CICY 6, $|\Gamma|=2$}
    \end{subfigure}
\end{figure*}


\setlength{\LTleft}{\fill}
\setlength{\LTright}{\fill}
\setlength{\LTcapwidth}{\linewidth}
\renewcommand{\arraystretch}{1}
\begin{longtable}{c | c | c | c | c | c | c}
\caption{Spectrum analysis summary. The meaning of $N$, $N_{S_5}$, $N_{G_{\mathbf{X}_3}}$ and $N_{\text{full}}$ is the same as in Table \ref{tab:detailed_counts}. In the `Solutions' column we show only the solutions with entries $|\mathbf{k}_a|\leq 2$. `Equivariance' shows solutions in the range $|\mathbf{k}_a|\leq 2$ that also satisfy \cref{eq:equivariantcond_trivial,eq:equivariantcond_nontrivial} for the equivariance of line bundles under the action of $\Gamma$. `Spectrum' counts the solutions satisfying the spectrum constraints in \eqref{eq:fullspectrum}. The line bundle cohomologies have been computed with \texttt{pyCICY} \cite{Larfors:2019sie}. We show in `Errors', the number of solutions for which \texttt{pyCICY} reported a warning that the computation was not reliable.} \label{tab:spectrum_stats} \\
\hhline{=|=|=|=|=|=|=}
\# & $h^{1,1}$ & $|\Gamma|$ & & Solutions ($|\mathbf{k}_a|\leq 2$) & Equivariance & Spectrum (Errors) \\
\hhline{=|=|=|=|=|=|=}
\endfirsthead
\multicolumn{7}{c}{{\tablename\ \thetable\ -- \textit{Continued from previous page}}} \\
\hhline{=|=|=|=|=|=|=}
\# & $h^{1,1}$ & $|\Gamma|$ & & Solutions ($|\mathbf{k}_5|\leq 2$) & Equivariance & Spectrum (Errors) \\
\hhline{=|=|=|=|=|=|=}
\endhead
\multicolumn{7}{r}{\textit{Continued on next page}} \\
\endfoot
\hhline{=|=|=|=|=|=|=}
\endlastfoot
\multirow{4}{*}{7862} & \multirow{4}{*}{4} & \multirow{4}{*}{4} & $N$ & 1586 & 132 & 0 (0) \\
 &  &  & $N_{S_5}$ & 24 & 2 & 0 (0) \\
 &  &  & $N_{G_{\mathbf{X}_3}}$ & 120 & 79 & 0 (0) \\
 &  &  & $N_{\text{full}}$ & 1 & 1 & 0 (0) \\ \hline
\multirow{4}{*}{7487} & \multirow{4}{*}{5} & \multirow{4}{*}{4} & $N$ & 105377 & 14394 & 1420 (0) \\
 &  &  & $N_{S_5}$ & 10501 & 1358 & 127 (0) \\
 &  &  & $N_{G_{\mathbf{X}_3}}$ & 7294 & 3118 & 806 (0) \\
 &  &  & $N_{\text{full}}$ & 172 & 103 & 52 (0) \\
\hline
\end{longtable}

\setlength{\LTleft}{\fill}
\setlength{\LTright}{\fill}
\setlength{\LTcapwidth}{\linewidth}
\renewcommand{\arraystretch}{1}
\begin{longtable}{c | c | c | c | c | c |  c | c | c | c | c | c | c}
\caption{
Counting of solutions found by the Hybrid LB-Explorer. The meaning of the columns is analogous to that in Table \ref{tab:detailed_counts}.} \label{tab:detailed_counts_CP} \\
\hhline{=|=|=|=|=|=====|=|=|=}
\multirow{2}{*}{\#} & \multirow{2}{*}{$h^{1,1}$} & \multirow{2}{*}{$|\Gamma|$} & \multirow{2}{*}{$|G_{\mathbf{X}_3}|$} & \multirow{2}{*}{Sol.} & \multicolumn{5}{|c|}{Seeds} & \multirow{2}{*}{Dup. $\left(\frac{N_\text{dup}}{N_\text{tot}}\%\right)$} & \multirow{2}{*}{Total} & \multirow{2}{*}{Equivariance} \\ \cline{6-10}
 &   &  &  &  & 42 & 43 & 44 & 45 & 46 &  &  & \\
\hhline{=|=|=|=|=|=|=|=|=|=|=|=|=}
\endfirsthead
\multicolumn{13}{c}{{\tablename\ \thetable\ -- \textit{Continued from previous page}}} \\
\hhline{=|=|=|=|=|=====|=|=|=}
\multirow{2}{*}{\#} & \multirow{2}{*}{$h^{1,1}$} & \multirow{2}{*}{$|\Gamma|$} & \multirow{2}{*}{$|G_{\mathbf{X}_3}|$} & \multirow{2}{*}{Sol.} & \multicolumn{5}{|c|}{Seeds} & \multirow{2}{*}{Dup. $\left(\frac{N_\text{dup}}{N_\text{tot}}\%\right)$} & \multirow{2}{*}{Total} & \multirow{2}{*}{Equivariance} \\ \cline{6-10}
 &   &  &  &  & 42 & 43 & 44 & 45 & 46 &  &  & \\
\hhline{=|=|=|=|=|=|=|=|=|=|=|=|=}
\endhead
\multicolumn{13}{r}{\textit{Continued on next page}} \\
\endfoot
\hhline{=|=|=|=|=|=|=|=|=|=|=|=|=}
\endlastfoot
\multirow{4}{*}{7862} & \multirow{4}{*}{4} & \multirow{4}{*}{2} & \multirow{4}{*}{24} & $N$ & 836 & 1521 & 1767 & 2434 & 2300 & 1124 (15.6\%) & 7222 & 1984 \checkmark \\
 &  &  &  & $N_{S_5}$ & 129 & 646 & 802 & 1389 & 1044 & 598 (20.9\%) & 2868 & 884 \checkmark \\
 &  &  &  & $N_{G_{\mathbf{X}_3}}$ & 143 & 383 & 277 & 589 & 429 & 260 (21.0\%) & 1237 & 582 \checkmark \\
 &  &  &  & $N_{\text{full}}$ & 16 & 164 & 96 & 286 & 183 & 158 (35.9\%) & 440 & 146 \checkmark \\ \hline
\multirow{4}{*}{7862} & \multirow{4}{*}{4} & \multirow{4}{*}{4} & \multirow{4}{*}{24} & $N$ & 7365 & 4787 & 8110 & 8218 & 7934 & 8664 (37.0\%) & 23403 & 2715 \checkmark \\
 &  &  &  & $N_{S_5}$ & 897 & 850 & 858 & 950 & 875 & 1023 (84.8\%) & 1206 & 88 \checkmark \\
 &  &  &  & $N_{G_{\mathbf{X}_3}}$ & 2619 & 2098 & 1741 & 1998 & 1769 & 2639 (66.7\%) & 3954 & 282 \checkmark \\
 &  &  &  & $N_{\text{full}}$ & 65 & 64 & 55 & 59 & 53 & 66 (91.7\%) & 72 & 10 \checkmark \\
\hline
\end{longtable}

\newpage
\begin{figure}[!htp]
\caption{Percentage of solutions as in Table \ref{tab:detailed_counts_CP} by seeds. Solutions found uniquely by Seeds 42, 43, 44, 45 46 are, respectively, shown in {\color{blue!60}{blue}}, {\color{red!60}{red}}, {\color{green!60}{green}}, {\color{orange!60}{orange}} and {\color{purple!60}{purple}}. Solutions that have been found by multiple seeds are shown in {\color{gray!60}{gray}}.}\label{fig:percentage_solutions_CP}
    \pgfplotsset{
        compat=1.18,
        every axis/.append style={
            width=\textwidth, 
            height=4.5cm, 
            ymin=0, ymax=110, ybar stacked, bar width=12pt,
            xtick={0,1,2,3}, xticklabels={$N$, $N_{S_5}$, $N_{G_{\mathbf{X}_3}}$, $N_{\text{full}}$},
            enlarge x limits=0.2, 
            legend style={cells={anchor=west}, legend pos=north east, font=\tiny},
            legend image code/.code={\draw[#1, draw=none] (0cm,-0.1cm) rectangle (0.2cm,0.1cm);},
            tick label style={font=\scriptsize},
            label style={font=\footnotesize}
        }
    }
    \pgfplotstableread{
Label S42 S43 S44 S45 S46 Dup
N 4.0 9.2 13.4 29.7 28.2 15.6
NS5 1.5 9.8 15.4 32.5 19.9 20.9
NGX3 1.6 10.3 1.8 39.9 25.4 21.0
Nfull 0.7 4.8 2.5 45.7 10.5 35.9
    }\dataAA
    \begin{subfigure}{0.32\textwidth}
        \centering
        \begin{tikzpicture}
            \begin{axis}[ylabel={Solutions (\%)},]
                \addplot [fill=blue!60, draw=black!80] table [y=S42, x expr=\coordindex] {\dataAA};
                \addplot [fill=red!60, draw=black!80] table [y=S43, x expr=\coordindex] {\dataAA};
                \addplot [fill=green!60, draw=black!80] table [y=S44, x expr=\coordindex] {\dataAA};
                \addplot [fill=orange!60, draw=black!80] table [y=S45, x expr=\coordindex] {\dataAA};
                \addplot [fill=purple!60, draw=black!80] table [y=S46, x expr=\coordindex] {\dataAA};
                \addplot [fill=gray!60, draw=black!80] table [y=Dup, x expr=\coordindex] {\dataAA};
            \end{axis}
        \end{tikzpicture}
        \caption*{CICY 7862, $|\Gamma|=2$}
    \end{subfigure}\hfill
    \pgfplotstableread{
Label S42 S43 S44 S45 S46 Dup
N 14.1 13.0 11.5 13.2 11.3 37.0
NS5 3.2 4.2 2.2 2.6 3.0 84.8
NGX3 10.3 13.3 2.5 4.3 2.8 66.7
Nfull 1.4 1.4 4.2 1.4 0.0 91.7
    }\dataAB
    \begin{subfigure}{0.32\textwidth}
        \centering
        \begin{tikzpicture}
            \begin{axis}[]
                \addplot [fill=blue!60, draw=black!80] table [y=S42, x expr=\coordindex] {\dataAB};
                \addplot [fill=red!60, draw=black!80] table [y=S43, x expr=\coordindex] {\dataAB};
                \addplot [fill=green!60, draw=black!80] table [y=S44, x expr=\coordindex] {\dataAB};
                \addplot [fill=orange!60, draw=black!80] table [y=S45, x expr=\coordindex] {\dataAB};
                \addplot [fill=purple!60, draw=black!80] table [y=S46, x expr=\coordindex] {\dataAB};
                \addplot [fill=gray!60, draw=black!80] table [y=Dup, x expr=\coordindex] {\dataAB};
            \end{axis}
        \end{tikzpicture}
        \caption*{CICY 7862, $|\Gamma|=4$}
    \end{subfigure}\hfill
    \pgfplotstableread{
Label S42 S43 S44 S45 S46 Dup
N 36.8 8.3 4.4 6.4 8.3 35.8
NS5 0.0 0.0 0.0 0.0 0.0 100.0
NGX3 15.6 9.4 0.0 3.1 0.0 71.9
Nfull 0.0 0.0 0.0 0.0 0.0 100.0
    }\dataAC
    \begin{subfigure}{0.32\textwidth}
        \centering
        \begin{tikzpicture}
            \begin{axis}[]
                \addplot [fill=blue!60, draw=black!80] table [y=S42, x expr=\coordindex] {\dataAC};
                \addplot [fill=red!60, draw=black!80] table [y=S43, x expr=\coordindex] {\dataAC};
                \addplot [fill=green!60, draw=black!80] table [y=S44, x expr=\coordindex] {\dataAC};
                \addplot [fill=orange!60, draw=black!80] table [y=S45, x expr=\coordindex] {\dataAC};
                \addplot [fill=purple!60, draw=black!80] table [y=S46, x expr=\coordindex] {\dataAC};
                \addplot [fill=gray!60, draw=black!80] table [y=Dup, x expr=\coordindex] {\dataAC};
            \end{axis}
        \end{tikzpicture}
        \caption*{CICY 7522, $|\Gamma|=4$}
    \end{subfigure}\hfill
    \pgfplotstableread{
Label S42 S43 S44 S45 S46 Dup
N 5.3 8.7 7.2 19.4 40.3 19.0
NS5 0.0 0.0 0.0 0.0 0.0 100.0
NGX3 0.0 3.3 1.1 17.8 33.3 44.4
Nfull 0.0 0.0 0.0 0.0 0.0 100.0
    }\dataAD
    \begin{subfigure}{0.32\textwidth}
        \centering
        \begin{tikzpicture}
            \begin{axis}
                \addplot [fill=blue!60, draw=black!80] table [y=S42, x expr=\coordindex] {\dataAD};
                \addplot [fill=red!60, draw=black!80] table [y=S43, x expr=\coordindex] {\dataAD};
                \addplot [fill=green!60, draw=black!80] table [y=S44, x expr=\coordindex] {\dataAD};
                \addplot [fill=orange!60, draw=black!80] table [y=S45, x expr=\coordindex] {\dataAD};
                \addplot [fill=purple!60, draw=black!80] table [y=S46, x expr=\coordindex] {\dataAD};
                \addplot [fill=gray!60, draw=black!80] table [y=Dup, x expr=\coordindex] {\dataAD};
            \end{axis}
        \end{tikzpicture}
        \caption*{CICY 7491, $|\Gamma|=4$}
    \end{subfigure}

    \pgfplotstableread{
Label S42 S43 S44 S45 S46 Dup
N 11.1 16.0 27.5 5.2 18.1 22.0
NS5 0.0 0.0 0.0 0.0 0.0 100.0
NGX3 5.6 5.6 30.0 3.3 3.3 52.2
Nfull 0.0 0.0 0.0 0.0 0.0 100.0
    }\dataAE
    \begin{subfigure}{0.32\textwidth}
        \centering
        \begin{tikzpicture}
            \begin{axis}[ylabel={Solutions (\%)},]
                \addplot [fill=blue!60, draw=black!80] table [y=S42, x expr=\coordindex] {\dataAE};
                \addplot [fill=red!60, draw=black!80] table [y=S43, x expr=\coordindex] {\dataAE};
                \addplot [fill=green!60, draw=black!80] table [y=S44, x expr=\coordindex] {\dataAE};
                \addplot [fill=orange!60, draw=black!80] table [y=S45, x expr=\coordindex] {\dataAE};
                \addplot [fill=purple!60, draw=black!80] table [y=S46, x expr=\coordindex] {\dataAE};
                \addplot [fill=gray!60, draw=black!80] table [y=Dup, x expr=\coordindex] {\dataAE};
            \end{axis}
        \end{tikzpicture}
        \caption*{CICY 7462, $|\Gamma|=4$}
    \end{subfigure}
\hfill
    \pgfplotstableread{
Label S42 S43 S44 S45 S46 Dup
N 12.0 15.3 12.8 6.9 35.0 17.9
NS5 0.0 0.0 0.0 0.0 0.0 100.0
NGX3 6.5 19.6 5.4 3.3 18.5 46.7
Nfull 0.0 0.0 0.0 0.0 0.0 100.0
    }\dataAF
    \begin{subfigure}{0.32\textwidth}
        \centering
        \begin{tikzpicture}
            \begin{axis}[]
                \addplot [fill=blue!60, draw=black!80] table [y=S42, x expr=\coordindex] {\dataAF};
                \addplot [fill=red!60, draw=black!80] table [y=S43, x expr=\coordindex] {\dataAF};
                \addplot [fill=green!60, draw=black!80] table [y=S44, x expr=\coordindex] {\dataAF};
                \addplot [fill=orange!60, draw=black!80] table [y=S45, x expr=\coordindex] {\dataAF};
                \addplot [fill=purple!60, draw=black!80] table [y=S46, x expr=\coordindex] {\dataAF};
                \addplot [fill=gray!60, draw=black!80] table [y=Dup, x expr=\coordindex] {\dataAF};
            \end{axis}
        \end{tikzpicture}
        \caption*{CICY 7435, $|\Gamma|=4$}
    \end{subfigure}\hfill
    \pgfplotstableread{
Label S42 S43 S44 S45 S46 Dup
N 29.6 0.1 0.0 0.7 33.2 36.4
NS5 18.9 0.2 0.1 1.3 44.8 34.6
NGX3 28.1 0.1 0.1 0.9 36.4 34.4
Nfull 16.5 0.2 0.2 1.1 49.4 32.8
    }\dataAG
    \begin{subfigure}{0.32\textwidth}
        \centering
        \begin{tikzpicture}
            \begin{axis}
                \addplot [fill=blue!60, draw=black!80] table [y=S42, x expr=\coordindex] {\dataAG};
                \addplot [fill=red!60, draw=black!80] table [y=S43, x expr=\coordindex] {\dataAG};
                \addplot [fill=green!60, draw=black!80] table [y=S44, x expr=\coordindex] {\dataAG};
                \addplot [fill=orange!60, draw=black!80] table [y=S45, x expr=\coordindex] {\dataAG};
                \addplot [fill=purple!60, draw=black!80] table [y=S46, x expr=\coordindex] {\dataAG};
                \addplot [fill=gray!60, draw=black!80] table [y=Dup, x expr=\coordindex] {\dataAG};
            \end{axis}
        \end{tikzpicture}
        \caption*{CICY 7403, $|\Gamma|=2$}
    \end{subfigure}
\hfill
    \pgfplotstableread{
Label S42 S43 S44 S45 S46 Dup
N 7.7 5.4 7.4 13.3 16.0 50.3
NS5 2.8 1.9 2.2 0.5 0.5 92.2
NGX3 4.4 2.2 3.9 8.1 11.6 69.9
Nfull 2.6 0.3 0.7 0.7 0.0 95.7
    }\dataAH
    \begin{subfigure}{0.32\textwidth}
        \centering
        \begin{tikzpicture}
            \begin{axis}[]
                \addplot [fill=blue!60, draw=black!80] table [y=S42, x expr=\coordindex] {\dataAH};
                \addplot [fill=red!60, draw=black!80] table [y=S43, x expr=\coordindex] {\dataAH};
                \addplot [fill=green!60, draw=black!80] table [y=S44, x expr=\coordindex] {\dataAH};
                \addplot [fill=orange!60, draw=black!80] table [y=S45, x expr=\coordindex] {\dataAH};
                \addplot [fill=purple!60, draw=black!80] table [y=S46, x expr=\coordindex] {\dataAH};
                \addplot [fill=gray!60, draw=black!80] table [y=Dup, x expr=\coordindex] {\dataAH};
            \end{axis}
        \end{tikzpicture}
        \caption*{CICY 7247, $|\Gamma|=3$}
    \end{subfigure}

    \pgfplotstableread{
Label S42 S43 S44 S45 S46 Dup
N 8.2 53.9 8.4 0.0 14.7 14.7
NS5 3.0 48.2 4.2 0.0 13.7 30.9
NGX3 7.1 38.0 5.4 0.0 21.4 28.0
Nfull 2.7 28.5 3.7 0.1 19.8 45.2
    }\dataAI
    \begin{subfigure}{0.32\textwidth}
        \centering
        \begin{tikzpicture}
            \begin{axis}[ylabel={Solutions (\%)},]
                \addplot [fill=blue!60, draw=black!80] table [y=S42, x expr=\coordindex] {\dataAI};
                \addplot [fill=red!60, draw=black!80] table [y=S43, x expr=\coordindex] {\dataAI};
                \addplot [fill=green!60, draw=black!80] table [y=S44, x expr=\coordindex] {\dataAI};
                \addplot [fill=orange!60, draw=black!80] table [y=S45, x expr=\coordindex] {\dataAI};
                \addplot [fill=purple!60, draw=black!80] table [y=S46, x expr=\coordindex] {\dataAI};
                \addplot [fill=gray!60, draw=black!80] table [y=Dup, x expr=\coordindex] {\dataAI};
            \end{axis}
        \end{tikzpicture}
        \caption*{CICY 7245, $|\Gamma|=2$}
    \end{subfigure}
\hfill
 \pgfplotstableread{
Label S42 S43 S44 S45 S46 Dup
N 2.9 0.0 4.0 72.0 2.7 18.4
NS5 2.2 0.0 1.9 88.5 2.0 5.4
NGX3 2.2 0.0 3.2 80.4 2.4 11.8
Nfull 1.5 0.0 1.5 91.4 1.6 3.9
    }\dataAJ
    \begin{subfigure}{0.32\textwidth}
        \centering
        \begin{tikzpicture}
            \begin{axis}
                \addplot [fill=blue!60, draw=black!80] table [y=S42, x expr=\coordindex] {\dataAJ};
                \addplot [fill=red!60, draw=black!80] table [y=S43, x expr=\coordindex] {\dataAJ};
                \addplot [fill=green!60, draw=black!80] table [y=S44, x expr=\coordindex] {\dataAJ};
                \addplot [fill=orange!60, draw=black!80] table [y=S45, x expr=\coordindex] {\dataAJ};
                \addplot [fill=purple!60, draw=black!80] table [y=S46, x expr=\coordindex] {\dataAJ};
                \addplot [fill=gray!60, draw=black!80] table [y=Dup, x expr=\coordindex] {\dataAJ};
            \end{axis}
        \end{tikzpicture}
        \caption*{CICY 6831, $|\Gamma|=2$}
    \end{subfigure}
\hfill
    \pgfplotstableread{
Label S42 S43 S44 S45 S46 Dup
N 0.0 100.0 0.0 0.0 0.0 0.0
NS5 0.0 100.0 0.0 0.0 0.0 0.0
NGX3 0.0 100.0 0.0 0.0 0.0 0.0
Nfull 0.0 100.0 0.0 0.0 0.0 0.0
    }\dataAK
    \begin{subfigure}{0.32\textwidth}
        \centering
        \begin{tikzpicture}
            \begin{axis}[]
                \addplot [fill=blue!60, draw=black!80] table [y=S42, x expr=\coordindex] {\dataAK};
                \addplot [fill=red!60, draw=black!80] table [y=S43, x expr=\coordindex] {\dataAK};
                \addplot [fill=green!60, draw=black!80] table [y=S44, x expr=\coordindex] {\dataAK};
                \addplot [fill=orange!60, draw=black!80] table [y=S45, x expr=\coordindex] {\dataAK};
                \addplot [fill=purple!60, draw=black!80] table [y=S46, x expr=\coordindex] {\dataAK};
                \addplot [fill=gray!60, draw=black!80] table [y=Dup, x expr=\coordindex] {\dataAK};
            \end{axis}
        \end{tikzpicture}
        \caption*{CICY 6828, $|\Gamma|=2$}
    \end{subfigure}
\hfill
    \pgfplotstableread{
Label S42 S43 S44 S45 S46 Dup
N 0.0 0.0 40.0 20.0 28.0 12.0
NS5 0.0 0.0 0.0 25.0 0.0 75.0
NGX3 0.0 0.0 31.6 21.1 26.3 21.1
Nfull 0.0 0.0 0.0 33.3 0.0 66.7
    }\dataAL
    \begin{subfigure}{0.32\textwidth}
        \centering
        \begin{tikzpicture}
            \begin{axis}[]
                \addplot [fill=blue!60, draw=black!80] table [y=S42, x expr=\coordindex] {\dataAL};
                \addplot [fill=red!60, draw=black!80] table [y=S43, x expr=\coordindex] {\dataAL};
                \addplot [fill=green!60, draw=black!80] table [y=S44, x expr=\coordindex] {\dataAL};
                \addplot [fill=orange!60, draw=black!80] table [y=S45, x expr=\coordindex] {\dataAL};
                \addplot [fill=purple!60, draw=black!80] table [y=S46, x expr=\coordindex] {\dataAL};
                \addplot [fill=gray!60, draw=black!80] table [y=Dup, x expr=\coordindex] {\dataAL};
            \end{axis}
        \end{tikzpicture}
        \caption*{CICY 6828, $|\Gamma|=4$}
    \end{subfigure}
\caption*{\hfill\textit{Continued on next page}}
\end{figure}

\begin{figure*}[!htp]
\ContinuedFloat
    \centering
    \pgfplotsset{
        compat=1.18,
        every axis/.append style={
            width=\textwidth, 
            height=4.5cm, 
            ymin=0, ymax=110, ybar stacked, bar width=12pt,
            xtick={0,1,2,3}, xticklabels={$N$, $N_{S_5}$, $N_{G_{\mathbf{X}_3}}$, $N_{\text{full}}$},
            enlarge x limits=0.2, 
            legend style={cells={anchor=west}, legend pos=north east, font=\tiny},
            legend image code/.code={\draw[#1, draw=none] (0cm,-0.1cm) rectangle (0.2cm,0.1cm);},
            tick label style={font=\scriptsize},
            label style={font=\footnotesize}
        }
    }
   
    \pgfplotstableread{
Label S42 S43 S44 S45 S46 Dup
N 0.0 0.0 0.0 100.0 0.0 0.0
NS5 0.0 0.0 0.0 100.0 0.0 0.0
NGX3 0.0 0.0 0.0 100.0 0.0 0.0
Nfull 0.0 0.0 0.0 100.0 0.0 0.0
    }\dataAM
    \begin{subfigure}{0.32\textwidth}
        \centering

        \caption*{CICY 6784, $|\Gamma|=2$}
    \end{subfigure}
    \pgfplotstableread{
Label S42 S43 S44 S45 S46 Dup
N 0.0 0.0 63.6 0.0 27.3 9.1
NS5 0.0 0.0 33.3 0.0 0.0 66.7
NGX3 0.0 0.0 60.0 0.0 20.0 20.0
Nfull 0.0 0.0 50.0 0.0 0.0 50.0
    }\dataAN
    \begin{subfigure}{0.32\textwidth}
        \centering
        %
        \caption*{CICY 6784, $|\Gamma|=4$}
    \end{subfigure}
\end{figure*}
\newpage

\setlength{\LTleft}{\fill}
\setlength{\LTright}{\fill}
\setlength{\LTcapwidth}{\linewidth}
\renewcommand{\arraystretch}{1}
\begin{longtable}{c | c | c | c | c | c | c | c | c | c | c | c | c}
\caption{Counting of the solutions for the theories involved in transfer learning within bucket. The meaning of the columns is analogous to that in Table \ref{tab:detailed_counts}, with `Src' and `Trg' referring respectively to the source and target CYs involved in the transfer learning.} \label{tab:detailed_counts_within_bucket} \\
\hhline{=|=|=|=|=|=|====|=|=|=}
\multirow{2}{*}{Src \#} & \multirow{2}{*}{Src $|G_{\IX_3}|$} & \multirow{2}{*}{Trg \#} & \multirow{2}{*}{Trg $|G_{\IX_3}|$} & \multirow{2}{*}{$h^{1,1}(\IX_3)$} & \multirow{2}{*}{$|\Gamma|$} & \multirow{2}{*}{Sol.} & \multicolumn{3}{c|}{Seeds} & \multirow{2}{*}{Dup. $\left(\frac{N_\text{dup}}{N_\text{tot}}\%\right)$} & \multirow{2}{*}{Total} & \multirow{2}{*}{Equivariance} \\ \cline{8-10}
 & & & & & & & 42 & 43 & 44 & & & \\
\hhline{=|=|=|=|=|=|=|=|=|=|=|=|=}
\endfirsthead
\multicolumn{13}{c}{{\tablename\ \thetable\ -- \textit{Continued from previous page}}} \\
\hhline{=|=|=|=|=|=|====|=|=|=}
\multirow{2}{*}{Src \#} & \multirow{2}{*}{Src $|G_{\IX_3}|$} & \multirow{2}{*}{Trg \#} & \multirow{2}{*}{Trg $|G_{\IX_3}|$} & \multirow{2}{*}{$h^{1,1}(\IX_3)$} & \multirow{2}{*}{$|\Gamma|$} & \multirow{2}{*}{Sol.} & \multicolumn{3}{c|}{Seeds} & \multirow{2}{*}{Dup. $\left(\frac{N_\text{dup}}{N_\text{tot}}\%\right)$} & \multirow{2}{*}{Total} & \multirow{2}{*}{Equivariance} \\ \cline{8-10}
 & & & & & & & 42 & 43 & 44 & & & \\
\hhline{=|=|=|=|=|=|=|=|=|=|=|=|=}
\endhead
\multicolumn{13}{r}{\textit{Continued on next page}} \\
\endfoot
\hhline{=|=|=|=|=|=|=|=|=|=|=|=|=}
\endlastfoot
\multirow{4}{*}{7447} & \multirow{4}{*}{120} & \multirow{4}{*}{7487} & \multirow{4}{*}{120} & \multirow{4}{*}{5} & \multirow{4}{*}{4}  & $N$ & 38561 & 40671 & 8246 & 19609 (29.0\%) & 67634 & 11506 \checkmark \\
 &  &  &  &  &   & $N_{S_5}$ & 8829 & 10873 & 7817 & 6951 (34.5\%) & 20176 & 4840 \checkmark \\
 &  &  &  &  &   & $N_{G_{\mathbf{X}_3}}$ & 1299 & 1710 & 304 & 1038 (50.4\%) & 2058 & 303 \checkmark \\
 &  &  &  &  &   & $N_{\text{full}}$ & 170 & 211 & 140 & 148 (47.9\%) & 309 & 65 \checkmark \\ \hline
\multirow{4}{*}{5452} & \multirow{4}{*}{8} & \multirow{4}{*}{7447} & \multirow{4}{*}{120} & \multirow{4}{*}{5} & \multirow{4}{*}{4}  & $N$ & 14146 & 17177 & 0 & 50 (0.2\%) & 31273 & 6568 \checkmark \\
 &  &  &  &  &   & $N_{S_5}$ & 8755 & 10499 & 0 & 213 (1.1\%) & 19041 & 4595 \checkmark \\
 &  &  &  &  &   & $N_{G_{\mathbf{X}_3}}$ & 2379 & 6581 & 0 & 270 (3.1\%) & 8690 & 1529 \checkmark \\
 &  &  &  &  &   & $N_{\text{full}}$ & 406 & 3162 & 0 & 221 (6.6\%) & 3347 & 617 \checkmark \\
\hline
\end{longtable}

\setlength{\LTleft}{\fill}
\setlength{\LTright}{\fill}
\setlength{\LTcapwidth}{\linewidth}
\renewcommand{\arraystretch}{1}
\begin{longtable}{c | c | c | c | c | c | c | c | c | c | c}
\caption{Efficacy metrics for transfer learning within bucket. The table compares the performance of a baseline model trained from scratch against a transfer learning model. The metrics are defined in Section \ref{sec:transferlearning}. \textit{\% Imp.} denotes the relative percentage improvement of the transfer policy.}\label{tab:efficacy_within_bucket} \\
\hhline{=|=|=|=|=|=|=|=|=|=|=}
\multirow{2}{*}{Src \#} & \multirow{2}{*}{Src $|G_{\IX_3}|$} & \multirow{2}{*}{Trg \#} & \multirow{2}{*}{Trg $|G_{\IX_3}|$} & \multirow{2}{*}{$h^{1,1}(\IX_3)$} & \multirow{2}{*}{$|\Gamma|$} & \multirow{2}{*}{{Seed}} & \multirow{2}{*}{{Metric}} & \multirow{2}{*}{{From Scratch}} & \multirow{2}{*}{{Transfer}} & \multirow{2}{*}{{\% Imp.}} \\
 & & & & & & & & & & \\
\hhline{=|=|=|=|=|=|=|=|=|=|=}
\endfirsthead
\multicolumn{11}{c}{{\tablename\ \thetable\ -- \textit{Continued from previous page}}} \\
\hhline{=|=|=|=|=|=|=|=|=|=|=}
\multirow{2}{*}{Src \#} & \multirow{2}{*}{Src $|G_{\IX_3}|$} & \multirow{2}{*}{Trg \#} & \multirow{2}{*}{Trg $|G_{\IX_3}|$} & \multirow{2}{*}{$h^{1,1}(\IX_3)$} & \multirow{2}{*}{$|\Gamma|$} & \multirow{2}{*}{{Seed}} & \multirow{2}{*}{{Metric}} & \multirow{2}{*}{{From Scratch}} & \multirow{2}{*}{{Transfer}} & \multirow{2}{*}{{\% Imp.}} \\
 & & & & & & & & & & \\
\hhline{=|=|=|=|=|=|=|=|=|=|=}
\endhead
\multicolumn{11}{r}{\textit{Continued on next page}} \\
\endfoot
\hhline{=|=|=|=|=|=|=|=|=|=|=}
\endlastfoot
\multirow{8}{*}{7447} & \multirow{8}{*}{120} & \multirow{8}{*}{7487} & \multirow{8}{*}{120} & \multirow{8}{*}{5} & \multirow{8}{*}{4}  & \multirow{2}{*}{42} & Zero-Shot Reward & 0.263 & 0.905 & +244.1\% \\
 & & & & &  &  & Episodes to First Solution & 901120 & 8192 & +99.1\% \\ \cline{7-11}
 & & & & &  & \multirow{2}{*}{43} & Zero-Shot Reward & 0.279 & 0.909 & +225.8\% \\
 & & & & &  &  & Episodes to First Solution & 974848 & 8192 & +99.2\% \\ \cline{7-11}
 & & & & &  & \multirow{2}{*}{44} & Zero-Shot Reward & 0.310 & 0.934 & +201.3\% \\
 & & & & &  &  & Episodes to First Solution & 917504 & 8192 & +99.1\% \\ \cline{7-11}
 & & & & &  & \multirow{2}{*}{Avg} & Zero-Shot Reward & 0.284 & 0.916 & +222.5\% \\
 & & & & &  &  & Episodes to First Solution & 931157 & 8192 & +99.1\% \\
\hline
\end{longtable}

\setlength{\LTleft}{\fill}
\setlength{\LTright}{\fill}
\setlength{\LTcapwidth}{\linewidth}
\renewcommand{\arraystretch}{1.2}
\begin{longtable}{c | c | c | c | c | c | c | c | c | c | c | c}
\caption{More structural metrics for transfer learning within bucket. The table compares the performance of a baseline model trained from scratch against a transfer learning model. The metrics are defined in Section \ref{sec:transferlearning}. \textit{\% Imp.} denotes the relative percentage improvement of the transfer policy. Entries with a dash (-) indicate that the metric could not be computed, because the model failed to reach 1000 unique solutions within the 5M episodes.}\label{tab:extra_tests_within_bucket} \\
\hhline{=|=|=|=|=|=|=|=|=|=|=|=}
\multirow{2}{*}{Src \#} & \multirow{2}{*}{Src $|G_{\IX_3}|$} & \multirow{2}{*}{Trg \#} & \multirow{2}{*}{Trg $|G_{\IX_3}|$} & \multirow{2}{*}{$h^{1,1}(\IX_3)$} & \multirow{2}{*}{$|\Gamma|$} & \multirow{2}{*}{{Seed}} & \multirow{2}{*}{{Mean Weight Dist}} & \multirow{2}{*}{{Sum $D_{KL}$}} & \multicolumn{2}{c|}{{Ep. to 1k Sols}} & \multirow{2}{*}{{\% Imp.}} \\ \cline{10-11}
 & & & & & & & & & Scratch & Transfer & \\
\hhline{=|=|=|=|=|=|=|=|=|=|=|=}
\endfirsthead
\multicolumn{12}{c}{{\tablename\ \thetable\ -- \textit{Continued from previous page}}} \\
\hhline{=|=|=|=|=|=|=|=|=|=|=|=}
\multirow{2}{*}{Src \#} & \multirow{2}{*}{Src $|G_{\IX_3}|$} & \multirow{2}{*}{Trg \#} & \multirow{2}{*}{Trg $|G_{\IX_3}|$} & \multirow{2}{*}{$h^{1,1}(\IX_3)$} & \multirow{2}{*}{$|\Gamma|$} & \multirow{2}{*}{{Seed}} & \multirow{2}{*}{{Mean Weight Dist}} & \multirow{2}{*}{{Sum $D_{KL}$}} & \multicolumn{2}{c|}{{Ep. to 1k Sols}} & \multirow{2}{*}{{\% Imp.}} \\ \cline{10-11}
 & & & & & & & & & Scratch & Transfer & \\
\hhline{=|=|=|=|=|=|=|=|=|=|=|=}
\endhead
\multicolumn{12}{r}{\textit{Continued on next page}} \\
\endfoot
\hhline{=|=|=|=|=|=|=|=|=|=|=|=}
\endlastfoot
\multirow{4}{*}{7447} & \multirow{4}{*}{120} & \multirow{4}{*}{7487} & \multirow{4}{*}{120} & \multirow{4}{*}{5} & \multirow{4}{*}{4}  & 42 & 0.04 & 5.69 & 1277952 & 57344 & +95.5\% \\
 & & & & &  & 43 & 0.04 & 4.45 & 1286144 & 57344 & +95.5\% \\
 & & & & &  & 44 & 0.07 & 37.63 & 1228800 & 311296 & +74.7\% \\
 & & & & &  & Avg & 0.05 & 15.92 & 1264298 & 141994 & +88.8\% \\ \hline
\multirow{3}{*}{5452} & \multirow{3}{*}{8} & \multirow{3}{*}{7447} & \multirow{3}{*}{120} & \multirow{3}{*}{5} & \multirow{3}{*}{4}  & 42 & 0.08 & 35.36 & 1253376 & 221184 & +82.4\% \\
 & & & & &  & 43 & 0.10 & 51.24 & 1261568 & 540672 & +57.1\% \\
 & & & & &  & Avg & 0.09 & 43.30 & 1257472 & 380928 & +69.7\% \\ \hline
\multirow{4}{*}{7447} & \multirow{4}{*}{120} & \multirow{4}{*}{6724} & \multirow{4}{*}{12} & \multirow{4}{*}{5} & \multirow{4}{*}{2}  & 42 & 0.09 & 10.86 & - & - & - \\
\hline
\end{longtable}

\setlength{\LTleft}{\fill}
\setlength{\LTright}{\fill}
\setlength{\LTcapwidth}{\linewidth}
\renewcommand{\arraystretch}{1}
\begin{longtable}{c | c | c | c | c | c | c | c | c | c | c | c | c | c | c}
\caption{Catastrophic forgetting metrics for transfer learning within bucket. 'Policy' defines the policy weights used, i.e. $\pi_{\theta_\text{src}}$ or $\pi_{\theta_\text{trg}}$, with 'On' being the evaluation environment, i.e. $\mathcal{E}_\text{src}$ or $\mathcal{E}_\text{trg}$, where they have been tested. 'Anom', 'Stab', 'Sum', 'Rng', and 'Pair' report the expected CRS in \eqref{eq:CRS} for each individual constraint. $\text{VSR}$ is the Validation Success Rate in \eqref{eq:VSR}.}\label{tab:forgetting_within_bucket} \\
\hhline{=|=|=|=|=|=|=|=|=|=|=|=|=|=|=}
\multirow{2}{*}{Src \#} & \multirow{2}{*}{Src $|G_{\IX_3}|$} & \multirow{2}{*}{Trg \#} & \multirow{2}{*}{Trg $|G_{\IX_3}|$} & \multirow{2}{*}{$h^{1,1}(\IX_3)$} & \multirow{2}{*}{$|\Gamma|$} & \multirow{2}{*}{Policy} & \multirow{2}{*}{On} & \multirow{2}{*}{Anom} & \multirow{2}{*}{Stab} & \multirow{2}{*}{Sum} & \multirow{2}{*}{Rng} & \multirow{2}{*}{Pair} & \multirow{2}{*}{VSR} & \multirow{2}{*}{Mean CRS} \\
 & & & & & & & & & & & & & & \\
\hhline{=|=|=|=|=|=|=|=|=|=|=|=|=|=|=}
\endfirsthead
\multicolumn{15}{c}{{\tablename\ \thetable\ -- \textit{Continued from previous page}}} \\
\hhline{=|=|=|=|=|=|=|=|=|=|=|=|=|=|=}
\multirow{2}{*}{Src \#} & \multirow{2}{*}{Src $|G_{\IX_3}|$} & \multirow{2}{*}{Trg \#} & \multirow{2}{*}{Trg $|G_{\IX_3}|$} & \multirow{2}{*}{$h^{1,1}(\IX_3)$} & \multirow{2}{*}{$|\Gamma|$} & \multirow{2}{*}{Policy} & \multirow{2}{*}{On} & \multirow{2}{*}{Anom} & \multirow{2}{*}{Stab} & \multirow{2}{*}{Sum} & \multirow{2}{*}{Rng} & \multirow{2}{*}{Pair} & \multirow{2}{*}{VSR} & \multirow{2}{*}{Mean CRS} \\
 & & & & & & & & & & & & & & \\
\hhline{=|=|=|=|=|=|=|=|=|=|=|=|=|=|=}
\endhead
\multicolumn{15}{r}{\textit{Continued on next page}} \\
\endfoot
\hhline{=|=|=|=|=|=|=|=|=|=|=|=|=|=|=}
\endlastfoot
\multirow{3}{*}{7447} & \multirow{3}{*}{120} & \multirow{3}{*}{7487} & \multirow{3}{*}{120} & \multirow{3}{*}{5} & \multirow{3}{*}{4}  & $\pi_{\theta_\text{src}}$ & $\mathcal{E}_\text{src}$ & 99.9\% & 87.2\% & 71.1\% & 93.0\% & 52.8\% & 0.7\% & 0.843 \\
 & & & & &  & $\pi_{\theta_\text{trg}}$ & $\mathcal{E}_\text{trg}$ & 98.1\% & 88.9\% & 76.0\% & 91.3\% & 60.9\% & 0.7\% & 0.859 \\
 & & & & &  & $\pi_{\theta_\text{trg}}$ & $\mathcal{E}_\text{src}$ & 97.9\% & 89.0\% & 75.3\% & 90.7\% & 60.6\% & 0.9\% & 0.871 \\ \hline
\multirow{3}{*}{5452} & \multirow{3}{*}{8} & \multirow{3}{*}{7447} & \multirow{3}{*}{120} & \multirow{3}{*}{5} & \multirow{3}{*}{4}  & $\pi_{\theta_\text{src}}$ & $\mathcal{E}_\text{src}$ & 100.0\% & 86.1\% & 27.4\% & 88.7\% & 52.1\% & 0.0\% & 0.708 \\
 & & & & &  & $\pi_{\theta_\text{trg}}$ & $\mathcal{E}_\text{trg}$ & 99.1\% & 88.4\% & 97.2\% & 98.1\% & 97.5\% & 1.1\% & 1.017 \\
 & & & & &  & $\pi_{\theta_\text{trg}}$ & $\mathcal{E}_\text{src}$ & 70.4\% & 74.3\% & 52.4\% & 55.1\% & 48.1\% & 0.0\% & 0.601 \\ \hline
\multirow{3}{*}{7447} & \multirow{3}{*}{120} & \multirow{3}{*}{6724} & \multirow{3}{*}{12} & \multirow{3}{*}{5} & \multirow{3}{*}{2}  & $\pi_{\theta_\text{src}}$ & $\mathcal{E}_\text{src}$ & 99.5\% & 84.3\% & 80.6\% & 93.9\% & 84.5\% & 0.3\% & 0.896 \\
 & & & & &  & $\pi_{\theta_\text{trg}}$ & $\mathcal{E}_\text{trg}$ & 99.2\% & 84.6\% & 28.0\% & 84.3\% & 50.0\% & 0.0\% & 0.690 \\
\hline
\end{longtable}

\setlength{\LTleft}{\fill}
\setlength{\LTright}{\fill}
\setlength{\LTcapwidth}{\linewidth}
\renewcommand{\arraystretch}{1}
\begin{longtable}{c | c | c | c | c | c | c | c | c | c | c | c}
\caption{Counting of the solutions for the theories involved in transfer learning across $\Gamma$. The meaning of the columns is analogous to that in Table \ref{tab:detailed_counts}, with `Src' and `Trg' referring respectively to the source and target CYs involved in the transfer learning.} \label{tab:detailed_counts_cross_gamma} \\
\hhline{=|=|=|=|=|====|=|=|=}
\multirow{2}{*}{\#} & \multirow{2}{*}{$h^{1,1}(\IX_3)$} & \multirow{2}{*}{$G_{\IX_3}$} & \multirow{2}{*}{Src $|\Gamma|$} & \multirow{2}{*}{Trg $|\Gamma|$} & \multirow{2}{*}{Sol.} & \multicolumn{3}{c|}{Seeds} & \multirow{2}{*}{Dup. $\left(\frac{N_\text{dup}}{N_\text{tot}}\%\right)$} & \multirow{2}{*}{Total} & \multirow{2}{*}{Equivariance} \\ \cline{7-9}
 & & & & & & 42 & 43 & 44 & & & \\
\hhline{=|=|=|=|=|=|=|=|=|=|=|=}
\endfirsthead
\multicolumn{12}{c}{{\tablename\ \thetable\ -- \textit{Continued from previous page}}} \\
\hhline{=|=|=|=|=|====|=|=|=}
\multirow{2}{*}{\#} & \multirow{2}{*}{$h^{1,1}(\IX_3)$} & \multirow{2}{*}{$G_{\IX_3}$} & \multirow{2}{*}{Src $|\Gamma|$} & \multirow{2}{*}{Trg $|\Gamma|$} & \multirow{2}{*}{Sol.} & \multicolumn{3}{c|}{Seeds} & \multirow{2}{*}{Dup. $\left(\frac{N_\text{dup}}{N_\text{tot}}\%\right)$} & \multirow{2}{*}{Total} & \multirow{2}{*}{Equivariance} \\ \cline{7-9}
 & & & & & & 42 & 43 & 44 & & & \\
\hhline{=|=|=|=|=|=|=|=|=|=|=|=}
\endhead
\multicolumn{12}{r}{\textit{Continued on next page}} \\
\endfoot
\hhline{=|=|=|=|=|=|=|=|=|=|=|=}
\endlastfoot
\multirow{4}{*}{7447} & \multirow{4}{*}{5} & \multirow{4}{*}{120} & \multirow{4}{*}{4} & \multirow{4}{*}{2}  & $N$ & 6078 & 13254 & 14242 & 5007 (18.2\%) & 27564 & 6621 \checkmark \\
 &  &  &  &   & $N_{S_5}$ & 1420 & 8158 & 11523 & 2805 (16.0\%) & 17511 & 4278 \checkmark \\
 &  &  &  &   & $N_{G_{\mathbf{X}_3}}$ & 212 & 586 & 226 & 217 (31.3\%) & 694 & 207 \checkmark \\
 &  &  &  &   & $N_{\text{full}}$ & 39 & 240 & 109 & 76 (25.9\%) & 294 & 86 \checkmark \\ \hline
\multirow{4}{*}{7447} & \multirow{4}{*}{5} & \multirow{4}{*}{120} & \multirow{4}{*}{2} & \multirow{4}{*}{4}  & $N$ & 0 & 3954 & 17184 & 88 (0.4\%) & 21050 & 3594 \checkmark \\
 &  &  &  &   & $N_{S_5}$ & 0 & 2033 & 9266 & 1523 (15.6\%) & 9776 & 2128 \checkmark \\
 &  &  &  &   & $N_{G_{\mathbf{X}_3}}$ & 0 & 1616 & 2299 & 403 (11.5\%) & 3512 & 568 \checkmark \\
 &  &  &  &   & $N_{\text{full}}$ & 0 & 138 & 571 & 115 (19.4\%) & 594 & 181 \checkmark \\
\hline
\end{longtable}

\setlength{\LTleft}{\fill}
\setlength{\LTright}{\fill}
\setlength{\LTcapwidth}{\linewidth}
\renewcommand{\arraystretch}{1}
\begin{longtable}{c | c | c | c | c | c | c | c | c | c}
\caption{Efficacy metrics for transfer learning across $\Gamma$. The table compares the performance of a baseline model trained from scratch against a transfer learning model. The metrics are defined in Section \ref{sec:transferlearning}. \textit{\% Imp.} denotes the relative percentage improvement of the transfer policy.}\label{tab:efficacy_cross_gamma} \\
\hhline{=|=|=|=|=|=|=|=|=|=}
\multirow{2}{*}{\#} & \multirow{2}{*}{$h^{1,1}(\IX_3)$} & \multirow{2}{*}{$G_{\IX_3}$} & \multirow{2}{*}{Src $|\Gamma|$} & \multirow{2}{*}{Trg $|\Gamma|$} & \multirow{2}{*}{{Seed}} & \multirow{2}{*}{{Metric}} & \multirow{2}{*}{{From Scratch}} & \multirow{2}{*}{{Transfer}} & \multirow{2}{*}{{\% Imp.}} \\
 & & & & & & & & & \\
\hhline{=|=|=|=|=|=|=|=|=|=}
\endfirsthead
\multicolumn{10}{c}{{\tablename\ \thetable\ -- \textit{Continued from previous page}}} \\
\hhline{=|=|=|=|=|=|=|=|=|=}
\multirow{2}{*}{\#} & \multirow{2}{*}{$h^{1,1}(\IX_3)$} & \multirow{2}{*}{$G_{\IX_3}$} & \multirow{2}{*}{Src $|\Gamma|$} & \multirow{2}{*}{Trg $|\Gamma|$} & \multirow{2}{*}{{Seed}} & \multirow{2}{*}{{Metric}} & \multirow{2}{*}{{From Scratch}} & \multirow{2}{*}{{Transfer}} & \multirow{2}{*}{{\% Imp.}} \\
 & & & & & & & & & \\
\hhline{=|=|=|=|=|=|=|=|=|=}
\endhead
\multicolumn{10}{r}{\textit{Continued on next page}} \\
\endfoot
\hhline{=|=|=|=|=|=|=|=|=|=}
\endlastfoot
\multirow{8}{*}{7447} & \multirow{8}{*}{5} & \multirow{8}{*}{120} & \multirow{8}{*}{4} & \multirow{8}{*}{2}  & \multirow{2}{*}{42} & Zero-Shot Reward & 0.263 & 0.731 & +177.9\% \\
 & & & &  &  & Episodes to First Solution & 1081344 & 16384 & +98.5\% \\ \cline{6-10}
 & & & &  & \multirow{2}{*}{43} & Zero-Shot Reward & 0.279 & 0.745 & +167.0\% \\
 & & & &  &  & Episodes to First Solution & 1089536 & 32768 & +97.0\% \\ \cline{6-10}
 & & & &  & \multirow{2}{*}{44} & Zero-Shot Reward & 0.310 & 0.738 & +138.1\% \\
 & & & &  &  & Episodes to First Solution & 974848 & 16384 & +98.3\% \\ \cline{6-10}
 & & & &  & \multirow{2}{*}{Avg} & Zero-Shot Reward & 0.284 & 0.738 & +159.9\% \\
 & & & &  &  & Episodes to First Solution & 1048576 & 21845 & +97.9\% \\
\hline
\end{longtable}

\setlength{\LTleft}{\fill}
\setlength{\LTright}{\fill}
\setlength{\LTcapwidth}{\linewidth}
\renewcommand{\arraystretch}{1.2}
\begin{longtable}{c | c | c | c | c | c | c | c | c | c | c}
\caption{More structural metrics for transfer learning across $\Gamma$. The table compares the performance of a baseline model trained from scratch against a transfer learning model. The metrics are defined in Section \ref{sec:transferlearning}. \textit{\% Imp.} denotes the relative percentage improvement of the transfer policy. Entries with a dash (-) indicate that the metric could not be computed, because the model failed to reach 1000 unique solutions within the 5M episodes.}\label{tab:extra_tests_cross_gamma} \\
\hhline{=|=|=|=|=|=|=|=|=|=|=}
\multirow{2}{*}{\#} & \multirow{2}{*}{$h^{1,1}(\IX_3)$} & \multirow{2}{*}{$G_{\IX_3}$} & \multirow{2}{*}{Src $|\Gamma|$} & \multirow{2}{*}{Trg $|\Gamma|$} & \multirow{2}{*}{{Seed}} & \multirow{2}{*}{{Mean Weight Dist}} & \multirow{2}{*}{{Sum $D_{KL}$}} & \multicolumn{2}{c|}{{Ep. to 1k Sols}} & \multirow{2}{*}{{\% Imp.}} \\ \cline{9-10}
 & & & & & & & & Scratch & Transfer & \\
\hhline{=|=|=|=|=|=|=|=|=|=|=}
\endfirsthead
\multicolumn{11}{c}{{\tablename\ \thetable\ -- \textit{Continued from previous page}}} \\
\hhline{=|=|=|=|=|=|=|=|=|=|=}
\multirow{2}{*}{\#} & \multirow{2}{*}{$h^{1,1}(\IX_3)$} & \multirow{2}{*}{$G_{\IX_3}$} & \multirow{2}{*}{Src $|\Gamma|$} & \multirow{2}{*}{Trg $|\Gamma|$} & \multirow{2}{*}{{Seed}} & \multirow{2}{*}{{Mean Weight Dist}} & \multirow{2}{*}{{Sum $D_{KL}$}} & \multicolumn{2}{c|}{{Ep. to 1k Sols}} & \multirow{2}{*}{{\% Imp.}} \\ \cline{9-10}
 & & & & & & & & Scratch & Transfer & \\
\hhline{=|=|=|=|=|=|=|=|=|=|=}
\endhead
\multicolumn{11}{r}{\textit{Continued on next page}} \\
\endfoot
\hhline{=|=|=|=|=|=|=|=|=|=|=}
\endlastfoot
\multirow{4}{*}{7447} & \multirow{4}{*}{5} & \multirow{4}{*}{120} & \multirow{4}{*}{4} & \multirow{4}{*}{2}  & 42 & 0.06 & 81.16 & 1589248 & 311296 & +80.4\% \\
 & & & &  & 43 & 0.07 & 94.40 & 1482752 & 319488 & +78.5\% \\
 & & & &  & 44 & 0.07 & 28.88 & 1425408 & 270336 & +81.0\% \\
 & & & &  & Avg & 0.06 & 68.15 & 1499136 & 300373 & +80.0\% \\ \hline
\multirow{3}{*}{7447} & \multirow{3}{*}{5} & \multirow{3}{*}{120} & \multirow{3}{*}{2} & \multirow{3}{*}{4}  & 43 & 0.07 & 12.07 & 1261568 & 4251648 & -237.0\% \\
 & & & &  & 44 & 0.09 & 26.45 & 1228800 & 204800 & +83.3\% \\
 & & & &  & Avg & 0.08 & 19.26 & 1245184 & 2228224 & -78.9\% \\ \hline
\multirow{4}{*}{6788} & \multirow{4}{*}{5} & \multirow{4}{*}{24} & \multirow{4}{*}{4} & \multirow{4}{*}{2}  & 42 & 0.06 & 49.95 & - & - & - \\
\hline
\end{longtable}

\setlength{\LTleft}{\fill}
\setlength{\LTright}{\fill}
\setlength{\LTcapwidth}{\linewidth}
\renewcommand{\arraystretch}{1}
\begin{longtable}{c | c | c | c | c | c | c | c | c | c | c | c | c | c}
\caption{Catastrophic forgetting metrics for transfer learning across $\Gamma$. 'Policy' defines the policy weights used, i.e. $\pi_{\theta_\text{src}}$ or $\pi_{\theta_\text{trg}}$, with 'On' being the evaluation environment, i.e. $\mathcal{E}_\text{src}$ or $\mathcal{E}_\text{trg}$, where they have been tested. 'Anom', 'Stab', 'Sum', 'Rng', and 'Pair' report the expected CRS in \eqref{eq:CRS} for each individual constraint. $\text{VSR}$ is the Validation Success Rate in \eqref{eq:VSR}.} \label{tab:forgetting_cross_gamma} \\
\hhline{=|=|=|=|=|=|=|=|=|=|=|=|=|=}
\multirow{2}{*}{\#} & \multirow{2}{*}{$h^{1,1}(\IX_3)$} & \multirow{2}{*}{$G_{\IX_3}$} & \multirow{2}{*}{Src $|\Gamma|$} & \multirow{2}{*}{Trg $|\Gamma|$} & \multirow{2}{*}{Policy} & \multirow{2}{*}{On} & \multirow{2}{*}{Anom} & \multirow{2}{*}{Stab} & \multirow{2}{*}{Sum} & \multirow{2}{*}{Rng} & \multirow{2}{*}{Pair} & \multirow{2}{*}{VSR} & \multirow{2}{*}{Mean CRS} \\
 & & & & & & & & & & & & & \\
\hhline{=|=|=|=|=|=|=|=|=|=|=|=|=|=}
\endfirsthead
\multicolumn{14}{c}{{\tablename\ \thetable\ -- \textit{Continued from previous page}}} \\
\hhline{=|=|=|=|=|=|=|=|=|=|=|=|=|=}
\multirow{2}{*}{\#} & \multirow{2}{*}{$h^{1,1}(\IX_3)$} & \multirow{2}{*}{$G_{\IX_3}$} & \multirow{2}{*}{Src $|\Gamma|$} & \multirow{2}{*}{Trg $|\Gamma|$} & \multirow{2}{*}{Policy} & \multirow{2}{*}{On} & \multirow{2}{*}{Anom} & \multirow{2}{*}{Stab} & \multirow{2}{*}{Sum} & \multirow{2}{*}{Rng} & \multirow{2}{*}{Pair} & \multirow{2}{*}{VSR} & \multirow{2}{*}{Mean CRS} \\
 & & & & & & & & & & & & & \\
\hhline{=|=|=|=|=|=|=|=|=|=|=|=|=|=}
\endhead
\multicolumn{14}{r}{\textit{Continued on next page}} \\
\endfoot
\hhline{=|=|=|=|=|=|=|=|=|=|=|=|=|=}
\endlastfoot
\multirow{3}{*}{7447} & \multirow{3}{*}{5} & \multirow{3}{*}{120} & \multirow{3}{*}{2} & \multirow{3}{*}{4}  & $\pi_{\theta_\text{src}}$ & $\mathcal{E}_\text{src}$ & 99.8\% & 86.3\% & 72.3\% & 93.0\% & 63.0\% & 0.4\% & 0.845 \\
 & & & &  & $\pi_{\theta_\text{trg}}$ & $\mathcal{E}_\text{trg}$ & 95.8\% & 87.0\% & 71.2\% & 88.9\% & 78.3\% & 0.8\% & 0.882 \\
 & & & &  & $\pi_{\theta_\text{trg}}$ & $\mathcal{E}_\text{src}$ & 95.7\% & 86.8\% & 7.3\% & 72.3\% & 72.8\% & 0.0\% & 0.670 \\ \hline
\multirow{3}{*}{7447} & \multirow{3}{*}{5} & \multirow{3}{*}{120} & \multirow{3}{*}{4} & \multirow{3}{*}{2}  & $\pi_{\theta_\text{src}}$ & $\mathcal{E}_\text{src}$ & 99.8\% & 86.3\% & 72.3\% & 93.0\% & 63.0\% & 0.4\% & 0.845 \\
 & & & &  & $\pi_{\theta_\text{trg}}$ & $\mathcal{E}_\text{trg}$ & 95.8\% & 87.0\% & 71.2\% & 88.9\% & 78.3\% & 0.8\% & 0.882 \\
 & & & &  & $\pi_{\theta_\text{trg}}$ & $\mathcal{E}_\text{src}$ & 95.7\% & 86.8\% & 7.3\% & 72.3\% & 72.8\% & 0.0\% & 0.670 \\ \hline
\multirow{3}{*}{6788} & \multirow{3}{*}{5} & \multirow{3}{*}{24} & \multirow{3}{*}{2} & \multirow{3}{*}{4}  & $\pi_{\theta_\text{src}}$ & $\mathcal{E}_\text{src}$ & 99.5\% & 82.7\% & 60.3\% & 95.0\% & 61.8\% & 0.0\% & 0.795 \\
 & & & &  & $\pi_{\theta_\text{trg}}$ & $\mathcal{E}_\text{trg}$ & 99.5\% & 84.2\% & 68.3\% & 93.8\% & 61.5\% & 0.0\% & 0.815 \\
\hline
\end{longtable}

\setlength{\LTleft}{\fill}
\setlength{\LTright}{\fill}
\setlength{\LTcapwidth}{\linewidth}
\renewcommand{\arraystretch}{1}
\begin{longtable}{c | c | c | c | c | c | c | c | c | c | c | c | c | c }
\caption{Counting of the solutions for the theories involved in transfer learning across $h^{1,1}(\IX_3)$. The meaning of the columns is analogous to that in Table \ref{tab:detailed_counts}, with `Src' and `Trg' referring respectively to the source and target CYs involved in the transfer learning.} \label{tab:detailed_counts_cross_h11} \\
\hhline{=|=|=|=|=|=|=|====|=|=|=}
\multirow{2}{*}{Src \#} & \multirow{2}{*}{Src $h^{1,1}(\IX_3)$} & \multirow{2}{*}{Src $|G_{\IX_3}|$} & \multirow{2}{*}{Trg \#} & \multirow{2}{*}{Trg $h^{1,1}(\IX_3)$} & \multirow{2}{*}{Trg $|G_{\IX_3}|$} & \multirow{2}{*}{$|\Gamma|$} & \multirow{2}{*}{Sol.} & \multicolumn{3}{c|}{Seeds} & \multirow{2}{*}{Dup. $\left(\frac{N_\text{dup}}{N_\text{tot}}\%\right)$} & \multirow{2}{*}{Total} & \multirow{2}{*}{Equivariance} \\ \cline{9-11}
 & & & & & & & & 42 & 43 & 44 & & & \\
\hhline{=|=|=|=|=|=|=|=|=|=|=|=|=|=}
\endfirsthead
\multicolumn{14}{c}{{\tablename\ \thetable\ -- \textit{Continued from previous page}}} \\
\hhline{=|=|=|=|=|=|=|====|=|=|=}
\multirow{2}{*}{Src \#} & \multirow{2}{*}{Src $h^{1,1}(\IX_3)$} & \multirow{2}{*}{Src $|G_{\IX_3}|$} & \multirow{2}{*}{Trg \#} & \multirow{2}{*}{Trg $h^{1,1}(\IX_3)$} & \multirow{2}{*}{Trg $|G_{\IX_3}|$} & \multirow{2}{*}{$|\Gamma|$} & \multirow{2}{*}{Sol.} & \multicolumn{3}{c|}{Seeds} & \multirow{2}{*}{Dup. $\left(\frac{N_\text{dup}}{N_\text{tot}}\%\right)$} & \multirow{2}{*}{Total} & \multirow{2}{*}{Equivariance} \\ \cline{9-11}
 & & & & & & & & 42 & 43 & 44 & & & \\
\hhline{=|=|=|=|=|=|=|=|=|=|=|=|=|=}
\endhead
\multicolumn{14}{r}{\textit{Continued on next page}} \\
\endfoot
\hhline{=|=|=|=|=|=|=|=|=|=|=|=|=|=}
\endlastfoot
\multirow{4}{*}{5302} & \multirow{4}{*}{6} & \multirow{4}{*}{720} & \multirow{4}{*}{7447} & \multirow{4}{*}{5} & \multirow{4}{*}{120} & \multirow{4}{*}{2}  & $N$ & 6846 & 4394 & 6859 & 3264 (22.0\%) & 14834 & 3516 \checkmark \\
 &  &  &  &  &  &   & $N_{S_5}$ & 4269 & 1748 & 1728 & 1790 (39.2\%) & 4569 & 1123 \checkmark \\
 &  &  &  &  &  &   & $N_{G_{\mathbf{X}_3}}$ & 106 & 82 & 153 & 94 (39.2\%) & 240 & 72 \checkmark \\
 &  &  &  &  &  &   & $N_{\text{full}}$ & 42 & 18 & 21 & 19 (42.2\%) & 45 & 12 \checkmark \\ \hline
\multirow{4}{*}{5302} & \multirow{4}{*}{6} & \multirow{4}{*}{720} & \multirow{4}{*}{7447} & \multirow{4}{*}{5} & \multirow{4}{*}{120} & \multirow{4}{*}{4}  & $N$ & 44750 & 45123 & 58014 & 34114 (33.1\%) & 102914 & 18752 \checkmark \\
 &  &  &  &  &  &   & $N_{S_5}$ & 6516 & 6491 & 6420 & 6286 (80.1\%) & 7843 & 1337 \checkmark \\
 &  &  &  &  &  &   & $N_{G_{\mathbf{X}_3}}$ & 1834 & 1813 & 3380 & 1831 (46.5\%) & 3934 & 750 \checkmark \\
 &  &  &  &  &  &   & $N_{\text{full}}$ & 138 & 111 & 105 & 112 (77.2\%) & 145 & 30 \checkmark \\
\hline
\end{longtable}

\setlength{\LTleft}{\fill}
\setlength{\LTright}{\fill}
\setlength{\LTcapwidth}{\linewidth}
\renewcommand{\arraystretch}{1}
\begin{longtable}{c | c | c | c | c | c | c | c | c | c | c | c}
\caption{Efficacy metrics for transfer learning across $h^{1,1}(\IX_3)$. The table compares the performance of a baseline model trained from scratch against a transfer learning model. The metrics are defined in Section \ref{sec:transferlearning}. \textit{\% Imp.} denotes the relative percentage improvement of the transfer policy.}\label{tab:efficacy_cross_h11} \\
\hhline{=|=|=|=|=|=|=|=|=|=|=|=}
\multirow{2}{*}{Src \#} & \multirow{2}{*}{Src $h^{1,1}(\IX_3)$} & \multirow{2}{*}{Src $|G_{\IX_3}|$} & \multirow{2}{*}{Trg \#} & \multirow{2}{*}{Trg $h^{1,1}(\IX_3)$} & \multirow{2}{*}{Trg $|G_{\IX_3}|$} & \multirow{2}{*}{$|\Gamma|$} & \multirow{2}{*}{{Seed}} & \multirow{2}{*}{{Metric}} & \multirow{2}{*}{{From Scratch}} & \multirow{2}{*}{{Transfer}} & \multirow{2}{*}{{\% Imp.}} \\
 & & & & & & & & & & & \\
\hhline{=|=|=|=|=|=|=|=|=|=|=|=}
\endfirsthead
\multicolumn{12}{c}{{\tablename\ \thetable\ -- \textit{Continued from previous page}}} \\
\hhline{=|=|=|=|=|=|=|=|=|=|=|=}
\multirow{2}{*}{Src \#} & \multirow{2}{*}{Src $h^{1,1}(\IX_3)$} & \multirow{2}{*}{Src $|G_{\IX_3}|$} & \multirow{2}{*}{Trg \#} & \multirow{2}{*}{Trg $h^{1,1}(\IX_3)$} & \multirow{2}{*}{Trg $|G_{\IX_3}|$} & \multirow{2}{*}{$|\Gamma|$} & \multirow{2}{*}{{Seed}} & \multirow{2}{*}{{Metric}} & \multirow{2}{*}{{From Scratch}} & \multirow{2}{*}{{Transfer}} & \multirow{2}{*}{{\% Imp.}} \\
 & & & & & & & & & & & \\
\hhline{=|=|=|=|=|=|=|=|=|=|=|=}
\endhead
\multicolumn{12}{r}{\textit{Continued on next page}} \\
\endfoot
\hhline{=|=|=|=|=|=|=|=|=|=|=|=}
\endlastfoot
\multirow{8}{*}{5302} & \multirow{8}{*}{6} & \multirow{8}{*}{720} & \multirow{8}{*}{7447} & \multirow{8}{*}{5} & \multirow{8}{*}{120} & \multirow{8}{*}{2}  & \multirow{2}{*}{42} & Zero-Shot Reward & 0.263 & 0.719 & +173.4\% \\
 & & & & & &  &  & Episodes to First Solution & 1081344 & 8192 & +99.2\% \\ \cline{8-12}
 & & & & & &  & \multirow{2}{*}{43} & Zero-Shot Reward & 0.279 & 0.705 & +152.7\% \\
 & & & & & &  &  & Episodes to First Solution & 1089536 & 8192 & +99.2\% \\ \cline{8-12}
 & & & & & &  & \multirow{2}{*}{44} & Zero-Shot Reward & 0.310 & 0.722 & +132.9\% \\
 & & & & & &  &  & Episodes to First Solution & 974848 & 8192 & +99.2\% \\ \cline{8-12}
 & & & & & &  & \multirow{2}{*}{Avg} & Zero-Shot Reward & 0.284 & 0.727 & +155.9\% \\
 & & & & & &  &  & Episodes to First Solution & 1003520 & 8192 & +99.2\% \\
\hline
\end{longtable}

\setlength{\LTleft}{\fill}
\setlength{\LTright}{\fill}
\setlength{\LTcapwidth}{\linewidth}
\renewcommand{\arraystretch}{1.2}
\begin{longtable}{c | c | c | c | c | c | c | c | c | c | c | c | c}
\caption{More structural metrics for transfer learning across $h^{1,1}(\IX_3)$. The table compares the performance of a baseline model trained from scratch against a transfer learning model. The metrics are defined in Section \ref{sec:transferlearning}. \textit{\% Imp.} denotes the relative percentage improvement of the transfer policy. Entries with a dash (-) indicate that the metric could not be computed, because the model failed to reach 1000 unique solutions within the 5M episodes.}\label{tab:extra_tests_cross_h11} \\
\hhline{=|=|=|=|=|=|=|=|=|=|=|=|=}
\multirow{2}{*}{Src \#} & \multirow{2}{*}{Src $h^{1,1}(\IX_3)$} & \multirow{2}{*}{Src $|G_{\IX_3}|$} & \multirow{2}{*}{Trg \#} & \multirow{2}{*}{Trg $h^{1,1}(\IX_3)$} & \multirow{2}{*}{Trg $|G_{\IX_3}|$} & \multirow{2}{*}{$|\Gamma|$} & \multirow{2}{*}{{Seed}} & \multirow{2}{*}{{Mean Weight Dist}} & \multirow{2}{*}{{Sum $D_{KL}$}} & \multicolumn{2}{c|}{{Ep. to 1k Sols}} & \multirow{2}{*}{{\% Imp.}} \\ \cline{11-12}
 & & & & & & & & & & Scratch & Transfer & \\
\hhline{=|=|=|=|=|=|=|=|=|=|=|=|=}
\endfirsthead
\multicolumn{13}{c}{{\tablename\ \thetable\ -- \textit{Continued from previous page}}} \\
\hhline{=|=|=|=|=|=|=|=|=|=|=|=|=}
\multirow{2}{*}{Src \#} & \multirow{2}{*}{Src $h^{1,1}(\IX_3)$} & \multirow{2}{*}{Src $|G_{\IX_3}|$} & \multirow{2}{*}{Trg \#} & \multirow{2}{*}{Trg $h^{1,1}(\IX_3)$} & \multirow{2}{*}{Trg $|G_{\IX_3}|$} & \multirow{2}{*}{$|\Gamma|$} & \multirow{2}{*}{{Seed}} & \multirow{2}{*}{{Mean Weight Dist}} & \multirow{2}{*}{{Sum $D_{KL}$}} & \multicolumn{2}{c|}{{Ep. to 1k Sols}} & \multirow{2}{*}{{\% Imp.}} \\ \cline{11-12}
 & & & & & & & & & & Scratch & Transfer & \\
\hhline{=|=|=|=|=|=|=|=|=|=|=|=|=}
\endhead
\multicolumn{13}{r}{\textit{Continued on next page}} \\
\endfoot
\hhline{=|=|=|=|=|=|=|=|=|=|=|=|=}
\endlastfoot
\multirow{4}{*}{5302} & \multirow{4}{*}{6} & \multirow{4}{*}{720} & \multirow{4}{*}{7447} & \multirow{4}{*}{5} & \multirow{4}{*}{120} & \multirow{4}{*}{2}  & 42 & 0.08 & 98.89 & 1589248 & 237568 & +85.1\% \\
 & & & & & &  & 43 & 0.06 & 114.61 & 1482752 & 278528 & +81.2\% \\
 & & & & & &  & 44 & 0.06 & 14.29 & 1425408 & 204800 & +85.6\% \\
 & & & & & &  & Avg & 0.07 & 44.96 & 1373525 & 157013 & +88.6\% \\ \hline
\multirow{4}{*}{5302} & \multirow{4}{*}{6} & \multirow{4}{*}{720} & \multirow{4}{*}{6724} & \multirow{4}{*}{5} & \multirow{4}{*}{12} & \multirow{4}{*}{2}  & 42 & 0.07 & 30.93 & - & - & - \\
 & & & & & &  & 43 & 0.08 & 32.47 & - & - & - \\
 & & & & & &  & 44 & 0.07 & 26.78 & - & - & - \\
 & & & & & &  & Avg & 0.07 & 30.06 & - & - & - \\
\hline
\end{longtable}

\setlength{\LTleft}{\fill}
\setlength{\LTright}{\fill}
\setlength{\LTcapwidth}{\linewidth}
\renewcommand{\arraystretch}{1}
\begin{longtable}{c | c | c | c | c | c | c | c | c | c | c | c | c | c | c | c}
\caption{Catastrophic forgetting metrics for transfer learning across $h^{1,1}(\IX_3)$. 'Policy' defines the policy weights used, i.e. $\pi_{\theta_\text{src}}$ or $\pi_{\theta_\text{trg}}$, with 'On' being the evaluation environment, i.e. $\mathcal{E}_\text{src}$ or $\mathcal{E}_\text{trg}$, where they have been tested. 'Anom', 'Stab', 'Sum', 'Rng', and 'Pair' report the expected CRS in \eqref{eq:CRS} for each individual constraint. $\text{VSR}$ is the Validation Success Rate in \eqref{eq:VSR}.} \label{tab:forgetting_cross_h11} \\
\hhline{=|=|=|=|=|=|=|=|=|=|=|=|=|=|=|=}
\multirow{2}{*}{Src \#} & \multirow{2}{*}{Src $h^{1,1}(\IX_3)$} & \multirow{2}{*}{Src $|G_{\IX_3}|$} & \multirow{2}{*}{Trg \#} & \multirow{2}{*}{Trg $h^{1,1}(\IX_3)$} & \multirow{2}{*}{Trg $|G_{\IX_3}|$} & \multirow{2}{*}{$|\Gamma|$} & \multirow{2}{*}{Policy} & \multirow{2}{*}{On} & \multirow{2}{*}{Anom} & \multirow{2}{*}{Stab} & \multirow{2}{*}{Sum} & \multirow{2}{*}{Rng} & \multirow{2}{*}{Pair} & \multirow{2}{*}{VSR} & \multirow{2}{*}{Mean CRS} \\
 & & & & & & & & & & & & & & & \\
\hhline{=|=|=|=|=|=|=|=|=|=|=|=|=|=|=|=}
\endfirsthead
\multicolumn{16}{c}{{\tablename\ \thetable\ -- \textit{Continued from previous page}}} \\
\hhline{=|=|=|=|=|=|=|==|=======}
\multirow{2}{*}{Src \#} & \multirow{2}{*}{Src $h^{1,1}(\IX_3)$} & \multirow{2}{*}{Src $|G_{\IX_3}|$} & \multirow{2}{*}{Trg \#} & \multirow{2}{*}{Trg $h^{1,1}(\IX_3)$} & \multirow{2}{*}{Trg $|G_{\IX_3}|$} & \multirow{2}{*}{$|\Gamma|$} & \multirow{2}{*}{Policy} & \multirow{2}{*}{On} & \multirow{2}{*}{Anom} & \multirow{2}{*}{Stab} & \multirow{2}{*}{Sum} & \multirow{2}{*}{Rng} & \multirow{2}{*}{Pair} & \multirow{2}{*}{VSR} & \multirow{2}{*}{Mean CRS} \\
 & & & & & & & & & & & & & & & \\
\hhline{=|=|=|=|=|=|=|=|=|=|=|=|=|=|=|=}
\endhead
\multicolumn{16}{r}{\textit{Continued on next page}} \\
\endfoot
\hhline{=|=|=|=|=|=|=|=|=|=|=|=|=|=|=|=}
\endlastfoot
\multirow{3}{*}{5302} & \multirow{3}{*}{6} & \multirow{3}{*}{720} & \multirow{3}{*}{7447} & \multirow{3}{*}{5} & \multirow{3}{*}{120} & \multirow{3}{*}{2}  & $\pi_{\theta_\text{src}}$ & $\mathcal{E}_\text{src}$ & 100.0\% & 95.9\% & 78.7\% & 96.2\% & 63.3\% & 7.3\% & 1.230 \\
 & & & & & &  & $\pi_{\theta_\text{trg}}$ & $\mathcal{E}_\text{trg}$ & 99.6\% & 88.4\% & 75.2\% & 93.0\% & 56.5\% & 0.8\% & 0.864 \\
 & & & & & &  & $\pi_{\theta_\text{trg}}$ & $\mathcal{E}_\text{src}$ & 98.4\% & 89.2\% & 22.6\% & 58.0\% & 37.7\% & 0.2\% & 0.620 \\ \hline
\multirow{3}{*}{5302} & \multirow{3}{*}{6} & \multirow{3}{*}{720} & \multirow{3}{*}{6724} & \multirow{3}{*}{5} & \multirow{3}{*}{12} & \multirow{3}{*}{2}  & $\pi_{\theta_\text{src}}$ & $\mathcal{E}_\text{src}$ & 100.0\% & 92.9\% & 84.3\% & 95.4\% & 67.4\% & 2.4\% & 0.997 \\
 & & & & & &  & $\pi_{\theta_\text{trg}}$ & $\mathcal{E}_\text{trg}$ & 99.9\% & 82.7\% & 20.6\% & 79.8\% & 46.3\% & 0.0\% & 0.651 \\
 & & & & & &  & $\pi_{\theta_\text{trg}}$ & $\mathcal{E}_\text{src}$ & 98.6\% & 85.3\% & 19.9\% & 66.7\% & 39.0\% & 0.0\% & 0.617 \\ \hline
\multirow{3}{*}{5302} & \multirow{3}{*}{6} & \multirow{3}{*}{720} & \multirow{3}{*}{6715} & \multirow{3}{*}{5} & \multirow{3}{*}{24} & \multirow{3}{*}{4}  & $\pi_{\theta_\text{src}}$ & $\mathcal{E}_\text{src}$ & 100.0\% & 96.0\% & 79.3\% & 96.2\% & 63.4\% & 8.1\% & 1.274 \\
 & & & & & &  & $\pi_{\theta_\text{trg}}$ & $\mathcal{E}_\text{trg}$ & 99.9\% & 81.4\% & 58.9\% & 89.2\% & 41.6\% & 0.0\% & 0.741 \\
\hline
\end{longtable}

\pagestyle{empty}
\setlength{\LTleft}{\fill}
\setlength{\LTright}{\fill}
\setlength{\LTcapwidth}{\linewidth}
\renewcommand{\arraystretch}{1}
\footnotesize
\begin{longtable}{c | c | >{\raggedright\arraybackslash}p{19cm}}
\caption{{CICYs used in our paper. We only write the freely acting symmetry $\Gamma$ we considered, despite some CICY might admit more \cite{Braun:2010vc,Lukas:2017vqp}. In $\langle . \rangle$, we write the action of the symmetry on the divisors basis $J_i$.}} \label{tab:CICYdataset}\\
\hhline{=|=|=}
\# & Data &  \\
\hhline{=|=|=}
\endfirsthead
\multicolumn{3}{c}%
{\tablename\ \thetable\ -- \textit{Continued from previous page}} \\
\hhline{=|=|=}
\# & Data &  \\
\hhline{=|=|=}
\endhead

\multicolumn{3}{r}{\textit{Continued on next page}} \\ 
\endfoot

\hhline{=|=|=} 
\endlastfoot

\multirow{10}{*}{\hypertarget{7862}{7862}} & Conf. & $\left[ \begin{array}{c|c} \mathbb{P}^1 & 2 \\ \mathbb{P}^1 & 2 \\ \mathbb{P}^1 & 2 \\ \mathbb{P}^1 & 2 \end{array} \right]^{4, 68}_{-128}$ \\ \cline{2-3}
 & $c_2(T\IX_3)$ & $\{24, 24, 24, 24\}$ \\ \cline{2-3}
 & $\mathcal{I}$ & $2 J_1 J_2 J_3 + 2 J_1 J_2 J_4 + 2 J_1 J_3 J_4 + 2 J_2 J_3 J_4$ \\ \cline{2-3}
 & \multirow{2}{*}{$G_{\IX_3}$} & $S_4$ \\
 & & $\langle (34), (23), (12) \rangle$ \\ \cline{2-3}
 & \multirow{2}{*}{$G_{\text{conf}}$} & $S_4$ \\
 & & $\langle (34), (23), (12) \rangle$ \\ \cline{2-3}
 & $\Gamma$ & $\ZZ_2$ $\coma$ $\ZZ_4$ \\
\end{longtable}

\end{landscape}

\bibliographystyle{JHEP}
\bibliography{references}

\end{document}